\documentclass[iop,revtex4,numberedappendix,twocolappendix]{emulateapj}
\usepackage{apjfonts}
\usepackage{epsfig,graphicx}
\usepackage{amssymb,amsmath}
\usepackage{latexsym}
\usepackage{multirow}
\usepackage{natbib}
\usepackage{lscape}
\usepackage[maxfloats=36]{morefloats}
\usepackage{longtable}
\usepackage{enumitem}
\usepackage{afterpage}  
\usepackage{comment}

\newcommand{\scriptsizeold}{}

\shorttitle{All-Sky Database of BHXBs}
\shortauthors{Tetarenko, B.E. et al.}

\submitted{Accepted to ApJS}

\begin{document}
\title{\textit{WATCHDOG:} A Comprehensive All-Sky Database of Galactic Black Hole X-ray Binaries}

\author{B.E. Tetarenko , G.R. Sivakoff, C.O. Heinke, and J.C. Gladstone}
\affil{Department of Physics, University of Alberta, CCIS 4-181, Edmonton, AB T6G 2E1, Canada}
 
\email{btetaren@ualberta.ca}   

\begin{abstract}

With the advent of more sensitive all-sky instruments, the transient Universe is being probed in greater depth than ever before.
Taking advantage of available resources, we have established a comprehensive database of black hole (and black hole candidate) X-ray binary (BHXB) activity between 1996 and 2015 as revealed by all-sky instruments, scanning surveys, and select narrow-field X-ray instruments aboard the INTErnational Gamma-Ray Astrophysics Laboratory (INTEGRAL),  Monitor of All-Sky X-ray Image (MAXI), Rossi X-ray Timing Explorer (RXTE), and Swift telescopes; the Whole-sky Alberta Time-resolved Comprehensive black-Hole Database Of the Galaxy or WATCHDOG.
Over the past two decades, we have detected 132 transient outbursts, tracked and classified behavior occurring in 47 transient and 10 persistently accreting BHs, and performed a statistical study on a number of outburst properties across the Galactic population.
We find that outbursts undergone by BHXBs that do not reach the thermally dominant accretion state make up a substantial fraction ($\sim$ 40\%) of the Galactic transient BHXB outburst sample over the past $\sim20$ years.
Our findings suggest that this ``hard-only'' behavior, observed in transient and persistently accreting BHXBs, is neither a rare nor recent phenomenon and may be indicative of an underlying physical process, relatively common among binary BHs, involving the mass-transfer rate onto the BH remaining at a low level rather than increasing as the outburst evolves.
We discuss how the larger number of these ``hard-only'' outbursts and detected outbursts in general have significant implications for both the luminosity function and mass-transfer history of the Galactic BHXB population.

\end{abstract}

\maketitle
%----------------------------------------------------------------------
\section{Introduction}\label{s:intro}

Black Hole X-ray Binaries (BHXBs) are interacting binary systems where X-rays are produced by material accreting from a secondary companion star onto a black hole (BH) primary.
Due to angular momentum in the system, accreted material does not flow directly onto the compact object, rather it forms a differentially rotating disk around the BH known as an accretion disk \citep{pringle72,ss73}.
While some material accretes onto the BH, a portion of this inward falling material may also be removed from the system via an outflow in the form of a relativistic plasma jet or an accretion disk wind  \citep{hjellmingwade71,blandford79,phinney82,whiteholt82,begelman83}.
For major reviews of BHXBs see \citet{tl97,ts96,chen97,mr06,rm06} and \citet{d07}.

The degree of BHXB variability separates them into transient and persistent sources.
Most transient BHXBs are Low Mass X-ray Binary (LMXB) systems where mass transfer occurs via Roche lobe overflow of a secondary companion with a mass $M_2 < 1 \,  M_{\odot}$ and spectral type A or later  \citep{white95}. A few transient BHXBs have high-mass donors (e.g., SAX J1819.2$-$2525; \citealt{or01}).

Transient X-ray binaries (XRBs) cycle between periods of quiescence and outburst.
The transient behavior is dependent upon the mass transfer rate onto the BH \citep{tl97}.
Transient XRBs generally spend most of their time in a quiescent state, characterized by long periods of time, lasting anywhere from a few months to decades. In quiescence, the system is exceptionally faint ($\sim10^{30} - 10^{33} {\rm \, erg \, s^{-1}}$) and very little material is transferred from the accretion disk onto the compact object  \citep{mr06}.
The transition to outburst occurs as a consequence of instabilities, both thermal and viscous in nature, developing in the accretion disk that cause more rapid mass transfer onto the BH and lead to bright X-ray emission \citep{meyer81,cannizzo95,kingrit8,lasota1}. As transients spend most of their time in a quiescent state at low luminosities their long-term mass transfer rates are usually low, on the order of $\sim 10^{-10} \, M_{\odot} {\rm \, yr^{-1}}$ \citep{king95}.

In contrast, there are two types of sources that persistently accrete for years. Typically, long-term persistent sources are High Mass X-ray Binaries (HMXBs) that spend most of their time in an X-ray bright (``outburst'') state. Here material is accreted from a massive companion, with $M_2\gtrsim M_{\rm BH}$ and a spectral type O or B, via a stellar wind. This massive companion drives a strong stellar wind resulting in long-term mass transfer rates as high as $10^{-6} \, M_{\odot} \, \rm{yr}^{-1}$ \citep{king95}. As these fast winds have small circularization radii, the outer radius of the disk is always ionized, allowing these systems to remain in a bright outburst state for long periods of time. On the other hand, some transient LMXBs can maintain bright outbursts for decades (e.g., GRS 1915+105; \citealt{ct94,deegan09}), and thus may be classified as persistent, at least over the timescales that we observe them.

The model used to explain the outburst mechanism in XRBs is referred to as the disk-instability model (DIM; \citealt{osaki1974,meyer81,cannizzo1985,cannizzo1993,lin1984,huang1989,mineshige1989}). While this model was originally developed to explain the mechanism behind dwarf novae outbursts in Cataclysmic Variables \citep{osaki1974}, the disc instability model, which we describe below, also seems to describe XRB outbursts reasonably well when the additional term of irradiation from the inner accretion disk in outburst is included.

In quiescence, the accretion disk is in a cool, neutral state. The quiescent disk is built up due to steady mass transfer from the counterpart star, either as a result of Roche lobe overflow in LMXBs or
winds in the case of HMXBs, and the temperature of the disk begins to rise.
Given the steep temperature dependence of the opacity,  eventually the disk temperature in the outer annuli will rise high enough to allow for ionization
of hydrogen to occur, and the disk will undergo a thermal limit cycle.

The cycle begins with the increase in temperature causing an increase in mass accretion rate through a particular annulus. The reason that this can occur is directly tied to hydrogen ionization causing a viscous instability within the disk. When the hydrogen is (at least partly) ionized, the magnetic field is locked into the disk.  As the disk rotates differentially, magnetic field lines that have radial extent are stretched, tending to slow down particles that are closer to the BH (thus making them fall inwards faster), and speed up those that are farther (making them move outwards).  Thus, the viscosity of the disk (i.e., the ability of the disk to move angular momentum around) increases dramatically; in the hot ionized state, material moves inward rapidly. The mechanism for the viscous instability described here is known as the magneto-rotational instability (MRI; \citealt{balhaw91}). See \cite{balhaw98} for a thorough review of this mechanism.

This process ultimately results in a heating wave that propagates inwards and/or outwards through the disc, causing a rapid infall of matter onto the compact object and in-turn an X-ray outburst. Eventually, as the disk is eaten away, the temperature and mass accretion rate through the disc are pulled down to a point where hydrogen is allowed to recombine, triggering the thermal instability once again, only in reverse, and the disk is allowed to return to a cool, neutral state once again.

To first order, the predictions of the disc instability model can explain many observable phenomena \citep{maccarone2014}. First, systems that have high enough mass transfer rates to keep disks fully ionized tend to be persistent and systems that have mass transfer rates below this threshold tend to be transient (see \citealt{coriat2012} and this work). Second, the positive correlation found between peak outburst luminosity and orbital period in transient XRBs (see \citealt{shab98,port2004,wu2010}; and this work) agrees roughly with the prediction that outburst peak luminosities should scale with the radius of the accretion disk \citep{kingrit8}. Third, observed outburst durations match relatively well with the viscous timescales of accretion disks in many BH systems (see \citealt{chen97} and this work). Lastly, we note that with the addition of tidal effects to the disk-instability (i.e., the tidal instability;\citealt{osaki1996,truss2002}), sources that show outbursts of varying amplitudes (e.g., ``super outbursts'' like Swift J1753.5$-$0127;\citealt{zur8,maccarone2013}) can also possibly be explained by this model. See for example \cite{maccarone2014} for a detailed discussion of XRB phenomenon associated with tidal interactions.

%\textbf{In the mass transfer instability model, hard X-rays from the compact object irradiate the subphotospheric layers of the counterpart star, causing them to expand. This expansion brings the atmosphere of the star into an unstable state, leading to sudden mass transfer. The quiescent accretion disk builds up as the mass transfer rate from the counterpart increases, causing the mass inflow from the outer regions of the disk onto the compact object to suddenly become enhanced, ultimately giving rise to an X-ray outburst. Eventually, the mass-loss instability will halt when the Lagrangian point is shielded by the disc, all the matter from the disc will be accreted onto the compact object, and the system will return to a quiescent state. }

However, there are a few observed phenomena that provide strong arguments for a mass-transfer instability occurring (i.e., variable mass-transfer from the counterpart star;\citealt{osaki1985,hameury1986,hameury1987,hameury1988,hameury1990}) and thus XRB outbursts cannot be accurately described by the disk-instabilty alone.
First, systems have been observed to undergo rapid flux variability on timescales of hours, too quickly to be described by the global disk instability and too strong to be the result of ``normal'' variability seen in XRBs (e.g., XTE J1819$-$254; \citealt{hj00b,or01}). Second, quiescent ultraviolet and optical flux are variable, presumably the result of variable mass transfer onto the accretion disk impact spot (e.g., 1A 0620$-$00; \citealt{can10,froning2011}). Therefore, given the observational evidence, it seems likely that some combination of disk, mass-transfer, and tidal instabilities is applicable to BHXBs at least some of the time \citep{maccarone2014}.

During the outburst state, BHXB light curves exhibit a range of morphological types that vary on both a source-by-source basis and between individual outbursts of the same source. While the most prominently observed type is the fast rise exponential decay (FRED) outburst, numerous other features including linear decays, plateaus, multiple peaks, and complex variability have been observed \citep{chen97}.

Notable variations in spectral and timing properties are also observed during an outburst, allowing a number of different accretion states to be defined. 
While X-ray accretion states have been known to exist since the early 1970's when \citet{tan72} first observed a global spectral change in Cygnus X$-$1, 
it was largely the multitude of population studies performed throughout the late 1990s and early 2000s (e.g.,  \citealt{tl97,chen97,mr06,rm06}) that propelled us beyond the largely phenomenological description of X-ray accretion states and into descriptions more firmly based on physical models (e.g., accretion disk, corona, and jet).

The launches of X-ray satellites with unparalleled capabilities like RXTE (1995), XMM-Newton (1999), Chandra (1999), and Swift (2004), have challenged, and continue to challenge, the prevailing views of X-ray accretion states in BHXBs. Access to large amounts of X-ray observations have made it possible to place observational constraints on accretion flows in strong gravity and has allowed for further theoretical understanding of these systems.
A variety of different models of the changing nature and geometry of accretion flows created over the last few decades have been developed to understand the wide variety and variability of emission observed from these systems.
From these models we now have an emerging picture that explains much of the behavior seen from BHXBs (e.g., \citealt{rm06}, \citealt{d07}). 

Currently, there are two theoretical stable accretion flow models that are generally thought to explain the majority of observed spectra.
The thermal disk black body spectral model, typically observed at low energies, is attributed to direct soft photons from a geometrically thin, optically thick disk \citep{ss73,mits84,makish86}.
While the hard Comptonized spectral model, typically observed at higher energies, is thought to come from a hot, geometrically thick, optically thin inner coronal flow existing above and around the inner disk.
The electrons within this flow are thought to (repeatedly) up-scatter a fraction of the lower energy disk photons, producing the observed smooth Comptonized spectrum extending to high energies \citep{thorneprice75,sunyaev1979}.
The structure of this flow is now most commonly associated with Advection Dominated Accretion Flows (ADAFs; \citealt{ich77,ny94}).
However, because the flow is thought to be more complex, other physical processes such as convection (CDAFs; \citealt{abig1}), magnetic fields (MDAFs; \citealt{meier5}), winds (ADIOS; \citealt{bb1999}), and jets (JDAFs; \citealt{falk4}) are necessary for a more realistic treatment.
We note that while a number of plausible alternatives for the origin of the Comptonized spectrum exist (e.g., see \citealt{malzac2009,plotkin2015}), we will focus only on the interpretation discussed above.

The behavioural pattern often observed during outburst \citep{maccarone2003,vadawale2003} involves the system cycling through a pattern of hard (dominated by Comptonized emission) and soft (dominated by  thermal emission) states, where the rise in luminosity at the start of outbursts occurs in the hard state. The peak of the outburst and the initial decline occurs in the soft state, while the final stages of the decline occur in the hard state.
This pattern, commonly referred to as the ``turtlehead'' or ``q-track'', can be clearly observed in a hardness-intensity diagram (HID; see Figure 1). 
 
The Hard (Comptonized) State (HCS) is characterized by spectra dominated by a power-law (Comptonized) component with a hard photon index of $\Gamma \sim1.5-1.7$ and a high energy cutoff at $\sim$100 keV, which may or may not be supplemented by a weak thermal component  \citep{done10}.
Observationally, the hard state is associated with radio detections of a flat to slightly inverted radio spectrum, thought to be the result of the presence of a compact, steady jet \citep{hjellmingwade71,tan72,fbg04,fe09,russ12}, involves low mass transfer rates, and is typically associated with lower Eddington-scaled luminosities \citep{done10}. 

In contrast, the Soft (Disc-Dominated) State (SDS) is characterized by spectra with a dominant disk component peaking at  $\sim 1$ keV accompanied by a weak power-law tail with $\Gamma\sim2$ that often extends past $\sim500$ keV and carries only a small portion of the power \citep{d07}.
Observationally, the soft state is associated with high mass transfer rates, is typically associated with higher Eddington-scaled luminosities, and lacks any persistent optically thick radio emission. The latter is thought to be a result of the quenching of the radio jet \citep{fbg04}.  

The SDS is associated with two different types of outflows.
The first, is optically thin jet ejecta that propagate out from the BH, which are typically detected at radio frequencies early in the SDS \citep{fbg04}.
Note that, even though the actual ejection event appears to occur before the SDS is fully reached \citep{mj12}, it is still included as an observational property of the SDS.
The second, is an accretion disk wind. While these winds have been seen recently in high resolution spectra (e.g., \citealt{lee2002,mil04,mi06,mi06c,miller2008,king2012,neil12,diaztrigo14}), evidence for their presence was identified well before high resolution X-ray spectroscopy had ever been done (e.g., V404 Cyg; \citealt{oos97}).
Originating from the outer disk, accretion disk winds have the ability to carry away large amounts of mass, sometimes on the order of, or larger than, the accretion rate onto the BH, $\Dot{M}_{\rm BH}$.
As such, these winds could be the mechanism behind the quenching of the radio jet in the soft state regimes \citep{neil09,ponti12}.

While it has been suggested that the two outflow regimes of the hard and soft states are most likely not connected by a simple rebalancing of the same outflow power, with the wind carrying more mass but less kinetic power then that of the jet, detailed calculations of quantities such as kinetic energy, mass, and momentum flux in these two types of outflows have not yet been carried out \citep{fendgal14}.
As a result the physical interaction between the winds, accretion flows, and jets in these systems are not fully understood. 
However, given the observationally suggested mass flux and power of these winds and their ubiquitous appearance only in the soft state, it stands to reason that both jets and winds are perhaps a fundamental component of the accretion phenomenon \citep{ponti12}.

The situation becomes far more complex during transitions between the hard and soft states.
This transitional stage, often collectively referred to as the Intermediate State (IMS), involves an increase in X-ray luminosity and a softer spectrum.
The softening of the spectrum is due to two effects that happen simultaneously; the appearance of a significant thermal disk component and the steepening of the hard power-law component to a photon index of $\Gamma \sim 2.0-2.5$.
Observationally, the IMS is associated with high mass accretion rates. The spectral behavior associated with this state can be observed at both low and high fractions of Eddington. 

\begin{figure}[t]%
%  \epsscale{1.2}
  \plotone{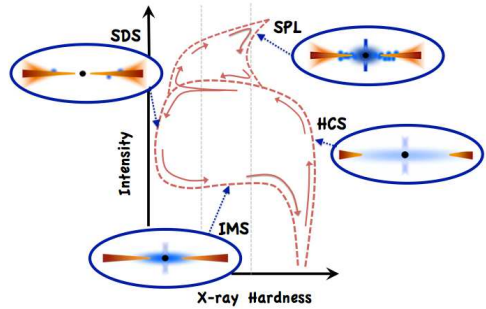}
  \caption{Schematic HID showing the ``turtlehead'' pattern often observed during a BHXB outburst. The source geometry thought to be producing the observed emission in each spectral state is depicted. HCS = hard comptonized state, IMS = intermediate state, SDS = soft disk-dominated state, and SPL = steep power-law. Source geometries taken from \citet{done10}.}%
  \label{fig:HID1}%
\end{figure}

While the majority of sources show the absence of a clear luminosity change during hard-soft and soft-hard spectral transitions \citep{macar05}, clear variations on this fundamental behavior have been observed.
The most noteworthy being the appearance of a steep power-law (SPL) state, in which rapid variations in luminosity accompany the softening and/or hardening of the source, resulting in the addition of a ``dragon horn'' like feature to the classic ``turtlehead'' pattern (see \citealt{mr06} and \citealt{d07} for a more detailed discussion and Figure \ref{fig:HID1} for a schematic representation).
In fact, evidence for SPL behavior is not just limited to a few cases, but can be seen in the brightest phases of many BHXBs, including GX 339$-$4 \citep{mi91,mr06,mo09}, GRS 1915+105 \citep{do04,r03}, GRO J1655$-$40 \citep{ku01,mr06,br06,du09}, 4U 1630$-$472 \citep{ab05},  and XTE J1550$-$564 \citep{mi01,ro03,mr06}.

While collectively the intermediate states are known to take place on relatively short time scales (hours to days), much of the physics occurring during this transitional stage is largely unknown \citep{rm06}.
The prevalent model (see \citealt{esin97,meyer2007}), stipulates that to transition between the HCS and SDS, the mass transfer rate must increase, causing the disk to move inwards.
Fewer seed photons are intercepted by the hot inner flow when the disk is truncated far from the BH, leading to a hard spectral component.
As the disk moves inwards further underneath the hot flow, the spectrum softens due to the disk component becoming significantly brighter (and more dominant) and a larger fraction of disk seed photons being intercepted, effectively cooling the corona via up-scattering.

The basic ``turtlehead'' pattern presented above has been modelled after numerous outbursts of GX 339$-$4 \citep{z04,fbg04,b05,hb05,de08,be09,mo09,co13,de13} and can be observed in a multitude of other sources \citep{mr06} including, but not limited to H 1743$-$322 \citep{z13,ch10}, GRO 1655$-$40 \citep{t99,br06}, 4U 1543$-$475 \citep{park04}, MAXI J1659$-$152 \citep{ku13}, and GRS 1739$-$278 \citep{bo00}. 

That being said, not all BHXB systems follow the basic ``turtlehead'' pattern during outburst.
A number of transient systems have been observed to undergo outbursts that do not involve any complete state transitions (i.e., ``hard-only'' outbursts). In this case the source either remains in the HCS \citep{hw94,hi00,b02,br01,br04,ar04,ss05,br10,si11,cu13} or only transitions as far as the IMS \citep{wm02,int02,ca09,fe11,r12,so12,z13,cu14} during outburst, never fully reaching the softer thermally dominant states.
This behavior is not just limited to transient systems but has also been exhibited by a fair number of persistently accreting systems as well. In their case, either spending long continuous periods of time in the HCS \citep{chur93,main99,de04,pott6,so12,sh13,fr14} or periodically undergoing ``incomplete'' state transitions \citep{pott3,so12}.

Despite the numerous advances made in the field over the past 45 years, we still do not have a theoretical framework to explain all the observational behavior exhibited by BHXBs.
Moreover, the physical parameter(s) that drives the critical instability that precipitates state transitions in BHXBs also remain largely unknown.
As such, being able to enumerate the frequency of outbursts occurring in BHXBs and quantitatively classify the wide range of behaviors exhibited during their outbursts is critical to furthering our understanding of the physical mechanisms driving mass accretion in BHXBs and in turn is a key step toward filling in the many gaps in our knowledge of how BHXBs form, accrete, and evolve.

To date, there exist numerous catalogues of XRBs (see \citealt{bm83,vp95,hmxb00,lmxb01,hmxb06,lmxb07}). However, with the advent of more sensitive all sky and scanning survey X-ray instruments allowing the transient X-ray Universe to be probed in greater depth, we are detecting a larger number of sources than ever before, culminating in the currently published catalogues\footnote{Note that we learned of the BlackCAT catalogue \citep{corralsantana2015} after submission of this manuscript.} quickly becoming dated.  

In addition to these catalogues, there also exists a copious amount of comprehensive reviews on BHXBs and X-Ray Novae (XRN; another name for BHXB outbursts) in the literature. For large scale reviews see \citet{tl97,ts96,mr06}, for a comprehensive study of X-ray properties of BHXBs in the pre-RXTE era see \citet{chen97} and for more recent global X-ray studies see \citet{gn06,rm06,du09,fe09,be09,gi09,yanyu14}. However, the majority of these studies only focus on the ``bright'' outburst events, and offer only a sampling of the sources exhibiting the well defined ``turtlehead'' behavior as observed by one telescope.
As such we have set out to build an update to the current picture in the form of a fully functioning modern BHXB database, accumulating the history of Galactic BH and BHC sources over the past 19 years (1996--2015). 

In Section 2 we outline the criteria for source inclusion in our sample, providing a detailed overview of the current state of the Galactic population on a source-by-source basis. Section 3 provides information on the data selection process and the inner workings of our custom pipeline, using a comprehensive algorithm built to discover, track, and quantitatively classify behavior, from which our database has been assembled. Section 4 presents the numerical and statistical results, and data products produced by the algorithm. Section 5 discusses the implications that both the larger number of ``hard-only'' outbursts and detected outbursts in general have on the long-term mass transfer rates and luminosity function of the Galactic population. Section 6 summarizes our findings for this all-sky study.

\subsection{Observational Techniques}
\label{sec:obstech}
In this subsection we provide a brief overview of the observational techniques that are used (and the current limitations of these techniques) for determining distance and compact object mass in XRB systems.

\subsubsection{Distance}
Here we outline four different methods (discussed in detail in \citealt{jonker2004}) to estimate distance to a XRB system. First, a direct, model-independent distance can be obtained using trigonometric parallax. While this method may provide the most accurate estimates of distance, for many sources this method is not feasible. XRBs may be (i) located several kpc away, and therefore require sub-milliarcsecond astrometry to measure their parallaxes\footnote{Very long baseline interferometry (VLBI) at radio wavelengths is currently the only technique available to do high-precision astrometric measurements \citep{mj9}.}, (ii) too faint to be detected at radio wavelengths, or (iii) located in the Galactic plane, where high-precision astrometry is next to impossible due to scatter broadening along the line of sight \citep{mj9}. Currently, there are only three trigonometric parallax distances to BH systems \citep{mj9,re11,reid2014}.

Second, distance can be estimated by comparing the absolute magnitude to the (dereddened) apparent magnitude of the counterpart star, taking into account a possible contribution from residual accretion \citep{jonker2004}. In this case, the absolute magnitude can be determined by either: (i) determining the spectral type by fitting the data (e.g., \citealt{marsh1994}), and assuming the absolute magnitude is that of a main-sequence star of this spectral type; or  (ii) determining radius, spectral type and luminosity class from the data, and then using the surface brightness of the determined spectral type (e.g. from \citealt{barnes1976,popper1980}), or a combination of the determined radius and effective temperature along with an appropriate bolometric correction, to estimate absolute magnitude. See for example, GS 2000+251 \citep{barr96b} or 1A 0620$-$00 \citep{gel01b}.

Third, limits can be placed on the distance to a source using the observed proper motions of approaching and receding jet ejections. See \cite{mirabelrodriguez99} for a detailed description of this method and \cite{hjr95} for an example of this method being employed for GRO J1655$-$40.

Lastly, the interstellar absorption properties of a source may also be used to determine distance. In this case, distance can be estimated by either: (i) using the correlation between the observed equivalent widths of interstellar absorption lines and diffuse interstellar bands and color excess \citep{herbig1995} and then converting the color excess to distance (e.g., see \citealt{beals1953}); or (ii) tracing the movement of individual gas clouds in velocity space through high resolution spectroscopic observations of interstellar absorption lines and associating them with distance by assuming the velocity is due to Galactic rotation. See for example GX 339$-$4 \citep{hi04}.

\subsubsection{Mass}
During quiescence, optical/infrared (OIR) observations of BHXBs permit detailed studies of the binary counterpart, allowing for the determination of key orbital parameters involved in the mass measurement of the compact object. Once the counterpart has been identified, the orbital period ($P_{\rm orb}$) can be measured from periodicity in either photometry (X-ray or optical/infrared), or radial velocity variations. Once the orbital period is known, radial velocity measurements of the secondary star can give the mass function,
\begin{equation}
f(M)=\frac{P_{\rm orb}}{2 \pi G}(K_2)^3 =\frac{M_1 \sin^3 i}{(M_1 + M_2)^2},
\end{equation}
where $K_2=v_2 \sin i$, the semi-amplitude of the radial velocity curve. The mass function gives a minimum mass of the compact object ($M_1$). A mass function greater then 3$M_{\odot}$ \citep{rhoades1974,kalb96} proves that the compact object is a BH. Next, either measuring $v_2 \sin i$ (the radial velocity of the compact object, by tracing the motion of the accretion disk), or independently measuring the mass of the counterpart star (for instance, by spectral typing and assuming it is a main-sequence star), allows for calculation of the mass ratio ($q=M_2/M_1$). Lastly, analysis of ellipsoidal variability in the photometric light curve provides a method to determine the inclination $i$. The ellipsoidal variability occurs as a result of gravitational distortion of the counterpart star, causing the projected area and average temperature seen by the observer to vary differently with orbital phase depending on the inclination. See for example \cite{chc6} for a detailed discussion of the entire process.

Recently, however, the accuracy of some existing mass measurements for individual BHXBs has been brought into question \citep{can10,kreid12}. It has been suggested that the most commonly used model to analyze ellipsoidal variability (i.e., the ``star-only'' model), which works under the assumption that sources of light not due to the star are negligible, may result in mass estimates that are inaccurate by substantial amounts \citep{kreid12}. This argument stems from the numerous observations of a non-stellar flux component contributing a significant fraction of the total flux of the system at both optical \citep{can10,zur02,orosz4} and IR \citep{hyn5,gel10} wavelengths, implying that OIR spectra of quiescent BHs are likely shaped by a number of competing emission mechanisms. 

Both direct thermal emission from the outer disc \citep{can10,gel10} and synchrotron emission from a relativistic jet \citep{gallo7,russ6} have been suggested as culprits for this flux contamination.
This non-stellar contribution is not modulated (as the counterpart star is) leading to flattening of the light curve (reducing the size of the modulation). This flattening tends to bias the inferred inclinations towards lower values, and can lead to differences of up to a factor of 2 in the derived mass of a compact object \citep{gel01b}. 
As such, orbital inclination is by far the largest source of systematic error involved in estimating the mass of the BH. In addition, as an accurate measurement of inclination relies on the ability to quantify the ratio of non-stellar flux to total flux, distinguish between the non-stellar emission mechanisms and characterize the degree to which these individual sources of emission contribute to the OIR light, it is also the most difficult orbital parameter to measure.

Fortunately, \cite{kreid12} recently published a study in which they characterize the systematic error caused by the effects discussed above. In particular, they build on the study of 1A 0620$-$00 quiescent light curves by
\cite{can08}, who define two separate states existent in the quiescent optical light curves, (i) passive: displaying minimum aperiodic variability, resulting in a stable light curve shape over short timescales; and
(ii) active: brighter, bluer, and more variable than the passive state, possibly as a result of increased accretion activity. 
Doing so has allowed them to use inclination estimates and light curves in the literature to assess the most probable value of inclination for 16 BHXB systems. 
We have made extensive use of this study in determining the orbital parameters used in our analysis.

\subsection{Black Hole Spectral and Temporal Signatures}
In this subsection we provide a list of BH spectral and temporal signatures that are often used to argue for the presence of a BH when the compact object mass (or mass function) is not known. This section will (i) define terms that will be used extensively throughout Section 2, and (ii) provide suggested references for further reading on the topics discussed.

As discussed in Section 1, the X-ray spectral shape associated with BHs is characterized by a soft, multi-color disc blackbody component and a hard power-law tail, in varying proportions. As the blackbody component, originating from the optically thick accretion disk, has a significantly lower characteristic temperature than the blackbody component originating from the surface in neutron star (NS) spectra \citep{mits84,done2003}, the soft blackbody characteristic of BHs is often referred to as ``ultra-soft''. For this reason, the observation of an ultra-soft+power-law spectrum is considered to be the X-ray spectral signature of a BH. For detailed discussion and examples see \cite{tl97} and \cite{mr06}. 

In addition, as most BH systems also exhibit rapid temporal variability (e.g., see \citealt{vanderk06}), the observation of low-frequency quasi-periodic oscillations (LFQPOs) can also be used in tandem with spectral characteristics to argue for the presence of a BH \citep{mr06}. There are three main types of low-frequency QPOs (LFQPOs), ranging in frequency from mHz to $\sim10$ Hz, commonly observed in BHXBs \citep{wijnands1999,casella2005}. The presence of each type  (A, B, or C) is closely correlated with different spectral states \citep{belloni2011}. For a detailed description of the behaviour of each type of LFQPO, see \cite{casella2005}.

Moreover, several BHXBs have also been shown to exhibit a strong correlation between X-ray and radio emission in the hard spectral state \citep{hann98,corb0,corb3}. The now established ``universal'' correlation, which takes the form of a non-linear power law, suggests a strong link between the compact radio jets and the accretion flows in these systems \citep{gallo2003}. BHXBs have been shown to (i) follow two distinct branches in the radio/X-ray ($L_R/L_X$) diagram, corresponding to the efficiency of the accretion flow in the hard spectral state \citep{gallo2012}, and (ii) be more radio bright (i.e., have much higher radio luminosities at a given X-ray luminosity) than NSs (e.g., see \citealt{corr11h}). As such, the position of a source in the $L_R/L_X$ plane can be used as evidence to argue for the presence of a BH in a binary system. See \cite{corr11h} and \cite{co13} and references therein for a more detailed discussion.

Lastly, the absence of observed type I X-ray bursts (i.e., the firm NS signature; \citealt{tl97}) can be used, in combination with BH spectral/timing signatures, to provide further evidence for the presence of a BH in the binary system. 

We note that in the past, an observed soft-hard spectral state transition was considered as a possible BH spectral signature \citep{tl97}. However, as this behavior is exhibited in NS systems as well (e.g., see \citealt{white95,tl97,gladstone2007,md14}), the observation of this state transition alone is not enough to confirm the nature of the compact object.

\section{Sample Selection}
\label{sec:sample}
\subsection{Source Selection Criteria}
We have compiled a sample of 77 XRB BHs and BHCs  in the Milky Way and Magellanic Clouds.
This sample has been built from
(i) the McClintock \& Remillard reviews \citep{mr06,rm06},
(ii) the most recent versions of the Low-Mass (LMXBCAT; \citealt{lmxb07}) and High-Mass (HMXBCAT; \citealt{hmxb06}) X-ray Binary Catalogues, 
(iii) the Swift/BAT Transient Monitor\footnote{http://swift.gsfc.nasa.gov/results/transients/} BH source list, and
(iv) sources listed in the Astronomers Telegram\footnote{http://www.astronomerstelegram.org/} (ATel) with a BH keyword that have not been shown to be a pulsar or Active Galactic Nuclei (AGN).
We note that some faint transients detected by Chandra, Swift, and XMM-Newton in the past $\sim$20 year period may not be included in our sample, even though they could very well be BHXB sources (e.g., \citealt{degenaar2009}).

Of the 77 sources included in our sample, 66 are classified as transient, 8 are known to be persistent, and the remaining 3, which are observed to be transient on long timescales (i.e., continuously bright for periods of $>2$ years in length), are treated as persistent. 

Our sample contains 21 dynamically confirmed BH sources, 18 LMXBs and 3 HMXBs.
In these cases, either the value of the mass function $f(M)$ far exceeds $\sim 3 \, M_{\odot}$, the widely agreed upon limit for the maximum stable mass of a neutron star (NS) in General Relativity \citep{rhoades1974,kalb96}, or dynamical studies have allowed for the measurement of a complete set of orbital parameters, namely $f(M)$, $q$, and $i$, and therefore a definitive estimate of BH mass, $M_{\rm BH}$.
The remaining 56 sources are BHCs, including 37 LMXBs, 6 HMXBs, and 14 undetermined systems.
These sources either lack radial velocity data, have no known optical/infrared (IR) counterpart, or, in some cases, have not been well studied at any wavelength.
Nevertheless, we can still hypothesize the nature of the primary in these systems based on X-ray spectral and timing behavior in tandem with radio characteristics \citep{mr06}.

At this point we must caution the reader. 
We have taken a very liberal approach in determining source membership in the BHC class, contrary to many previous compilations.
As such, there are sources that are far more likely to contain a BH primary than others.
Nevertheless, we believe the few discrepancies we may have are justified in the interest of providing a complete sample of the BH and BHC systems in the Galaxy.

Given our liberal identification of BHCs, we have divided our sample into three  classes.
Class A contains 21 dynamically confirmed BHs.
Class B contains  BHC sources with BH-like spectra (e.g., an ultra-soft spectrum, absence of X-ray bursts) and QPO/timing properties characteristic of BHs, and/or correlated radio/X-ray behavior typical of BHs (i.e., micro-quasar/relativistic jet behavior). 
Class C  are most likely Galactic XRBs, but only have weak evidence for a BH primary. 

We note that the McClintock \& Remillard reviews \citep{mr06,rm06} also use quantitative letter grade labels to classify how likely it is that the system in question contains a BH. To avoid confusion, we remind the reader that their letter grade definitions, which take into account both the amount of observations available as well as X-ray/radio characteristics displayed by the source, differ from our own definitions. In their classification scheme, Class A through C correspond to most-likely to least likely BHCs, synonymous to our Class B and C.

We do not include IGR J06074+2205, IGR J17586$-$2129, or IGR J17354$-$3255 in our sample. All of these sources meet requirements ii and iv above.
However, (i) the only known information on IGR J06074+2205 is the spectral type of the optical counterpart, a B0.5Ve star \citep{ha05,tom2006},
(ii) based on its spectrum and probable optical counterpart, IGR J17586$-$2129 is most likely an obscured HMXB \citep{kr9,sf9}, and
(iii) IGR J17354$-$3255 is likely a Super Giant Fast X-ray Transient (SFXT; \citealt{sg11}).  

\subsection{A Census of Galactic Black Holes \& Black Hole Candidates}
In the following Section, in ascending order of right ascension (RA), we provide:
(i) a brief summary of the X-ray discovery;
(ii) an outline of optical/IR, radio, and X-ray detections;
(iii) an overview of the outburst history/long-term behavior;
(iv) a summary of spectral and timing characteristics exhibited during outburst;
(v) a discussion of the past estimates and currently accepted orbital parameters found through dynamical studies of the system;
(vi) a justification of BH or BHC status, and
(vii) an indication of our assigned BH certainty class, within parentheses in the subsection headers.

For a summary of primary source information, orbital parameters, and binary system information see Tables \ref{table:primaryBH} and \ref{table:binaryBH}.

\subsubsection{XTE J0421+560 (C)}
XTE J0421+560 was discovered by the All-Sky Monitor (ASM) aboard RXTE, when it underwent its first and only outburst in 1998 \citep{sm98}.
The outburst peaked first in X-rays followed by optical and radio wavelengths, a behavior commonly associated with XRN.
Several soft X-ray flares were observed after the initial flare \citep{front98}.
The optical counterpart, CI Cam, is a B0-2 supergiant Be star (assuming a BH or NS accretor; \citealt{rob02}), so the system is a HMXB.

The distance is quite uncertain, with distance estimates ranging from 1 kpc \citep{barsukova2006} to $>10$ kpc \citep{rob02}, and thus the X-ray luminosity is poorly constrained. This large uncertainty in luminosity has impacted accurate determination of the nature of both the mass donor and the accretor \citep{bart13}.
If the distance is $>$2 kpc, the resulting X-ray luminosity would indicate the presence of a NS or BH \citep{be99b}, while a closer distance would suggest a white dwarf accretor \citep{orl02,ish04} and a B$4_{\rm III-IV}$ counterpart \citep{barsukova2006}.
Despite the uncertainty in its nature, we include this system in our BHC sample because its X-ray properties are similar to known BHXBs.

\subsubsection{GRO J0422+32 (A)}
GROJ0422+32 is an LMXB discovered in 1992 by BATSE on the Compton Gamma-ray Observatory (CGRO) when it underwent a FRED type outburst \citep{pa92}.
The spectrum was well described by a hard power-law \citep{sun94} and timing analysis revealed properties commonly associated with the HCS \citep{vdh99b}, indicating that this was a ``hard-only'' outburst.
The X-ray and optical light curves of this outburst were similar to 1A 0620$-$00, except that they showed two unusual optical mini-outbursts following the main burst in 1993, which were not detected in hard X-rays \citep{sh97,cal5}.
A series of mini-outbursts, with a recurrence period of $\sim$120 days, were also detected in 1997 \citep{iy97}.
 
Rigorous determination of binary parameters showed this system to contain a BH primary \citep{fi95,cas95b} and an M1--4V optical counterpart \citep{har99,webb2}.
The detection of a radio counterpart and evolution of its spectrum is discussed by \citet{shrad4}. 
We adopt the distance derived by \citet{gh03} of $2.49\pm0.30$ kpc.

There have been numerous discussions of the inclination of GRO J0422+32.
\citet{orb95,gh03,rey7,fi95} all estimate $i>45^{\circ}$.
However, (i) the light curve of \citet{orb95} exhibits shape changes between two different nights and a difference in mean $I$ magnitude of 0.05 suggesting it is probably in the active state (i.e., variable and poorly-defined light curve shape as a result of contributions from the accretion disk and/or jet; see \citealt{can10}),
(ii) \citet{gh03} and \citet{rey7} find similar mean magnitudes in the $H$ and $K$ light curves, but because there is evidence for IR disk contamination in the former and no detection of ellipsoidal variability in the light curves of the latter, the conflicting results suggest that both authors' light curves are in the active state, and,
(iii) while \citet{fi95} find an $i=48 \pm 3^{\circ}$, which is consistent with the above three results, they assume a normal mass M2V secondary.

As discussed in \citet{kreid12}, only lower limits on the inclination may be derived if the light curves are in the active state, or if the secondary star is assumed to have a mass equal to or less than that of a main sequence star with the same radius.
The latter limitation stems from the fact that the relationship between mass, radius, and surface temperature for a star which fills its Roche-lobe may be very different from that of a spherical star.
Thus, if $q$ is known, assuming a standard mass for the secondary star's spectral type (i.e., an upper limit) will result in overestimated $M_2$, and therefore an overestimated $M_{\rm BH}$.
If $M_{\rm BH}$ is overestimated, then $i$ will be underestimated (for a fixed $q$ and $f(M)$).
 
Several authors measure $i<45^{\circ}$, including \citet{cas95,cal6,beek7}.
However, all three make use of binned light curves.  
The act of binning light curves may flatten their shape, therefore implying a lower inclination. 
We therefore adopt the value of $i=63.7^{\circ}$ calculated by \citet{kreid12} by adjusting the $i=45^{\circ}$ value assuming the source is active.
We adopt a $M_{\rm BH}$  from this corrected inclination.
 
\subsubsection{4U 0538$-$641 (A) --- LMC X-3}
Commonly known as LMC X-3, it was discovered by the UHURU satellite in 1971 \citep{le71}.
LMC X-3 is one of three known persistently accreting, dynamically confirmed BHs in the Milky Way and Magellanic Clouds \citep{mr06}.
From the large mass function, the absence of X-ray eclipses, and an estimated mass of the B3V optical counterpart, \citet{cow83} calculated a lower limit on the mass, confirming the presence of a BH \citep{kuip88,or14}. 

LMC X-3 almost continuously maintains itself in a bright state like persistent HMXB wind-fed systems, but the BH is actually fed by Roche lobe overflow, similar to transient systems \citep{stein14}.
The X-ray spectrum is nearly constantly thermal and disk-dominated, spending most of its time in the SDS \citep{trev88,e93,no01} with occasional transitions to the HCS \citep{wilm01,smale12}.
Due to its location in the Large Magellanic Cloud (LMC), the distance is well constrained \citep{cow83}. 

Recently, \citet{or14} established a new dynamical model for the system.
More high quality radial velocity data (yielding an improved determination of $K$-velocity) and an accurate measurement of the projected rotational velocity for the companion star, along with a much larger array of ellipsoidal light curves, have yielded a more precise $M_{\rm BH}, q$ and $i$ than previous studies. 

\subsubsection{4U 0540$-$697 (A) --- LMC X-1}
Commonly known as LMC X-1,  
it is among the three known persistently accreting, dynamically confirmed BHs in the Milky Way and Magellanic Clouds \citep{mr06}.
Spectroscopic studies of the optical counterpart \citep{hut83,hut87} established it as a firm BHC.
However,  the presence of a BH in this system was not confirmed until an accurate measurement of the mass function was made \citep{haa01}. 

Similar to LMC X-3, LMC X-1 has been observed to spend most of its time in the SDS \citep{e93,no01}.
Given its location in the LMC  \citep{cow83}, the distance to the system is well determined \citep{free1}. 
 
\citet{or09} have improved the dynamical model of \citet{hut87}.
High and medium resolution spectra have reduced the uncertainty in radial velocity amplitude $K$ by a factor of 6 and produced a secure value of the rotational line broadening.
Additionally, they present the first optical light curves and infrared magnitudes and colors of LMC X-1, allowing strong constraints on the inclination of the system, as well as the temperature and radius of the companion star. 

\subsubsection{1A 0620$-$00 (A)}
The X-ray transient 1A 0620$-$00 was discovered by the Ariel V Satellite during an outburst in 1975 \citep{e75}.
A radio counterpart was detected almost two weeks into the outburst, remaining visible for approximately one week \citep{owen76}.
\citet{kul99} collected data from the 1975 outburst and found multiple (jet) ejections with expansion velocities in excess of $0.5 c$, a behavior associated with BHXB systems \citep{mr06}.
A previous outburst, occurring in 1917, was discovered retrospectively on photographic plates at the Harvard College Observatory \citep{each76}. 

\citet{oke77} identified the optical counterpart as a K5 dwarf.  
A radial velocity study of the bright, low-mass counterpart led to the measurement of the mass function, solidifying 1A 0620$-$00 as a strong BHC \citep{mcr86}.

Many attempts have been made to measure inclination of the system: \citet{has93} found $i>62^{\circ}$; \citet{shab94} found $i=36.7^{\circ}$; \citet{marsh1994} found $i=37^{\circ}$; \citet{gel01a} found $i=40.8^{\circ}$; and \citet{frr01} found a wide range of inclinations ($38^{\circ}<i<75^{\circ}$) corresponding to different epochs of data.
Overall inclination estimates remained inconsistent at best.
The inconsistency stems from the highly asymmetric ellipsoidal variations in the light curve \citep{leb98} and contamination of the K-star flux by light from the accretion disk \citep{neil08}.

\citet{can10} rectify these issues, making use of an extensive data set spanning a decade.
They restrict their sample to light curves that are in the passive state (i.e., minimal aperiodic variability from accretion disk contributions) and those where the non-stellar luminosity (NSL) fraction and magnitude calibration are well constrained. They also require each light curve to maintain the same shape over its duration.
As a result of fitting an 11 parameter model to the 8 remaining light curves, they estimate a weighted average of $i=51^{\circ}\pm0.9^{\circ}$.
We assume this value to be unbiased by systematic error \citep{kreid12}, and use this inclination along with $M_{\rm BH}$ inferred from it in our calculations.
We also adopt the distance estimated by \citet{can10}, using their dynamical model, in our analysis. 

\subsubsection{GRS 1009$-$45 (A)}
GRS 1009$-$45 was discovered  by the GRANAT/WATCH ASM \citep{lep93} and by BATSE aboard CGRO \citep{h93}.
It was shown to have an ultra-soft spectrum typical of BHXB systems \citep{kani93}.
\citet{delv93} discovered a blue optical counterpart that showed a spectral type of G5--K7 \citep{del97,fi99}.
Further optical photometry revealed a secondary outburst 6 months later, followed by a series of ``mini-outbursts'',  reminiscent of BH XRN systems like GRO J0422+32 \citep{baiorz95}. 

While \citet{fi99} were able to obtain a mass function and mass ratio, their estimate of inclination assumed that the secondary is a K7--K8 star that is not under massive.
\citet{shab96} performed an ellipsoidal variability analysis; however, because the light curve shows clear evidence of a significant non-stellar contribution and the star-only model fit yields a large $\chi^2$, the light curve is likely active. 

The inclination found by \citet{shab96} conflicts with the estimate by \citet{fi99}, and the spectral type of the secondary is not clear.
Assuming it is a G5V star, as suggested by \citet{del98}, \citet{kreid12} find a lower limit on inclination of $i=62^{\circ}$, consistent with their corrected estimate.
We therefore adopt this inclination in our calculation of $M_{\rm BH}$.
\citet{hi05} discusses their preference for the distance estimate of $3.82$ kpc by \citet{gh02}, which we also adopt.

\subsubsection{XTE J1118+480 (A)}
XTE J1118+480 was discovered by RXTE/ASM in 2000 as a weak, slowly rising X-ray source \citep{rem00}.
Both optical \citep{uem20,uem20b} and radio \citep{pool00} counterparts were quickly found.
Strong low-frequency variability and a spectrum dominated by a hard power-law component extending past 100 keV \citep{rev00c}, accompanied by a persistently inverted radio spectrum \citep{hi00}, suggested it was a BHC in the HCS \citep{fend01}. 

This outburst lasted $\sim$7 months and exhibited some complex behavior.
After the first peak in January of 2000 \citep{rem00,wren2000}, the source decayed only to re-brighten to a plateau state (of similar brightness to the first peak; \citealt{uem20b}) for $\sim$ 5 months \citep{br10}.
XTE J1118+480 remained in the HCS for the duration of the outburst, making it a ``hard-only'' outburst source \citep{br04,zur06}. 

Its second outburst was discovered in the optical \citep{zur05} and confirmed by X-ray and radio observations \citep{rem05,pool5} in 2005.
RXTE would confirm that the source again remained in the HCS for the duration of the outburst \citep{sm05,zur06}.
However, the 2005 event exhibited behavior more typical of a soft X-ray transient, with short lived jet ejections, a more dominant disk component to the spectrum, and a FRED type light curve \citep{br10}.

Dynamical measurements establishing a very large mass function ($>6 \, M_{\odot}$) confirmed a BH primary in this system \citep{mc01b,wagn1}.
Observations in quiescence have allowed refinement of the system parameters \citep{gel6,gh2008}; however, estimates of inclination have proven more challenging.
While the consensus is that XTE J1118+480 has a high inclination, accurate measurements are challenging due to strong superhump modulation (for a discussion of superhumps in LMXBs, see \citealt{has01}) in addition to ellipsoidal variability \citep{zur02}, and a large and variable NSL fraction \citep{wagn1}.
The inclination measurements by \citet{wagn1}, \citet{mc01}, \citet{zur02} and \citet{gel6} all lie in the range  $68^{\circ}<i<82^{\circ}$.
While these estimates are all consistent with each other, none are free of significant systematic errors \citep{kreid12}.
Therefore, we adopt the full range of inclinations in our calculation of $M_{\rm BH}$.
We adopt the distance derived by \citet{gel6} using their established dynamical model.

\subsubsection{GS 1124$-$684 (A)}
The X-ray nova GS 1124$-$684 was discovered by the GINGA ASM  \citep{mak91} and by the GRANAT/WATCH ASM  \citep{lb1991} in 1991.
It was observed extensively from radio to hard X-rays \citep{k92}.
Its X-ray spectra and decay timescales similar to 1A 0620$-$00 \citep{delv91,k92}, a known BH \citep{mcr86}, established it as a firm BHC.
The orbital period and radial velocity curve of the secondary by \citet{rem92} established it as a dynamically confirmed BH.
\citet{or96} refined the determination of the mass function, spectral type of the secondary, and inclination.

This estimate of inclination is both higher and less precise than the estimate in \citet{gel01}.
However, as there is clear evidence for non-stellar flux in the IR \citep{gel10} when the source was active, the difference may be due to the non-stellar contributions.
Even though the source may have been passive during the \citet{gel01} observations, this does not guarantee a negligible NSL fraction.
The \citet{or96} estimate is also higher than that of \citet{shab94}.
However, the \citet{shab94} best-fit inclination has a large $\chi^2$, suggested to be the result of incorrect sky subtraction.
Following \citet{kreid12}, we adopt the inclination range given in \citet{or96} of $54^{\circ}-65^{\circ}$ in our calculation of $M_{\rm BH}$. 

Using infrared photometry, corrected for reddening, \citet{gel01} find a distance of $\sim$ 5.1 kpc. 
\citet{gel1} refined their estimate with simulations to include error bars.
\citet{hi05} prefer this estimate, and we adopt it in our calculations.

\subsubsection{IGR J11321$-$5311 (C)}
The transient hard X-ray source IGR J11321$-$5311 was discovered by ISGRI aboard INTEGRAL in 2005 during a short flare lasting $\sim 3.5$ hours \citep{kr05}.
During this time the source exhibited a very hard spectrum ($\Gamma \sim0.55$), with no evidence for a break up to 300 keV leading \citet{sg07} to suggest that it was probably a magnetar.
However, \citet{kr05} suggest that the spectrum is reminiscent of a BHXB.
Therefore, we include this source in our sample as a possible BHC. 

\subsubsection{MAXI J1305$-$704 (B)}
The X-ray transient MAXI J1305$-$704 was discovered by the Gas Split Camera (GSC) aboard MAXI in 2012 \citep{sato12}.
This source was proposed to be a BHXB based on X-ray and optical spectra as well as light curve and hardness variations over time \citep{gr12,ken12b,suw12,mor13}.
\citet{ken12d,ken12c} observed dip like features and \citet{mill12a,mill12b} observed possible absorption line features.
These dips and absorption profiles provide a strong indication that this source has a large inclination \citep{shid13}.

\citet{shid13} identified a 9.74 hr orbital period from the recurrence interval between absorption dips and inferred an inclination between $60^{\circ}<i<75^{\circ}$,  most likely $\sim75^{\circ}$ as the source shows dips but no eclipses.
A precise value of inclination has not yet been determined.
 
\subsubsection{ Swift J1357.2$-$0933 (A)}
Swift J1357.2$-$0933, a new Galactic BHC \citep{cas11}, was discovered by the Swift/BAT in January of 2011 when it went into outburst \citep{kr11b}. 
The source remained in the HCS for the duration of the outburst, classifying it as a ``hard-only'' outburst source \citep{arm13}.
Both the X-ray spectrum \citep{kr11c} and the magnitude difference (between quiescence and outburst) of the detected optical counterpart \citep{rau11} pointed to an LMXB nature. 
 
Photometry from the  Sloan Digital Sky Survey (SDSS) suggested an M4 counterpart \citep{rau11,shab13}.
The distance is debated; \citet{corr13} suggest 1.5 kpc, but \citet{shab13} compare the estimated magnitude of the companion (not detected) with the expected magnitude of an M4.5V star to infer a possible distance range from 0.5--6 kpc.
Given this information, we choose to use 1.5 kpc as a lower limit and 6 kpc as an upper limit on the distance to this system.
 
\citet{arm13} calculate the peak luminosity for the outburst (for a distance of 1.5 kpc) to be $L_{\rm X}=1.1 \times 10^{35} \rm{ergs}$ $\rm{s}^{-1}$, making it the only confirmed BH Very Faint X-ray Transient\footnote{VFXTs are classified as systems with a peak (2$-$10 keV) $L_{X} \sim 10^{34}-10^{36} \rm{erg} \rm{s}^{-1}$ \citep{wij6}.} (VFXT; \citealt{arm13,corr13}).  
  
\citet{corr13} established $M_{\rm BH}>3.6 \, M_{\odot}$, dynamically confirming this as a BH.
\citet{corr13} also find an orbital period of 2.8 hours and presented an observation of intense dips in the optical light curve, which they explained as toroidal structure in the inner region of the disk, seen at high inclinations ($i \geq 70^{\circ}$), moving outward as the outburst progressed. This implies we may be observing the system close to edge on.
In addition, \citet{shab13} find evidence for quiescent optically thin synchrotron emission, which they discuss could possibly arise from a jet in the system.
 
\subsubsection{ GS 1354$-$64 (A)}
In 1987 the GINGA ASM discovered GS 1354$-$64 in outburst \citep{ma87}.
The X-ray spectrum was well fit with soft disk black body and hard power-law components \citep{kit90} typical of X-ray transient outbursts, and suggestive of a BH nature \citep{br01}.
The position of GS 1354$-$64 is consistent with two transient sources, Cen X-2 \citep{fran71} and MX 1353$-$64 \citep{mark79}, which were observed in outburst in 1967 and 1972, respectively.
Both sources show different X-ray spectral properties than the 1987 outburst of GS 1354$-$64.
If all three were in fact the same source, then GS 1354$-$64 must show at least four different spectral states \citep{kit90}.
\citet{br01} argue that this is not unfeasible as sources such as GX 339$-$4 routinely show multiple state behaviors during outburst \citep{mr06}.
We therefore assume all three are in fact the same source. 
 
GS 1354$-$64 was again observed in outburst in 1997 \citep{rev00a,br01}.
During this time, optical \citep{ct97}, IR \citep{sor97}, and radio \citep{fend97c} counterparts were detected. The radio source was too faint to detect extended structure.
However, analysis of the radio spectrum showed a weak flat synchrotron spectrum, which suggested possible mass ejections in the form of a jet \citep{br01}.
This source has recently (2015 June) gone into outburst for the fifth time, where it has been detected across X-ray \citep{millerjm2015,millerjm2015b}, optical \citep{russell2015}, and radio \citep{coriat2015} wavelengths.
 
While the first and third outbursts of GS 1354$-$64 show very soft X-ray spectra \citep{br01,br04}, the second and fourth events display spectra dominated by a hard power-law, typical of XRBs in the hard state, indicating the source did not reach the softer states in these cases (i.e., ``hard-only'' outbursts; \citealt{rev00a,br01}). 

\citet{ca04} obtained the first radial velocity curve of the optical counterpart, BW Cir, identified its spectral type, mass ratio, period, and in turn a mass function of $f(M)=5.75 \pm 0.30 \, {M}_{\odot}$, confirming a BH primary.
\citet{cas9} infer a lower limit on the distance of 25 kpc (from the companion's luminosity) and estimate an upper limit of 61 kpc (assuming a $10 \, M_{\odot}$ BH) from calculations by \citet{kit90}.
We take the distance to be characterized by a uniform distribution between 25 and 61 kpc for the purpose of our analysis. 
 
While \citet{cas9} have multi-wavelength photometry and spectroscopy between 1995 and 2003 of the source, the data is characterized by strong aperiodic variability without discernible ellipsoidal modulation. Therefore, no lower limit on inclination can be found.
From the spectral type and eclipse limits \citet{kreid12} find $27.2^{\circ}<i<80.8^{\circ}$
 We take $80.8^{\circ}$ as the upper limit on the inclination of GS 1354$-$64 to calculate a lower limit on the mass of the system.
 
\subsubsection{ 1A 1524$-$62 (B)} 
The X-ray transient 1A 1524$-$62 was discovered by Ariel V \citep{p74} and has been observed in outburst twice.
It is considered to be a BHC due to its ultra-soft spectrum, bi-modal spectral behavior, and absence of type I X-ray bursts. 
  
The 1990 outburst was observed in both the hard \citep{barr92} and soft (ROSAT All-Sky Survey; RASS) X-rays.
The soft X-ray spectrum could be fit equally well by a cool black body or a power-law.
The presence of an ultra-soft component to the spectrum could not be ruled out \citep{barret95}.
However, the outburst was insufficient to trigger the soft X-ray ASMs (WATCH and Ginga) and Ginga provided an upper limit consistent with the ROSAT detection \citep{barret95,br04}.
For these reasons we include this outburst as a possible ``hard-only'' outburst \citep{br04}. 
 
\citet{mur77} identify a possible optical counterpart.  
From similarities with 1A 0620$-$00, \citet{mur77} estimate a distance of $>$ 3 kpc, while \citet{van84} propose 4.4 kpc assuming $M_{v}=1.0$ and E(B$-$V)=0.7. We adopt 3 kpc as a lower limit, set the upper limit to 8 kpc and take 4.4 kpc as the most likely value.

\subsubsection{ Swift J1539.2$-$6227 (B)} 
Swift J1539.2$-$6227 was discovered by Swift/BAT in 2008 \citep{kr8b}.
\citet{kr11} present the complete evolution of spectral and timing properties during the outburst, including the rise of the disk component in the SDS and power density spectra signatures of transitions between the HCS and SDS. 
These features, coupled with a lack of observed pulsations, establish the source as a possible BHC \citep{kr13}.

\citet{tor9} performed optical spectroscopy and found a possible optical counterpart with a blue continuum.
No Balmer lines, He II 4686\AA $\,$ or Bowen blend emission were detected.
\citet{kr11} suggest that the lack of emission features in the outburst spectrum paired with the faintness of the source in quiescence points to a low mass main sequence or degenerate donor star companion to the compact accretor.

\subsubsection{ MAXI J1543$-$564 (B)}   
MAXI J1543$-$564 was discovered by the GSC aboard MAXI in 2011 \citep{ne11}.
Type-C QPOs, and an observed decrease in fractional rms and hardness ratios and steepening of the photon index during the outburst led \citet{md11} to classify the source as a BHC.
\citet{st12} present a full spectral and timing analysis.
 \citet{mj11} detect a radio counterpart with an optically thin spectrum. 
 
Three possible optical/IR counterparts have positions consistent with the XRT, ATCA, and Chandra error circles \citep{russ11,rau11b,roj11,chak11}. 
None of these three candidates show any variability \citep{russ11,rau11b,roj11}, leaving the optical counterpart unknown.

\subsubsection{4U 1543$-$475 (A)}  
4U 1543$-$475 is a recurrent X-ray transient discovered in 1971 \citep{m72}, and also observed in outburst in 1983 \citep{kit84}, 1992 \citep{harm92}, and 2002 \citep{park04,kale05}.
4U 1543$-$475 displayed classical Soft X-ray Transient (SXT) behavior during the first, second, and fourth outbursts \citep{m72,kit84,park04}.
However, hard X-ray observations during the third event reveal a power-law spectrum \citep{harm92}.
Unfortunately, there are no soft X-ray observations of this source during the 1992 outburst. 
Following \citet{br04} we include this outburst as a possible ``hard-only'' outburst.
Overall, the observation of a wide array of spectral features make 4U 1543$-$475 a strong BHC \citep{or98}.

4U 1543$-$475 is one of the few sources that has near-simultaneous photometry and spectroscopy \citep{kreid12}.
The optical counterpart, IL Lupi, was discovered by \citet{ped83} and classified as spectral type A2V \citep{chev92}.
A radio counterpart was detected by \citet{hun02}. 

4U 1543$-$475 has been subject to many detailed dynamical studies \citep{or98,or02,or03}, 
confirming its BH primary.
However, when estimating inclination, \citet{or98} include the mass ratio as a free parameter, which results in a large source of error \citep{kreid12}.

A more precise inclination measurement is in a conference proceeding \citep{or02}, however, there is no formal published record of the light curve.
We therefore follow \citet{kreid12} in taking the inclination estimates from \citet{or98} as the boundaries for a uniform distribution.
A precise measurement of distance is found in \citet{oz10}.

\subsubsection{XTE J1550$-$564 (A)}  
The Galactic microquasar\footnote{A microquasar is an XRB system that is known to reach near Eddington luminosity and launch discrete jet ejecta, similar to the behavior of AGN.} XTE J1550$-$564 was discovered by the RXTE ASM \citep{sm98b}.
Shortly after discovery, optical \citep{orosz98b}, and radio \citep{camp8} counterparts were detected along with a superluminal ejection observed in the radio \citep{hann01}.  
Observations of rapid X-ray variability, hard spectrum, and the absence of X-ray bursts or pulsations suggested a BHC \citep{cui1999}. 

Later, \citet{or02b} confirmed the BH nature of the primary.
It has been seen in outburst five times: 1998/1999 \citep{s2000,r02,ku04}, 2000 \citep{ro04,kale01,mi01,tom01a}, 2001 \citep{tom01}, 2001/2002 \citep{b02} and 2003 \citep{ss05,ar04}.
Radio observations during the 2001/2002 outburst by \citet{cor02} confirmed the presence of a hard state jet spectrum. 

The first two outbursts showed the typical ``turtlehead'' pattern through BH states.
Complete spectral and timing analysis for the 1998/1999 and 2000 outbursts can be found in \citet{s2000,homan01,cui1999,rem99b} and \citet{tom01a,mi01,kale01,b02}, respectively.
The last three have been shown to be under-luminous ``hard-only'' outbursts \citep{tom01,swank2002,b02,cor02,ss05}.

\citet{or11b} provide an improved dynamical model including $P_{ \rm orb}$, $f(M)$, $q$, and $i$.
They have determined an inclination using photometry and spectroscopy over 7 years of data.
However, they use NSL fractions determined at a different time than the photometry measurements, which can produce unreliable inclination measurements \citep{kreid12}.
\citet{or11b} acknowledge the uncertainty and fit a model, which includes a disk and four free parameters, using eight different combinations of light curves and NSL fractions.
They find a reasonably narrow possible range in inclination of $57.7^{\circ}<i<77.1^{\circ}$.
Following \citet{kreid12}, we adopt an isotropic distribution over this range for our inclination, a mass ratio uniformly distributed over the range given in \citet{or11b}, and use these values to calculate a $M_{\rm BH}$.

\subsubsection{4U 1630$-$472 (B)}  

The recurrent X-ray transient 4U 1630$-$472 was discovered by the VELA 5B and UHURU satellites \citep{pr86,j76}.
Over the last 46 years, 4U 1630$-$472 has undergone 23 outbursts occurring quasi-regularly ($\sim$ 600--700 days; \citealt{kul97}) and exhibiting a wide range of complex outburst behavior (see Table \ref{table:outhistBH} for a complete list of references for each outburst). 

Highly polarized radio emission was observed in 1998, confirming the presence of jets \citep{hje99}.
The most recent studies of the radio jets in 4U 1630$-$472 discuss possible baryonic matter content within the jets \citep{diaz13,neil14}.
An additional outflow, in the form of an accretion disk wind, has also been detected in this source \citep{ponti12,diaztrigo14}.

No optical counterpart is known, most likely due to its high reddening and crowded field \citep{parm86} resulting in difficulty performing optical and infrared studies.
While no compact object mass is known, \citet{mr06} classify it as a very likely class ``A''  BHC.
It is considered likely to contain a BH based on spectral properties \citep{parm86} and fast timing behavior \citep{kul97b}.

\subsubsection{XTE J1637$-$498 (C)} 
The X-ray transient XTE J1637$-$498 was discovered by PCA aboard RXTE during a regular scan of the Galactic bulge and ridge regions in 2008 \citep{mark08}.
During this time, the source was also detected as a weak source with Swift/BAT (Private Communication with H. Krimm).
\citet{wij8} obtained an X-ray spectrum described by an absorbed power-law with a photon index of  $\sim$1.5, which is consistent with the source being a LMXB, but due to the large errors on the spectral parameters they stress that other types of systems cannot be excluded. 
\citet{curran11} identify a possible optical/IR counterpart.
Given the uncertain nature of this object, 
we include XTE J1637$-$498 in our sample as a possible BHC.

\subsubsection{XTE J1650$-$500 (A)} 
XTE J1650$-$500 is a soft X-ray transient that was discovered by the ASM aboard RXTE in 2001 \citep{rem01}.
The X-ray spectrum \citep{mar2001}, power density spectra \citep{revsun1,wij01}, evolution through the ``turtlehead'' pattern of accretion states \citep{ros4,tom4}, and observation of QPOs \citep{homan3,kale03} during the outburst confirmed the source to be a BHC. 

The optical counterpart was discovered by \citet{ct01}, confirmed by \citet{groot1,aug01}, and later classified as a star of spectral type G5--K4III \citep{orosz4}.
The radio counterpart was discovered by \citet{groot1}, and \citet{co04} observed the source at radio frequencies for the duration of the outburst, finding evidence for the existence of a steady compact jet.
An additional outflow, in the form of an accretion disk wind, has also been detected in this source \citep{mil04,ponti12}.

Further optical observations of the source revealed the $f(M)$, $P_{\rm orb}$, and $i$ for the system \citep{orosz4}.
While \citet{orosz4} are able to determine a lower limit on the inclination of $i>50^{\circ}$, by fitting a star only model with photometry obtained between May and August of 2003, \citet{kreid12} suggests that the source was active during this time, as there was more scatter in the light curve than one would expect from photometric errors alone. Therefore they use this lower limit to calculate their own corrected inclination.
We adopt their estimate of $i=75.2^{\circ}$.

\subsubsection{XTE J1652$-$453 (B)} 
The transient source XTE J1652$-$453 was discovered in 2009 during PCA monitoring of the Galactic region \citep{mar9}.
Further observations showed a quickly rising flux and an X-ray spectrum that evolved from a soft disk blackbody  \citep{marsw9,marbe9} to a hard power-law \citep{corr99}, which suggested a BHC \citep{han9}.
For complete spectral and timing analysis of the outburst, see \citet{han9} and \citet{hi11}. 

Near-IR observations show at least two possible counterparts ($<$1.8 \arcsec from each other) within the XRT and ATCA error circles \citep{calv9,marbe9,rey9}.
While the ATCA observation favors the fainter of the two as the probable counterpart, further near IR observations have detected no significant variability in this source, suggesting that it may be an unrelated interloper star \citep{tor9b}.
No conclusive argument has been made regarding the true counterpart.

A radio counterpart was detected by \citet{calv9}, whose observations indicated emission from the decay of an optically thin synchrotron event associated with the activation of XTE J1652$-$453.

\subsubsection{GRO J1655$-$40 (A)} 
GRO J1655$-$40 was discovered in 1994 by BATSE aboard CGRO \citep{harm95}. 
Radio jets travelling with apparent superluminal motion were discovered  \citep{ting95,hjr95},  
allowing for a precise measurement of distance to the source\footnote{Note that the jets of GRO J1655-40 only give a geometric distance because they are seen to precess. Otherwise, two-sided jets can only give upper limits on distances. See Section \ref{sec:obstech} and \cite{mirabelrodriguez99}.} \citep{hjr95}. 

GRO J1655$-$40 underwent additional outbursts in 1996/1997, exhibiting complex multi-peak behavior \citep{zhang7,s1999,rem99c}, and 2005 \citep{saito6,shap7,br06,caballero07,joinet08,migliari2007,diaztrigo2007}, in which variable radio emission was detected again \citep{rupen5a,rupen5b,rupen5c,br06}. 

An additional outflow, in the form of an accretion disk wind, has also been detected in this source \citep{mi06c,miller2008}. The magnetically driven wind in GRO J1655$-$40 \citep{kallman9} has the largest known mass loss rate of all BH sources in which such winds have been detected \citep{ponti12}.
 
Studies of the orbital parameters in full quiescence published by \citet{orb97}, \citet{green1} and \citet{beer2}, all of which are consistent, provided $P_{\rm orb}$ and $M_{\rm BH}$, and confirmed the BH nature of the primary. 

However, \citet{kreid12} suggested that the distance of $3.2$ kpc from \citet{beer2} is more accurate than that of \citet{green1}.  We adopt the inclination and other orbital parameters from \citet{beer2}, assuming it was passive during their observation.   

\subsubsection{MAXI J1659$-$152 (B)} 
MAXI  J1659$-$152 was detected by Swift/BAT \citep{man10a} and MAXI \citep{ne10} in 2010. 
Optical spectroscopy proved the Galactic origin of the source and its X-ray binary classification \citep{deug10,kaur12}.
It was established as a BHC by observations of fast timing behavior, similar to other BH transients \citep{kall1,ken11,md11b,yam12}. 
Its orbital period of $\sim$ 2.4 hours \citep{kul10,ken11,ku13} makes MAXI J1659$-$152 the shortest period BHXB source known \citep{ken10}. 

 \citet{ku13} constrain its inclination between $65^{\circ}<i<80^{\circ}$, from the obscuration of $\sim$90\% of the total emission on cyclical timescales, as well as the absence of eclipses.
 Mass estimates range from 2.2--$20 \, M_{\odot}$ \citep{ken11,yam12,shap12}.  

Distance estimates range from 1.6--4.2 kpc \citep{mj11a} and 8.6 kpc \citep{yam12}.  
We take a range from 1.6 to 8 kpc in our analysis. 
A possible optical counterpart  \citep{kong10,kong12} has a spectral type between M2 and M5 \citep{mj11a,kong12,ku13}.

\subsubsection{GX 339$-$4 (A)} 
The Galactic X-ray binary GX 339$-$4, discovered in 1972 by the MIT X-ray detector aboard the Orbiting Solar Observatory (OSO) 7 satellite \citep{ma73}, is the most extensively studied transient Galactic BHXB
system \citep{z04}.
Over 43 years, GX 339$-$4 has undergone 20 outbursts in which the entire array of spectral accretion states have been observed \citep{be99}. 
GX 339$-$4 has undergone numerous hard state outbursts, making it a ``hard-only'' outburst source \citep{r98,kong02,bu12,bell13}.
For a complete list of references for each outburst see Table \ref{table:outhistBH}. 

The secondary star is not clearly detected during quiescence, with most of the observed optical emission originating from the accretion disk.
Its LMXB nature has been inferred from the upper limits on the luminosity of this companion star \citep{shab01}.
The system was classified as a BHC \citep{zdz8,sunr0} from its spectral and temporal characteristics.
Fluorescence spectroscopy of NIII and He II emission lines during outburst, formed on the donor star surface due to X-ray irradiation, allowed \citet{hi03} to measure $P_{\rm orb}$ and put an upper limit on $q$ and a lower limit on the mass function of $> 2 \, M_{\odot}$.

The radio source associated with GX 339$-$4 was discovered by \citet{sood4} in 1994.
\citet{wilm99} argued that the radio emission  
could come from a compact self-absorbed jet. 
The picture of the radio jet existing only in the hard states, and being quenched in the soft states \citep{fend99a,fbg04}, and disk-jet coupling implied from observed X-ray-Radio correlations \citep{hann98,corb0,corb3,markoff3,homan5,co13}, arose from numerous observations of GX 339$-$4. 

An additional outflow, in the form of an accretion disk wind, has also been detected in this source \citep{mil04,ponti12}.

\citet{hi04} (based on optical spectra) argue that GX 339$-$4 is located beyond the Galactic tangent point (implying a lower limit of $\geq 6$ kpc) and favor $d\geq15$ kpc.
However, \citet{z04} prefer GX 339$-$4 to be in the Galactic bulge, and use optical/IR data to estimate a favored distance of 8 kpc, a result still consistent with the lower limit given by \citet{hi04}.
We adopt the \citet{z04} estimate. 

\subsubsection{H 1705$-$250 (A)} 
H 1705$-$250 was discovered by the ASMs aboard the Ariel V \citep{g78} and HEAO 1 satellites \citep{k77}.
The optical counterpart, discovered on plates taken at the Anglo-Australian Telescope and UK Schmidt Telescope \citep{long77,g78}, is a star of spectral type K3--7V \citep{har97}. 

The light curve behavior and observed soft \citep{g78} and hard \citep{wilrot83} components to the spectrum resembled that of other SXTs \citep{martin5}. 
This evidence, along with a dynamical mass function measurement of $f(M)=4.86\pm 0.13 \, M_{\odot}$ \citep{fi97} led to the confirmation of a BH primary. 

\citet{martin5} made the first inclination measurement of H 1705$-$250 to be $48^{\circ} < i <51^{\circ}$.
\citet{rem96} obtained a conflicting result of $i>60^{\circ}$.
However, the former only show folded light curves and the latter analyzed a light curve which exhibits uneven maxima.
Due to the uncertainty as to whether the source was active or passive during these observations, we agree with \citet{kreid12} and adopt the \citet{martin5} $i=48^{\circ}$ as a lower limit on the inclination, yielding an upper limit on $M_{\rm BH}$.
As the source does not eclipse, is not a dipper or an accretion disk corona source, we estimate a lower limit on $M_{\rm BH}$ by taking $i=80^{\circ}$.
The distance to the source is quoted in \citet{barr96b}.

\subsubsection{IGR J17091$-$3624 (B)} 
\label{subsubsec:IGR17091}
IGR J17091$-$3624 was discovered by INTEGRAL in 2003 \citep{ku03}.
Spectral analysis of the outburst revealed typical state transition behavior (i.e., ``turtlehead''), where the X-ray emission softened as the outburst progressed \citep{lr2003,cap06}.
IGR J17091$-$3624 has been classified as a probable BHC \citep{lr2003}. 

An archival search of TTM-KVANT and BeppoSAX Wide Field Camera (WFC) data revealed previous outbursts in 1994, 1996, and 2001 \citep{rev03,int03,cap06}. 
The radio counterpart showed a flux increase over two weeks and an inverted spectrum, characteristic of a compact jet \citep{ca09b}.
An accretion disk wind has also been detected \citep{king2012}.

Two more outbursts have been reported, in 2007, showing spectral behavior typical of a BHC \citep{ca09b}, and 2011--2013 \citep{cap12}.
The most recent outburst has been well studied across the X-ray regime \citep{krc11,rod11b,del11,cap11,cap12}. 
Radio observations again revealed evidence for a self-absorbed compact jet \citep{tor11,corb11,rod11a} as well as discrete jet ejections \citep{rod11a}.
This outburst, unlike the others, showed peculiar pseudo periodic flare-like events (``heartbeats'') \citep{alt11a} closely resembling those observed in GRS 1915+105 \citep{alt11b,alt11,bell2000}. 

\subsubsection{IGR J17098$-$3628 (B)} 
The transient IGR J17098$-$3628 was discovered by INTEGRAL in 2005 \citep{greb05}, and remained detectable until 2007.
The source was also briefly detected in 2009 \citep{prat2009}.
Its evolution in brightness and spectral shape \citep{greb05b,ca09b} formed the basis for its BHC classification \citep{greb07}. 
Probable radio \citep{rup5} and optical \citep{st05} counterparts have been detected for this source.

\subsubsection{SAX J1711.6$-$3808 (C)}
The transient SAX J1711.6$-$3808 was discovered by the WFC aboard BeppoSAX in 2001 \citep{int01}.
The Galactic latitude, flux, spectral, and timing properties confirmed the XRB nature of SAX J1711.6$-$3808, while the lack of observation of type I X-ray bursts and lack of coherent oscillations along with the appearance of a broad Fe-K emission feature  suggested that the primary could be a BH \citep{int02,wm02}.
\citet{mr06} classify the system with a grade ``B'' likelihood of harbouring a BH. 
 
The spectrum, which was dominated by a Comptonized continuum, and timing properties of SAX J1711.6$-$3808 indicate that it never left the HCS during outburst, making it one of the ``hard-only'' outburst sources \citep{int02}. 
 
The search for an optical counterpart has proven difficult due to large extinction \citep{int02}.
 
\subsubsection{Swift J1713.4$-$4219 (C)}
The transient Swift J1713.4$-$4219 was discovered by Swift/BAT in 2009 \citep{krimm2009b}.
Unfortunately, due to Sun constraints the source could not be observed by Swift/XRT or UVOT, and was only visible with PCA for 3 days.
The PCA spectrum was well fit with a power-law of photon index $\Gamma=1.68$, and timing analysis revealed strong, aperiodic variability in the power spectrum, both of which point to a possible BH transient in the HCS \citep{kr13}.
For this reason, we include Swift J1713.4$-$4219 in our sample as a possible BHC.

\subsubsection{XMMSL1 J171900.4$-$353217 (C)}
The hard X-ray transient XMMSL1 J171900.4-353217 was discovered in an XMM-Newton slew \citep{read10}.  \citet{mar10} argued for its association with XTE J1719$-$356, a faint transient seen by PCA aboard RXTE in the same month \citep{mar10}. 
Decreases in flux \citep{armd10} and re-brightening events \citep{armd10b} were observed in the source over the course of a few months, further indicating its transient nature. 
As no definitive arguments have been made regarding the nature of the compact object, we include XMMSL1 J171900.4$-$353217 in our sample as a possible BHC.

\subsubsection{XTE J1719$-$291 (C)}
XTE J1719$-$291 was discovered by the RXTE/PCA bulge scans in 2008 \citep{mark8}.
Over $\sim$ 46 days, it showed both flux decreases and rebrightening events \citep{mark8,deg08,deg08b}.
Later, \citet{deg08c} (with Swift/XRT observations) observed large X-ray variability during an outburst that lasted almost two months. 

\citet{grein8} found a possible optical counterpart, deriving a spectral type K0V or later. \citet{arm11} put constraints on the orbital period ($0.4<P_{\rm orb}<12$ hrs) and calculated a long-term mass transfer rate of $\sim 3.7 \times 10^{-13} \, M_{\odot} \rm{yr}^{-1}$ if it does contain a BH.

Assuming a distance of 8 kpc, \citet{arm11} estimate that XTE J1719$-$291 would have had a 2--10 keV peak luminosity of $7 \times 10^{35} \rm{erg} \rm{s}^{-1}$ during its 2008 outburst, therefore classifying the system as a VFXT.

To date, no conclusive evidence is available about the nature of the accretor. As such, we include XTE J1719$-$291 in our sample as a possible BHC.

\subsubsection{GRS 1716$-$249 (GRO J1719$-$24; B) --- X-ray Nova Ophiuchi}
X-ray Nova Ophiuchi (GRS 1716$-$249) was discovered in 1993 by SIGMA aboard the GRANAT satellite \citep{b93} and BATSE aboard CGRO \citep{h93a}.
 Follow up observations discovered the optical and radio counterparts, derived the distance, and suggested that GRS 1716$-$249 was a LMXB system \citep{del94}. 
 
The X-ray spectrum was comparable to Cyg X-1 in the hard state \citep{rem98b}, and the power spectra showed a QPO which varied in frequency \citep{vdhoo96}.
This confirmed that GRS 1716$-$249 never left the HCS during the 1993 outburst, and solidified it as a ``hard-only'' outburst source \citep{br04}. 
 
In early 1995, Mir/Kvant detected five slow-rise, fast decay hard X-ray flares, ``mini-outbursts'' resembling those of the SXTs, GRS 1009$-$45 and GRO J0422+32 \citep{mas96}.
\citet{hj96} found the relation between high energy X-ray and radio emission was very similar (i.e., radio emission follows the peak or onset of decay in X-ray flares seen between 20--200 keV) to that observed in GRO J1655$-$40 and GRS 1915+105 \citep{fost96}, implying the presence of a jet linked to state changes in the accretion disk.
 
\citet{mas96} estimate a period of $\sim14.7$ hours and a lower limit on the mass of the primary from the super hump period\footnote{There are two methods for estimating a lower limit on compact object mass (in close binary systems) when superhump behavior is present. If only the superhump period $P_{\rm sh}$ is known, $(M_{\rm BH}/M_{\odot}) > 0.33 P_{\rm sh}$ hr. If both $P_{\rm sh}$ and $P_{\rm orb}$ are known, $(M_{\rm BH}/M_{\odot}) \gtrsim 0.01 (P_{\rm orb}/\Delta P)$ where $\Delta P \equiv (P_{\rm sh}-P_{\rm orb})/P_{\rm orb}$. For detailed analysis and application of both methods, see \citet{mine92}.} to be $>4.9 \, M_{\odot}$. 
 
\subsubsection{XTE J1720$-$318 (B)}
XTE J1720$-$318 was discovered by the ASM aboard RXTE undergoing an XRN-like outburst in 2003 \citep{rem03}.
Its X-ray spectrum included a 0.6 keV thermal component and a hard tail.
Both the spectral characteristics and source luminosity were typical of a BH in the soft state \citep{mar03a}.
\citet{cb4} perform a detailed spectral analysis of the outburst and conclude that XTE J1720$-$318 is a BHC. 
 
The radio counterpart was discovered by \citet{rup3c,orbrien3}.
\citet{br5} observed an unresolved radio source during the rise phase of the outburst, which reached a peak approximately coincident with the X-ray light curve.
Through study of the spectral indices, they conclude that at least two ejection events occurred, similar to behavior observed in XTE J1859+226.
Following a period in which the radio source was not detected, the source again switched on as XTE J1720$-$318 transitioned back into the hard state. 
 
\subsubsection{XTE J1727$-$476 (C)}
The X-ray transient XTE J1727$-$476 was discovered by RXTE \citep{lev5} and INTEGRAL \citep{tur5} in 2005.
Observations yielded a soft spectrum, reminiscent of a BHXB in outburst \citep{lev5,ken05}.
We therefore include XTE J1727$-$476 in our sample as a possible BHC.
An optical counterpart was discovered by \citet{mait5}.

\subsubsection{IGR J17285$-$2922 (C)}
IGR J17285$-$2922 was discovered with INTEGRAL in 2004 \citep{w04}.
Based on the characteristic evolution of the spectrum from soft \citep{marsw10} to hard \citep{bar05} observed during the $\sim$ 2 week outburst and its location in the Galactic bulge, IGR J17285$-$2922 was suggested to be an LMXB undergoing transient activity \citep{bar05}. Given the lack of type I X-ray bursts and the relative hardness of the spectrum, \citet{bar05} tentatively suggested the system may be harbouring a BH. 

Assuming a distance of 8 kpc, the peak luminosity of this outburst was $\sim 8 \times 10^{35} \rm{erg} \rm{s}^{-1}$, classifying it as a VFXT \citep{si11}. 

Renewed activity in the transient identified as XTE J1728$-$295 was observed in 2010 \citep{marsw10}.
During this time, the source was also detected by Swift/BAT (although never officially reported), reaching a peak of $\sim16$ mCrab on August 28--29, coincident with the RXTE detection (Private Communication with H. Krimm).
\citet{turler10} confirmed that IGR J17285$-$2922 and XTE J1728$-$295 were the same source.
Using XMM and INTEGRAL, \citet{si11} were able to obtain the first broadband spectrum, dominated by a power-law with a slope consistent with the canonical range for BHXBs in the HCS ($\Gamma \sim 1.5-1.7$; \citealt{be09}).
IGR J17285$-$2922 remained in the HCS for the entire 2010 outburst, making it a ``hard-only'' outburst source and further suggesting a BH nature \citep{si11}. 

A possible optical counterpart is known \citep{russ10,tor10,kong10b}.

\subsubsection{GRS 1730$-$312 (C)}
The X-ray source GRS 1730$-$312 was discovered by SIGMA aboard GRANAT \citep{chur94} and the TTM telescope aboard Mir-Kvant \citep{bor94} in 1994.
Both \citet{bo95} and \citet{v96} observed that the source peak luminosity, contribution of the hard and soft spectral components, and hard to soft state transition during outburst, resemble the other BH sources GS 1124$-$684 \citep{ebis94} and 1A 0620$-$00 \citep{rick75}, and therefore classified the source as a BHC.

\subsubsection{KS 1732$-$273 (B)}
GS 1732$-$273 was discovered by the GINGA satellite during scanning observations and was first designated GS 1734$-$275 \citep{ma88}.  
The spectrum was fit well with a black body model similar to those of BHXBs \citep{yk90}. 

The source position was redetermined, now finding agreement with KS 1732$-$273 \citep{vp95,lmxb01,yn04}, discovered by Mir-Kvant in 1989 \citep{int91} and 1RXS J173602.0-272541, discovered with the RASS in 1990 \citep{vog99}.  
Given the ultra-soft spectrum and transient behavior as evidence, \citet{yn04} classify the source as a BHC. 

\subsubsection{IGR J17379$-$3747 (C)}
IGR J17379$-$3747, originally designated XTE J1737$-$376, was discovered by RXTE/PCA in 2004.
The outburst lasted $\sim$9 days.
Although not published formally, it appeared on the PCA bulge scan webpage\footnote{http://asd.gsfc.nasa.gov/Craig.Markwardt//galscan/main.html}. The source position coincided with IGR J17379$-$3747, a weak hard X-ray source reported in the 3rd INTEGRAL catalog \citep{bir07}. Given limits on its positional accuracy, \citet{mar08} concluded that XTE J1737$-$376 and IGR J17379$-$3747 were most likely the same source.

This source was again detected by PCA, INTEGRAL, and BAT in 2008 \citep{mar08}. Swift/XRT observations revealed a hard spectrum with photon index $\Gamma \sim1.78$, consistent with the RXTE/PCA spectrum \citep{kr8}.
The observed spectrum suggests a BHC.
In addition, \citet{curran11} identify a possible optical/NIR counterpart. 

\subsubsection{GRS 1737$-$31 (C)}
GRS 1737$-$31 was discovered by SIGMA aboard GRANAT in March of 1997 \citep{su97}.
It was also found in RXTE and BeppoSAX data \citep{marshsm97,cui97b,he7}.
Both the hard spectrum of the source \citep{su97} and the observed chaotic variability \citep{cui97} were similar to properties observed by XRN and BH source Cyg X-1 \citep{sut79,sun2,vik94,tr99}. 

Observations from \citet{tr99} and \citet{cui97} revealed a hard power-law spectrum.
Further observations using BeppoSAX and ASCA \citep{he7,ueda7} confirmed this hard spectrum $\sim$2 weeks after the initial detection.
Based on spectral and temporal properties, it was suggested that GRS 1737$-$31 was a distant XRN and BHC in the HCS \citep{su97,cui97,tr99}. 
GRS 1737$-$31 appears to have never left the hard state for the duration of the 1997 outburst, making it a ``hard-only'' outburst source \citep{br04}. 

No optical or radio observations have ever been published \citep{br04}.

\subsubsection{GRS 1739$-$278 (B)}
The X-ray source GRS 1739$-$278 was discovered by SIGMA aboard GRANAT in 1996 \citep{p96}.
The outburst was observed by Mir-Kvant \citep{bor96}, ROSAT \citep{grein97}, GRANAT \citep{v97}, and RXTE \citep{tak96}. 

A radio counterpart was discovered with Very Large Array (VLA) data  \citep{durx96,hj96b} and an optical counterpart was discovered by \citet{mart97} to be either a luminous early/middle B type main sequence star or a middle G/early K giant star. 

\citet{bo98} and \citet{v97} find the light curve behavior, optical and radio observations, evolution of the spectrum and spectral characteristics correspond to the SDS and SPL states of BHCs, and \citet{bo00} observe QPOs, present when the source was in the SPL and SDS, allowing GRS 1739$-$278 to reliably be classified as a BHC and soft XRN. )\
 
In 2014 March, GRS 1739$-$278 was again detected by Swift \citep{krimm2014} and INTEGRAL \citep{fil14} where, like the 1996 outburst, it completed the ``turtlehead'' BHXB pattern through the spectral states.

\subsubsection{1E 1740.7$-$2942 (B) --- The Great Annihilator}
The micro-quasar 1E 1740.7$-$2942, located near the Galactic centre, was discovered by Einstein in 1984 \citep{hg84}.
Its hard X-ray emitting nature was first reported by \citet{sk87}.
Given the spectral shape of its soft $\gamma$-ray emission and the similarities to Cyg X-1, \citet{su91} classified it as a BHC and the strongest persistent source in the Galactic centre region.
Its micro-quasar classification came with the discovery of a double-sided radio emitting jet \citep{mirbel2}.
Its radio emission has been found to be variable and correlated with the X-ray flux \citep{paul91}. 

1E 1740.7$-$2942 has been suggested as a possible source of electron-positron annihilation due to an observed high energy spectral feature with GRANAT \citep{bou91,su91c}, hence the common name ``The Great Annihilator''. However, near simultaneous observations by CGRO \citep{jung95} and BATSE \citep{smith96} and high energy observations by INTEGRAL \citep{bou09} could never confirm this feature. 

1E 1740.7$-$2942 is one of only three BHCs that not only remain persistently near their maximum luminosity but also spend most of their time in the HCS \citep{chur93,main99,de04}, with an X-ray spectrum usually described by an absorbed power-law with photon index $\Gamma \sim 1.4-1.5$ \citep{gallof2} and high energy cutoff \citep{sid99,nata14}. Occasionally the source has been observed to make the transition to the softer states \citep{su91c,de04}. 

Due to a source environment characterized by a high concentration of dust and high column density ($\sim10^{23} \rm{cm}^{-2}$), optical identification is difficult \citep{gallof2} and its nature as a HMXB or LMXB, inclination, and distance remain unknown.
However, the high absorption, position near the Galactic centre and presence of bipolar jets, all favor a distance of $\sim$8.5 kpc and disfavor a face on geometry \citep{nata14}.  

Periodic modulation has been detected and interpreted as an $P_{\rm orb}\sim 12.7$ days, suggesting the object could have a red giant companion \citep{sm2002}. 
However, \citet{mart10} have reported a candidate IR counterpart that would exclude the red giant companion possibility.
A much longer periodicity has also been reported by \citet{ogb1}, thought to be related to cyclic transitions between a flat and warped disk, similar to what is observed in Cyg X-1 and LMC X-3.

\subsubsection{Swift J174510.8$-$262411 (B) --- Swift J1745$-$26}
The transient Swift J174510.8$-$262411 (or Swift J1745$-$26) was discovered by Swift/BAT in 2012 \citep{cg12}. 
During this time a variable IR counterpart was identified \citep{rau12b} and an optical counterpart with H$\alpha$ emission was discovered \citep{md13}.
\citet{md13} argue a $P_{\rm orb} \leq 21$ hours and a spectral type of the counterpart as A0 or later.

Spectral and timing observations (by Swift and INTEGRAL) suggest that Swift J1745$-$26 was an LMXB BH system that never left the hard states for the duration for the outburst \citep{bell12,grebsun12,tom12,vovk12,sbar13,kr13}, making Swift J1745$-$26 a ``hard-only'' outburst source \citep{cu14}.
A radio counterpart was detected by \citet{mjs12} and both \citet{corb12} and \citet{corr13a} find a spectral index suggestive of optically-thick synchrotron emission from a partially self-absorbed compact jet.
See \citet{cu14} and \citet{tetarenkoa2015} for complete analysis of the evolution of the jet throughout the outburst from radio through sub-mm frequencies.

\subsubsection{IGR J17454$-$2919 (C)}
The new Galactic source IGR J17454$-$2919 was discovered by the INTEGRAL/JEM-X in October of 2014 \citep{chenevez14} and labelled a new transient XRB source \citep{chenevez2014b}. 
Spectral and timing analysis with NuSTAR revealed a hard power-law index, high energy cutoff, and level of variability consistent with that of an accreting BHXB in the hard state \citep{tendulkar2014}.
This result paired with the fact that no evidence for bursts or pulsations were found make the source a probable BHC \citep{tendulkar2014}.

\subsubsection{1A 1742$-$289 (B)}
The transient source 1A 1742$-$289 was discovered by the Ariel V satellite in 1975 \citep{ey75}.
\citet{brand76} observe similarity in spectral behavior and timescales with transient BH source 4U 1543$-$475 indicating that 1A 1742$-$289 contains a compact object (based on high X-ray luminosity) with variable mass transfer from a low mass companion of spectral type M--K.
\citet{dav76} find a radio counterpart and observe changes in intensity at radio and X-ray wavelengths in 1A 1742$-$289 similar to behavior observed in Cyg X-1 and 1A 0620$-$00 \citep{e75,owen76}.
\citet{mae6} give an estimate of $P_{\rm orb} \sim 8.4$ hours. 
While the nature of the compact object is not known, the source is included in the BHC list of \citet{mr06} and therefore is included in our sample.

\subsubsection{CXOGC J174540.0-290031 (B)}
The new transient source CXOGC J174540.0-290031, located only 0.1 pc in projection from Sgr A*, was discovered by Chandra in 2004 \citep{muno2005}. This source has also been detected by XMM-Newton \citep{belanger2005}. The 7.9 hour orbital modulation detected in the Chandra X-ray light curve and upper limit on the magnitude of the infrared counterpart of $K>16$, coupled with its low peak X-ray luminosity, suggest that it is an LMXB that is being viewed nearly edge-on \citep{muno2005b}. While the bright radio outburst observed (coincident with the X-ray) suggests that it contains a BH. See \cite{bower2005} for a detailed analysis of the radio emission.

\subsubsection{H 1743$-$322 (B)}
H 1743$-$322 was discovered during a bright outburst by the Ariel V \citep{k77} and HEAO--1 \citep{dox77} satellites in 1977.
The source was classified as a BHC based on its very soft spectrum \citep{wm1984}. 

H 1743$-$322 was detected in 1984 with EXOSAT \citep{r99}, in 1996 with Mir-Kvant \citep{e00}, and again in 2003 by INTEGRAL \citep{rev03a} and RXTE \citep{marsw3}.
In 2003, QPOs of typical BHC frequencies were observed \citep{homan03b} and the system followed the typical ``turtlehead'' pattern, transitioning through hard and soft accretion states \citep{cap5,mc09}. 

Detections of the radio \citep{rupen3ee} and optical \citep{st03} counterparts followed quickly thereafter.
H1743$-$322 is classified as a micro-quasar, as jets have been detected at both radio and soft X-ray wavelengths \citep{rupen4,co05}.
There exists numerous studies of this source at radio wavelengths \citep{kale06,jonm10,corr11h,mj12}.
An additional outflow, in the form of an accretion disk wind, has also been detected in this source \citep{mi06c,ponti12}.

Since 2003, H 1743$-$322 has undergone 13 observed outbursts, 3 of which where the source never reached the softer states, we label H 1743$-$322 as a ``hard-only'' outburst source. 
Despite being one of the most well studied BHXBs in the Galaxy, no dynamical confirmation has ever been made on the system \citep{mo10}.
We adopt the distance estimated by \citet{co05} from the proper motion of the jet, $10.4\pm2.9$ kpc, for the purpose of our analysis.

\subsubsection{XTE J1748$-$288 (B)}
XTE J1748$-$288 was discovered in 1998 by the ASM aboard RXTE \citep{sm98bb} and BATSE aboard CGRO \citep{harm98}.
General spectral and timing properties and their evolution during the outburst were typical of BH XRN \citep{rev00}.
\citet{rev00} and \citet{br7} suggest that the outburst began in the SPL state.
As the outburst continued, the source then passed through the SDS and made the transition back to the HCS. 
In addition, an iron emission line was detected by \citet{kot00} and \citet{mi01b}. 

The optically thin radio counterpart was discovered soon after by \citet{hj98}, confirmed to be associated with XTE J1748$-$288 \citep{hj98d,fend98b} and resolved by the VLA \citep{rup98}.
Follow-up work revealed a jet with a velocity $>0.93c$ \citep{hj98e}, making XTE J1748$-$288 only the third known Galactic source that has displayed superluminal motion \citep{br7}.

\subsubsection{IGR J17497$-$2821 (C)}
IGR J17497$-$2821 was discovered with ISGRI onboard INTEGRAL in 2006 \citep{sold6}.
Assuming a distance of 8 kpc, \citet{ku06b} estimated a peak   2--200 keV luminosity of $\sim 10^{37} \rm{erg} \rm{s}^{-1}$.
Given this luminosity and a source position closely projected towards the Galactic center, this strongly suggests that IGR J17497$-$2821 was an XRB \citep{ro07}. Spectral analysis and the FRED type light curve observed further implied that the source was a BHC in the HCS  \citep{ku06,w07}.
The source never left the HCS during this outburst, classifying it as a ``hard-only'' outburst source \citep{ro07,w07,pai09}. 

Given a refined Chandra position, \citet{pai07} identified  two possible optical/IR counterparts consistent with a B-type star (i.e., HMXB) or a K-giant (which would make the system a symbiotic LMXB).
However, given that neither is consistent with the standard LMXB light curve of this source, it is probable that the true counterpart has not yet been identified.
No radio counterpart has been found for this source \citep{ro07}.

\subsubsection{SLX 1746$-$331 (B)}
SLX 1746$-$331 was discovered by the SpaceLab 2 X-ray Telescope in 1985 \citep{sk90} and detected by the RASS in 1990 \citep{mot98}.
Both \citet{sk90} and \citet{wvp96} speculated that SLX 1746$-$331 may harbour a BH based on both its transient nature and its soft spectrum.
Further evidence was added to this claim when the source was once again detected in 2003 by the RXTE/PCA bulge scan, finding a very soft spectrum modelled by black body emission at $\sim1.3$ keV \citep{mar03,remlev3} and by INTEGRAL/ISGRI that found an additional hard component to the spectrum that only contributed at most $\sim 10$ \% of the flux \citep{ll03a}.
Spectral and timing analysis showed that the source had made the transition from the soft state back to the hard state, commonly associated with BHCs \citep{homan3}. 

SLX 1746$-$331 has since been detected twice more, once in 2007/2008 by both RXTE/PCA \citep{mar7a} and INTEGRAL/JEM-X \citep{ku8} and once in late 2010 by MAXI \citep{oz11}.
Both outbursts showed the typical very soft spectrum, commonly associated with BHCs in the soft state.

\subsubsection{XTE J1752$-$223 (B)}
The X-ray transient XTE J1752$-$223 was discovered by the ASM aboard RXTE in 2009 \citep{mar9b}.
RXTE, MAXI, and Swift monitoring suggested that the source was a BHC as variability was indicative of an imminent state transition \citep{nak9,mar9c,remil9,shap09,shap10a,shap10}.
Later works agreed \citep{md10,st11,reis11,nak12b}. 

Optical/NIR observations revealed a possible counterpart  \citep{tor9c,tor9d}.
However, given the crowded field \citep{russell2010}, it remains unclear whether this is the true counterpart.
\citet{br8} discovered the radio counterpart with a flat spectrum consistent with a compact jet in the hard spectral state.
The source appeared to stay in the hard state for an extended period of time and finally transitioned to the softer states \citep{hom10,cur10,nak10,shap10a,shap10,chun3}.
Radio observations of ejection events, often associated with state changes, supported this result \citep{br10b,y10b,yang11}.
There has also been evidence to suggest that XTE J1752$-$223 contains X-ray jets \citep{yang11}.
For in-depth radio analysis see \citet{brock13}.
Using correlations between spectral and variability properties with GRO J1655$-$40 and XTEJ1550$-$564, \citet{shap10} estimated a distance of $3.5\pm0.4$ kpc and a BH mass of $9.6\pm0.9 \, M_{\odot}$, which we adopt for the purpose of our analysis. 

\subsubsection{Swift J1753.5$-$0127 (B)}
Swift J1753.5$-$0127 is an X-ray transient discovered in outburst by Swift/BAT in 2005 \citep{pa05}.
Soon after, the source was detected in UV, optical, NIR and radio bands \citep{still5a,halp5,torr5,fend5}.
The radio observations by \citet{fend5} indicated likely compact jet activity.
The hard X-ray spectrum, up to $\sim$ 600 keV (no NS system has been detected past $\sim$ 200 keV; \citealt{cb07}), and the detection of QPOs \citep{morg5}, has provided strong evidence that the system harbours a BH \citep{cb07}. 
 
The system has not returned to quiescence since 2005, and therefore we treat Swift J1753.5$-$0127 as a persistent source in our analysis.
It is usually seen in the HCS, making it a ``hard-only'' outburst source \citep{so12,sh13,fr14}.
However, it has been observed to make occasional transitions to the intermediate states before returning back to the hard state, similar to the incomplete state transitions of Cyg X-1 \citep{so12}.
 
Swift J1753.5$-$0127 follows the lower track in the X-ray/radio luminosity plane \citep{cb07}, along with an increasing number of BH sources \citep{corr11h,co13}, and it could be the BH with the second shortest orbital period of $\sim$3.2 hours, according to \citet{zur8}.
For complete multi-wavelength analysis, see \citet{fr14,cb07} and for spectral and timing analysis, see \citet{mi06b,so12,mos13}. 

\subsubsection{XTE J1755$-$324 (C)}
XTE J1755$-$324 was discovered by the ASM aboard RXTE in 1997 \citep{rem97}.
The spectrum was fit with a multicolor disk black body at $\sim$0.7 keV and a hard power-law tail extending to $\sim$20 keV \citep{rem97}.
Given the spectrum, the FRED type light curve behavior \citep{rev98}, typical of XRN \citep{ts96}, and the fact that no Type-I X-ray bursts or pulsations were observed throughout the course of the outburst \citep{gold9}, the source was suggested as a good BHC \citep{gold9}.
For complete X-ray spectral analysis, see \citet{rev98} and \citet{gold9}. \cite{gold9} did not detect any quiescent X-ray emission coincident with the position of the source in the ROSAT All-Sky Bright Source Catalogue. No radio counterpart has ever been found \citep{ogley7}.

\subsubsection{H 1755$-$338 (B)}
H 1755$-$338 was discovered by the UHURU satellite when it was active in 1970 \citep{j77}.
It had an unusually soft spectrum \citep{wm1984,white84} and a hard X-ray tail \citep{p95}, which was suggestive of a BHC \citep{karr6}.
H 1755$-$338 shows X-ray dips indicating a high inclination and a $P_{\rm orb} \sim4.4$ days \citep{white84,mason85}.
The source was still active in 1993 \citep{church97}, but in quiescence in 1996 \citep{rob96}.
Therefore the source likely remained active for at least 23 years.
We classify this source as persistent \citep{karr6}. 

The distance to the source is likely $>4$ kpc, as the optical counterpart, which was identified during outburst by \citet{mc78}, was not detected in quiescence \citep{wach8} and $<9$ kpc, suggested by the low level of visual extinction \citep{mason85}.
\citet{ang3} found a linear structure in the X-ray, roughly symmetric, about the position of the source and extending outwards by $\sim 3\arcmin$, suggesting the presence of X-ray jets, see \citet{park05} and \citet{karr6}.

\subsubsection{GRS 1758$-$258 (B)}
The hard X-ray source GRS 1758$-$258 was discovered with GRANAT in 1990 \citep{ma90,su91}.
GRS 1758$-$258 displays a hard power-law spectrum with photon indices $\Gamma \sim1.4$--$1.9$ and a high energy cutoff above $\sim$100 keV \citep{kuz9,main99,lin2000} and strong short term variability up to 10 Hz \citep{smith97,lin2000}, making it one of only three BHCs that not only remain persistently near their maximum luminosity, with the exception of a few dim states that can last up to several months \citep{pott6}, but also spend most of their time in the HCS.
Sometimes a weak soft excess is seen in the spectrum, observed in conjunction with a slightly reduced X-ray flux, thought to be characteristic of the source transitioning into the intermediate states \citep{mer94,mer97,lin2000,heindl2}. 

From the X-ray properties and radio double sided jet structure \citep{rod92}, GRS 1758$-$258 is classified as a micro-quasar \citep{pott6}.
An additional outflow, in the form of an accretion disk wind, has also been detected in this source \citep{ponti12}. 

Three possible counterparts have been identified, a K0 III giant and two main sequence A stars \citep{mart98,eik01,roth2}. The former has been suggested as the most likely counterpart given the $P_{\rm orb} \sim 18.5$ days \citep{sm2002}. 

Numerous observational campaigns, in both the soft \citep{mer94,mer97,smith97,main99,sm2001,sm2002,sm2002b,pott6} and hard \citep{gilf3,kuz9,pott6} X-rays have been undertaken over the years, as well as a multi-wavelength study by \citet{lin2000}. 

\subsubsection{XTE J1812$-$182 (C) --- XMMU J181227.8$-$181234}
The X-ray transient XTE J1812$-$182 (or XMMU J181227.8$-$181234) was discovered by XMM-Newton in outburst in 2003.
After reprocessing of data, this source was found in the RXTE/ASM data as well. \cite{cac6} searched the two Micron All-Sky Survey (2MASS) All-Sky Catalog of Point Sources and the USNO-B1.0 catalogue within 6\arcsec \ of the source position and found no counterpart. They reason that this is most likely due to the large absorption ($\sim10^{23} \, {\rm cm^{-2}}$) in the direction of the source.

The spectrum of the source is fit equally well with an absorbed power-law ($\Gamma\sim2.5$) or a multi-color disk blackbody ($\sim2$ keV) and no pulsations were detected in the timing analysis.
A color-color diagram, along with the high absorption found in the direction of the source suggests an HMXB system.
However, the power-law spectral index is more typical of an LMXB \citep{cac6}. 

The source was detected again in 2008 with RXTE/PCA, and confirmed to be XMMU J181227.8$-$181234 \citep{markw8,tor8b}.
The spectrum was again consistent with a highly absorbed power-law.
While \citet{cac6} speculated that the source was an HMXB, \citet{markw8} interpreted the spectrum and variability behavior as being a BHC in a soft state.
Following \citet{markw8}, we include XTE J1812$-$182 in our sample as a possible BHC.

\subsubsection{IGR J18175$-$1530 (C) --- XTE J1817$-$155}

The hard X-ray transient IGR J18175$-$1530 was discovered by INTEGRAL in 2007 \citep{pai07a} and also detected during RXTE/PCA scans of the region and designated XTE J1817$-$155 \citep{mar7}.
\citet{che7} discuss the detection of a radio source that may be associated with IGR J18175$-$1530.

\subsubsection{XTE J1817$-$330 (B)}
XTE J1817$-$330 was discovered by RXTE in 2006 \citep{re6} and shown to have a very soft spectrum, dominated by the accretion disk component, typical of transient BHCs in the soft state \citep{sala7}
Both the radio counterpart \citep{rup6} and a probable optical \citep{torr6} counterpart were identified.
\citet{sala7} suggest a spectral type of K--M for the optical counterpart.
An outflow, in the form of an accretion disk wind, has also been detected in this source \citep{ponti12}.

At the peak of the outburst, the source was in the SDS and then later transitioned back to the HCS as the source intensity gradually decreased \citep{gi8,roy11}.
QPOs associated with the intermediate states have also been detected in this source \citep{hom6b,roy11}.
For complete analysis of QPOs present during the outburst of XTE J1817$-$330, see \citet{sri13}. 

\subsubsection{XTE J1818$-$245 (B)}
XTE J1818$-$245 was discovered by the ASM aboard RXTE in 2005.
The hardness ratio indicated a very soft spectrum, typical of BHCs \citep{lev5aa} and no pulsations were detected \citep{mar5}.
Soon thereafter, the optical \citep{st05b} and radio \citep{rup5b} counterparts were discovered.
The spectral type of the optical counterpart could not be identified as the optical emission was found to be dominated by the accretion disk \citep{zur11}. 

Spectral parameters showed behavior typical of the SDS and intermediate states seen in BHXBs, including the usual decrease in disk temperature, increase in inner disk radius and decrease in disk flux as the high-energy flux became stronger, and radio flares associated with discrete ejecta \citep{cb09}.
Based on the above analysis and the observed light curve behavior, \citet{cb09} concluded that XTE J1818$-$245 is most likely a LMXB and a BHC.

They estimate a distance of 2.8--4.3 kpc using the model developed by \citealt{shab98} and the outburst light curve properties of the source. We assume a uniform distribution in this range for the purposes of our analysis.

\subsubsection{SAX J1819.3$-$2525 (A) --- V4641 Sgr}
In 1999 the transient SAX J1819.3$-$2525 was discovered by BeppoSAX \citep{int99} and RXTE \citep{mar99} with a position consistent with variable star V4641 Sgr.
Its optical \citep{stub99} and X-ray \citep{sm99aa,sm99bb} flux increased rapidly and then began to decline within two hours.
Emission lines found in both optical and infrared spectra, during this bright X-ray flare \citep{ayp99,lill99,djor99,charles99}, were reminiscent of accretion onto a compact object.

Shortly thereafter, the radio counterpart was discovered by \citet{hje99b}. This counterpart which declined on timescales of hours to days \citep{hje99b,hje99c} and showed ejecta moving with relativistic motion \citep{hje99d}, which led to SAX J1819.3$-$2525 being classified as a possible micro-quasar.
For a discussion of the rapid X-ray variability occurring at super-Eddington luminosities during this flare, see \citet{wv00} and \citet{rev02}.

Since 1999, major outbursts have been observed in 2000 \citep{hj00}, 2002 \citep{uem04}, 2003 \citep{bux3,mb06}, 2004 \citep{swa4} and 2015 \citep{yoshit2015}.
Each outburst was much shorter than typical compact transient systems \citep{mb06}.
Weaker flare-like activity has been reported in 2000 \citep{hj00,uem04}, 2007 \citep{cm07}, 2008/2009 \citep{yam08}, 2010/2011 \citep{yam10a,yam10bb,yam10} and 2014 \citep{tac14}.
An outburst in 1978 was found on photographic plates at the Sternberg Astronomical Institute \citep{barsuk14}.

Optical spectroscopy and photometry during quiescence allowed \citet{or01} to measure a $f(M)=2.74 \pm 0.12 \, M_{\odot}$ and $P_{\rm orb}\sim$2.8 days , estimate a distance between 7.4 and 12.3 kpc and mass between $\sim8.7$--$11.7 \,M_{\odot}$, and classify V4641 Sgr as a B9III star, in turn making SAX J1819.3$-$2525 a confirmed BH \citep{rev02}. 

Its optical companion, has a mass estimated at $\sim 5.5$--$8.1 \, M_{\odot}$ \citep{or01}.
However, \citet{mac11} argue the maintenance of the LMXB label for two reasons.
The first, mass transfer occurs via Roche lobe overflow in the system. The second, the optical counterpart, is not more massive than the probable BH in the system.

More recently, \citet{macdonald2014} have compiled and subsequently separated 10 years of data on this source into passive and active states. They find the passive state data to be dominated by ellipsoidal variations and  stable in the shape and variability of the light curve. By fitting ellipsoidal models to the passive state data they find an improved set of dynamical parameters including $i = 72.3\pm4.1^{\circ}$, $M_{BH} = 6.4\pm0.6 M_{\odot}$, and an updated distance of $6.2\pm0.7$ kpc.

\subsubsection{MAXI J1836$-$194 (B)}
MAXI J1836$-$194 was discovered in August of 2011 simultaneously by MAXI \citep{ne11b} and Swift/BAT \citep{fe11}.
Follow-up observations led to the discovery of the optical \citep{ken11b} and radio \citep{mjss11} counterparts.
The relatively strong radio and IR emission observed was associated with a jet \citep{mjss11,trush11}. 

\citet{ss11} classified the source as a BHC from a spectrum consistent with a power-law of photon index $\Gamma \sim1.8$, the presence of an iron line, and a transition from the HCS to the IMS with RXTE/PCA.
\citet{fe11} found that the source never made the transition from the HCS to the SDS, thereby classifying MAXI J1836$-$194 as a ``hard-only'' outburst source (also see \citealt{r12}).
For full multi-wavelength analysis of the outburst see \citet{rus13} and \citet{r14}.

Using optical/UV signatures of an accretion disk (H$\alpha$, HeII 4686) found in the optical spectra, \citet{r14b} infer a plausible inclination range for the system between $4^{\circ}$ and $15^{\circ}$. In addition, \citet{r14b} are also able to place mass and radius constraints on the counterpart, and in-turn derive an upper limit on the orbital period of $<4.9$ hours, using stellar evolution models and the quiescent luminosity limits found from optical photometry.
 
\subsubsection{Swift J1842.5$-$1124 (C)}
Swift J1842.5$-$1124 was discovered with Swift/BAT in 2008 \citep{kr8c} and studied in detail \citep{racus8,kr8d,kr8e}.
The spectrum of the source was fit with a black body ($\sim0.9$ keV) and power-law model ($\Gamma \sim1.5$), where the black body component only contributed $\sim6$ \% of the total 2--40 keV flux.
Strong QPOs near $\sim0.8$ Hz were also observed \citep{mar08b}.
A QPO at 8 Hz and a hard spectrum suggested that the source was transitioning from the HCS to the SDS \citep{kr13}. 

As the hard X-ray peak preceded the soft X-ray peak by $\sim10$ days in the light curve, a behavior also seen in BH sources Swift J1539.2$-$6227 \citep{kr11} and GRO J1655$-$40, \citep{br06}, suggested the system was a BHC \citep{kr13}.
Swift J1842.5$-$1124 underwent a later, very weak outburst in February of 2010.
Follow-up optical/near IR observations performed in 2008 revealed a possible candidate counterpart \citep{tor8c}. 

\subsubsection{EXO 1846$-$031 (C)}
EXO 1846$-$031 was discovered by EXOSAT in 1985 \citep{parmwh85}.
\citet{p93} observe an X-ray spectrum well fit with a multi-color disk blackbody and power-law component extending to $\sim$25 keV as well as significant variability in this hard component, suggesting that EXO 1846$-$031 is a BHC. 
While a search for the optical counterpart was performed, no significant conclusions were made due to technical issues that affected the source positions derived \citep{p93}.

\subsubsection{IGR J18539+0727 (C)}
The hard X-ray transient IGR J18539+0727 was discovered by INTEGRAL in 2003 \citep{ll03}.
\citet{lr2003} observe an X-ray spectrum fit well with a power-law and a fluorescent line at $\sim$6.4 keV, typical of XRBs in the HCS \citep{gil99}.
IGR J18539+0727 shows strong flux variability on timescales of tenths to tens of seconds and a break frequency in the power spectrum at $<0.1$ Hz.
Although most NS systems do not demonstrate a break with this low of a frequency \citep{wv99}, \citet{linar7} have observed an accreting milli-second pulsar that exhibited BH-like X-ray variability, including a break frequency below 0.1 Hz.
As such, alone these power spectra properties should not be taken as strong evidence for a BH primary.
Given the spectrum and observed properties in the power spectra of the source, \citet{lr2003} suggest  IGR J18539+0727 is a BHC.

\subsubsection{XTE J1856+053 (C)}
XTE J1856+053 was discovered by RXTE/PCA in 1996 \citep{marsh96}.
The RXTE/ASM light curve showed two peaks, separated by $\sim$ 4.5 months.
The first in April displayed a symmetric shape, and the second in September displayed a FRED pattern \citep{rem99}.
Both outbursts are classified as successful,especially when the HIDs are also considered \citep{sala8}.

In 2007, XTE J1856+053 was again detected by RXTE \citep{levrel7}.
Much like the 1996 outburst, two peaks were observe but this time they were separated by only a few weeks and therefore are counted as only one outburst \citep{sala8}.
\citet{sala8} observed that the X-ray spectrum of the 2007 outburst was dominated by emission from the accretion disk, consistent with a BHC in the soft state \citep{mr06}.
Our algorithm detected a fourth outburst in 2009.
This short outburst, lasting $\sim 24$ days, is not found in the literature. 

More recently, in March 2015, XTE J1856+053 was observed in outburst again, this time undergoing a short-lived ``hard-only" outburst.
Originally detected by MAXI  \citep{suzuki2015}, only basic spectral analysis from Swift/XRT is available for this outburst \citep{sanna2015}.
Similar to the 1996 and 2007 events, a second outburst, occurring $\sim$70 days after the onset of the first outburst, was also observed.
During this outburst, which was brighter than the first, the source appears to have entered the soft state \citep{negoro2015b}.

\citet{sala8} classify XTE J1856+053 as a LMXB based on its non-detection at IR wavelengths, therefore ruling out a massive companion. They also suggest a BH primary based on the low temperature of the accretion disk and their rough estimate of an accretor mass range between $1.3$--$4.2 \, M_{\odot}$.

\subsubsection{XTE J1859+226 (A)}
XTE J1859+226 was discovered by the ASM aboard RXTE in 1999 \citep{wood9}.
Follow-up RXTE/PCA observations exhibited a hard power-law spectrum and the existence of QPOs of frequency 0.45 Hz \citep{mar99b}.
BATSE observations confirmed the hard spectrum extending up to $\sim$200 keV and revealed that the hard X-ray flux peaked while the soft X-ray flux was still rising \citep{mcwil9}.
A series of soft flares were seen \citep{fock0}, in which QPOs were detected, at 6--7 Hz and 82--187 Hz \citep{cui2000}.
Further analysis of the outburst \citep{far13,cas4,brock2} suggested a likely BH primary in the system. 

Radio \citep{pool97} and optical \citep{garn99} counterparts were discovered.
\citet{brock2} found a series of radio ejections occurred simultaneously with spectral hardening of the source, suggesting a disk/jet connection.
\citet{garnq0} and \citet{sf0} searched the optical photometry finding a potential $P_{\rm orb} \sim 9.2$ hours.
\citet{fic99} determine a mass function for XTE J1859+226, indicating the BH nature of the primary. 

More recently, \citet{corr11} perform optical photometry and spectroscopy of XTE J1859+226 and find an $P_{ \rm orb} \sim 6.6$ hours, a companion spectral type of K5--7V, and a $f(M)=4.5\pm0.6 \, M_{\odot}$ which, while lower than the original estimate, still requires a BH primary in the system.
\citet{corr11} fit a star-only model to find an $i=60^{\circ}$.
As their data is consistent with the passive state \citep{kreid12},  we adopt the \citet{corr11} value for the inclination and use it to calculate $M_{\rm BH}$.

The distance to XTE J1859+226 remains problematic.
\citet{zur02} estimate 11 kpc based on the brightness of the outburst and the quiescent optical counterpart.
However, this estimate is based on assumed values for both the orbital period and companion spectral type \citep{fic99}. 
\citet{mar2001a} use a combination of spectral and timing information to estimate $\sim$5--13 kpc and \citet{hi22}, using fits to optical-UV spectral energy distribution, find an estimate of $\sim$4.6--8.0 kpc.
Due to the uncertainty that still exists in the system parameters, we follow \citet{hi05} in believing that the distances estimated during outburst are more reliable and thus adopt their distance estimate of $8\pm3$ kpc.

\subsubsection{XTE J1901+014 (C)}
XTE J1901+014 was discovered by the ASM aboard RXTE in 2002 \citep{r02a}.
This outburst lasted between 2 minutes and 3 hours.  
Later, when reanalyzing RXTE/ASM data, \citet{r02a} found a previous outburst of the source occurring in 1997, lasting between 6 minutes and 8 hours. 

\citet{kar7} performed spectral and timing analysis on the 1997 and 2002 outbursts, arguing that they are not type I X-ray bursts.
They fit the spectrum with a power-law of photon index $\Gamma \sim2.3$, finding no cutoffs at energies between 20--30 keV (typical of pulsars; \citealt{fil05}), or any emission lines in the spectrum.
They conclude that such a spectrum, indicates that the primary is in fact a BH.
\citet{kar7} suggest that the outbursts are similar in spectral and timing properties to those observed in Galactic BH source SAX J1819.3$-$2525 \citep{stubp99} and to a lesser degree, the outbursts of fast transients like SAX J1818.9$-$1703 \citep{grebsun05}, concluding that XTE J1901+014 may be a fast X-ray transient containing a BH. 

XTE J1901+014 once again became active in 2006.
\citet{kar8} found a power-law spectrum consistent with accreting XRBs.
They failed to find the optical counterpart, estimating upper limits on the magnitude of $\sim$23.5 in the $r^{\prime}$ band and $\sim$24.5 in the $I$ band. 
In 2010, Swift/BAT detected XTE J1901+014 in outburst for the fourth time. 
his outburst lasted at least 2.5 minutes.
The spectrum was consistent with the other three outbursts \citep{krimm10}.
It is interesting to note that this source is generally detected as a low-level persistent source in the BAT survey\footnote{See http://swift.gsfc.nasa.gov/results/bs70mon/SWIFT\_J1901.6p0129}.

\subsubsection{XTE J1908+094 (B)}
XTE J1908+094 was discovered serendipitously in observations of SGR 1900+14 with RXTE/PCA in 2002.
The spectrum was consistent with an absorbed power-law of photon index $\Gamma \sim1.6$ and no pulsations were detected suggesting that XTE J1908+094 was an XRB containing a BH primary \citep{woods2}.
Subsequent BeppoSAX observations \citep{int02c} confirmed the hard spectrum, extending up to $\sim250$ keV, and the high Galactic absorption. Detailed spectral and timing analysis by \citet{int02b} and \citet{gog4}, which confirmed that the source passed through both the HCS and SDS during outburst, agreed with \citet{woods2} that XTE J1908+094 is a BHC.
A second outburst from this source was detected in 2013/2014 \citep{kr13a,kr13c,mj13,corr13b}.

The radio counterpart was discovered by \citet{rup2}.
\citet{jon4} analyzed simultaneous X-ray and radio observations during the outburst decay and discuss the X-ray/radio correlation. 

A likely near IR counterpart was detected \citep{wagner2002,garnavich2002,chaty2002}.
Once the source had returned to quiescence \citet{chat6} performed NIR observations with CFHT, finding two possible counterparts (separated by 0.8\arcsec) within the X-ray error circle.
They postulate that the companion star could be either (i) an intermediate/late type main-sequence star of spectral type A--K, located between 3--10 kpc; or (ii) a late-type main-sequence star of spectral type K or later, located between 1--3 kpc.
They favor the former due to an independently determined lower limit on distance of 3 kpc derived by \citet{int02b} from the peak bolometric flux.
We follow the suggestion by \citet{chat6} and take the distance to be in the range 3--10 kpc.

\subsubsection{Swift J1910.2$-$0546 (C) --- MAXI J1910$-$057}
Swift J1910.2$-$0546 was simultaneously discovered by Swift/BAT \citep{krimm12} and MAXI (as MAXI J1910$-$057; \citealt{us12}) in 2012.
\citet{raugs12} detected the optical/NIR counterpart.
\citet{ll12} and  \citet{casar12}  report possible periodic variations in the optical light curve, which could be attributed to orbital variations, of $\sim 2.2$ hours, and $\sim 4$ hours, respectively.

The complex light curve \citep{kr13} of the source, as well as spectral analysis from MAXI \citep{kim12,nak12} and INTEGRAL \citep{king12}, show progression through state transitions and the mirrored behavior of the hard and soft X-ray flux.
For this reason \citet{kr13} tentatively suggest that Swift J1910.2$-$0546 is a BHC.

\subsubsection{SS 433 (C)}
\label{subsubsec:SS433s}
The Galactic micro-quasar SS 433 was first discovered in a survey of stars exhibiting H$\alpha$ mission \citep{stephenson77} and later identified as a variable X-ray source in 1978 by the UHURU satellite \citep{mar80}. SS433 is unique among XRBs. The main property that distinguishes it from other ``normal'' XRBs, is that the system remains in a continuous regime of supercritical accretion onto the compact object. Basically all photometric and spectroscopic properties of the system are determined by the accretion disk and its orientation. See \cite{fabrika2004} for an extensive review of the system.

In the X-ray, SS 433 is a weak source, generally not observed past $\sim$30 keV \citep{nan5}.
The highly erratic spectral and temporal behavior of the system combined with the internal complexity of the system \citep{brink89} makes the nature of the compact object and companion uncertain \citep{fab90}.

The basic picture involves an evolved binary undergoing extensive mass-transfer.
The secondary feeds an enlarged accretion disk around a compact object (NS or BH; \citealt{blun10}).
Some of this mass is directed through the disk toward the oppositely facing relativistic jets.
We observe red and blue shifted optical lines, indicating material is accelerated by the jets \citep{fabr79,mil79,gies02}.

 \citet{mar84} successfully fit a precessing jet model to these lines, finding that the jets moved near constant velocity of $\sim0.26c$ and had a precession periodicity of 162.15 days.
 Radio imaging of the source, showed twin processing jets with structure on scales ranging from milliarcseconds to arcseconds \citep{hjon81,ver87,ver93,fej88}.
 The system is known to have a 13 day orbital period.
 This photometric periodicity, originally identified as orbital by \citet{crampton80,crampton81}, is further supported by both the observation of eclipses at 13 days \citep{cherep81}, and the X-ray dimming observed during the optical eclipses \citep{stewart87}.

SS 433 is believed to be a binary system consisting of a compact object and an O or B type star \citep{mar84}.
Extensive arguments for the nature of the compact object being a BH  \citep{zwit89,fab90,lopez06,hillwig2008,kubota2010,goranskij11} and a NS  \citep{fi88,deo91,goranskij11} have been made.
The currently favored distance to the source is 5.5 kpc.
Originally estimated by \citet{hjon81} (and later \citealt{bb2004}) using the proper motion of the jets, this distance is now also supported by observations of HI absorption occurring as a result of a gas cloud interacting with the related supernova remnant W50 \citep{lockman2007}.

\subsubsection{GRS 1915+105 (A)}
\label{subsubsec:GRS1915}
GRS 1915+105 was discovered in 1992 by the WATCH ASM aboard GRANAT \citep{ct92} and has remained active ever since \citep{bellalt13}.
The system exhibits very peculiar behavior, in the form of complex structured variability, such as QPOs ranging in frequency from $10^{-3}$ Hz to $67$ Hz and patterns of dips and rapid transitions between high and low intensity in the light curve \citep{grein96,morg97,bell2000,klein2,fb2004,hann05}.
Currently, there are 14 known variability classes \citep{bell2000,klein2,hann05}.
Modelling of this X-ray variability has led to major insights into the connection between the accretion disk and relativistic jets in XRBs \citep{bello97,klein2}.
Its luminosity is estimated to be near Eddington \citep{mirabelrodriguez99}. 
 
The probable optical counterpart was discovered by \citet{boer96} and the radio counterpart was found with the VLA \citep{mirbel3}.
Further radio monitoring revealed structures travelling at superluminal speed \citep{mirbel4}, making GRS 1915+105 the first superluminal source in the Galaxy.
GRS 1915+105 was suggested to harbour a BH based on its similarity with GRO J1655$-$40, the second Galactic source to exhibit superluminal motion for which the dynamical mass estimate proved the presence of a BH \citep{bail95}.
GRS 1915+105 also exhibits a second type of outflow in the form of an accretion disk wind \citep{ponti12}. 
Studies of the accretion disk wind in GRS 1915+105 have suggested that these winds could act as the jet suppression mechanism in the soft states \citep{neil09}.

The binary system parameters for GRS 1915+105 remained elusive, despite extensive observational effort \citep{ct96,eik8,mirbel7,ma20b,har01,gr01b}, until \citet{gr01} obtained a radial velocity curve, $P_{\rm orb}$ and therefore $f(M)$. 

Inclination has been estimated based on the orientation of the jets ($i=70^{\circ}\pm2^{\circ}$ from \citealt{mirbel4} or $i=66^{\circ}\pm2^{\circ}$ from \citealt{fend99}).
These parameters, along with the mass ratio estimated by \citet{harg4}, allowed for a dynamical mass estimate confirming the BH accretor.
A trigonometric parallax measurement by \citet{reid2014} on the system yielded a direct distance measurement and a revised estimate for BH mass.

\subsubsection{4U 1956+350 (A) --- Cyg X-1}

Cygnus X-1, one of the brightest X-ray sources in the sky,  was discovered in the X-rays by the UHURU satellite in 1971 \citep{tan72}.  
The X-ray emission has been shown to exhibit strong variability on timescales from millliseconds to months \citep{pr83,mi89}.  
 
Numerous spectral and timing studies over the years \citep{holt79,ling83,be90,kit90b,barr90,uber91,mi92,gi97,zdz2,pott3,wilm06,gi10,grin13} have shown Cyg X-1 to spend most of its time in the HCS, resulting in its X-ray spectrum never fully being disk dominated \citep{grin13}. Cyg X-1 often undergoes ``incomplete'' state transitions, never fully transitioning to the softer states \citep{pott3}.
This extended hard state of Cygnus X$-$1 shows weak and persistent radio emission, which has been resolved to be a steady jet \citep{stir1}.
In fact, the ``incomplete'' transitions exhibited by this source are thought to be connected to the radio jet, as jet activity is thought to be quenched in the soft accretion states \citep{tan72}.

With this being said, the source has been occasionally observed in the soft state (e.g., see \citealt{grinberg2011,yamada13,jourdain14,grinberg14}) and there is evidence for the presence of a weak compact jet in the softer states of Cyg X-1 \citep{rushton2012}. For a complete list of X-ray studies see Table \ref{table:outhistBH}. For studies of Cyg X-1 at radio wavelengths, see \citet{hj71,hj73,tan72,stir1,gleiss4,fbg04,fend6}.

This system is known to contain a O9.7Iab type supergiant companion \citep{gies86}, which orbits around a compact object with a period of $\sim 5.6$ days \citep{holt79}.
Over the past 44 years, many estimates on the mass of the compact object have been made \citep{or11}.
While there exist several low-mass models \citep{trim73,bolt75}, all conventional models, which assume an O-type supergiant companion \citep{pac74,gies86,nin87,cab9}, find a large (and uncertain) mass of the compact object exceeding $\sim3 \, M_{\odot}$, therefore confirming a BH primary. \citep{kalb96}. 
 
The large range of these mass estimates is mainly due to the large uncertainty in distance to the source \citep{re11}.
With the more recent trigonometric parallax distance measurement calculated by \citet{re11}, \citet{or11} was able to build a complete improved dynamical model for Cyg X-1, including $M_{\rm BH}$, $q$ and $i$. We make use of this dynamical model for the purposes of our analysis. 

\subsubsection{4U 1957+115 (C)}
4U 1957+115 was discovered by UHURU in 1973 \citep{gn74}.
Despite having an X-ray brightness that is comparable to (and occasionally larger) than the well-studied persistent BH sources LMC X-1 and LMC X-3, little is known about the nature of the system \citep{now8}. 

Optical spectra reveal a power-law continuum with H$\alpha$, H$\beta$ and He II 4686\AA $\,$ emission lines \citep{cow88,shab96b}, typical of systems dominated by an accretion disk \citep{thor87}. 
Long-term variations in the optical light curve have revealed modulations (interpreted as evidence for an accretion disk with a large outer rim (possibly due to a warp) that is seen close to edge on; \citealt{hakk99}) with a 9.33 hour period, generally believed to be the orbital period of the system \citep{thor87}. 

The short orbital period, indicative of a late-type main sequence star as a companion, and the absence of X-ray eclipses, has allowed for an upper limit estimate of inclination between $70^{\circ}$ and $75^{\circ}$, which is consistent with the model of optical variability \citep{hakk99}. 

4U 1957+115 was first classified as a possible BHC in 1984 when EXOSAT observations revealed a very soft spectrum \citep{ricci95}.
Observations by \citet{now8}, who analyzed the complete set of available data from RXTE, and more recently, \citet{now12}, reveal that the X-ray spectrum is a pure disk spectrum $\sim$85\% of the time, with the remaining $\sim$15\% involving some non-thermal component. 

This dominant soft spectrum, coupled with the observed low fractional variability \citep{now99,wij02,now8}, both characteristic of the soft state, make 4U 1957+115  one of only three BHCs that not only remains persistently in outburst but also spends most of its time in the soft disk-dominated accretion state \citep{now8}.
Further evidence for this behavior is implied from the recent radio non-detection \citep{russ11b}, as jet production is believed to be quenched in the soft state.
While 4U 1957+115 may not show evidence for a relativistic jet, this source has been observed to exhibit an outflow in the form of an accretion disk wind \citep{ponti12}.
As there exists no dynamical mass or distance measurements for the system, X-ray and optical observations, analyzed at different times, have been used to argue whether the compact object is a BH \citep{wij02,now8,now12,mait13,gomez15} or NS \citep{yac93,ricci95,rob12}.
Regardless of the uncertainty, we include this system in our sample as a possible BHC.

\subsubsection{GS 2000+251 (A)}
GS 2000+251 was discovered by the ASM aboard GINGA in 1988 \citep{t89}.
The source has been observed to exhibit spectral and temporal characteristics similar to other X-ray nova systems believed to contain BH primaries \citep{vp95,tl97}.
\citet{hj88} found a transient radio source associated with GS 2000+25 exhibiting a spectrum that was fit well with a synchrotron model, similar to the radio emission observed in 1A 0620$-$00 \citep{owen76}, suggesting the possibility of a radio jet in the system.
Optical photometry after this outburst revealed that the system had a $\sim$8.3 hour orbital period \citep{chev93}.
Dynamical measurements first obtained by \citet{fi95} and later improved upon by \citet{cas95} and \citet{har96} revealed a $f(M)=5.01\pm0.12$, subsequently confirming the BH nature of the primary.
The distance to this source is estimated by \citet{barr96b}.

\citet{iou4} has performed the most extensive study of ellipsoidal variability.
They measure an inclination of $54^{\circ}<i<60^{\circ}$, which is consistent with estimates by \citet{cal6} and \citet{beek6}.
However, while \citet{kreid12} conclude that the source was passive during their observations, they suggest their inclination measurement is depressed due to the binning of the light curves, which can slightly decrease the amplitude of ellipsoidal variations.
Following \citet{kreid12} we adopt $55^{\circ} < i < 65^{\circ}$ from \citet{cal6}, and assume a uniform distribution across this range to calculate $M_{\rm BH}$.

\subsubsection{XTE J2012+381 (C)}
The transient X-ray source XTE J2012+381 was discovered with the ASM aboard RXTE in 1998 during outburst \citep{rem98}. The light curve exhibited FRED like behavior \citep{chen97} and the X-ray spectrum was well described by a combination multicolor disk black body ($T\sim0.76$ keV) and power-law with photon index $\Gamma\sim2.9$ \citep{w98}, characteristic of BHXBs \citep{mr06}. In addition, extensive spectral and timing analysis of this outburst reveal the source exhibited the typical ``turtlehead'' evolution seen in BHXBs, transitioning between the hard and soft states \citep{v00,ca01}. The radio counterpart was discovered by \citet{hj98b}, while \citet{hy99} identified a probable optical counterpart with a faint red star heavily blended with a brighter foreground star.  Given the light curve behavior, spectral and timing characteristics, XTE J2012+381 is considered a BHC.

\subsubsection{GS 2023+338 (A) --- V404 Cyg}

GS 2023+338 (V404 Cyg) was discovered with the GINGA satellite in 1989 during outburst \citep{ma89}. V404 Cyg is one of the most well-known transient X-ray sources due to both its high X-ray luminosity and levels of variability at many different wavelengths in outburst and quiescence \citep{tl97,hi02}. The many X-ray observations \citep{k89,int92,mi92,mi93} that exist have shown that, despite its high X-ray luminosity, no soft component exists in the spectrum \citep{oos97}, suggesting that V404 Cyg is a ``hard-only'' outburst source \citep{br04}. While this fact has been disputed by \citet{z99}, who claim that the source spent a short period of time in the soft state, we consider the 1989 event as a possible ``hard-only'' outburst. Two additional outbursts of this source have been found on photographic plates at the Sonnenberg Observatory \citep{r87}. V404 Cyg began its fourth recorded outburst in June 2015 \citep{negoroh15,kuulkerse15}.

The source has also been detected at radio \citep{hjh89} wavelengths. The persistent radio emission displays a flat spectrum, which is indicative of a self-absorbed synchrotron jet \citep{gallo5,mj8}. In addition, another type of outflow in the form of an accretion disk wind has also been observed in this source \citep{oos97}. The optical counterpart was identified by \citet{wagn89} as Nova Cygni 1938. Further optical observations led to a calculation of the mass function by \citet{casc92}, which was later refined by \citet{casc94} to be $f(M)=6.08\pm0.06 \, M_{\odot}$, confirming the BH nature of the primary. \citet{shab94a} modelled ellipsoidal variations of the source obtaining a $\sim 6.5$ day orbital period. More recently, \citet{mj9} obtained a distance to V404 Cyg of $d=2.39\pm0.14$ kpc using trigonometric parallax, making this the first accurate parallax distance measurement to a BH system. We make use of this distance, which is significantly lower than the previously accepted values, for the purposes of our analysis.

The light curve of GS 2023+338 exhibits strong aperiodic variability making it difficult to obtain precise inclination measurements for the system. \citet{wagn2} and \citet{shab94} constrained the inclination to $50^{\circ} < i < 80^{\circ}$ and $45^{\circ} < i <83^{\circ}$, respectively. \citet{wagn2} obtained their lower limit from observations of Balmer spectral lines and the upper limit based on the lack of eclipses. While \citet{shab94} obtain their range by fitting a star only model, which provided a poor fit due to incorrect color correction. In addition, while \citet{san6} fit a star only model to the IR data, the authors note hour time scale variability in the light curve, suggesting the source is most likely in the active state. We therefore adopt the inclination calculated by \citet{kreid12}, which makes use of the inclination estimate from \citet{san6} of $i>62^{\circ}$ to obtain a corrected inclination estimate to calculate a $M_{ \rm BH}$.

\subsubsection{4U 2030+40 (B) --- Cyg X-3}
The X-ray source Cyg X-3 was discovered in 1967 \citep{giacconi66}. It is the only known Galactic XRB containing a compact object and Wolf-Rayet counterpart \citep{vank1993,vank96,fenderhp1999}. Despite being an HMXB, Cyg X-3 has an unusually short orbital period of only 4.8 hours \citep{vank96}. The distance to the system is estimated at $\sim$7--9 kpc \citep{dickey83,predehl2000,ling2009}.

Cyg X-3 is a persistently accreting X-ray source that can be observed in both the hard and soft spectral states \citep{hjalm2008,szostek2008,szostek2008b,koljonen2010}. In addition, it is also the brightest radio source among XRBs \citep{mccollough99}, exhibiting both strong radio flares as well as resolved jets \citep{watanabe1994,marti2001,miod2001}. The true nature of the compact object in the system remains uncertain (e.g., BH or NS; \citealt{hanson2000,vilhu2009,zdz2013}) due to the lack of reliable estimates for the mass function and inclination. However, a BH is favored over a NS given that (i) the X-ray spectral evolution resembles the typical ``turtlehead'' pattern through BH accretion states, and (ii) the radio/X-ray correlation closely corresponds to that found in BHXB systems \citep{szostek2008b}.

\subsubsection{MWC 656 (B) --- HD 215227}
In 2010, the emission line Be star MWC 656 was found within the error circle of the gamma-ray source AGL J2241+4454 with the AGILE satellite \citep{lucarelli10,williams10}. \cite{williams10} noted the similarity of its spectral properties and rapid H$\alpha$ variations (unusually fast for Be stars) to the gamma-ray binary LSI +61$^{\circ}$303 and analyzed available optical photometry, finding a variation with a period of 60 days, which they interpreted as an orbital modulation of the flux from the disk surrounding the Be star. The distance to the star is known to be $2.6\pm0.6$ kpc \citep{williams10}.

Performing a radial velocity study, \cite{casares2012} test the binary hypothesis, estimate an inclination of $i>66^{\circ}$ and a compact object mass. Later, with new optical observations \cite{casar14}  improve upon the previous radial velocity curve and determine a more precise set of binary parameters including a spectral classification of the Be companion of B1.5-B2 III, a mass of the companion star of 10--16 $M_{\odot}$, and a compact object mass of 3.8--6.9 $M_{\odot}$. This evidence makes MWC 656 the first know Be XRB to contain a BH. 

The source has also been detected at X-ray \citep{munar14} and radio wavelengths \citep{dzib2015}. Both radio and X-ray luminosities of the source agree with the behaviour of BHXBs in the hard and quiescent state, and the source occupies the same region of the radio/X-ray plane that the faintest known BHs do (e.g., 1A0620$-$00 and XTE J1118+480), making it a strong BHC.

\section{Data Selection and Analysis}
 To construct the database we have incorporated data from the All-Sky Monitors (ASMs), Galactic Bulge Scan Surveys, and select narrow-field X-ray instruments available on four separate telescopes, making it possible to study nearly two decades of behavior exhibited by the Galactic BHXB population. Table \ref{table:telescopes} presents specific details on the telescopes and instruments used in this study. Data for this work has been acquired from the
the INTEGRAL Galactic Bulge Monitoring Program\footnote{http://integral.esac.esa.int/BULGE/},
the MAXI Database\footnote{http://maxi.riken.jp/top/},
 the High Energy Astrophysics Science Archive Research Center (HEASARC) Online Service provided by the NASA/Goddard Space Flight Center\footnote{http://heasarc.gsfc.nasa.gov/}, Craig Markwardt's (RXTE/PCA) Galactic Bulge Survey Webpage\footnote{http://asd.gsfc.nasa.gov/Craig.Markwardt//galscan/main.html}, and 
Swift/BAT Transient Monitor\footnote{http://swift.gsfc.nasa.gov/results/transients/}.
 
\subsection{The All-Sky Monitors}
The ASMs are indispensable in the study of XRBs as they (i) provide near real-time coverage of large portions of the X-ray sky across both hard and soft X-ray energies, (ii) operate on short  timescales, on the order of $\sim$ 1 day or less, allowing them to track short term changes in behavior in known sources as well as discover new sources, and (iii) accumulate vast databases of activity, which can be used to track outbursts, study evolution and state transitions, and derive a long term history for numerous sources (e.g., \citealt{mr06,kr13}).

\subsubsection{The Rossi X-ray Timing Explorer (RXTE) ASM} 
The RXTE \citep{s98} satellite was perhaps the most important vehicle for the study of transient phenomena in the last decade due to the wide-sky coverage of the ASM, high sensitivity of the PCA and its overall fast response time \citep{mr06}. The ASM \citep{lev96}  aboard RXTE, made up of three wide-field proportional counters, operated in the 1.5--12 keV band from 1996--2012 and had the ability to cover $\sim$ 90\% of the sky every orbit, which took about 90 minutes, with a sensitivity between $\sim$10--20 mCrab (integrating all orbits over a full day)\citep{mr06}.

\subsubsection{The Swift Burst Alert Telescope (BAT)} 
The BAT X-ray monitor \citep{kr13} has provided near real time, wide-field (2 steradians) coverage of the X-ray sky in the 15--150 keV energy range, with an energy resolution of 5\% at 60 keV, since 2005. The BAT has the ability to observe 80--90\% of the sky every day with a sensitivity of 16 mCrab (integrating scans over 1 day) and arcminute positional accuracy. One of the key characteristics of the Swift satellite is the ability to ``swiftly'' ($\lesssim 90$ s) and autonomously repoint itself after detection by BAT to bring the source within the field of view of the sensitive narrow-field X-ray and UV/optical instruments that are also on board the observatory.

\subsubsection{The Monitor of All-Sky Image (MAXI) Telescope} 
MAXI \citep{ma09}, mounted on the International Space Station (ISS) has the ability to scan 85\% of the sky every 92 minutes (one orbit/rotation period of the ISS) with its wide field of view providing near real-time coverage with a positional accuracy of $<6$ arcminutes and a daily sensitivity of 9 mCrab. The Gas Slit Camera (GSC; \citealt{mih11}) detector, one of the two ASMs aboard MAXI, contains a proportional counter that covers the 2--20 keV energy band with its large detection area (5000 $\rm{cm}^2$) and an energy resolution of 18\% at 6 keV.

\subsection{The Scanning Surveys}
The scanning surveys we make use of observe the Galactic Bulge, a region rich in bright variable high-energy X-ray sources, regularly during all visible periods, and provide high sensitivity long-term light curves of numerous X-ray sources that supplement the all-sky coverage.

\subsubsection{The INTErnational Gamma-Ray Astrophysics Laboratory (INTEGRAL) Galactic Bulge Monitoring Program} 
The INTEGRAL \citep{wi03,ku07} Monitoring Program has provided periodic scans of the Galactic Bulge since 2005. Data is taken approximately every 3 days (the length of one orbit) and is provided in the form of single observations, consisting of 7 pointings in a hexagonal pattern of spacing 2 degrees, and lasting $\sim$1.8 ksec total. The INTEGRAL satellite contains three coded mask imagers. One of which is the Integral Soft Gamma-Ray Imager (IBIS/ISGRI; \citealt{ubertini03}), which has a primary energy range of 20--60 keV, an energy resolution of 8\% at 100 keV, and a field of view of 29 square degrees. The other two are the Joint European X-Ray Monitor (JEM-X; \citealt{lund03}) X-ray detectors, which have a primary energy range of 3--35 keV, an energy resolution of 9\% and 13\% at 30 keV, and a circular field of view of diameter 10 degrees. Note that within each observation, only one JEM-X unit is used at a time. Each observation covers 0.1\% of the sky with JEM-X and 2\% of the sky with ISGRI, with a sensitivity of 3--9 mCrab.

\subsubsection{The Proportional Counter Array (PCA) Galactic Bulge Scan} 
The Galactic Bulge Scan Survey, which used the PCA \citep{j96,sm01} aboard RXTE (see Section \ref{sec:PCA_NFI}), provided periodic scans of the Galactic bulge region in the 2.5--10 keV energy band between 1999 and 2011. Each scan covered $\sim8$\% of the sky with a sensitivity of $\sim$3--7 mCrab (estimated from real observations; see Table \ref{table:telescopes} for details).

\afterpage
{
\renewcommand{\thefootnote}{\alph{footnote}}
\renewcommand\tabcolsep{3.0pt}
\tabletypesize{\scriptsize}
\begin{longtable*}{lllclllcccccr}
\caption[Telescope/Instrument Technical Details]{Telescope/Instrument Technical Details}  \\
\hline \hline \\[-2ex]
   \multicolumn{1}{l}{Telescope} &
   \multicolumn{1}{l}{Instruments} &
     \multicolumn{1}{l}{Type} &
          \multicolumn{1}{c}{Sky} &
     \multicolumn{1}{c}{$t_{\rm active}$$^a$} &
    \multicolumn{1}{l}{Energy Band} &
    \multicolumn{1}{c}{Reported} &        
        \multicolumn{1}{c}{Data$^b$} &   
                \multicolumn{1}{c}{1-Day} &  
            \multicolumn{1}{c}{Crab$^c$}&   
                        \multicolumn{1}{c}{Flux$^d$}&   
   \multicolumn{1}{c}{Band$^e$}&
      \multicolumn{1}{r}{Ref.$^n$}\\
      & &&\multicolumn{1}{c}{Coverage}& \multicolumn{1}{c}{(MJD)}&\multicolumn{1}{c}{(keV)} & \multicolumn{1}{c}{Units }&\multicolumn{1}{c}{Type} &\multicolumn{1}{c}{Sensitivity}&  \multicolumn{1}{c}{Conversion} &\multicolumn{1}{c}{Conversion} &\multicolumn{1}{c}{Usage}&  \\
            & &&& &&&&\multicolumn{1}{c}{(3$\sigma$; mCrab)}&  \multicolumn{1}{c}{(crabs)} &\multicolumn{1}{c}{(erg $\rm{cm}^{-2} \, \rm{s}^{-1}$)} &&  \\[0.5ex] \hline
   \\[-1.8ex]
\endfirsthead
  \\[-1.8ex] \hline \\[-0.9ex]          
     \multicolumn{13}{p{0.98\textwidth}}{\hangindent=1ex$^a$The time period over which the instrument has been active.} \\
     \multicolumn{13}{p{0.98\textwidth}}{\hangindent=1ex$^b$Data type collected from the original source. All orbital data was converted to daily averaged data by calculating a weighted mean count rate per day.}\\
     \multicolumn{13}{p{0.98\textwidth}}{\hangindent=1ex$^c$The average photon flux for the Crab Nebula in a given energy band.}\\ 
     \multicolumn{13}{p{0.98\textwidth}}{\hangindent=1ex$^d$Equivalent of 1 crab in flux units (erg $\rm{cm}^{-2}\rm{s}^{-1}$) assuming the Crab Nebula has a spectrum $I(E) = 9.7(\pm1)^{-1.10(\pm.03)} \rm{keV} \rm{cm}^{-2} \rm{s}^{-1}  \rm{keV}^{-1}$ \citep{tods74}.} \\ 
     \multicolumn{13}{p{0.98\textwidth}}{\hangindent=1ex$^e$Indicates whether the energy band in question is used for outburst detection (D), quantitative outburst classification (C), or spectral fitting (F).}\\
     \multicolumn{13}{p{0.98\textwidth}}{\hangindent=1ex$^f$Energy bands manufactured (after crab conversion per band) via an addition process.}\\ 
     \multicolumn{13}{p{0.98\textwidth}}{\hangindent=1ex$^g$The energy range of the PCA varied over the instrument lifetime due to gain changes (e.g., see \citealt{shaposhnikov2012,garcia2014}). The PCA Bulge Scan energy range was roughly 2.5--10 keV with slow 5\% changes as the detector gain varied over the mission (e.g., see \citealt{cart13}).} \\ 
     \multicolumn{13}{p{0.98\textwidth}}{\hangindent=1ex$^h$There are two JEM-X units, each with a separate crab conversion. The column format given is J1/J2.}\\ 
     \multicolumn{13}{p{0.98\textwidth}}{\hangindent=1ex$^i$Crab rates for HEXTE and PCA pointed observation bands are found by calculating the weighted mean count rate of the Crab Nebula using the available HEASARC mission-long products.}\\
     \multicolumn{13}{p{0.98\textwidth}}{\hangindent=1ex$j$Calculated using the 16 (PCA), 29 (ISGRI) and 6 (JEM-X) square degree fields of view quoted for the Galactic Bulge scans. See [2] and [12].}\\ 
     \multicolumn{13}{p{0.98\textwidth}}{\hangindent=1ex$^k$ Scaled to 1-day given that HEXTE was capable of measuring a 100 mCrab X-ray source to 100 keV or greater in $10^3$ live seconds.} \\
      \multicolumn{13}{p{0.98\textwidth}}{\hangindent=1ex$^l$Scan sensitivity (3$\sigma$) for the PCA Bulge scans was typically 7 mCrab for real observations (e.g., \citealt{heinke2010}, Fig. 3). \citet{markwardt06} mentions a nominal 1$\sigma$ sensitivity for scans of 0.5--1 mCrab. Given that he indicates flux detections down to 2--3 mCrab and the possibility that the limiting flux was better in some areas, we estimate a 3$\sigma$ scan sensitivity of 3--7 mCrab.} \\
           \multicolumn{13}{p{0.98\textwidth}}{\hangindent=1ex$^m$3$\sigma$ sensitivity for one observation lasting $\sim1.8$ ks.}\\
     \multicolumn{13}{p{0.98\textwidth}}{\hangindent=1ex$^n${\bf References.} ---
     [1]~\citet{wi03};
     [2]~\citet{ku07};
     [3]~\citet{ubertini03};
     [4]~\citet{lund03};
     [5]~\citet{mih11};
     [6]~\citet{hiroi11};
     [7]`\citet{sugiz11};
     [8]~\citet{lev96};
     [9]~\citet{mr06};
     [10]~\citet{rothschild98};
     [11]~\citet{j96};
     [12]~\citet{sm01};
     [13]~\citet{markwardt06};
     [14]~\citet{kr13}
     }
\endlastfoot
INTEGRAL&ISGRI&scan&2\%/obs$^j$&53419--&$18-40$& \rm{ct} $\rm{s}^{-1}$&orbital&3--9$^m$&208&$8.93 \times 10^{-9}$&D,C,F&1--3\\ 
&&&&&$40-100$&&&&100&$9.41 \times 10^{-9}$&D,F&\\[0.1cm]
&JEM-X&scan&0.1\%/obs$^j$&53419--&$3-10$& \rm{ct} $\rm{s}^{-1}$&orbital&3--9$^m$& 97/103$^h$&$1.58 \times 10^{-8}$&D,C,F&1,2,4\\ 
&&&&&$10-25$&&&&29/27$^h$&$1.08 \times 10^{-8}$&D,F&\\[0.25cm]

MAXI&GSC&all-sky&95\%/day&55058.5--&$4-10$&\rm{ph} $\rm{cm}^{-2}\rm{s}^{-1}$&daily&9& 1.24&$1.18 \times 10^{-8}$&D,C,F&5--7\\
&&&&&$4-20$$^f$&&&&\nodata&$2.01 \times 10^{-8}$&D,F&\\[0.25cm]
RXTE&ASM&all-sky&$\sim$90\%/orbit&50088--55924&$3-5$ (B)& \rm{ct} $\rm{s}^{-1}$&orbital&$\sim$10--20&23.3&$6.93 \times 10^{-9}$&D,C,F&8,9\\
&&&&&$5-12$ (C)&&&&25.4&$1.11 \times 10^{-8}$&D,C,F&\\ 
&&&&&$3-12$$^f$ (B+C)&&&&\nodata&$1.80 \times 10^{-8}$&D,C,F&\\[0.1cm]
&HEXTE&pointed&\nodata&50088--55924&$15-30$& \rm{ct} $\rm{s}^{-1}$&orbital&$\sim1$$^k$&17.8$^i$&$7.94 \times 10^{-9}$&D,C,F&10\\[0.1cm]
&PCA$^g$&scan&8\%/scan$^j$&51214--55869&$2.5-10$& \rm{ct} $\rm{s}^{-1}$$5 \, \rm{PCU}^{-1}$&orbital&3--7$^l$&13930&$1.84 \times 10^{-8}$&D,C,F&11--13\\ 
&&pointed&\nodata&50088--55924&$2-4$& \rm{ct} $\rm{s}^{-1}$$ \, \rm{PCU}^{-1}$&&$\sim$1&321$^i$&$9.71 \times 10^{-9}$&D,F&\\ 
&&&&&$4-9$&&&&637$^i$&$1.05 \times 10^{-8}$&D,C,F&\\ 
&&&&&$9-20$&&&&220$^i$&$9.57 \times 10^{-9}$&D,C,F&\\[0.25cm]
Swift&BAT&all-sky&80--90\%/day&53414--&$15-50$&\rm{ct} $\rm{cm}^{-2}\rm{s}^{-1}$&daily&16& 0.22&$1.34 \times 10^{-8}$&D,C,F&14
\label{table:telescopes}
\end{longtable*}
\renewcommand{\thefootnote}{\arabic{footnote}}
\renewcommand\tabcolsep{5.0pt}
}

\subsection{Narrow-Field Instruments}
We also supplement our database with pointed observations from the PCA and HEXTE instruments aboard RXTE. For details on the reduction of the PCA and HEXTE pointed observations see Section 3.6.

\subsubsection{The High Energy X-ray Timing Experiment (HEXTE)} 
HEXTE \citep{rothschild98} aboard RXTE provided high energy coverage of the X-ray sky in the 15--250 keV energy band, with an energy resolution of 15\% at 60 keV, from 1996--2012. HEXTE consisted of two clusters of detectors, each of which contained four NaI(Tl)/CsI(Na) phoswich scintillation counters. Each cluster had the ability to ``rock'' along mutually orthogonal directions, providing background measurements 1.5 or 3.0 degrees away from the source every 16 to 128 s. Overall, HEXTE was capable of measuring a 100 mCrab X-ray source to 100 keV or greater in $10^3$ live seconds. The field of view per cluster was 1 degree, and all 8 detectors in both clusters covered a net open area of 1600 ${\rm cm^2}$. 

\subsubsection{The Proportional Counter Array (PCA)} 
\label{sec:PCA_NFI}
The PCA \citep{j96} aboard RXTE consisted of an array of 5 proportional counters, which operated in the 2--60 keV\footnote{The energy range of the PCA varied over the instrument lifetime due to gain changes (see, e.g. \citealt{shaposhnikov2012,garcia2014}).} range, with an energy resolution of 18\% at 6 keV, between 1996 and 2012. It had a total collecting area of 6500$\rm{cm}^2$, a field of view of 16 square degrees and a sensitivity of 1mCrab (estimated from observations; see \citealt{heinke2010}).

\subsection{Calibration with the Crab Nebula}

We use the Crab Nebula to effectively probe the X-ray sky over a wide energy range and long consecutive time periods using the X-ray emission from multiple telescopes/instruments. The Crab Nebula is often used as
an X-ray calibration source due to the fact that it has been observed to be a bright, approximately steady X-ray source producing a constant spectrum. We make use of the now accepted ``canonical'' simple power-law spectrum, originally estimated by \citet{tods74}, of the form,
\begin{equation}
I(E)=(9.7 \pm 1.0)E^{(-1.10\pm0.03)} \rm{keV}\, \rm{cm}^{-2}\, \rm{s}^{-1}\, \rm{keV}^{-1}
\end{equation}
valid in the 2--50 keV range. Note, in this form, the index on the power-law is quoted as $\alpha=\Gamma-1$. See \citet{kirsh5} for a review of past estimates of this spectrum. Using crabs as a baseline unit of flux not only allows us to calculate approximate count rate equivalences in each energy band, therefore giving us the ability to directly compare data from a particular source across telescopes and instruments (i.e., light curves, hardness ratios etc.), but also gives us a straightforward method for converting between count rate and flux in a given energy band (via integration of the known spectrum over the given band).

That being said, our analysis clearly relies heavily on the assumption that the Crab is a steady X-ray source in any given band. Specifically the count rate over the instrumental area (or flux density in counts $\rm{cm}^{-2} \rm{s}^{-1}$) is assumed constant. This assumption brings with it two separate issues. First, being that the Crab is variable, more so at higher energies then lower energies \citep{wilhog10}, which will subtly affect Swift/BAT, INTEGRAL, and HEXTE data. The measured variations reach 10\% at most (and no more than 4\% in one year), which would lead to smaller flux errors than those produced by the current distance uncertainties in most of our sources. Since correcting for these small variations in the flux of the Crab would be quite difficult,  we neglect them in our analysis.

Second, assuming that the spectrum of each of our sources is Crab-like induces errors in the flux computations. As such, we have investigated what the effect on the unabsorbed flux (in a given band) is, if we were to assume a harder photon index ($\Gamma=1.2$ 
versus $\Gamma=2.1$) or a 1 keV blackbody, using PIMMS\footnote{https://heasarc.gsfc.nasa.gov/docs/software/tools/pimms.html}. 
Within each (fairly narrow) X-ray energy band we find that assuming a Crab spectral shape gives errors on the flux of no more then 20\% for a flat power-law versus a blackbody, and typically $\lesssim10$\%, which justifies our approach in assuming the Crab spectral shape for our analysis.

\subsection{Other Data Issues}
Inspection of the MAXI data suggests that it contains some remaining calibration issues . These problems are most obvious in the MAXI soft (2--4 keV) band, where the Crab light curve, displays oscillatory behavior.
Given that many X-ray telescopes regularly monitor the Crab and this behavior has not been reported elsewhere, as well as that these oscillations are observed in a number of other bright sources, this behavior is unlikely to be physical in origin.
As including this band would lead to a large systematic uncertainty, the MAXI soft band is not used in our analysis.
In addition, we also omit analysis of just the MAXI hard (10--20 keV) band based on its relatively poor signal-to-noise ratio; however we include data from this band for combined analysis with the MAXI medium (4--10 keV) band.

As we are studying sources within the Galaxy across a wide energy range, X-ray absorption by the ISM becomes important, especially in the soft X-ray regime ($<2$ keV). 
LMXBs in the Galaxy show a range of $N_H$, from $\sim3\times10^{20}$ to $\sim5\times10^{22}$ atoms cm$^{-2}$ (e.g., \citealt{lmxb07}), with only a few objects (e.g., 1E 1740.7$-$2942) having $N_H$ near or above $10^{23}$ cm$^{-2}$.  This leads to a turnover in the low-energy spectrum, which for a Crablike spectrum occurs around 0.5 keV for $10^{21}$\,cm$^{-2}$, 1.2 keV for $10^{22}$\,cm$^{-2}$, 2.2 keV for $5\times10^{22}$\,cm$^{-2}$, and 3 keV for $10^{23}$\,cm$^{-2}$.  At $N_H=5\times10^{22}$\,cm$^{-2}$, 
the flux in the 1.5--3 keV energy range is 0.3 of the unabsorbed amount. As such, we choose to omit the RXTE/ASM A band (1.5--3 keV).

\subsection{Reduction of RXTE Pointed Observations}
\label{subsec:rxte}
To make use of the vast array of pointed observations available in the RXTE public archive, we first collected the available Mission-Long Data Products\footnote{https://heasarc.gsfc.nasa.gov/docs/xte/recipes/mllc\_start.html} for individual sources.
The Mission-Long Data Products, created via the standard data products (StdProds) from the PCA and HEXTE, combine all available good pointed RXTE observations (no scans or slews) of an individual source (i.e., all observation IDs across all available proposals and the many sequential observing years of the RXTE mission).
Each ObsID provides one data point for each energy band.

For PCA, a single data point in three separate energy bands (2--4, 4--9, and 9--20 keV) is calculated by taking the average rate of the background-subtracted PCA Std2 mode for that particular observation.
For HEXTE, a single data point in the 15--30 keV energy band is calculated by taking the average rate of the background-subtracted HEXTE cluster B archive mode. 
 
However, only 31 sources within our sample have available Mission-Long Data Products.
As such, we have made use of existing Perl scripts within the Heasoft Software Package\footnote{http://heasarc.nasa.gov/lheasoft/} to emulate the reduction and subsequent production of the mission-long background subtracted light curve data for the remainder of the sources in our sample. 

For HEXTE data, we used the \textit{hxtlcurv}\footnote{https://heasarc.gsfc.nasa.gov/docs/xte/recipes/hexte.html\#script} script to create background-subtracted light curves in the 15--30 keV band, with 16 second time bins, from the cluster B archive mode data.
For PCA data, we made use of the \textit{rex}\footnote{https://heasarc.gsfc.nasa.gov/docs/xte/recipes/rex.html} script to extract and create background subtracted light curves in the 2--4, 4--9, and 9--20 keV bands from the PCA Std2 data.
Following reduction via the specified scripts, we implemented the same procedure as outlined above, whereby long-term light curves are created by combining single data points representing each individual observation.

Note that, after 2009 December 14 (MJD = 55179), HEXTE cluster B no longer had the ability to ``rock''. As such the cluster remained in an off-source position and data from this cluster was only used for background estimation until the end of the RXTE mission. In addition, during the time period of 2009 December 14 to 2010 March 29 (MJD = 55179--55284) the telemetry values that indicated the cluster position incorrectly indicated the cluster was on-source in the header files, when it was actually in an off-source position\footnote{https://heasarc.gsfc.nasa.gov/docs/xte/whatsnew/newsarchive\_2010.html}. For these reasons we do not include any HEXTE cluster B data occurring after MJD = 55179 in our analysis.

\subsection{The Algorithm: Outburst Discovery, Tracking, Classification, and Analysis}
Data from the seven instruments listed in Table \ref{table:telescopes} are run through a custom pipeline composed of a comprehensive algorithm built to discover, track, and quantitatively classify outburst behavior.
The products produced via this algorithm can then be used to analyze the details of outburst behavior, including duty cycles, recurrence times, total energy released during outburst, long-term outburst rates, state transitions, luminosity functions, and mass transfer rates of Galactic BHXB systems (see Section~\ref{sec:results}). This algorithm consists of a seven stage process; data cleaning, detection, sensitivity selection, X-ray hardness computation, spectral fitting, luminosity function and mass-transfer rate estimation, and empirical classification.

\subsubsection{Data Cleaning}
\label{subsubsec:clean}
The purpose of data cleaning is two-fold.
As we have found the background subtraction performed by the surveys to be inadequate and the quoted errors to be underestimated across all four telescopes, it becomes necessary to perform a further background subtraction on the data and include an additional factor in the treatment of the errors.
The following analysis is performed on daily averaged data.
If the data is collected in the form of orbital data, it is first converted to daily average data by calculating the weighted mean count rate (and uncertainty) over each day of data available (see \citealt{beving3}).
 
In the case of the transient sources, the algorithm must first be able to identify times when a source was in quiescence for each energy band and instrument combination.
To do so, it begins by calculating a background rate $R_{\rm bg}$ using an iterative bi-weight\footnote{Here the bi-weight refers to a robust statistic for determining the central location of a distribution and quantifying the statistical dispersion in a set of data. See \citet{beers90} for a detailed discussion. } method.
This method, which makes use of $\sigma$-clipping, allows for a determination of both the location and scale (i.e., mean and standard deviation) of all the long-term light curve data for the source.
This task is performed in part using the \textit{astLib.astStats}\footnote{Created by Matt Hilton and Steven Boada and available on SourceForge: http://astlib.sourceforge.net.}  python module, which makes use of methods described in \citet{beers90} to provide robust estimations of location and scale (i.e., robust equivalents of mean and standard deviation).
From here, quiescence is defined as times in which the count rate is less then $3\sigma$ above the background rate, $R_{\rm bg}$.

Generally, if the quoted errors $\sigma_{\rm quo}$ are correct, you would expect the quiescent data to follow a Gaussian distribution \citep{beving3}.
However, this is not observed in our data sets, which becomes a serious problem when the algorithm begins to identify individual outbursts (see Section ~\ref{subsubsec:detect}).
To remedy this problem, an \textit{ad hoc} method is employed, defining a correction factor $C_{\sigma}$ as the standard deviation of a Gaussian distribution fit to the quiescent non-detection data.
$C_{\sigma}$ then acts as a multiplicative factor, scaling up the quoted errors appropriately to be $\sigma_{\rm corr}=C_{\sigma} \sigma_{\rm quo}$. 
 
To ensure that a transient source is not in fact being detected during times that the algorithm has defined as quiescence (which would effectively decrease the value of our errors), we have performed numerous trials comparing the algorithm results to pointed observations with more sensitive instruments found in the literature for the three most well studied recurrent transient sources in the Galaxy: 4U 1630$-$472, GX 339$-$4, and H 1743$-$322.
 
While the above analysis is adequate for the transient sources, the persistent sources must be handled differently.
In this case, the algorithm makes use of background estimates and error corrections (i.e., the standard deviation of the distribution of observed quiescence count rates) reflective of the local transient source population.
Here it takes the average background rate and average error correction for all the transient sources within a $16^{\circ}$ radius of each persistent source, per energy band, to be the background rate and error correction for each persistent source, respectively.
The $16^{\circ}$ radius was specifically chosen to ensure that at least two transient sources are used in the background/error correction estimation of each persistent source.
The LMC sources are the obvious exception to this criterion as there are no nearby transients from our sample.
In their case the algorithm makes use of all transient sources in our sample to compute the background/error correction estimates.

\subsubsection{Detection}
\label{subsubsec:detect}
The detection stage begins by first performing a second background subtraction (via the method outlined in Section~\ref{subsubsec:clean}) on the now error-corrected data, followed by differentiating the data into two separate catagories, (i) detection: in which the count rate $R_c\geq R_{\rm bg}+3\sigma_{\rm corr}$; and (ii) non-detection: in which $R_c<R_{\rm bg}+3\sigma_{\rm corr}$. Note that negative count rates, which happen on occasion due to over-subtraction of the background are included in the non-detections. From here the algorithm produces lists of individual outbursts detected, in every given energy band (see Table  \ref{table:outhistBH}), based on the minimum criteria that there must be at least 2 detections occurring every 8 days to be counted as an outburst.
We set a minimum number of detections to eliminate a large number of spurious ``outbursts''.
This criterion is justified given that we do not currently know of any outbursts from BH transients lasting less than 2 days.
Lists of outbursts detected per energy band are then combined, creating outburst lists per instrument, then per telescope, and lastly into a final outburst list for each individual source, taking into account detections from all four telescopes in our data sample (where available). 

The detection stage of the algorithm is equipped to deal with 
(i) situations in which large gaps in the data exist (largely due to Sun constraints), there is a lack of continuous daily coverage (e.g., survey instruments), and there is known down time (e.g., Space Shuttle docked at the ISS affecting MAXI coverage); and
(ii)  complicated non-trivial behavior exhibited during outburst such as double (or multiple) peak features (e.g., XTE J1550$-$564; \citealt{ku04}), extended flare-like activity (e.g., 4U 1630$-$472; \citealt{t05}), and prolonged outburst periods (e.g., GX 339$-$4; \citealt{z04}).

To deal with the complicated non-trivial behavior the algorithm repeats the data cleaning (Section~\ref{subsubsec:clean}) and detection (this Section) stages on both weekly (8-day averaged) and monthly (24-day averaged) data with the minimum criteria for an outburst being at least 2 detections occurring every 24 days and at least 2 detections occurring every 72 days, respectively.
This is followed by combining the produced results (1-day average, 8-day average, and 24-day average) into a final list of outbursts detected in an individual source.

\afterpage{
\renewcommand{\thefootnote}{\alph{footnote}}
\renewcommand\tabcolsep{10pt}
\tabletypesize{\footnotesize}
\begin{longtable}{lcccr}
\caption[Empirical Outburst Classification Criteria]{Empirical Outburst Classification Criteria}  \\
\hline \hline \\[-2ex]
   \multicolumn{1}{c}{Telescope$^a$} &
   \multicolumn{1}{c}{Hard Band} &
   \multicolumn{1}{c}{Soft Band} &
   \multicolumn{1}{c}{$C_{\rm soft}$$^b$} &
   \multicolumn{1}{c}{$C_{\rm hard}$$^b$}\\
   \multicolumn{1}{c}{ID} &
   \multicolumn{1}{c}{(keV)} &
   \multicolumn{1}{c}{(keV)} &
   &
   \\[0.5ex]\hline
   \\[-3ex]
   \label{table:HRtable}
\endfirsthead
  \\ \hline \\[-1.0ex]  
  \multicolumn{5}{p{0.44\textwidth}}{\hangindent1ex$^a$SM:~Swift~\&~MAXI,\par\hspace{1ex}\hangindent=1ex SR:Swift~\&~RXTE/ASM,\par\hspace{1ex}\hangindent=1ex SRp:~Swift~\&~RXTE/PCA,\par\hspace{1ex}\hangindent=1ex SI:~Swift~\&~INTEGRAL/\mbox{JEM-X},\par\hspace{1ex}\hangindent=1exII:~INTEGRAL/ISGRI~\&~INTEGRAL/\mbox{JEM-X},\par\hspace{1ex}\hangindent=1ex HR:~RXTE/HEXTE~\&~RXTE/ASM,\par\hspace{1ex}\hangindent=1ex HRpp:~RXTE/HEXTE~\&~RXTE/PCA,\par\hspace{1ex}\hangindent=1ex RRpp:~RXTE/PCA~\&~RXTE/PCA.}\\
  \multicolumn{5}{p{0.44\textwidth}}{\hangindent1ex$^b$$H_{X}$ boundaries defining the HCS and SDS (See Sections \ref{subsubsec:HR} and \ref{subsubsec:classify}).}  
\endlastfoot
SM&15--50&\phantom{0.}4--10&0.2846&0.3204\\[0.15cm]
SR&15--50&\phantom{0.}3--12&0.1646&0.2675\\[0.15cm]
SRp&15--50&2.5--10&0.5597&0.8601\\ [0.15cm]
SI&15--50&\phantom{0.}3--10&0.3884&0.5751\\[0.15cm]
RR&\phn5--12&\phantom{0.}3--5\phn&0.3843&0.4220\\  [0.15cm]
II&18--40&\phantom{0.}3--10&0.3579&0.5449\\  [0.15cm]
HR&15--30&\phantom{0.}3--12&0.3717&0.6890\\  [0.15cm]
HRpp&15--30&\phantom{0.}4--9\phn&0.2246&0.4938\\  [0.15cm]
RRpp&\phn9--20&\phantom{0.}4--9\phn&0.4620&0.6433\\[-1.0ex]
\end{longtable}
 \renewcommand{\thefootnote}{\arabic{footnote}}
 \renewcommand\tabcolsep{5pt}
 }

\subsubsection{Sensitivity Selection}
\label{subsubsec:selection}
Given the variable data quality at times, to ensure that the outburst detector (Section~\ref{subsubsec:detect}) is catching ``real'' outbursts rather than artificial flare-like/dip-like profiles, which can be the product of unexpected background increases/decreases (i.e., Sun glints) or instrumental errors, the algorithm implements a sensitivity limit.
  
During the sensitivity selection stage, a weighted mean method (see \citealt{beving3}) is used to calculate the mean count rate ($\mu_{\rm rate}$) and error on the mean count rate ($\sigma_{\rm rate}$) per outburst.
This calculation is followed by the application of a $\sigma$-clip, whereby only outbursts with $\mu_{\rm rate}>10\sigma_{\rm rate}$  are considered ``real'' outbursts.
The $10\sigma_{\rm rate}$ limit has been determined to be the optimal value through numerous trials comparing the results from the outburst detector to literature detections of the well studied transient sources 4U 1630$-$472, GX 339$-$4, and H 1743$-$322. In the hardness computation stage the algorithm computes hardness ratios using nine different combinations of energy bands, each of which are listed in Table \ref{table:HRtable}.

\subsubsection{X-Ray Hardness Computation}
\label{subsubsec:HR}

The algorithm assumes the form of the hardness ratio $H_{X}$ to be the hard band flux density (in crab units) divided by the soft band flux density (in crab units) and computes the quantity using a Markov Chain Monte-Carlo (MCMC) method via the \textit{emcee} python module \citep{mack12}.
As this problem involves two-dimensional uncertainties, the standard linear formulation (i.e., $y=mx+b$) is ill-suited.
As such, the algorithm makes use of an alternative model, originally proposed by \citet{hogg10}, which involves parameterizing the slope $m$ in terms of the angle $\theta$ that the generative locus (line) makes with the $x$-axis.

Adapting this technique, it parametrizes $H_{X}$ (slope $m$) in terms of hard band flux density ($y$-variable) and soft band flux density ($x$-variable) such that,
\begin{equation}
H_{X}=\tan\theta=\frac{f_{\rm hard}}{f_{\rm soft}},
\end{equation}
 
\noindent yielding a log likelihood of the form \citep{hogg10},
\begin{equation}
\ln \mathcal{L}=K- \sum_{i=1}^{N} \frac{(f_{\rm hard}\cos \theta-f_{\rm soft}\sin \theta)^2}{\sigma_{\rm soft}^2\sin^{2}\theta+\sigma_{\rm hard}^2\cos^{2}\theta}.
\label{eq:loglikelyHX}
\end{equation}
Initialization is performed using the following prescription,
\begin{equation}
x_0=(\pi/4)+(0.1)r,
\end{equation}
where $r$ is a random number in the range $(0,1]$, (E.~Rosolowsky, private communication) in combination with a 100 step ``burn in'' phase.

After likelihood maximization is performed on $\theta$ (via Equation \ref{eq:loglikelyHX} ), a probability distribution function (PDF) of $H_{X}$ is obtained by taking the tangent of the resulting PDF for $\theta$ found via the MCMC algorithm.
The algorithm takes the median of the resulting distribution to be the accepted value of $H_{X}$ at the particular time $t$, and defines the $1\sigma$ confidence interval as the upper and lower limits on $H_{X}$. 

Within our data sets there exist three separate situations.
On any given day $t$ there may be a
(i) detection in both hard and soft bands;
(ii) detection in the hard band and non-detection in the soft band, or;
(iii) detection in the soft band and non-detection in the hard band.

In case (i), both upper and lower limits on $H_{X}$ are tabulated. 
However, in cases (ii) and (iii) the situation is more complex.
Here we must be particularly cautious as there is a possibility for large errors on the non-detection data.
As such, the algorithm disregards data with large errors by applying a $\sigma$-cut.
Here only non-detection data points in which the errors are within $2\sigma$ of the mean error value for the given band are considered valid\footnote{This mean and error on this mean are calculated via the bi-weight method discussed in Section~\ref{subsubsec:clean}.}.
Note that this $\sigma$-cut is not applied to detection data (i.e., times in which we have detections in both the hard and soft bands).
In a case (ii) situation, where there is only a hard detection, only lower limits on $H_{X}$ are tabulated.
This indicates the source is most likely in the HCS, assuming similar sensitivity in hard and soft bands.
Similarly, in a case (iii) situation, where there is only a soft detection, only upper limits on $H_{X}$ are tabulated.
This indicates the source is most likely in the SDS.

The last part of the hardness computation stage involves a calculation of a total hardness range for each outburst, followed by the placement of a ``classification flag'' on each outburst indicating whether or not it meets the ``minimum data requirement''.
This requirement indicates whether or not the algorithm has enough data on the outburst to later confidently classify it (via procedure outlined in Section~\ref{subsubsec:classify}).
To receive a ``classification flag'' the outburst must consist of at least 5 data points in which $H_{X}$ has been computed.
In addition to cases where there are not enough $H_{X}$ data points, an outburst will also fail to receive a ``classification flag''  when data are only available in one energy band (making the calculation of $H_{X}$ impossible).

\subsubsection{Spectral Fitting}
\label{subsubsec:spectra}
By modelling each day's flux of a BHXB as a combination of a soft disk black body spectral component and hard Comptonized spectral component, assuming a Crab-like spectrum in each given energy band and a known distance (from the literature), it is possible to obtain X-ray luminosity $L_{\rm X}$ (in any energy band) for a source on a given day $t$.
To accomplish this task we use an algorithm that relies on X-ray hardness to determine the relative dominance of the disk and Comptonized spectral components in the spectrum.
Here we have assumed a standard two component spectral model in {\sc Xspec}, {\sc diskbb+comptt}, representing
(i) the soft disk component of the spectrum with a $T_{in}=1$ keV  multi-color disk black body, typical of BHXBs in the SDS \citep{mr06}, and
(ii) the hard Componized component with the analytical {\sc comptt} model corresponding to a plasma temperature of $T_{e}=50$ keV, a soft disk photon temperature of $T_{in}=1$ keV, and an optical depth $\tau=1.26$, which has been calculated to roughly match a typical hard state photon index of 1.7 for the 3--20 keV range, as is often found in BHXBs \citep{mr06,done10}.

Several sources have been shown to soften as they fade into quiescence (e.g., see \citealt{plotkin2013,wijnands2015}), but we are not capable of following a source into quiescence with the instruments that we use (i.e., our sensitivity limit is a few times $10^{35}$ ${\rm erg \, s^{-1}}$; see Figure \ref{fig:poplum1a}). Thus, this trend is not relevant to our analysis.

\afterpage{
\renewcommand{\thefootnote}{\alph{footnote}}
\renewcommand\tabcolsep{5.0pt}
\tabletypesize{\footnotesize}
\begin{longtable}{llcc}
\caption[Spectral Fitting Constants]{Spectral Fitting Constants}  \\
\hline \hline \\[-2ex]
   \multicolumn{1}{c}{Instrument} &
   \multicolumn{1}{c}{Energy Band} &
   \multicolumn{1}{c}{{\sc diskbb} flux $m_1$$^a$} &
     \multicolumn{1}{c}{{\sc comptt} flux $m_2$$^b$} \\
&\multicolumn{1}{c}{(keV)}&\multicolumn{1}{c}{($\rm{erg} \, \rm{cm}^{-2} \, \rm{s}^{-1}$)} & \multicolumn{1}{c}{($\rm{erg} \,  \rm{cm}^{-2} \, \rm{s}^{-1}$)}\\[0.5ex] \hline
   \\[-4ex]
   \label{table:specfit}
\endfirsthead
  \\[-1.8ex] \hline \\[-1ex]   
    \multicolumn{4}{l}{$^a$the flux of the {\sc diskbb} model in the energy band specified.}\\      
        \multicolumn{4}{l}{$^b$the flux of the {\sc comptt} model in the energy band specified.}\\  
\endlastfoot
ASM&3--5 (B)&$4.08\times10^{-12}$&$5.08\times10^{-8}$ \\[0.1cm]
ASM&5--12 (C)&$1.49\times10^{-12}$&$1.43\times10^{-7}$ \\[0.1cm]
ASM&3--12 (B+C)&$5.57\times10^{-12}$&$1.94\times10^{-7}$\\[0.1cm]
BAT&15--50&$4.99\times10^{-16}$&$3.23\times10^{-7}$ \\[0.1cm]
GSC&4--10&$2.92\times10^{-12}$&$1.36\times10^{-7}$ \\[0.1cm]
GSC&4--20&$2.95\times10^{-12}$&$2.80\times10^{-7}$ \\[0.1cm]
HEXTE&15--30&$4.99\times10^{-16}$&$1.69\times10^{-7}$ \\[0.1cm]
ISGRI&18--40&$4.24\times10^{-17}$&$2.12\times10^{-7}$ \\[0.1cm]
ISGRI&40--100&$7.29\times10^{-26}$&$7.06\times10^{-8}$ \\[0.1cm]
\mbox{JEM-X}&3--10&$5.55\times10^{-12}$&$1.60\times10^{-7}$ \\[0.1cm]
\mbox{JEM-X}&10--25&$2.95\times10^{-14}$&$2.00\times10^{-7}$ \\[0.1cm]
PCA&2.5--10&$9.83\times10^{-12}$&$1.80\times10^{-7}$ \\[0.1cm]
PCA&2--4&$6.91\times10^{-12}$&$4.42\times10^{-8}$ \\[0.1cm]
PCA&4--9&$2.88\times10^{-12}$&$1.17\times10^{-7}$ \\[0.1cm]
PCA&9--20&$7.01\times10^{-14}$&$1.63\times10^{-7}$ \\
\end{longtable}
\renewcommand{\thefootnote}{\arabic{footnote}}
\renewcommand\tabcolsep{5pt}
}

Each day's flux for a chosen source and energy band can then be modelled as,
\begin{equation}
f_{X} = a \, m_1 + b \, m_2,
\label{eq:fluxfit}
\end{equation}
where $m_1$ and $m_2$ are the flux of the {\sc diskbb} and {\sc comptt} models for the energy band in question, respectively.
Provided that there is one energy band available that can act as a hard band, one available energy band that can act as a soft band, and that at least one of these bands exhibits a detection of the source on the day in question, the algorithm estimates the flux of a source on this day in the chosen energy band using an MCMC method, via the \textit{emcee} python module \citep{mack12}, to fit for the normalization parameters $a$ and $b$.
For a list of the $m_1$ and $m_2$ constants corresponding to each energy band used in this study see Table \ref{table:specfit}.
In this case, the log likelihood takes the form,
\begin{equation}
\ln \mathcal{L}=K -  \sum_{i=1}^{N} \frac{(y_i - a \, m_1 - b\, m_2)^2}{\sigma_{y_i}^2}.
\end{equation}
Initialization is accomplished by starting the MCMC ``walkers'' in a very compact grid in parameter space around the point that is expected to be close to the maximum probability point, in combination with the implementation of a 500 step ``burn in'' phase.

The algorithm computes where this compact grid is located in any particular case by fitting for the $a$ and $b$ parameters from Equation \ref{eq:fluxfit} using a bounded least-squares fitting algorithm (\textit{leastsqbound}\footnote{Created by Jonathan J. Helmus, \textit{leastsqbound} is a modified version of the \textit{scipy.optimize.leastsq} module that allows input of bounds on each fit parameter. Constraints are enforced by using an unconstrained internal parameter list, which is transformed into a constrained parameter list, using non-linear functions. The source code is available via GitHub:https://github.com/jjhelmus/leastsqbound-scipy/.}), as bounds are necessary (both $a$ and $b$ must be greater than zero) to obtain a ``physical'' flux (i.e., non-negative). 

Once $a$ and $b$ are obtained from \textit{leastsqbound}, a suitable range for each parameter is found by performing a simple grid search, splitting the parameter space from $0 \rightarrow a$ and $0 \rightarrow b$ into four equal fractional sections  (i.e., 1/8, 3/8, 5/8, and 7/8 multiples of the parameter). The grid point with the minimum $\Delta \chi^2$ is applied as a symmetric error for each parameter.

After likelihood maximization is performed, the algorithm takes the median and $1\sigma$ confidence intervals of the resulting PDFs created via the MCMC algorithm as the accepted value, and upper and lower limits on each of the normalization parameters $a$ and $b$, respectively.

Once $a$ and $b$ are found for a given day, the algorithm can then compute the flux in the 2--50 keV band on that day. This band limited flux is then converted to bolometric flux (0.001--1000 keV) by multiplying each component (disk and comptonized) by a derived bolometric correction from the {\sc Xspec} models. Lastly, the disk and comptonized flux (and in turn luminosity, $L_{\rm X}$) components are obtained via, 
\begin{equation}
f_{\rm bol} = a \, m_{1,\rm bol} + b \, m_{2,\rm bol} = f_{\rm disk} + f_{\rm comp},
\end{equation}
and a disk fraction, $d_f=f_{\rm disk}/f_{\rm bol}$, is computed.

We stress that this algorithm assumes one spectral model is valid for all systems. 
For example, if the disk was in fact hotter then expected (i.e., when a source is in the SDS), then the algorithm would overestimate the Comptonized component and underestimate the thermal disk component in the spectrum.
Similarly, if the disk was cooler then expected (i.e., when a source is in the HCS), then the algorithm would underestimate the Comptonized component and overestimate the thermal disk component in the spectrum.
This assumption also breaks down when a source has a power-law index $\Gamma$ that is significantly steeper then the chosen $\Gamma=1.7$ value. This situation could occur if a source enters a SPL state or other anomalous high luminosity variability states (e.g., GRS 1915+105 and IGR J17091$-$3624; see Sections \ref{subsubsec:GRS1915}
and \ref{subsubsec:IGR17091} for references).

To determine which outbursts are likely affected by these problems we
(i) observe the evolution of the disk fraction over each algorithm-classified outburst, looking for deviations from the expected trend (i.e., lower disk fraction in the HCS and higher disk fraction in the SDS), and
(ii) check for appearances (via spectral/timing analysis in the literature) of SPL state behavior (which could be a possible explanation for uncharacteristic behavior in the disk fraction) in outbursts that deviate from the expected disk fraction trend.
Overall, this allows us to quantify the problem and eliminate affected outbursts from our analysis (see Sections \ref{subsec:peaklum}--\ref{subsec:mdothist} and \ref{subsec:peaklum_porb} for a detailed discussion on the impact of this problem).

\subsubsection{Luminosity Function and Mass-Transfer Rate Estimation}
\label{subsubsec:lfmdot}
With the computed bolometric X-ray luminosities $L_{\rm bol}$, the algorithm next builds an X-ray Luminosity Function (XLF), obtains a time-averaged bolometric luminosity (over the last 19 years), and derives a mass-transfer history for a source.

For transient systems, the algorithm assumes $L_{\rm bol}=0$ during all quiescent (non-detection) periods. It then uses the days in which data was available (and thus those days  an estimate of $L_{\rm bol}$ exists via Section \ref{subsubsec:spectra}) during outburst to interpolate a $L_{\rm bol}$ for the missing days in between. Using the $L_{\rm bol}$ estimates for every day a source was in outburst, the algorithm creates a complete transient XLF.
To accomplish this task, the algorithm checks if an  $L_{\rm bol}$ exists for every day during an outburst.
If no estimate exists, the algorithm takes the nearest days bracketing the missing day(s) on either side (call them $t_1$ and $t_2$), that have estimates and performs a linear interpolation using an MCMC method via the \textit{emcee} python module \citep{mack12}.

While the log likelihood may take the standard linear form in the linear interpolation MCMC method, the situation is more complex due to the fact that the flux values have asymmetric errors. We thus add  a conditional statement  to the likelihood function.
If the fit (the flux calculated from fitted parameters $m$ and $b$) obtains a flux value greater than that of the input data point (i.e., the fit is above the data), the upper error bar is used in $\ln \mathcal{L}$.
In contrast, if the fit obtains a flux value less than that of the input data point (i.e., the fit is below the data), the lower error bar is used in $\ln \mathcal{L}$ (E.~Rosolowsky, private communication).
This asymmetric error situation complicates the MCMC initialization procedure.
To set the initialization of the ``walkers'' in the linear interpolation MCMC method the algorithm again makes use of the idea of starting the ``walkers'' in a very compact grid in parameter space around the point that is expected to be close to the maximum probability point in combination with the implementation of  a 500 step ``burn in'' phase. 
However, to compute the compact grid, the procedure presented in Section \ref{subsubsec:spectra},which makes use of the bounded least squares algorithm, must be modified.
Performing a Monte-Carlo sampling for each data point (flux $f_X$) in the range bracketed by its asymmetric error bars, allows the algorithm to compute the 1$\sigma$ confidence intervals of the resulting distributions and in turn a suitable range for each parameter to create the compact grid.

After likelihood maximization, the algorithm takes the median of the PDF distributions created via the MCMC algorithm as the accepted values of  $m$ and $b$ and the $1\sigma$ confidence intervals as the upper and lower limits on each parameter.
Knowing the values of $m$ and $b$ across the time interval $t_1 \rightarrow t_2$, it can then interpolate $f_{\rm bol}$ (and in turn calculate $L_{\rm bol}$) for the days missing estimates in this time interval and as such calculate a time-averaged bolometric luminosity over the last 19 years via, 
\begin{equation}
L_{\rm bol,avg}=\frac{\sum_{i=1}^{N} L_{\rm{bol},i}}{t_{\rm tot}}.
\label{eq:timeavg}
\end{equation}

The situation is more involved for the persistent systems as, unlike transient systems, many of these sources never fully return to quiescence. Instead persistent sources occasionally transition to a state, often occurring in between long bright outburst periods, in which a low-flux level is maintained.
To deal with this situation the algorithm first estimates a detection limit, $L_{\rm det}$, for each particular source, by fitting the non-detection data (using the procedure outlines in Section ~\ref{subsubsec:spectra}) to find the lowest non-detection luminosity upper limit.
Whenever $L_{\rm bol}>L_{\rm det}$, then $L_{\rm bol}$ is assumed to be the true luminosity on that day. 

In contrast, when $L_{\rm bol}<L_{\rm det}$ the algorithm is equipped to deal with three separate cases: when a source, on a particular day, in a given energy band, is detected
(i) above where the source is normally detected during that time period (e.g., when the Sun is near the Galactic Center or when a nearby bright source goes into outburst),
(ii) below where the source is normally detected during that time period (e.g., a clear decrease in luminosity), or
(iii) where the source is normally detected during that time period within error (e.g., no discernible increase/decrease in luminosity).

\afterpage{
\renewcommand{\thefootnote}{\alph{footnote}}
\renewcommand\tabcolsep{5pt}
\tabletypesize{\footnotesize}
\begin{longtable}{llcr}
\caption[Calibration Source Details]{Calibration Source Details}  \\
\hline \hline \\[-2ex]
   \multicolumn{1}{l}{Source Name} &
   \multicolumn{1}{c}{Outburst} &
     \multicolumn{1}{c}{Calibration$^a$} &
          \multicolumn{1}{r}{References$^b$} \\
&\multicolumn{1}{c}{ID} & \multicolumn{1}{c}{Type}&\\[0.5ex] \hline
   \\[-4ex]
   \label{table:calibS}
\endfirsthead
\multicolumn{4}{l}{{\ldots Continued from Previous Page}}\\
\hline \hline \\[-2ex]
   \multicolumn{1}{l}{Source Name} &
   \multicolumn{1}{c}{Outburst} &
     \multicolumn{1}{c}{Calibration$^a$} &
          \multicolumn{1}{r}{References$^b$} \\
&\multicolumn{1}{c}{ID} & \multicolumn{1}{c}{Type}&\\[0.5ex] \hline
   \\[-4ex]
\endhead
  \\[-1.8ex] \hline  \\
  \multicolumn{4}{l}{{Continued on Next Page\ldots}} \\
\endfoot
  \\[-1.8ex] \hline \\[-1ex]
      \multicolumn{4}{p{0.44\textwidth}}{\hangindent1ex$^a$States whether the outburst was used to calibrate the hardness limits for a successful (S) or hard-only (H) outburst.}\\
      \multicolumn{4}{p{0.44\textwidth}}{\hangindent1ex$^b${\bf References.} --- [1]~\citet{z04}, [2]~\citet{be99}, [3]~\citet{homan5}, [4]~ \citet{b06}, [5]~\citet{bu12}, [6]~\citet{mo09}, [7]~\citet{de13}, [8]~\citet{cap5}, [9]~\citet{mi06}, [10]~\citet{kale06}, [11]~\citet{mc09}, [12]~\citet{cap06b}, [13]~\citet{ca10},  [14]~\citet{z13}, [15]~\citet{ca09}, [16]~\citet{mo10}, [17]~\citet{corr11}, [18]~\citet{de13a}, [19]~\citet{fe11}, [20]~\citet{r12}, [21]~\citet{r14}, [22]~\citet{hi00}, [23]~\citet{mc01}, [24]~\citet{br10}, [25]~\citet{front00}, [26]~\citet{rev00c}, [27]~\citet{br01}, [28]~\citet{br04}, [29]~\citet{rev00a},  [30]~\citet{ro07}, [31]~\citet{w07}, [32]~\citet{pai09}, [33]~\citet{s2000}, [34]~\citet{r02}, [35]~\citet{ku04}, [36]~\citet{tom01}, [37]~\citet{b02}, [38]~\citet{ss05}, [39]~\citet{ar04}, [40]~\citet{int02},  [41]~\citet{si11}, [42]~\citet{kr13}, [43]~\citet{cu14}} 
\endlastfoot
GX 339$-$4&1996--1999&S&1,2 \\
&2002/2003&S&3 \\
&2004/2005&S&4 \\
&2006&H&5 \\
&2006/2007&S&5,6 \\
&2009--2011&S&7 \\[0.15cm]
H 1743$-$322&2003&S&3,8--11 \\
&2004&S&12 \\
&2005&S&12 \\
&2007/2008&S&13,14 \\
&2008&H&15,16 \\
&2010&S&14,17,18 \\[0.15cm]
MAXI J1836$-$194&2011/2012&H&19--21 \\[0.15cm]
XTE J1118+480&1999/2000&H&22--26 \\[0.15cm]
GS 1354$-$64&1997/1998&H&27--29 \\[0.15cm]
IGR J17497$-$2821&2006&H&30--32 \\[0.15cm]
XTE J1550$-$564&1998/1999&S&33--35\\
&2001&H&36\\
&2001/2002&H&37\\
&2003&H&38,39\\[0.15cm]
SAX J1711.6$-$3808&2001&H&40 \\[0.15cm]
IGR J17285$-$2922&2010&H&41 \\[0.15cm]
Swift J174510.8$-$262411&2012/2013&H&42,43
\end{longtable}
\renewcommand{\thefootnote}{\arabic{footnote}}
\renewcommand\tabcolsep{5pt}
}

To quantify the limits on $L_{\rm bol}$ when a source is below $L_{\rm det}$, we begin by estimating a local sensitivity limit $f_{\rm sl}$ in each available energy band for the day $t$ in question.
This is done by finding the weighted mean (and corresponding error) of the local data points corresponding to the closest two days bracketing $t$ on either side, in each available instrument/energy band (in crab units).
From here we then compare the flux (in crabs) of the source on the day in question, $f_{\rm t}$, to this sensitivity limit in that band.
If at least one instrument/energy band has a lower limit of $f_{\rm t}$ greater then the upper limit of $f_{\rm sl}$ (i.e., case i), then we use the linear interpolation method, described above for transients, to estimate $L_{\rm bol}$ of the source on this day.
If none of the data satisfies the above condition, then the algorithm checks if any of the available data show $f_{\rm t}$ consistent (within error) of $f_{\rm sl}$ (i.e., case iii).
If at least one instrument/energy band satisfies this condition then we assume $f_{\rm t}=f_{\rm sl}$ on that day for each available energy band, and then use these new estimates for flux to fit for $L_{\rm bol}$ using the spectral fitting algorithm discussed in Section~\ref{subsubsec:spectra}.
Lastly, if all available instrument/energy bands have an upper limit of $f_{\rm t}$ less then the lower limit of $f_{\rm sl}$ (i.e., case ii), then we assume the flux on that day is a uniform distribution between zero and the lower limit of $f_{\rm sl}$ in each available energy band, and then fit for $L_{\rm bol}$ using the spectral fitting algorithm discussed in Section~\ref{subsubsec:spectra}.
We stress that, if all requirements for spectral fitting are not met on any particular day (e.g., we do not have at least one energy band which can act as a soft band and one energy band which can act as a hard band, see Section \ref{subsubsec:spectra} for detail) or no data exists on this day, then the algorithm will default to using the linear interpolation method, described above for transients, to estimate $L_{\rm bol}$ of the source using the closest available luminosity estimates bracketing the day in question.

Because we are dealing with non-detection data, there is a possibility of large errors in flux.
To deal with this possibility we use the same method used in our X-ray hardness computation algorithm, whereby we test whether a band limited flux has an error greater then $2\sigma$ of the mean error value for the given band (see Section \ref{subsubsec:HR}).
If this condition is satisfied in a given band, we ignore the data and instead interpolate the flux on this day using the algorithm described above for the transient sources and the local flux data points corresponding to the closest two days bracketing the day in question on either side.
Using this interpolated flux we then follow the the same three step procedure described previously to estimate $L_{\rm bol}$.

Following the estimation of $L_{\rm bol}$ on every day over the past 19 year period, a time-averaged bolometric luminosity is calculated for each persistent source using Equation \ref{eq:timeavg}. Lastly, using this calculated $L_{\rm bol,avg}$ for all transient and persistent sources, the algorithm estimates a long-term averaged mass-transfer rate for each individual source via,

\begin{equation}
\Dot{M}_{\rm avg}=\frac{L_{\rm bol,avg}}{c^2 \eta},
\end{equation}
where $\eta$ denotes accretion efficiency. 

To provide an accurate estimate of $\Dot{M}_{\rm avg}$, the algorithm uses Monte-Carlo simulations to take into account the uncertainties which come into the calculation in the form of errors in distance ($d$), and spectral modelling (fit parameters $a$ and $b$).
Accretion efficiency ($\eta$) is fixed at 0.10, corresponding to radiatively efficient flow through a thin accretion disk \citep{frank2}
Once again, the algorithm takes the median of the resulting distribution of $\Dot{M}_{\rm avg}$ to be the accepted value and the $1\sigma$ confidence interval as the upper and lower limits on this value. 

We note that if the true accretion efficiency differs from the chosen value of 0.1 at any point during outburst (e.g., during the lower-luminosity hard state, which is thought to be dominated by radiatively inefficient flows; \citealt{ny94,meyer4,knev14}), then the resulting $\Dot{M}_{\rm avg}$ will be affected to a certain degree.
See Section ~\ref{subsec:mdot_porb} for a thorough discussion of the factors that have the potential to affect the long term $\Dot{M}$ balance of BH systems.

\subsubsection{Empirical Classification}  
\label{subsubsec:classify}
In the final stage, the algorithm makes use of the empirical hardness ratio ($H_{X}$) parameter to categorize outburst behavior into one of three classes: ``successful'', ``indeterminate'', and ``hard-only''. 
 
The classification procedure begins by differentiating data for each outburst into hard, soft, and intermediate states based on critical hard $C_{\rm hard}$ and critical soft $C_{\rm soft}$ hardness values.
As these critical values will differ depending on the telescopes involved in $H_{X}$, the algorithm makes use of 10 calibration sources (found in Table \ref{table:calibS}) to set these baseline critical values.
These calibration sources have been specifically chosen based on the criteria that they have exhibited (proven via spectral and/or timing analysis) either ``hard-only''  outbursts, or a combination of ``successful'' and ``hard-only'' outbursts over the last 19 years.
The literature classification is then used to find the baseline critical values for each of the nine $H_{X}$ combinations. See Table \ref{table:HRtable} for the critical values corresponding to each $H_{X}$ combination.

The criteria for a source to be in the soft state requires at least one upper error bar on $H_X$, $\sigma_{\rm H,high}<C_{\rm soft}$.
In turn for a source to be in the hard state, all lower error bars on $H_X$, $\sigma_{\rm H,low}>C_{\rm hard}$.
If the observation does not fall into either category, then the source is classified to be in an intermediate state. 
 
 $C_{\rm hard}$ was found by taking the minimum $\sigma_{\rm H,low}$ for each ``hard-only'' calibration outburst, followed by finding the absolute minimum of these values across all calibration sources, yielding the softest a source can be while still remaining in the hard state.
 $C_{\rm soft}$ is found by taking the minimum $\sigma_{\rm H,high}$ for each ``successful'' calibration outburst (thus fulfilling the minimum requirement for a source to reach the soft state that at least one $\sigma_{\rm H,high}<C_{\rm soft}$), followed by finding the absolute maximum of these values across all calibration sources, yielding the hardest a source can be while still being in the soft state. 
 
There are multiple telescope pairs involved in this process (between 1--9 separate pairs).
As such, each pair of observations (two bands involved in $H_X$) will indicate whether the source is currently hard, soft, or intermediate. The algorithm must be able to take into account all possible combinations of hard, soft, and intermediate classifications and logically combine them to classify the outburst as a whole.
If a classification flag exists on the outburst the algorithm will perform the following procedure.
If all observations during the outburst classify the state of the source as hard, then the outburst is classified as ``hard-only''. If any observations during the outburst classify the state is soft, then the outburst is classified as ``successful''.
If neither of these conditions are met and/or if no classification flag exists, the outburst is classified ``indeterminate'', which indicates that the outburst was detected by the algorithm but that there was not enough data available to confidently determine whether or not the source reached the soft state during the outburst.

We recommend caution when interpreting both outburst classification and accretion state when only RXTE/ASM hardness ratios (i.e., 5--12 keV/3--5 keV; ``RR'') are available. We have found that when using the ratio of 3--5 keV and 5--12 keV RXTE/ASM fluxes, if the disk is hotter than we would typically assume for a source in the soft state, then the hardness ratio will mimic a ratio typical of the HCS, leading to mis-classification of the accretion state of the source on a given day and in-turn the possibility for the outburst as a whole being classified wrong by our algorithm. We have also found that if an unusually power-law dominant SPL state occurs, then the algorithm will incorrectly label this state as the HCS as we do not have the ability to differentiate the SPL state from the HCS state due to the limited available spectral information.
If this situation occurs, it is possible for our ``successful''/``hard-only'' dichotomy to be confused.
See Section ~\ref{subsec:failed} for specific examples of sources that display this behavior and further discussion on the impact these issues have on our ``successful'' and ``hard-only'' outburst detection rates.

\afterpage{
\renewcommand{\thefootnote}{\alph{footnote}}
\renewcommand\tabcolsep{5.0pt}
\tabletypesize{\small}
\begin{longtable*}{llccccr}

\caption[Transient Outburst Rate per Instrument]{Transient Outburst Rate per Instrument}  \\

\hline \hline \\[-2ex]
   \multicolumn{1}{c}{Telescope} &
   \multicolumn{1}{c}{Instrument} &
     \multicolumn{1}{c}{Type} &
          \multicolumn{1}{c}{$t_{\rm active}$$^a$} &
                    \multicolumn{1}{c}{$f_{\rm collect}$$^b$} &
                              \multicolumn{1}{c}{Outbursts} &
          \multicolumn{1}{c}{Rate$^c$} \\
&&& \multicolumn{1}{c}{(yrs)}&& \multicolumn{1}{c}{Detected}&\multicolumn{1}{c}{($\rm{yr}^{-1}$)}\\[0.5ex] \hline
  \\[-4ex]
   \label{table:outrate}
\endfirsthead

\multicolumn{7}{c}{{\tablename} \thetable{} -- Continued} \\[0.5ex]
\hline \hline \\[-2ex]
   \multicolumn{1}{c}{Telescope} &
   \multicolumn{1}{c}{Instrument} &
     \multicolumn{1}{c}{Type} &
          \multicolumn{1}{c}{$t_{\rm active}$$^a$} &
                    \multicolumn{1}{c}{$f_{\rm collect}$$^b$} &
                              \multicolumn{1}{c}{Outbursts} &
          \multicolumn{1}{c}{Rate$^c$} \\
&&& \multicolumn{1}{c}{(yrs)}&& \multicolumn{1}{c}{Detected}&\multicolumn{1}{c}{($\rm{yr}^{-1}$)}\\[0.5ex] \hline
   \\[-1.8ex]
\endhead

  \\[-1.8ex] \hline  \\
  \multicolumn{7}{l}{{Continued on Next Page\ldots}} \\
\endfoot

  \\[-1.8ex] \hline \\[-1ex]
%  \multicolumn{7}{p{0.55\columnwidth}}{{Note. --}} \\
  \multicolumn{7}{p{0.55\columnwidth}}{\hangindent=1ex$^a$Amount of time the instrument has been active.} \\        
  \multicolumn{7}{p{0.55\columnwidth}}{\hangindent=1ex$^b$fraction of time the instrument was taking data over the total time active.} \\     
  \multicolumn{7}{p{0.55\columnwidth}}{\hangindent=1ex$^c$detection rate of the instrument between 1996 January 6 -- 2015 May 14 (50088--57156), quoted with 1$\sigma$ Gehrels errors.} \\  
  \multicolumn{7}{p{0.55\columnwidth}}{\hangindent=1ex$^d$As no all-sky monitoring survey was done by HEXTE, no outburst rate is calculated.}\\        
  \multicolumn{7}{p{0.55\columnwidth}}{\hangindent=1ex$^e$JEM-X only covers a fraction of the sky that ISGRI does.} \\   

\endlastfoot
INTEGRAL&JEM-X$^e$&scan&10.23&0.19&16&$8.16^{+2.59}_{-2.02}$\\ [0.1cm]
&ISGRI&scan&10.23&0.25&24&$10.12^{+2.44}_{-2.00}$\\[0.4cm]
MAXI&GSC&all-sky&5.74&1.0&28&$4.87^{+1.11}_{-0.92}$\\[0.4cm]
RXTE&ASM&all-sky&15.98&0.92&76&$5.17^{+0.66}_{-0.59}$\\[0.1cm]
&HEXTE$^d$&pointed&-&-&38&-\\ [0.1cm]
&PCA&scan/pointed&12.74&0.59&105&$14.06^{+1.51}_{-1.37}$\\ [0.4cm]
Swift&BAT&all-sky&10.25&0.92&62&$6.67^{+0.95}_{-0.84}$
\end{longtable*}
     \renewcommand{\thefootnote}{\arabic{footnote}}
}

\section{Results}
\label{sec:results}
\subsection{Outburst History}
Combining the outburst detector, tracker, and empirical classification tools of the algorithm with an exhaustive literature review we have compiled a complete outburst  history for the Galactic (transient and persistent) BHXB population encompassing over 50 years of activity and including over 200 outbursts in 66 transient sources and the long-term activity of 11 persistent sources.

Table \ref{table:outhistBH} presents, on a source by source basis,
(i) a complete outburst list,
(ii) the beginning and end times of outbursts (in MJD) detected by the algorithm, or the outburst year if record of the outburst is only found in the literature, and a unique outburst ID,
(iii) literature and (where applicable) algorithm classification,
(iv) a list of instruments that have detected each outburst, and
(v) an extensive list of references.
Any discrepancies existing between the literature classification and algorithm classification and/or instances in which the algorithm did not detect an outburst found in the literature, are discussed on a source by source basis in the footnotes of Table \ref{table:outhistBH}.

\subsection{Outburst Detection Rates}
\label{subsec:rates}
Calculating the overall (and instrument specific) outburst rate for transient BHXB events in the Galaxy is a non-trivial task as it depends on the sky coverage, lifetime, and limiting sensitivity of instruments launched at different times \citep{chen97}.
To address these issues we begin by estimating outburst rates per year, with quoted 1$\sigma$ Gehrels errors \citep{geh86}, for each individual instrument (see Table \ref{table:outrate}).

We attempt to quantify sky coverage and instrument lifetime by using only the time in which data was being taken by the instrument  ($f_{\rm collect}*t_{\rm active}$), rather than the total time the instrument was active ($t_{\rm active}$) in the calculation of this rate.

Due to Sun constraints, most X-ray instruments do not point near the Galactic Bulge region (where most of our sources are located) around December every year. To take into account these Sun constraints in the calculation of $f_{\rm collect}$ for Swift/BAT and RXTE/ASM, we assume that for one month per year of operation the instrument in question was not actively observing, yielding $f_{\rm collect}=0.92$ in both cases. Given that MAXI has no Sun constraint, we assume that the GSC aboard MAXI has taken data daily for the full time period it has been active (i.e., $f_{\rm collect}=1.0$).

The instruments involved in the scanning surveys (as well as those that have pointed observations available) only take data in short consecutive intervals resulting in $f_{\rm collect}\ll1.0$. To calculate $f_{\rm collect}$ for PCA, ISGRI, and \mbox{JEM-X} we begin by parsing through all available data on an individual source (including the scanning survey and/or pointed observations), checking if data is available from that instrument on a 8 day time scale. If the instrument observed the source at least once in any particular 8-day time period, that full time period is counted toward the total time the instrument was observing. 
If the algorithm finds a time gap greater than 8 days in which the instrument has no observations of the source, the total duration of this gap is calculated and subsequently not counted toward the total time the instrument was observing.
Once the total observation time is calculated for each transient source, an average over all transient sources is calculated. This average is then used to calculate $f_{\rm collect}$ for the instrument in question.

\begin{figure}%
\plotone{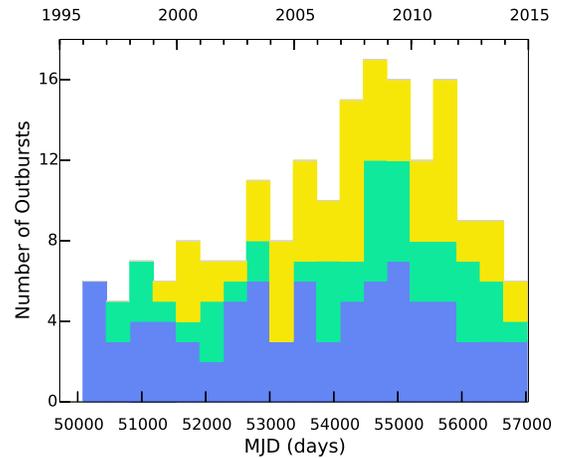}
    \caption{Histogram showing transient BHXB outbursts detected by the algorithm between 1996 January 1 -- 2015 May 14. Time bins are one year in length and colors correspond to outburst classification; blue for successful, green for hard-only, and yellow for indeterminate.}%
    \label{fig:outhist1}%
\end{figure}

We do not attempt to take into account the differing limiting sensitivities between the instruments used in this study (e.g., PCA is the most sensitive soft instrument and INTEGRAL is the most sensitive hard instrument).

Overall, our algorithm has detected 132 outbursts occurring in 47 transient BHXBs between 1996 January 6 -- 2015 May

\afterpage{
\renewcommand\tabcolsep{5pt}
\renewcommand{\thefootnote}{\alph{footnote}}
\tabletypesize{\footnotesize}
\begin{longtable*}[t]{lcccccl}
\caption[Hard-only Outburst Ratios as Revealed by X-ray Hardness]{Hard-only Outburst Ratios as Revealed by X-ray Hardness}  \\
\hline \hline \\[-2ex]
   \multicolumn{1}{c}{Telescope$^a$} &
   \multicolumn{1}{c}{Hard Band} &
     \multicolumn{1}{c}{Soft Band} &
          \multicolumn{1}{c}{Successful} &
                    \multicolumn{1}{c}{Hard-only} &
                    \multicolumn{1}{c}{Indeterminate} &
          \multicolumn{1}{c}{Hard-only$^b$} \\
\multicolumn{1}{c}{ID} &\multicolumn{1}{c}{(keV)} & \multicolumn{1}{c}{(keV)}& \multicolumn{1}{c}{Detected}& \multicolumn{1}{c}{Detected}& \multicolumn{1}{c}{Detected}&\multicolumn{1}{c}{Ratio}\\[0.5ex] \hline
   \\[-1.8ex]
\endfirsthead

\multicolumn{7}{c}{{\tablename} \thetable{} -- Continued} \\[0.5ex]
\hline \hline \\[-2ex]
   \multicolumn{1}{c}{Telescope$^a$} &
   \multicolumn{1}{c}{Hard Band} &
     \multicolumn{1}{c}{Soft Band} &
          \multicolumn{1}{c}{Successful} &
                    \multicolumn{1}{c}{Hard-only} &
                                        \multicolumn{1}{c}{Indeterminate} &
          \multicolumn{1}{c}{Hard-only$^b$} \\
\multicolumn{1}{c}{ID} &\multicolumn{1}{c}{(keV)} & \multicolumn{1}{c}{(keV)}& \multicolumn{1}{c}{Detected}& \multicolumn{1}{c}{Detected}& \multicolumn{1}{c}{Detected}&\multicolumn{1}{c}{Ratio}\\[0.5ex] \hline
   \\[-1.8ex]
\endhead

  \\[-1.8ex] \hline  \\
  \multicolumn{7}{l}{{Continued on Next Page\ldots}} \\
\endfoot

  \\[-1.8ex] \hline \\ 
       \multicolumn{7}{p{0.60\columnwidth}}{\hangindent=1ex$^a$%
HR:~RXTE/HEXTE \& RXTE/ASM,
HRpp:~RXTE/HEXTE \& RXTE/PCA,\par\hspace{1ex}\hangindent=1ex
II:~INTEGRAL/ISGRI \& INTEGRAL/\mbox{JEM-X},
RR:~RXTE/ASM \& RXTE/ASM,\par\hspace{1ex}\hangindent=1ex
RRpp:~RXTE/PCA \& RXTE/PCA,
SI:~Swift \& INTEGRAL/\mbox{JEM-X},
SM:~Swift \& MAXI,
SR:~Swift \& RXTE/ASM,
SRp:~Swift \& RXTE/PCA.} \\
       \multicolumn{7}{p{0.60\columnwidth}}{\hangindent=1ex$^b$ratio of hard-only to successful plus hard-only, with 1$\sigma$ binomial Gehrels errors, valid between 1996 January 6 -- 2015 May 14 (50088--57156).}\\[-1ex]
\endlastfoot
HR&15--30&\phantom{0.}3--12&28&\phn5&14&$0.15^{+0.090}_{-0.064}$\\ [0.1cm] 
HRpp&15--30&\phantom{0.}4--9\phn&21&\phn6&20&$0.22^{+0.11}_{-0.084}$\\ [0.1cm]
II&18--40&\phantom{0.}3--10&\phn3&\phn6&\phn3&$0.67^{+0.18}_{-0.22}$\\ [0.1cm] 
RR&\phn5--12&\phantom{0.}3--5\phn&27&14&19&$0.34^{+0.090}_{-0.081}$\\  [0.1cm]
RRpp&\phn9--20&\phantom{0.}4--9\phn&31&\phn7&25&$0.18^{+0.085}_{-0.065}$\\ [0.1cm]
SI&15--50&\phantom{0.}3--10&\phn2&\phn5&\phn5&$0.71^{+0.18}_{-0.26}$\\[0.1cm]
SM&15--50&\phantom{0.}4--10&10&\phn9&\phn7&$0.47^{+0.14}_{-0.13}$\\[0.1cm]
SR&15--50&\phantom{0.}3--12&18&\phn5&\phn7&$0.22^{+0.12}_{-0.090}$\\[0.1cm]
SRp&15--50&2.5--10&14&14&29&$0.50^{+0.11}_{-0.11}$\\ [0.1cm]
\hline \\[-1.8ex]
All&-&-&52&32&47&$0.38^{+0.060}_{-0.056}$
\label{table:failedrate}
\end{longtable*}
\renewcommand{\thefootnote}{\arabic{footnote}}
\renewcommand\tabcolsep{5pt}
}

\noindent 14 (see Figure \ref{fig:outhist1}).
Taking the above mentioned factors into consideration, we estimate that with the current suite of instruments in space we are detecting $\sim$4--12 BHXB transient events every year, more than a factor of three larger than in the pre-RXTE era \citep{chen97}.

Note that on 2015 May 14 (the cutoff date for our analysis in this paper), 4U 1630$-$472 and GX339$-$4 were both in the decay stage of their most recent outbursts (and had already reached the soft state during their outburst).
As such, these outbursts are included in all of our analyses with the exception of the outburst duration statistics.
In addition, during the months of June and July (2015), H 1743$-$322, GS 1354$-$64, GS 2023+338 (V404 Cyg), and SAX J1819.3$-$2525 (V4641 Sgr) were all observed in outburst again.
While these outbursts occurred after the cut-off date, and therefore are not included in our analysis, we still make an effort to provide an up-to-date list of references for them. See the relevant subsections of Section ~\ref{sec:sample} and Tables \ref{table:primaryBH} and \ref{table:outhistBH}.

\subsection{``Hard-only'' Outburst Behaviour}
\label{subsec:failed}
Using the empirical classification tools of the algorithm, we have been able to classify the behavior exhibited during (i) 92 of the 132 total transient outbursts detected and (ii) the bright periods of 10 of the 11 persistently accreting BH sources, over the last 19 years.

In contrast to the picture found in much of the large-scale population studies in the literature on state-changing versus ``hard-only'' BHXB behavior (e.g., \citealt{zhang07,du09}), we find that the outbursts undergone by BHXBs that do not complete the ``turtlehead'' pattern (i.e.,undergo a state change), failing to transition from the HCS to the SDS, the so-called ``hard-only'' outbursts, make up $\sim$ 40\% (i.e., $0.38^{+0.060}_{-0.056}$) of all outbursts occurring in Galactic transient BHXBs in the past 19 years.
Table \ref{table:failedrate} presents the ``hard-only'' ratio, quoted with 1$\sigma$ binomial Gehrels errors \citep{geh86}, computed for each of the nine separate hardness ratios used in the algorithm, while Table~\ref{table:failedovertime} presents the ``hard-only'' ratio as it changes over time.
With these numbers, it is clear that the ``hard-only'' outbursts represent a (surprisingly) substantial contribution to the total outburst sample.

By splitting the 19 year period we studied into logical segments defined by the addition/loss of each instrument, we find that after Swift and INTEGRAL were turned on the rate of observed outbursts increased by $\sim1.5$ and the ``hard-only'' ratio increased by 
$\sim15\%$ (see Table \ref{table:failedovertime}). While this suggests that the ``hard-only'' outburst ratio may have increased when the threshold for detection was lowered (i.e., since ``hard-only'' outbursts are faint, higher sensitivity instruments would detect more ``hard-only'' outbursts), the difference in ``hard-only'' ratios is only significant at the $1.7\sigma$ level. Moreover, through an extensive literature search (as presented in Table \ref{table:outhistBH}), we find a near constant appearance of these ``hard-only'' outbursts over the last $\sim50$ years. Our findings suggest the ``hard-only'' behavior is neither a rare nor recent phenomenon.

We postulate that the steady appearance of these ``hard-only'' outbursts over the last $\sim50$ years is indicative of an underlying physical process that is relatively common.
Such a physical process would likely involve the mass-transfer rate onto the BH remaining at a low level rather than increasing as the outburst evolves, resulting in no state transition to the softer states occuring.
To test this theory we must be able to determine whether or not a ``hard-only'' outburst reaches the luminosity regime where the transition to the soft state tends to happen (i.e., are all ``hard-only'' outbursts faint, as the detection ratios above tentatively suggest).
We have therefore compared the peak Eddington scaled luminosities of all algorithm classified ``hard-only'' outbursts for which our spectral fitting algorithm succeeded (see Section~\ref{subsec:peaklum} and Table \ref{table:lumdata}), to our estimated mean HCS-SDS transition luminosity ($\sim0.11 \, L_{\rm edd}$; see Section~\ref{subsec:transition}).
We find that all of these outbursts (with the exception of GS 1354$-$64) either have
 (i) upper limits on their Eddington scaled peak outburst luminosities that are $<0.11 \, L_{\rm edd}$, or
 (ii) have Eddington scaled peak luminosities consistent within error of the $<0.11 \, L_{\rm edd}$ regime.
Overall, this suggests that ``hard-only'' behavior may indicate that a source did not reach the required mass transfer rate needed to transition to the soft state.
In the case of GS 1354$-$64, the distance is poorly constrained (25--61 kpc).
If we were to place this system at our assumed standard Galactic value (i.e., a uniform distribution between 2 and 8 kpc), its Eddington scaled luminosity would be consistent within error of the $<0.11 \, L_{\rm edd}$ regime.

\afterpage{\renewcommand{\thefootnote}{\alph{footnote}}
\renewcommand\tabcolsep{5pt}
\tabletypesize{\footnotesize}
\begin{longtable*}[t]{lcccccc}
\caption[Detected Hard-only Outburst Ratio over Time]{Detected Hard-only Outburst Ratio over Time} \\
\hline \hline \\[-2ex]
   \multicolumn{1}{c}{Time Segment} &
   \multicolumn{1}{c}{Time Period} &
     \multicolumn{1}{c}{$N_S$} &
          \multicolumn{1}{c}{$N_H$} &
                    \multicolumn{1}{c}{$N_I$} &
                    \multicolumn{1}{c}{Hard-only} &
                    \multicolumn{1}{c}{Outburst Rate$^a$} \\
&\multicolumn{1}{c}{(MJD)} & & && \multicolumn{1}{c}{Ratio} & \multicolumn{1}{c}{(yr$^{-1}$)}\\[0.5ex] \hline
   \\[-4ex]
   \label{table:failedovertime}
\endfirsthead
  \\[-1.8ex] \hline \\   
      \multicolumn{7}{p{0.6\columnwidth}}{Note. -- $N_S$, $N_H$,  and $N_I$ represent the number of successful, hard-only, and indeterminate outbursts detected during the corresponding time period. The hard-only ratio is defined as the ratio of hard-only to the sum of successful and hard-only outbursts for the given time period and is quoted with 1$\sigma$ binomial Gehrels errors. Present = 2015 May 14 (57156).}\\
            \multicolumn{7}{p{0.6\columnwidth}}{\hangindent=1ex$^a$No corrections have been applied to this outburst rate.}\\        
\endlastfoot
ASM \underline{ON}--PCA \underline{ON}&50088--51214&10&4&1&$0.286^{+0.169}_{-0.131}$&$\phn4.86^{+1.61}_{-1.24}$\\[0.1cm]
PCA \underline{ON}--BAT/INTEGRAL \underline{ON}&51214--53414&14&6&13&$0.300^{+0.136}_{-0.112}$&$\phn5.48^{+1.13}_{-0.95}$\\[0.1cm]
BAT/INTEGRAL \underline{ON}--GSC \underline{ON}&53414--55058&15&12&19&$0.444^{+0.114}_{-0.109}$&$10.21^{+1.74}_{-1.50}$\\ [0.1cm]
GSC \underline{ON}--ASM \underline{OFF}&55058--55924&\phn8&\phn6&\phn8&$0.429^{+0.167}_{-0.155}$&$\phn9.28^{+2.43}_{-1.96}$\\[0.1cm]
ASM \underline{OFF}--Present& 55924--Present&\phn5&\phn4&\phn6&$0.444^{+0.213}_{-0.198}$&$\phn4.44^{+1.47}_{-1.13}$\\[0.1cm] \hline\\[-1.8ex]
Before BAT/INTEGRAL \underline{ON} & 50088--53414&24&10&13&$0.294^{+0.098}_{-0.084}$&$\phn5.16^{+0.87}_{-0.75}$\\[0.1cm]
After BAT/INTEGRAL \underline{ON} & 53414--Present&28&22&34&$0.440^{+0.081}_{-0.078}$&$\phn8.19^{+0.99}_{-0.89}$\\[0.1cm] \hline\\[-1.8ex]
Total& 50088--Present&52&32&47&$0.380^{+0.060}_{-0.056}$&$\phn6.76^{+0.64}_{-0.59}$
\end{longtable*}
     \renewcommand{\thefootnote}{\arabic{footnote}}
\renewcommand\tabcolsep{5pt}
}

This being said, additional factors need to be addressed before such a strong claim against selection biases are made. These factors include (i) the effect that distance could have on the outburst behaviors that we are able to observe, as an increase in sensitivity could largely increase the distance range within which we could observe the same outburst behaviors,
(ii) the significance of individual instrument performance on outburst detection rates over time (e.g., RXTE/ASM detected significantly fewer outbursts towards the end of its life in 2011--2012 in comparison to during its earlier operation; \citealt{yanyu14}), and
(iii) the change in sensitivity of each instrument between soft and hard X-rays (i.e., while RXTE ASM and PCA included high sensitivity to soft state X-rays, INTEGRAL/ISGRI and Swift/BAT are only sensitive to hard X-rays, INTEGRAL/\mbox{JEM-X} has a relatively small field-of-view, and the only band in MAXI we found useful is not sensitive below 4 keV).

Lastly, we find, through our ability to track the accretion state of a source throughout an outburst via the algorithm (see Section~\ref{subsec:transition} and Tables \ref{table:statetransitions} and \ref{table:statetransitions2}), that this particular class of behavior is not limited to the transient systems, but is also exhibited by a number of persistently accreting systems. Rather than a ``hard-only outburst'', this behavior manifests in the form of long continuous periods spent in the HCS (in the case of Cyg X-1, 1E 1740.7$-$2942, and SS 433) or periodic ``incomplete'' state transitions (in the case of GRS 1758$-$258 and Swift J1753.5$-$0127). See Sections~\ref{subsec:peaklum} and \ref{subsec:xlfs} for further discussion on these behaviors.

\begin{figure*}%
\plotone{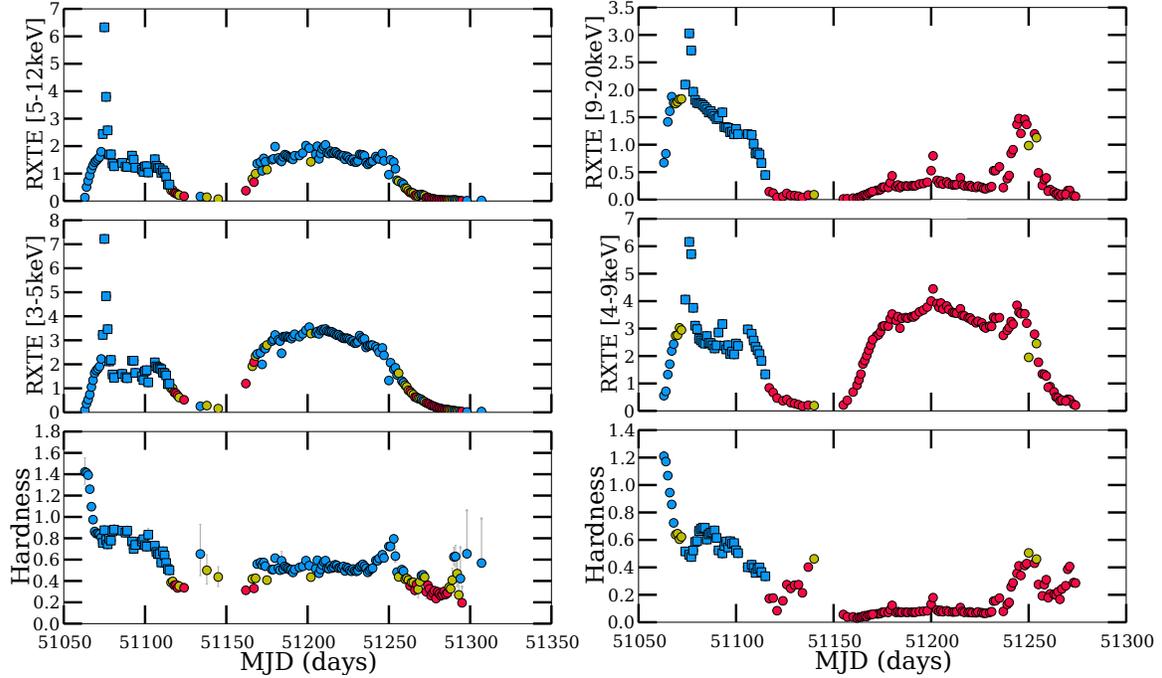}
 \caption{Analysis of the 1998/1999 outburst of XTE J1550$-$564 with RXTE/ASM (left) and RXTE/PCA (right) displaying the limitations of using RXTE/ASM data alone in discriminating between hard and soft states. The colors represent the accretion state: HCS (blue), SDS (red), and IMS (yellow). Square shapes depict days that spectral/timing analysis in the literature has classified as a SPL state. According to spectral analysis presented in \citet{s2000}, this outburst can be divided into two halves. During the first half the source was in the hard state (51063--51072), followed by a SPL state (51074--51115), in agreement with both the ASM and PCA hardness ratio presented here. During the second half of the outburst, the source began transitioning (51115), reached the soft state (51150-51249), and then made the reverse transition back to the hard state (51249 onward). While the PCA hardness ratios are in agreement with the spectral evolution of the second half, the ASM hardness ratios tell a different story. Between 51200 and 51250 the ASM hardness ratio indicates a hard state even though the spectrum during this period is clearly thermally dominated.}%
     \label{fig:1RXTE}%
\end{figure*}

As discussed in Section~\ref{subsubsec:classify}, there is a possibility that some ``successful'' outbursts may be mis-classified as ``hard-only'' when using RXTE/ASM data alone if (i) the disk is hotter then we typically expect in the soft state, or (ii) the source enters into an unusual power-law dominated SPL state, which may be mislabelled as a HCS.
To investigate whether case (i) (see Figure \ref{fig:1RXTE} for example of this behavior) has caused any mis-classifications we analyzed all 92 algorithm classified outbursts, looking for those that had only RXTE/ASM available, and then used spectral studies in the literature to confirm or refute that the algorithm-classified accretion states reached during the outburst in question.
In doing so, we found that this was a serious issue affecting the group of outbursts detected prior to the beginning of the Swift, INTEGRAL, and MAXI missions. This group amounted to over a third of the algorithm classified outbursts.
To circumvent the issue we adjusted our classification scheme to include all available pointed observations from PCA and HEXTE (see Section~\ref{subsec:rxte} for details).
With the addition of the new data, only 13 of the algorithm-classified outbursts had only RXTE/ASM coverage.
Among these 13 outbursts, we found no discrepancies between literature and algorithm classifications.
Note that, while this behavior persists over the full 19 year period of our analysis, the availability of data from multiple telescopes after 2005 has allowed for the possibility of outbursts being classified with multiple instruments/energy bands.
As such, the ASM-analysis alone in these cases has had no effect on the final outburst classification.

\begin{figure*}%
\plotone{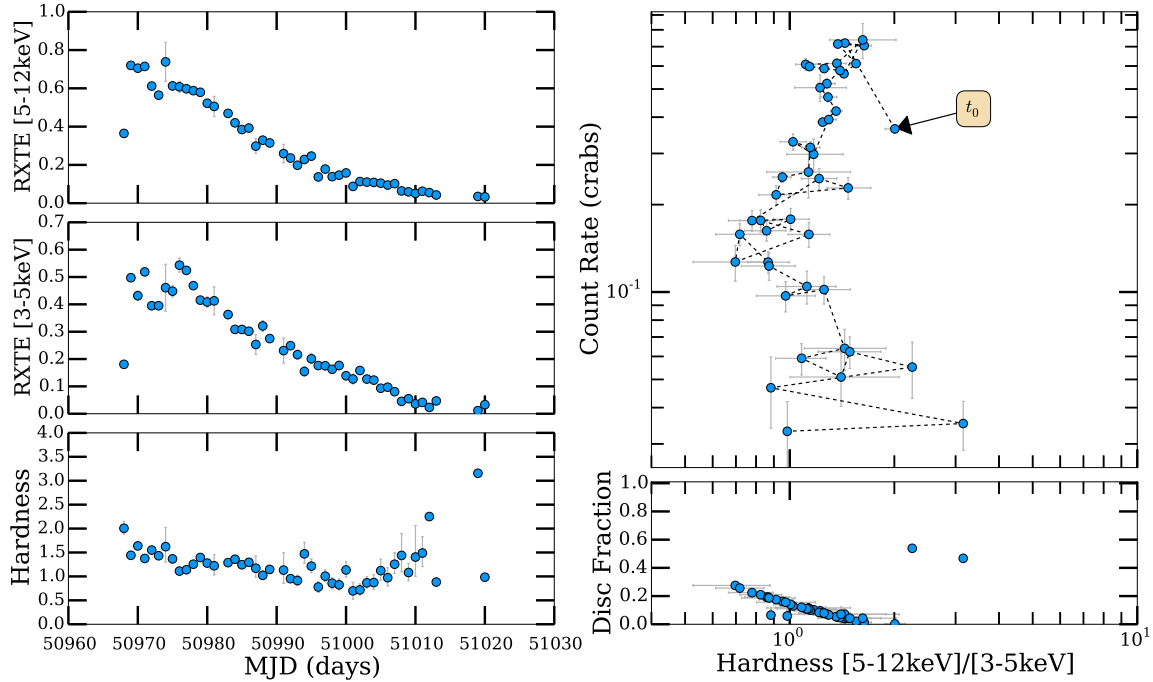}
 \caption{Analysis of the 1998 outburst of XTE J1748$-$288 (top). Colors represent accretion state: HCS (blue), SDS (red) and IMS (yellow). Square points are days in which spectral/timing studies from the literature have indicated a SPL state \citep{rev00}. Here, an SPL state with an unusually dominant power-law component causes a ``successful'' outburst to be classified as ``hard-only'' (compare to Figure~\ref{fig:1bSPL}.)}%
    \label{fig:1SPL}%
    \label{fig:1aSPL}%
\end{figure*}

\begin{figure*}%
\plotone{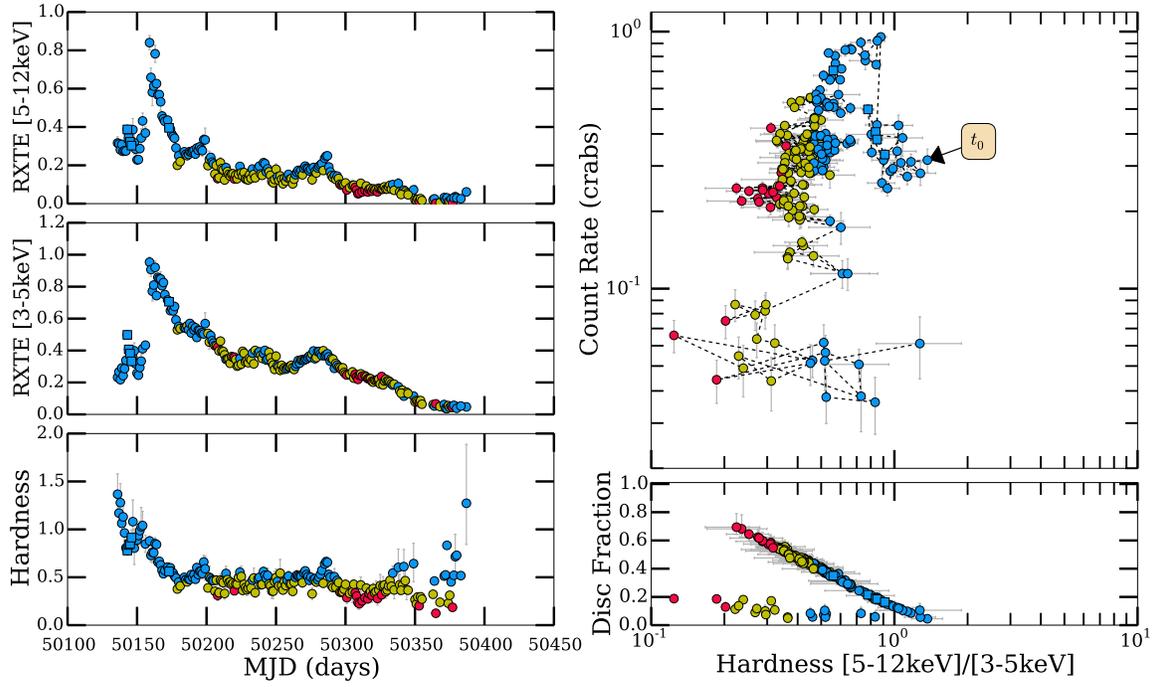}
 \caption{Analysis of the 1996 outburst of GRS 1739$-$278 (bottom) with RXTE/ASM. Colors represent accretion state: HCS (blue), SDS (red) and IMS (yellow). Square points are days in which spectral/timing studies from the literature have indicated a SPL state \citep{bo98}. Here, an SPL state with an unusually dominant power-law component has no effect on the outburst classification (compare to Figure~\ref{fig:1aSPL}).}%
    \label{fig:1bSPL}%
\end{figure*}

To investigate whether case (ii) has caused any outburst mis-classifications, we have compared our algorithm and literature classifications for all 92 outbursts in our sample and cross referenced any differences with the proven appearance of a SPL state via spectral/timing studies in the literature.
In doing so we find one instance that meets this criteria, the 1998 outburst of XTE J1748$-$288 (see Figures~\ref{fig:1aSPL} and \ref{fig:1bSPL}).
During this outburst \citet{rev00} were able to distinguish between the SPL, SDS, and HCS using a combination of spectral and timing analysis.
In doing so they found an abnormally bright hard component (power-law contributed $\gtrsim80$\% of the 3--25 keV flux) during the observations in the SPL state.

\begin{figure*}%
\plotone{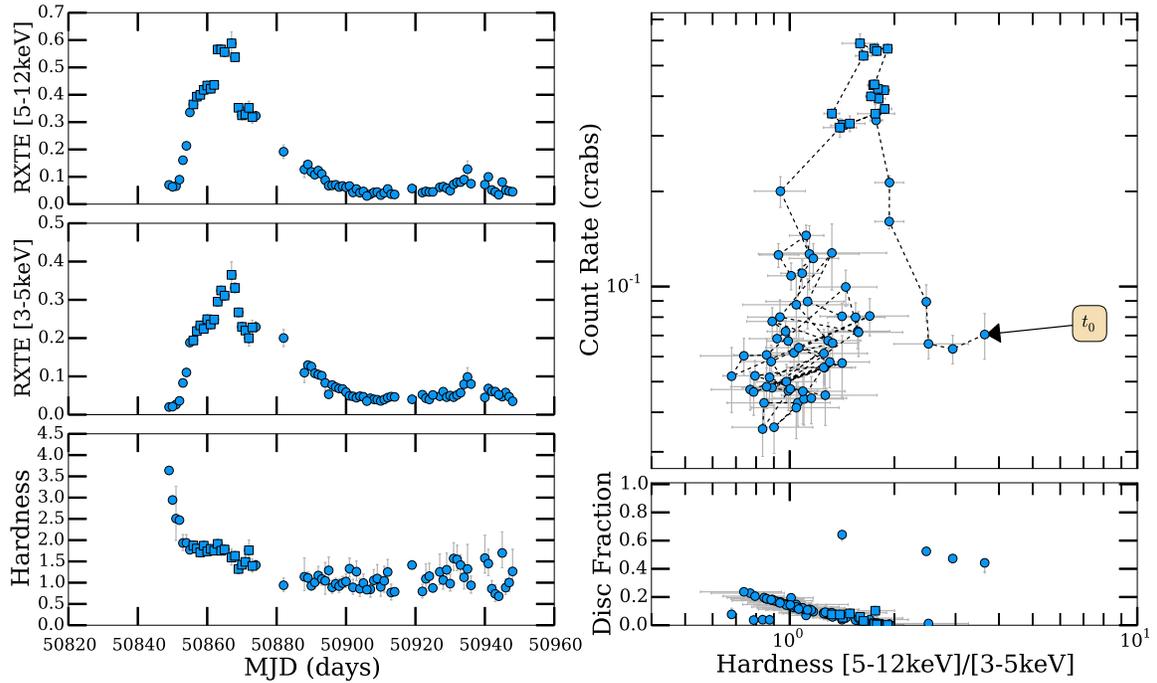}
\caption{Analysis of the 1998 outburst of 4U 1630$-$472 with RXTE/ASM. Colors represent accretion state: HCS (blue), SDS (red) and IMS (yellow). Square points are days in which spectral/timing studies from the literature have indicated a SPL state \citep{trul01}. An SPL state with an unusually dominant power-law component can cause SPL observations to be classified as HCS with RXTE/ASM (compare to Figure~\ref{fig:2bSPL}).}%
    \label{fig:2SPL}%
    \label{fig:2aSPL}%
\end{figure*}

\begin{figure*}%
\plotone{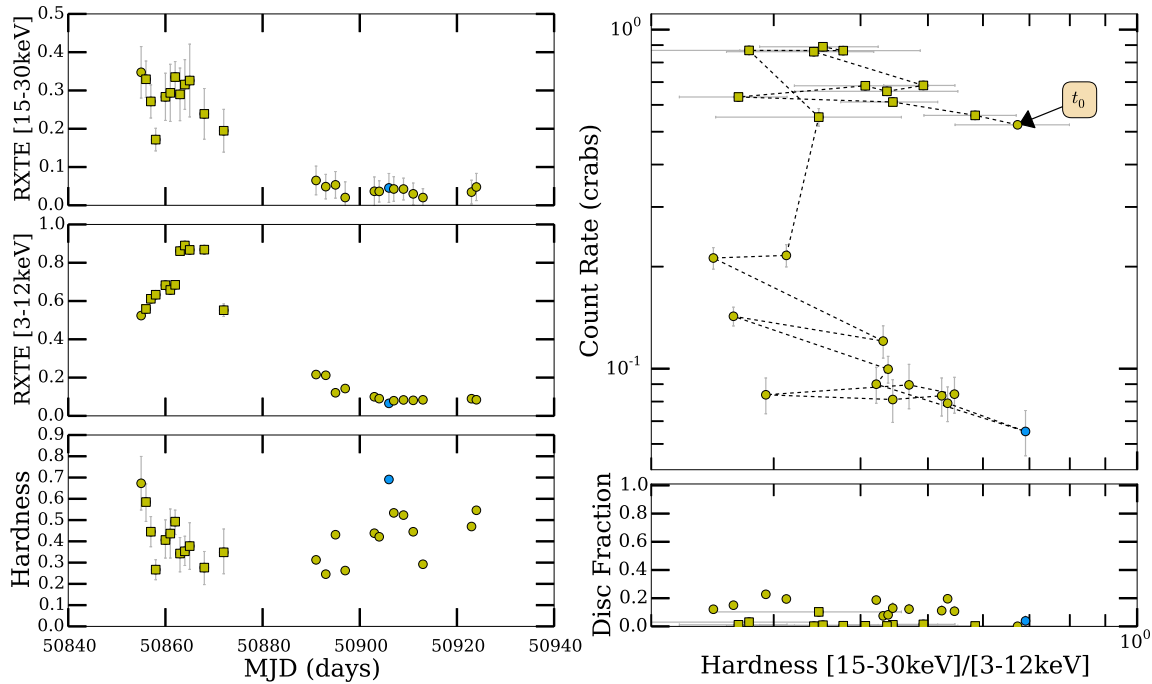}
 \caption{Analysis of the 1998 outburst of 4U 1630$-$472 with RXTE/HEXTE and RXTE/ASM. Colors represent accretion state: HCS (blue), SDS (red) and IMS (yellow). Square points are days in which spectral/timing studies from the literature have indicated a SPL state \citep{trul01}. The combination of RXTE/HEXTE and RXTE/ASM will properly classify an SPL state with an unusually dominant power-law component
 as an IMS (compare to Figure~\ref{fig:2aSPL}).}%
    \label{fig:2bSPL}%
\end{figure*}

This type of behavior is not an anomalous spectral feature exclusive to this source.
It has been observed during the SPL state in a handful of sources including GS 1124$-$684 \citep{k92,ebis94}, GRS 1730$-$312 \citep{bo95,trudol96}, GRS 1739$-$278 \citep{bo98}, and 4U 1630$-$472 \citep{trul01}.
In addition to XTE J1748$-$288, our algorithm classifies this power-law dominant SPL state behavior as the HCS in RXTE/ASM data on a number of other occasions (e.g., GRS 1739$-$278 and 4U 1630$-$472; see Figures~\ref{fig:1aSPL}--\ref{fig:2bSPL}), though the mis-classification of these few single observations does not affect the final outburst classification as a whole in any of these cases.

Lastly, we note that many of the algorithm detected outbursts that have been classified as ``hard-only'' are also under sampled, which brings with it the possibility that we may be missing the soft state due to the lack of coverage. 



We have checked all 32 outbursts that have been classified as ``hard-only'' by the algorithm against  available spectral/timing information in the literature, confirming that 19 of these outbursts have been classified as ``hard-only'' in the literature.
If we assume as a worst case scenario that the remaining 13 algorithm classified ``hard-only'' outbursts are in fact ``successful'' outbursts, the ratio obtained would still be within the quoted error range (see Table \ref{table:failedrate}).
While we may not be able to confidently rule out a source's presence in the soft state in many of these ``hard-only'' cases, the 1$\sigma$ Gehrels errors provides a very conservative range on the ``hard-only'' outburst ratio.

\subsection{Outburst Duration, Recurrence Rate and Duty Cycles}
\label{subsec:outburststats}
\begin{figure}[t]%
\plotone{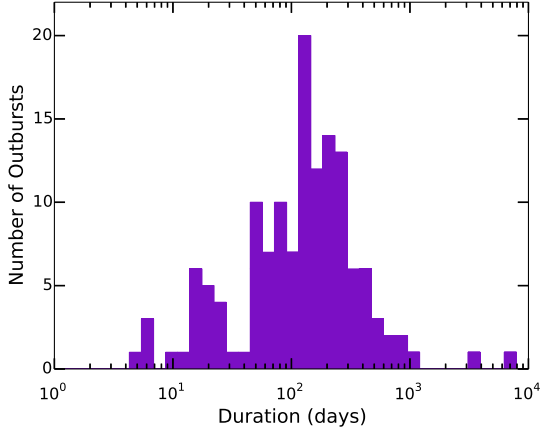}
    \caption{Distribution of the duration of transient (and long-term transient) outbursts detected by the algorithm between 1996 January 6 -- 2015 May 14. Data is distributed into 50 equal size bins between 1 and $10^4$ days with a mean of $250$ days.}%
    \label{fig:outdur1}%
\end{figure}

We have calculated the duration of 130 of the 132 (only excluding the ongoing (as of May 2015) outbursts of 4U 1630$-$472 and GX 339$-$4) transient outbursts and the 2 outbursts from the long-term transients detected by the algorithm (see Figure \ref{fig:outdur1} and Table \ref{table:transienthistory}).
We find that the mean outburst duration for the Galactic transient (and long-term transient) BHXB population is $\approx250$ days. When comparing the outburst durations of ``successful'' (i.e., state transitions have occurred) versus ``hard-only'' outbursts, we find mean outburst durations of $\approx247$ and $\approx391$ days, respectively. In addition, we test whether or not the durations of ``successful''  and ``hard-only''  outbursts are systematically different by performing a two sample KS-test. We find a p-value of $3.9\times 10^{-4}$, providing clear statistical evidence that the durations of ``successful'' and ``hard-only'' outbursts do not arise from the same parent distribution.

Using these durations and the number of outbursts detected, we have estimated the duty cycle for the 47 transient and 2 long-term transient sources in which the algorithm has detected at least one outburst in the last 19 years (See Table \ref{table:transienthistory}).
The duty cycle, defined as the fraction of its lifetime that a transient source has spent in outburst, is an important parameter needed to understand both the luminosity functions and binary evolution of these types of systems (e.g, \citealt{belz04,fragos08,fragos09}).
As such, being able to quantify the range of duty cycles exhibited by the transient BH population is of crucial importance.

\begin{figure}[h!]%
\plotone{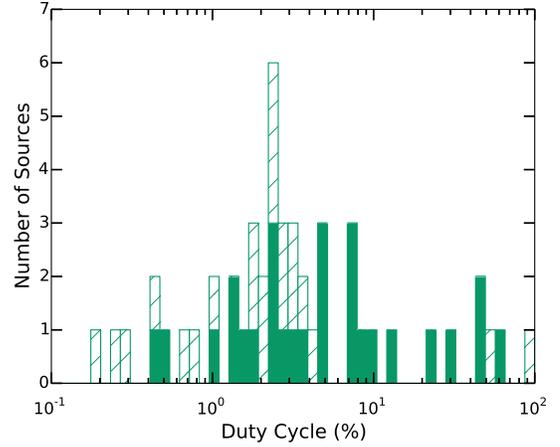}
    \caption{Distribution of the duty cycles for the Galactic transient (and long-term transient) BHXB population between 1996 January 6 -- 2015 May 14. Data is distributed into 50 equal size bins between 0.1\% and 100\% with a mean of $10$\% and median of 2.7\% (including upper limits). Green-colored sources have undergone more than one outburst in the time period. The cross hatch region represent those sources for which we only have upper limit estimates on the duty cycle due to only one outburst being detected in the last 19 years.}%
    \label{fig:outdutycycle1}%
\end{figure}

Figure \ref{fig:outdutycycle1} shows the distribution of duty cycles, calculated by taking the total time each source has spent in outburst and dividing by the total observation time (1996 January 6 -- 2015 May 14; $7068$ days), for all 47 transient and 2 long-term transient BH sources.
Here the dividing line between outburst and quiescence is defined as a count rate (in crabs) greater then the background rate plus $3\sigma$ for each individual source (see Sections \ref{subsubsec:clean} and \ref{subsubsec:detect} for thorough discussion).
We find a wide distribution of duty cycles exhibited by the transient population.
Including upper limits we find duty cycles ranging from 0.20--100\%, with an mean value of 10\%, a median value of 2.7\%, and that there is no observable systematic difference in duty cycle between those sources that have exclusively undergone ``successful'' outbursts, exclusively undergone ``hard-only'' outbursts, or have undergone a combination of the two types of outbursts over the last 19 years.

We note that many of the sources in our sample have only undergone one outburst during the 19 year time period, making the duty cycle we estimated only an upper limit on the true value.
While the long term evolutionary history of these transient systems may not be fully represented by the 19 years of behavior we have cataloged, our analysis can at least provide order of magnitude estimates for their duty cycles, which are still exceedingly helpful input into the theoretical modelling of luminosity functions and binary evolution codes \citep{yanyu14}.

We note that in making this argument we are assuming that the outburst recurrence times for these systems do not exceed $\sim200$ years  (an order of magnitude longer than our observation history).
To determine if this assumption is valid we begin with an analytical relationship between orbital period and recurrence time.
Using the Disc Instability Model (e.g., see \citealt{meyer81,cannizzo95,kingrit8,lasota1}) and assuming recurrence time can be estimated as the time required to fill the disk to its maximum mass $M_{\rm max}$, \citet{menou99} find,
\begin{equation}
t_{\rm recur}\lesssim\frac{M_{\rm max}}{\Dot{M}_{\rm BH}}=2.64 \times 10^{21} \, \alpha^{-0.85} \, \Dot{M}_{\rm BH}^{-1} \, M_{\rm BH}^{-0.37} \, R_{10}^{2.11} 
\label{eq:trecur}
\end{equation}
where $\alpha$ is the viscosity parameter, $M_{\rm BH}$ is the mass of the BH, $\Dot{M}_{\rm BH}$ is the mass transfer rate onto the BH, and $R_{10}$ is the disk outer radius in units of $10^{10}$ cm. Following \citet{menou99}, we assume $R_{10}$ is approximately 70\% of the Roche-lobe equivalent radius and use the formula for the Roche lobe equivalent radius given by \citet{pac71} to approximate $R_{10}\sim15 \, M_{\rm BH}^{1/3} \, P_{\rm orb}^{2/3}$. 

From Equation \ref{eq:trecur}, it is clear that a large $P_{\rm orb}$ would be needed to give a long recurrence time.
For a system with a $10 \, M_{\odot}$ BH,  and an $\Dot{M}_{\rm BH}$ typical for longer period ($>10$ hr) systems of $\sim10^{-9} \, M_{\odot} {\rm yr ^{-1}}$ (see Figure \ref{figure:mdotporb}), a $P_{\rm orb}\sim 286$ hours would be required to give a $t_{\rm recur}\sim200$ years.
Using the typical radius-period relation \citep{frank2}, a system with an $P_{\rm orb}\gtrsim286$ hours (and therefore a $t_{\rm recur}\gtrsim200$ years) would have to contain a giant companion of spectral type K7 or later.
None of the known spectral types in our sample are red and large enough to give such a long recurrence time, except for GRS 1915+105, which is thought to have a $t_{\rm recur}\sim10^4$ years \citep{deegan09}. 

While companion stars as large as the one in GRS 1915+105 can be ruled out in many of our systems with this argument, it can not be used to rule out long recurrence times in systems where we have no serious limits on the companion.
With this being said, one could also turn this argument around and say that a long recurrence time will also imply a long outburst time, so short outbursts will suggest short recurrence times.
Examining the remainder of the transient population, for which we have no information on companion spectral type or orbital period, we find all outburst durations are $\lesssim3$ years.
We therefore believe that the assumption that $t_{\rm recur}$ does not exceed $\sim200$ years for the systems in our sample is a reasonable one.

\begin{figure}[t]%
\plotone{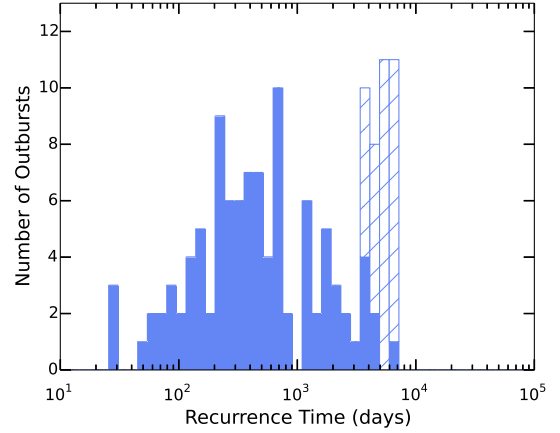}
    \caption{Distribution of the outburst recurrence times for Galactic transient (and long-term transient) BHXB population between 1996 January 6 -- 2015 May 14. Data is distributed into 50 equal size bins between 10 and $10^5$ days. Blue-colored sources have undergone more than one outburst in the time period and the cross hatch region represents those sources for which we only have a lower limit estimate on the recurrence rate due to only one outburst being detected in the last 19 years. When a source has undergone more then one outburst, the median of the recurrence times between outbursts for the source is plotted. Not including the lower limits, we find a median recurrence time for the transient population of $\approx414$ days.}%
    \label{fig:outrecur1}%
\end{figure}

In addition, we have also calculated the recurrence times between outbursts (over the last 19 years) for these 47 transient and 2 long-term transient sources (see Table \ref{table:transienthistory}).
Figure \ref{fig:outrecur1} displays the distribution of outburst recurrence times over the 19 year time period, calculated by finding the time difference between the beginning of each outburst detected in a 

\noindent particular source.
In the case where a source has only undergone one outburst in the last 19 years, a lower limit is estimated by taking the time difference between
(i) the beginning of our analysis (1996 January 6 -- MJD = 50088) and the beginning of the outburst in the case where the outburst occurred closer to end of our analysis (2015 May 14 -- MJD = 57156), or
(ii) the end of our analysis and the beginning of the outburst in the case where the outburst occurred closer to the beginning of our analysis.
Taking into account all detected recurrent events (i.e., not including lower limits) and ignoring the possibility that we may have missed events between some of the recorded outbursts, we find the minimum, maximum, and median recurrence time exhibited by the transient population to be approximately 29, 6589, and 414 days, respectively.

We note that there is a possibility that our distribution is not the true distribution due to two contributing factors.
First, there could be (and are) outbursts that our algorithm has not detected (see footnotes of Table \ref{table:outhistBH}).
 This would increase the count of short recurrence times.
Second, the many sources that have only one recorded outburst (and in turn only a lower limit on recurrence time) may go into outburst in the future. This would effectively increase the count of long recurrence times.

\subsection{State Transitions and Transition Luminosities}
\label{subsec:transition}
In addition to classification, our algorithm makes use of the X-ray hardness ratio to track a source as it transitions through varying combinations of the three accretion states, during  outburst and/or periods of continuous activity on a day-by-day basis.
In Tables \ref{table:statetransitions} and \ref{table:statetransitions2} we present the results of tracking 46 transient and 10 persistent sources for which we have sufficient data available.
Each outburst (or period of long-term activity) has been differentiated into three separate stages:
(i) rise, corresponding to times preceding the outburst peak (in luminosity);
(ii) decline, corresponding to times following the outburst peak (in luminosity); and
(iii) transition, corresponding to time periods in which the source changes accretion state, either between soft and hard, or just to an intermediate state and back.

\afterpage{
\tabletypesize{\footnotesize}
\renewcommand{\thefootnote}{\alph{footnote}}
\renewcommand\tabcolsep{5pt}
\begin{longtable*}{lccccccccr}

\caption{Activity of the Transient (and long-term transient) Galactic BHXB Population from 1996--2015}  \\

\hline \hline \\[-2ex]
   \multicolumn{1}{l}{Source Name (s)} &
   \multicolumn{1}{c}{Successful$^a$} &
     \multicolumn{1}{c}{Indeterminate$^a$} &
     \multicolumn{1}{c}{Hard-only$^a$} &
    \multicolumn{1}{c}{Total$^a$} &
    \multicolumn{1}{c}{$t_{\rm quies}$$^b$} &        
        \multicolumn{1}{c}{$t_{\rm out}$$^b$} &   
            \multicolumn{1}{c}{$<t_{\rm recur}>$$^c$}&   
                \multicolumn{1}{c}{Duty Cycle$^d$}&
   \multicolumn{1}{c}{States$^e$}\\
      &\multicolumn{1}{c}{Outbursts} & \multicolumn{1}{c}{Outbursts} & \multicolumn{1}{c}{Outbursts}&\multicolumn{1}{c}{Outbursts} & \multicolumn{1}{c}{(days) }&\multicolumn{1}{c}{(days)} &  \multicolumn{1}{c}{(days)}&\multicolumn{1}{c}{(\%)} &\multicolumn{1}{c}{Achieved}\\[0.5ex] \hline
   \\[-4ex]
\label{table:transienthistory}
\endfirsthead

\multicolumn{10}{c}{{\tablename} \thetable{} -- Continued} \\[0.5ex]
\hline \hline \\[-2ex]
   \multicolumn{1}{l}{Source Name (s)} &
   \multicolumn{1}{c}{Successful$^a$} &
     \multicolumn{1}{c}{Indeterminate$^a$} &
     \multicolumn{1}{c}{Hard-only$^a$} &
    \multicolumn{1}{c}{Total$^a$} &
    \multicolumn{1}{c}{$t_{\rm quies}$$^b$} &        
        \multicolumn{1}{c}{$t_{\rm out}$$^b$} &   
            \multicolumn{1}{c}{$<t_{\rm recur}>$$^c$}&   
                \multicolumn{1}{c}{Duty Cycle$^d$}&
   \multicolumn{1}{c}{States$^e$}\\
      &\multicolumn{1}{c}{Outbursts} & \multicolumn{1}{c}{Outbursts} & \multicolumn{1}{c}{Outbursts}&\multicolumn{1}{c}{Outbursts} & \multicolumn{1}{c}{(days) }&\multicolumn{1}{c}{(days)} &  \multicolumn{1}{c}{(days)}&\multicolumn{1}{c}{(\%)} &\multicolumn{1}{c}{Achieved}\\[0.5ex] \hline
   \\[-1.8ex]
\endhead

  \\[-1.8ex] \hline  \\
  \multicolumn{10}{l}{{Continued on Next Page\ldots}} \\
\endfoot

  \\[-1.7ex] \hline \\[-1.8ex] 
           \multicolumn{10}{p{0.95\columnwidth}}{\hangindent=1ex NOTE 1.-- This table is also available in machine readable format online at the Astrophysical Journal and on the WATCHDOG website - http://astro.physics.ualberta.ca/WATCHDOG/.}\\
         \multicolumn{10}{p{0.95\columnwidth}}{\hangindent=1ex NOTE 2.-- The sources not included in this table because we have not detected at least one outburst in the time period between 1996 January 6 -- 2015 May 14, 2015 are as follows: GROJ0422+32, 1A0620$-$00, GRS1009$-$45, GS1124$-$684, IGRJ11321$-$5311, 1A1524$-$62, H1705$-$250, XTEJ1719$-$291, GRS1716$-$249, GRS1730$-$312, KS1732$-$273, 1A1742$-$289, CXOGC J174540.0-290031, EXO1846$-$031, IGRJ18539+0727, XTEJ1901+014, GS2000+251, GS2023+338, MWC 656.}\\
         \multicolumn{10}{p{0.95\columnwidth}}{\hangindent=1ex $^a$The number of successful, indeterminate, hard-only and total outbursts detected and classified by the algorithm.}\\
         \multicolumn{10}{p{0.95\columnwidth}}{\hangindent=1ex $^b$Days spent in quiescence ($t_{\rm quies}$) and outburst ($t_{\rm out}$) calculated in the time period 50088.0--57156.0 (1996 January 6 -- 2015 May 1), starting when our data coverage began with RXTE.}\\
         \multicolumn{10}{p{0.95\columnwidth}}{\hangindent=1ex $^c$Median outburst recurrence time. A lower limit (indicated by $>$) is given when only one outburst has been detected. See Section~\ref{subsec:outburststats} for the method used for calculation.}\\
         \multicolumn{10}{p{0.95\columnwidth}}{\hangindent=1ex $^d$Transient duty cycle. For details on the method used for calculation see Section~\ref{subsec:outburststats}.}\\
         \multicolumn{10}{p{0.95\columnwidth}}{\hangindent=1ex $^e$States achieved by each source during outbursts. ``N/A'' refers to sources where we do not have enough data to determine the state.}
          
\endlastfoot

XTEJ0421+560             & \phn0 & \phn0 & 1 & \phn1 & 7019          & \phn\phn49 & >6271           & \phn\phn0.69        & HCS\\[0.015cm]
XTEJ1118+480             & \phn0 & \phn1 & 1 & \phn2 & 6861          & \phn207    & \phantom{>}2809 & \phn\phn2.9\phn     & HCS\\[0.015cm]
MAXIJ1305$-$704          & \phn0 & \phn1 & 0 & \phn1 & 6888          & \phn180    & >5922           & \phn\phn2.6\phn     & IMS\\[0.015cm]
SWIFTJ1357.2$-$0933      & \phn0 & \phn0 & 1 & \phn1 & 6992          & \phn\phn76 & >5489           & \phn\phn1.1\phn     & HCS\\[0.015cm]
GS1354$-$64              & \phn0 & \phn0 & 1 & \phn1 & 6912          & \phn156    & >6442           & \phn\phn2.2\phn     & HCS\\[0.015cm]
SWIFTJ1539.2$-$6227      & \phn1 & \phn0 & 0 & \phn1 & 6894          & \phn174    & >4704           & \phn\phn2.5\phn     & HCS,SDS,IMS\\[0.015cm]
MAXIJ1543$-$564          & \phn1 & \phn0 & 0 & \phn1 & 6889          & \phn179    & >5082           & \phn\phn2.5\phn     & HCS,SDS,IMS\\[0.015cm]
4U1543$-$475             & \phn1 & \phn0 & 0 & \phn1 & 6934          & \phn134    & >5848           & \phn\phn1.9\phn     & HCS,SDS,IMS\\[0.015cm]
XTEJ1550$-$564           & \phn2 & \phn0 & 3 & \phn5 & 6539          & \phn529    & \phantom{>0}464 & \phn\phn7.5\phn     & HCS,SDS,IMS\\[0.015cm]
4U1630$-$472             & 10    & \phn0 & 0 & 10    & 3924          & 3144       & \phantom{>0}714 & \phn44\phantom{.00} & HCS,SDS,IMS\\[0.015cm]
XTEJ1637$-$498           & \phn0 & \phn1 & 0 & \phn1 & 7054          & \phn\phn14 & >4615           & \phn\phn0.20        & N/A\\[0.015cm]
XTEJ1650$-$500           & \phn1 & \phn0 & 0 & \phn1 & 6851          & \phn217    & >5007           & \phn\phn3.1\phn     & HCS,SDS,IMS\\[0.015cm]
XTEJ1652$-$453           & \phn1 & \phn0 & 0 & \phn1 & 6910          & \phn158    & >4912           & \phn\phn2.2\phn     & HCS,SDS,IMS\\[0.015cm]
GROJ1655$-$40            & \phn2 & \phn0 & 0 & \phn2 & 6562          & \phn506    & \phantom{>}3486 & \phn\phn7.2\phn     & HCS,SDS,IMS\\[0.015cm]
MAXIJ1659$-$152          & \phn1 & \phn0 & 0 & \phn1 & 6910          & \phn158    & >3698           & \phn\phn2.2\phn     & HCS,SDS,IMS\\[0.015cm]
GX339$-$4                & \phn6 & \phn0 & 4 & 10    & 3602          & 3466       & \phantom{>0}501 & \phn49\phantom{.00} & HCS,SDS,IMS\\[0.015cm]
IGRJ17091$-$3624         & \phn1 & \phn1 & 0 & \phn2 & 6118          & \phn950    & \phantom{>}1445 & \phn13\phantom{.00} & HCS,SDS,IMS\\[0.015cm]
IGRJ17098$-$3628         & \phn0 & \phn1 & 0 & \phn1 & 7034          & \phn\phn34 & >3710           & \phn\phn0.48        & N/A\\[0.015cm]
SAXJ1711.6$-$3808        & \phn0 & \phn0 & 1 & \phn1 & 6929          & \phn139    & >5229           & \phn\phn2.0\phn     & HCS\\[0.015cm]
SWIFTJ1713.4$-$4219      & \phn0 & \phn1 & 0 & \phn1 & 7051          & \phn\phn17 & >5054           & \phn\phn0.24        & N/A\\[0.015cm]
XMMSL1J171900.4$-$353217 & \phn0 & 13    & 3 & 16    & 4906          & 2162       & \phantom{>0}154 & \phn31\phantom{.00} & HCS\\[0.015cm]
XTEJ1720$-$318           & \phn1 & \phn0 & 0 & \phn1 & 6880          & \phn188    & >4512           & \phn\phn2.7\phn     & HCS,SDS,IMS\\[0.015cm]
XTEJ1727$-$476           & \phn1 & \phn0 & 0 & \phn1 & 7015          & \phn\phn53 & >3551           & \phn\phn0.75        & HCS,SDS,IMS\\[0.015cm]
IGRJ17285$-$2922         & \phn0 & \phn1 & 1 & \phn2 & 6901          & \phn167    & \phantom{>}2148 & \phn\phn2.4\phn     & HCS\\[0.015cm]
IGRJ17379$-$3747         & \phn0 & \phn1 & 1 & \phn2 & 7038          & \phn\phn30 & \phantom{>}2868 & \phn\phn0.42        & HCS\\[0.015cm]
GRS1737$-$31             & \phn0 & \phn0 & 1 & \phn1 & 6968          & \phn100    & >6659           & \phn\phn1.4\phn     & HCS\\[0.015cm]
GRS1739$-$278            & \phn2 & \phn0 & 0 & \phn2 & 6374          & \phn694    & \phantom{>}3510 & \phn\phn9.8\phn     & HCS,SDS,IMS\\[0.015cm]
SWIFTJ174510.8$-$262411  & \phn0 & \phn0 & 1 & \phn1 & 6783          & \phn285    & >6090           & \phn\phn4.0\phn     & HCS\\[0.015cm]
IGRJ17454$-$2919         & \phn0 & \phn1 & 0 & \phn1 & 7049          & \phn\phn19 & >6853           & \phn\phn0.27        & N/A\\[0.015cm]
H1743$-$322              & \phn6 & \phn2 & 5 & 13    & 5515          & 1553       & \phantom{>0}304 & \phn22\phantom{.00} & HCS,SDS,IMS\\[0.015cm]
XTEJ1748$-$288           & \phn0 & \phn0 & 1 & \phn1 & 6991          & \phn\phn77 & >6195           & \phn\phn1.1\phn     & HCS\\[0.015cm]
IGRJ17497$-$2821         & \phn0 & \phn0 & 1 & \phn1 & 6970          & \phn\phn98 & >6188           & \phn\phn1.4\phn     & HCS\\[0.015cm]
SLX1746$-$331            & \phn2 & \phn1 & 0 & \phn3 & 6551          & \phn517    & \phantom{>}1619 & \phn\phn7.3\phn     & HCS,SDS,IMS\\[0.015cm]
XTEJ1752$-$223           & \phn1 & \phn0 & 0 & \phn1 & 6736          & \phn332    & >5004           & \phn\phn4.7\phn     & HCS,SDS,IMS\\[0.015cm]
SWIFTJ1753.5$-$0127      & \phn0 & \phn1 & 0 & \phn1 & 3427          & 3641       & >3627           & \phn52\phantom{.00} & HCS,IMS\\[0.015cm]
XTEJ1755$-$324           & \phn1 & \phn0 & 0 & \phn1 & 6944          & \phn124    & >6519           & \phn\phn1.8\phn     & HCS,SDS,IMS\\[0.015cm]
XTEJ1812$-$182           & \phn1 & \phn0 & 1 & \phn2 & 6955          & \phn113    & \phantom{>}2250 & \phn\phn1.6\phn     & HCS,SDS,IMS\\[0.015cm]
IGRJ18175$-$1530         & \phn0 & \phn1 & 0 & \phn1 & 7035          & \phn\phn33 & >4240           & \phn\phn0.47        & N/A\\[0.015cm]
XTEJ1817$-$330           & \phn1 & \phn0 & 0 & \phn1 & 6816          & \phn252    & >3656           & \phn\phn3.6\phn     & HCS,SDS,IMS\\[0.015cm]
XTEJ1818$-$245           & \phn1 & \phn0 & 0 & \phn1 & 6943          & \phn125    & >3581           & \phn\phn1.8\phn     & HCS,SDS,IMS\\[0.015cm]
SAXJ1819.3$-$2525        & \phn1 & 19    & 4 & 24    & 2911          & 4157       & \phantom{>0}221 & \phn59\phantom{.00} & HCS,SDS,IMS\\[0.015cm]
MAXIJ1836$-$194          & \phn0 & \phn0 & 1 & \phn1 & 6707          & \phn361    & >5565           & \phn\phn5.1\phn     & HCS\\[0.015cm]
SWIFTJ1842.5$-$1124      & \phn1 & \phn0 & 0 & \phn1 & 6841          & \phn227    & >4542           & \phn\phn3.2\phn     & HCS,SDS,IMS\\[0.015cm]
XTEJ1856+053             & \phn3 & \phn1 & 0 & \phn4 & 6718          & \phn350    & \phantom{>}1523 & \phn\phn5.0\phn     & HCS,SDS,IMS\\[0.015cm]
XTEJ1859+226             & \phn1 & \phn0 & 0 & \phn1 & 6844          & \phn224    & >5719           & \phn\phn3.2\phn     & HCS,SDS,IMS\\[0.015cm]
XTEJ1908+094             & \phn1 & \phn1 & 0 & \phn2 & 6446          & \phn622    & \phantom{>}2415 & \phn\phn8.8\phn     & HCS,SDS,IMS\\[0.015cm]
SWIFTJ1910.2$-$0546      & \phn1 & \phn0 & 0 & \phn1 & 6799          & \phn270    & >5859           & \phn\phn3.8\phn     & HCS,SDS,IMS\\[0.015cm]
GRS1915+105              & \phn1 & \phn0 & 0 & \phn1 & \phn\phn\phn0 & 7068       & >7068           & 100\phantom{.00}    & HCS,SDS,IMS\\[0.015cm]
XTEJ2012+381             & \phn1 & \phn0 & 0 & \phn1 & 6867          & \phn201    & >6216           & \phn\phn2.8\phn     & HCS,SDS,IMS
\end{longtable*}
\renewcommand\tabcolsep{5pt}
\renewcommand{\thefootnote}{\arabic{footnote}}
}

In addition to stage differentiation, an outburst (or period of long-term activity) is also separated into three possible accretion states where the data will allow: HCS, SDS, or IMS.
Due to the limited spectral information in broad-band count rates, while we observe the likely signature of the SPL state (i.e., the ``dragon horn'' in the HIDs) occurring in a number of sources over the last 19 years, we do not have the ability to empirically differentiate it from the other three accretion states using hardness ratio, and flux, alone.

By using independent spectral/timing studies in the literature we are able to make note of two interesting observations regarding the behavior of sources which enter this state.
First, we observe that this ``dragon horn'' feature appears in two different shapes in our HIDs, either
(i) curling backwards (i.e., a significant increase in hardness coupled with a moderate increase in luminosity, followed by a softening and moderate decrease in luminosity of the source) or (ii) standing closeto straight up (i.e., a near constant hardness and rapid luminosity change during the time period).
Second, we observe that SPL behavior spans the hardness ratios that our algorithm associates with the HCS, SDS, and IMS.  
See Figures \ref{fig:1aSPL}--\ref{fig:2bSPL}, and \ref{fig:44SPL}--\ref{fig:33bSPL} for examples of the aforementioned behavior as well as Sections \ref{subsubsec:classify} and \ref{subsec:failed} for a discussion on how the appearance of the SPL states affects our outburst classification algorithm. 

In addition to single states represented, Tables \ref{table:statetransitions} and \ref{table:statetransitions2} also present transitions of two different forms, ``(state name one) $\rightarrow$ (state name two)'' and ``(state name one) $\rightarrow$ IMS $\rightarrow$ (state name two)''.
The first depicts the  full transitions between hard and soft states, while the second describes the attempted transitions (or erratic ``jumps'')  between one of the two principal states and the IMS.
It is important to note that as we only have daily time resolution, HCS to SDS transitions taking less than 1 day to complete are not included here.
Lastly, Tables \ref{table:statetransitions} and \ref{table:statetransitions2} also indicate times when we do not have adequate information to define the state of the system. 
The ``-'' symbol is used to indicate situations where
(i) only one energy band is available during the time period, which is adequate for outburst detection but not for classification via $H_{X}$, or
(ii) no data is available during the particular time period.

\begin{figure}[h!]%
\plotone{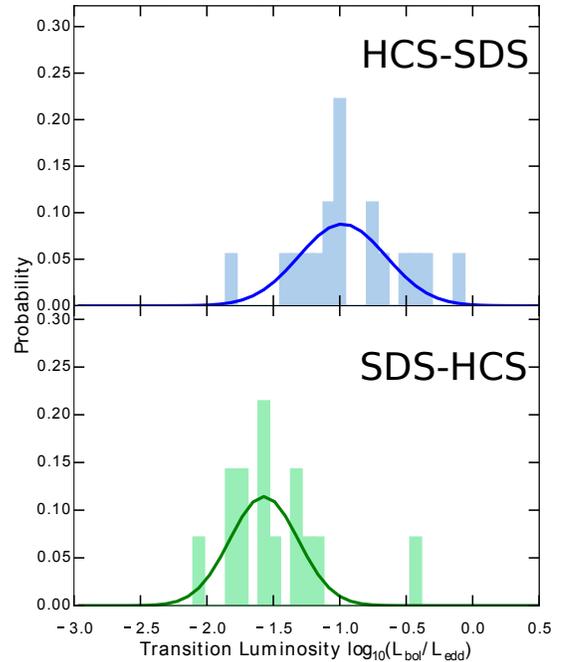}
    \caption{Probability distribution of the HCS-SDS (top) and SDS-HCS (bottom) transition luminosities in Eddington units for the Galactic transient BHXB population. The histograms include only those outbursts that the spectral fitting algorithm has not failed on and only complete HCS-SDS and SDS-HCS state transitions. The mean and standard deviation of the log luminosities for the HCS-SDS and SDS-HCS transitions, found via a Monte-Carlo method (see text), are $(\mu,\sigma)=(-0.94,0.409)$ and $(\mu,\sigma)=(-1.50,0.369)$, respectively. We overplot Gaussian distributions (blue and green lines) with these parameters for illustrative purposes.}%
    \label{fig:outtransL1}%
\end{figure}

\begin{figure*}[t]%
\plotone{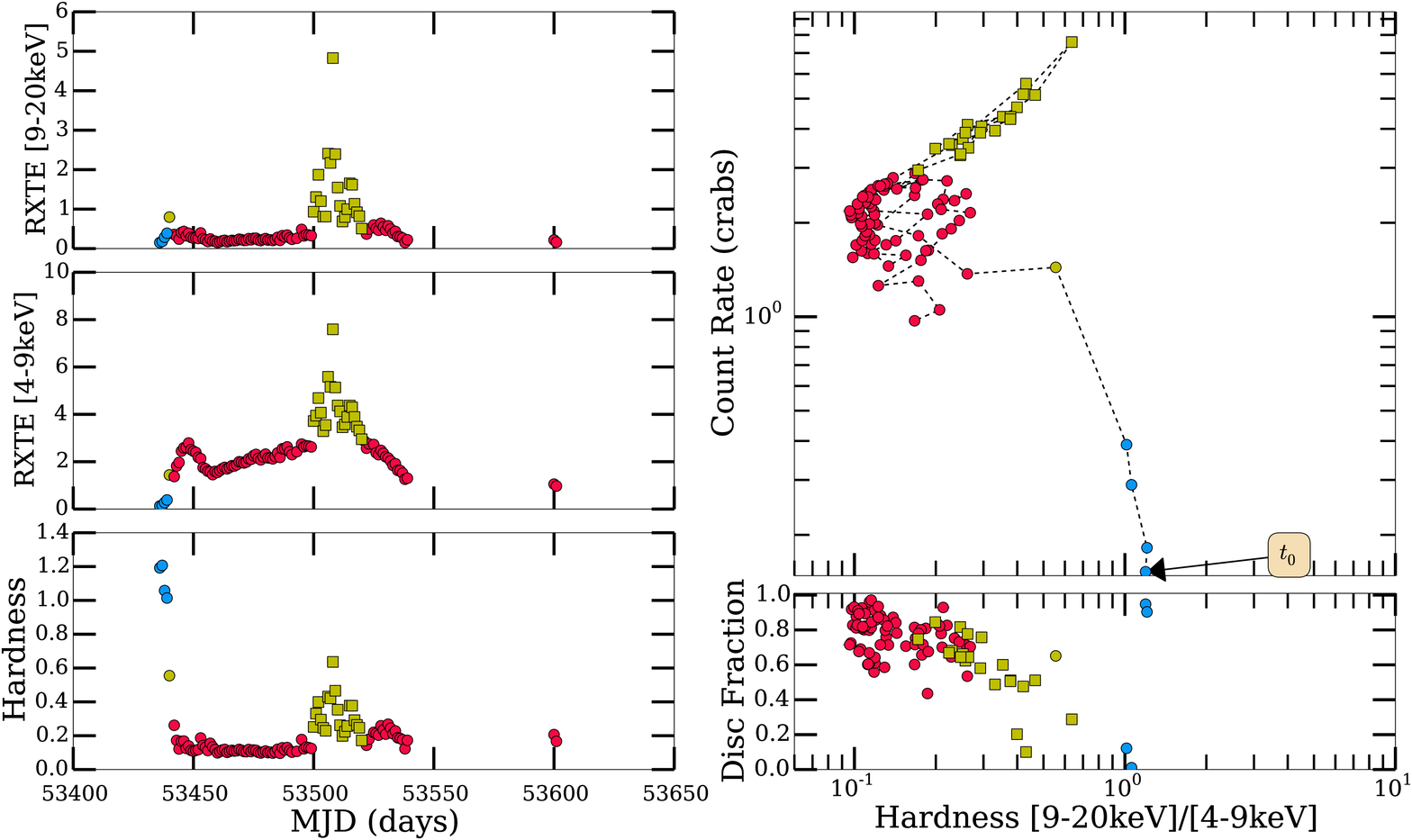}
 \caption{Analysis of the 2005 outburst of GRO J1655$-$40 with RXTE/PCA. Colors represent accretion state: HCS (blue), SDS (red) and IMS (yellow). Square points are days in which spectral/timing studies from the literature have indicated an SPL state \citep{br06} . Observe (i) how the SPL state can be classified by our algorithm as a HCS, SDS, IMS or any combination of the three and (ii) the first of two shapes the SPL state takes (i.e., curls backward).}%
    \label{fig:44SPL}%
\ \\
\plotone{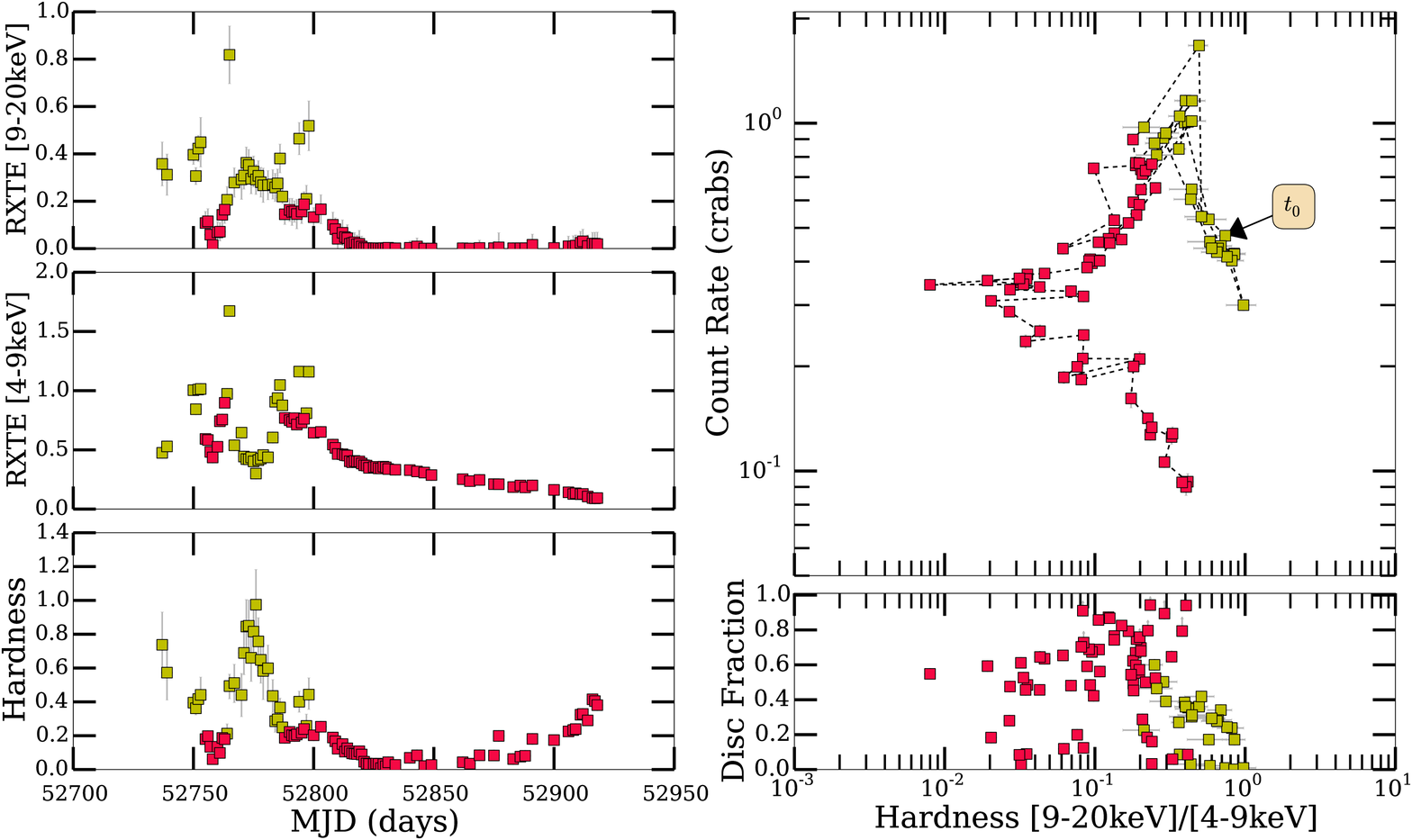}
 \caption{Analysis of the 2003 outburst of H 1743$-$322 with RXTE/ASM and RXTE/PCA. Colors represent accretion state: HCS (blue), SDS (red) and IMS (yellow). Square points are days in which spectral/timing studies from the literature have indicated a SPL state \citep{mc09}. Observe (i) how the SPL state can be classified by our algorithm as a HCS, SDS, IMS or any combination of the three and (ii) the second of two shapes the SPL state takes (i.e., stands straight up).}%
    \label{fig:33SPL}%
    \label{fig:33aSPL}%
\end{figure*}

\begin{figure*}%
\plotone{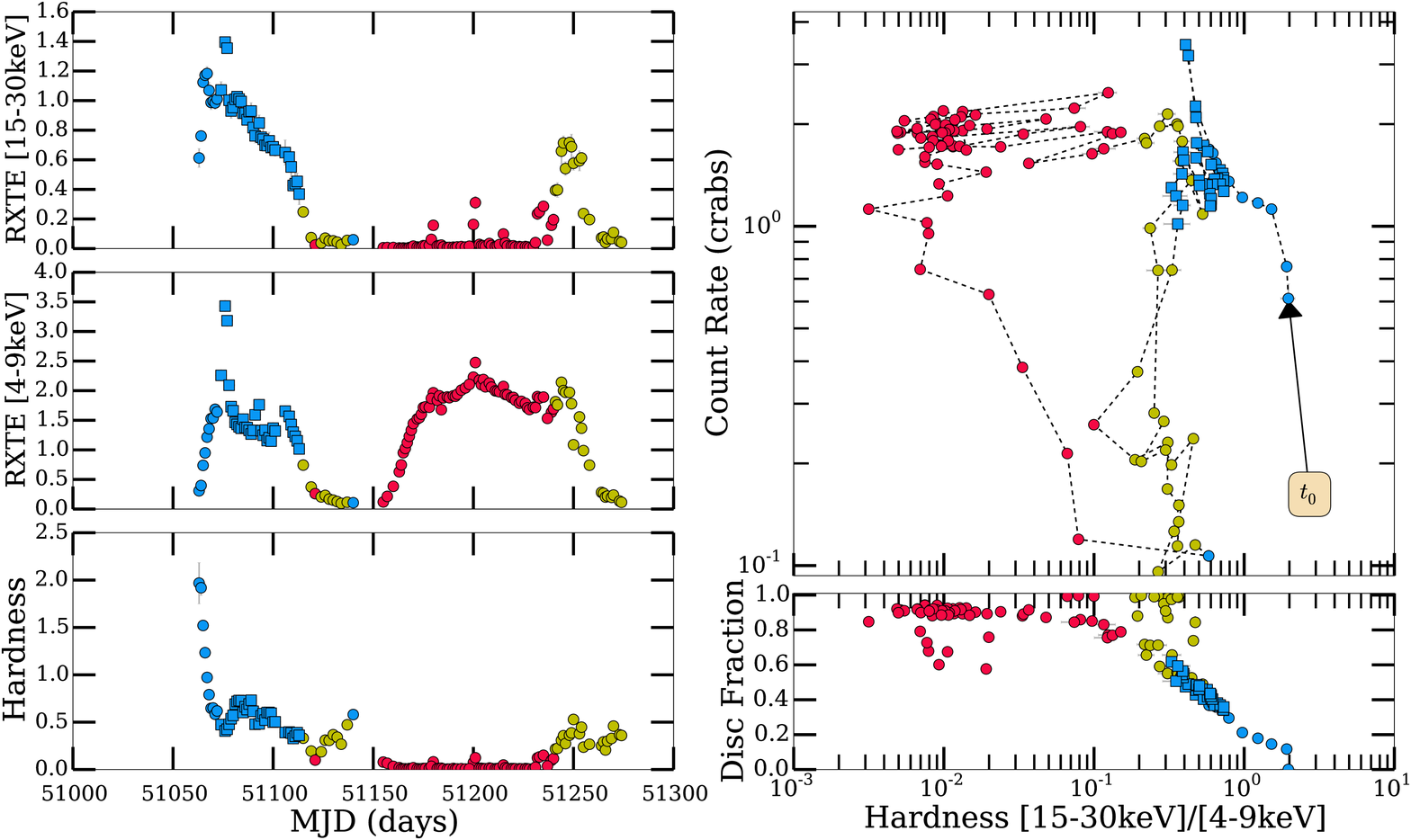}
 \caption{Analysis of the 1998/1999 outburst of XTE J1550$-$564 with RXTE/HEXTE and RXTE/PCA. Colors represent accretion state: HCS (blue), SDS (red) and IMS (yellow). Square points are days in which spectral/timing studies from the literature have indicated an SPL state \citep{s2000}. Observe (i) how the SPL state can be classified by our algorithm as a HCS, SDS, IMS or any combination of the three and (ii) the second of two shapes the SPL state takes (i.e., stands straight up). In comparison to Figures \ref{fig:44SPL} and \ref{fig:33aSPL},  the SPL state occurs before the transition to the SDS.}%
    \label{fig:33bSPL}%
\end{figure*}

We have also estimated the luminosity of transient BHXB systems during both the forward (HCS-SDS) and reverse (SDS-HCS) state transitions.
In Figure~\ref{fig:outtransL1} we plot the probability distribution of transition luminosities in Eddington units ($L_{\rm bol}/L_{\rm edd}$). Using luminosities in Eddington units allows us to accurately compare across sources and take into account both uncertainties in distance to a source as well as BH mass. See Section \ref{subsec:peaklum} for a discussion of how the distances are estimated for the sources without
distance estimates in the literature.
We are careful to include only those outbursts in which our spectral fitting algorithm has not failed and only instances within each of these outbursts where a source made a complete hard-soft or soft-hard transition.
We do not include the erratic ``jumps'' between the intermediate states and the hard and soft states in this analysis.

The transition luminosity of a particular source during an outburst is estimated by finding the days in which the source was undergoing the specific state transition (see Table \ref{table:statetransitions}) and calculating a weighted mean (and error) of the $L_{\rm bol}/L_{\rm Edd}$ estimates during this time period. To take into account the errors in the estimated transition luminosities we make use of the Monte Carlo method presented in \cite{du09}, whereby we randomly select a value of $L_{\rm bol}/L_{\rm Edd}$ from a Gaussian distribution with 1$\sigma$ values derived from the propagated error in the transition luminosity estimate for each state transition in our sample. We then use this to estimate the underlying probability distribution and display this in Figure \ref{fig:outtransL1}.

Rather than fit a Gaussian distribution to the probability distribution (despite a large number of Monte Carlo simulations, the precision of the distribution is relatively poor given the small number of total sources) as has been done in previous work, we have performed more distribution agnostic characterizations of the hard-soft transition luminosities and the soft-hard transition luminosities. For each Monte Carlo simulation we measure the mean and standard deviation of the log luminosities. The hard to soft transition is described by $\mu_{log L_{\rm Edd}} = -0.94^{+0.15}_{-0.16}$ and  $\sigma_{log L_{\rm Edd}} = 0.409^{+0.085}_{-0.033}$, while the soft to hard transition is described by $\mu_{log L_{\rm Edd}} = -1.50^{+0.15}_{-0.14}$ and  $\sigma_{log L_{\rm Edd}} = 0.369^{+0.033}_{-0.022}$. For illustrative purposes, we overplot Gaussian distributions given by the median $\mu_{log L_{\rm Edd}}$ and $\sigma_{log L_{\rm Edd}}$ (Shapiro-Wilk tests can not reject the hypothesis that either population is drawn from a normal distribution; $p_{\rm H \rightarrow S} = 0.95$ and $p_{\rm S \rightarrow H} = 0.086$). A non-parametric Wilcoxon rank-sum test of each simulation implies that the median (two-sided) probability that the hard-to-soft transition luminosities and the soft-to-hard transition luminosities are drawn from the same distribution is $5.5\times10^{-4}$ ($3.5\sigma$).

A number of similar studies analyzing transition luminosities in BHXBs exist.
\citet{du09} used disk fraction luminosity diagrams (DFLDs) to determine transition luminosities for 25 BHXB systems, and found a mean HCS-SDS transition luminosity ($\sim0.3 \, L_{\rm edd}$) that is significantly larger than our estimate; however, their range was comparable to our estimate (0.03--$1 \, L_{\rm edd}$).
\citet{gn06}, using RXTE/ASM data and HIDs, find a HCS-SDS transition luminosity comparable to our result, though they quote a range that is significantly narrower (0.01--$0.28 \, L_{\rm edd}$).

In contrast, our mean SDS-HCS transition luminosity of $\sim0.03 \, L_{\rm edd}$ is consistent with both the estimates by \citet{du09} ($\sim0.03 \, L_{\rm edd}$) and \citet{macar03} ($\sim0.02\,  L_{\rm edd}$), the latter who tabulated transition luminosities using available data of 6 BHXBs from the literature.
However, both \citet{du09}  and \citet{macar03} find a significantly narrower range (0.05--$0.10 \, L_{\rm edd}$ and 0.01--$0.04 \,L_{\rm edd}$, respectively) for the SDS-HCS transition luminosity, when compared to our results.

\subsection{Peak Outburst Luminosity}
\label{subsec:peaklum}
We analyzed the peak luminosities for individual transient outbursts and long-term activity in the persistent sources.
Table \ref{table:lumdata} presents peak luminosities in the HCS, SDS, and for the outburst as a whole, a deconvolution of the outburst into total time spent in the HCS, SDS, and in transition, and an estimate of total energy released during outburst (discussed further in Section~\ref{subsec:etotal}) for each of the 92 classified algorithm detected outbursts. If there are no available distance and/or BH mass estimates for a source, luminosity analysis is performed assuming a distance corresponding to a uniform distribution between 2 and 8 kpc and a BH mass that is sampled from the \cite{oz10} mass distribution. Note that the BH mass estimates are only used for scaling the peak luminosity by the Eddington luminosity. 

This inferred distance was chosen by taking the central 90\% range of distances spanned by the 18 individual systems with known mass measurements (i.e., the dynamically confirmed BHs in class A and/or the most likely BHCs in class B), allowing us to sample from a range in distance in our analysis (e.g., bolometric luminosity, luminosity functions, long-term mass transfer rates) that we would expect the majority of Galactic BH systems to lie.
Using this criterion the closest and farthest systems, 1A 0620$-$00 at $\sim$1 kpc and GS 1354$-$64, which if the estimated distance (inferred from X-ray observations) is correct would make it the only known BH source in the Galactic halo, are excluded.

By choosing this uniform distribution in distance, we are implicitly assuming that whether or not follow-up observations have been performed on a source to determine distance/mass is more or less random. However, one could argue that the sources without distance/mass estimates are most likely to be those sources that are optically the faintest and behind the most reddening, which is more likely for sources on the far side of the Galaxy. In addition, if one were to look at a map of locations of XRBs with distance estimates (e.g., \citealt{jonker2004}), they seem to be concentrated on the near side of the Galaxy. 
On the other hand, (i) many large scale population studies like this work use similar assumptions for distance (e.g., \citealt{du09}), and (ii) we separate sources with and without known distances, and provide individual results, in all analyses involving luminosity in this work. As such, the reader should either (i) retain the caveat that, for our full sample, truncating uncertain distance distributions at 8 kpc may potentially cause inferred luminosity estimates to be skewed lower than the true luminosities for those sources, or (ii) focus only on the ``known distance'' parts of our analysis.

We perform several analyses with the peak outburst luminosities.
First, we observe a clear demonstration (also see Figure \ref{figure:lpeakporb}) of the under-luminous nature of a ``hard-only'' outburst (or long-term period spent in the HCS), indicated by the sub-Eddington peak luminosities, when compared to the ``successful''  outbursts (or long-term persistent ``turtlehead'' pattern behavior), as expected. 

Second, we determine the state in which the peak luminosity of an outburst occurs.
In the standard picture of outburst evolution \citep{maccarone2003,vadawale2003}, the peak luminosity is expected in the SDS, when the mass transfer rate is high enough to shift the truncation radius of the disk in towards the ISCO. However, a fair number of outbursts that display this (``turtlehead'') outburst behavior appear to exhibit a peak in the HCS.
However, upon further analysis, we find that this HCS peak behavior can be explained away by
(i) the fact that the bolometric corrections are poorly known,
(ii) missing coverage of the soft state peaks due to Sun constraints near the month of December, or
(iii) the breakdown of our spectral modelling algorithm in certain situations where the true spectral shape of the source does not match the assumed spectral shape of our model (e.g., the presence of a SPL state).

\begin{figure}%
\plotone{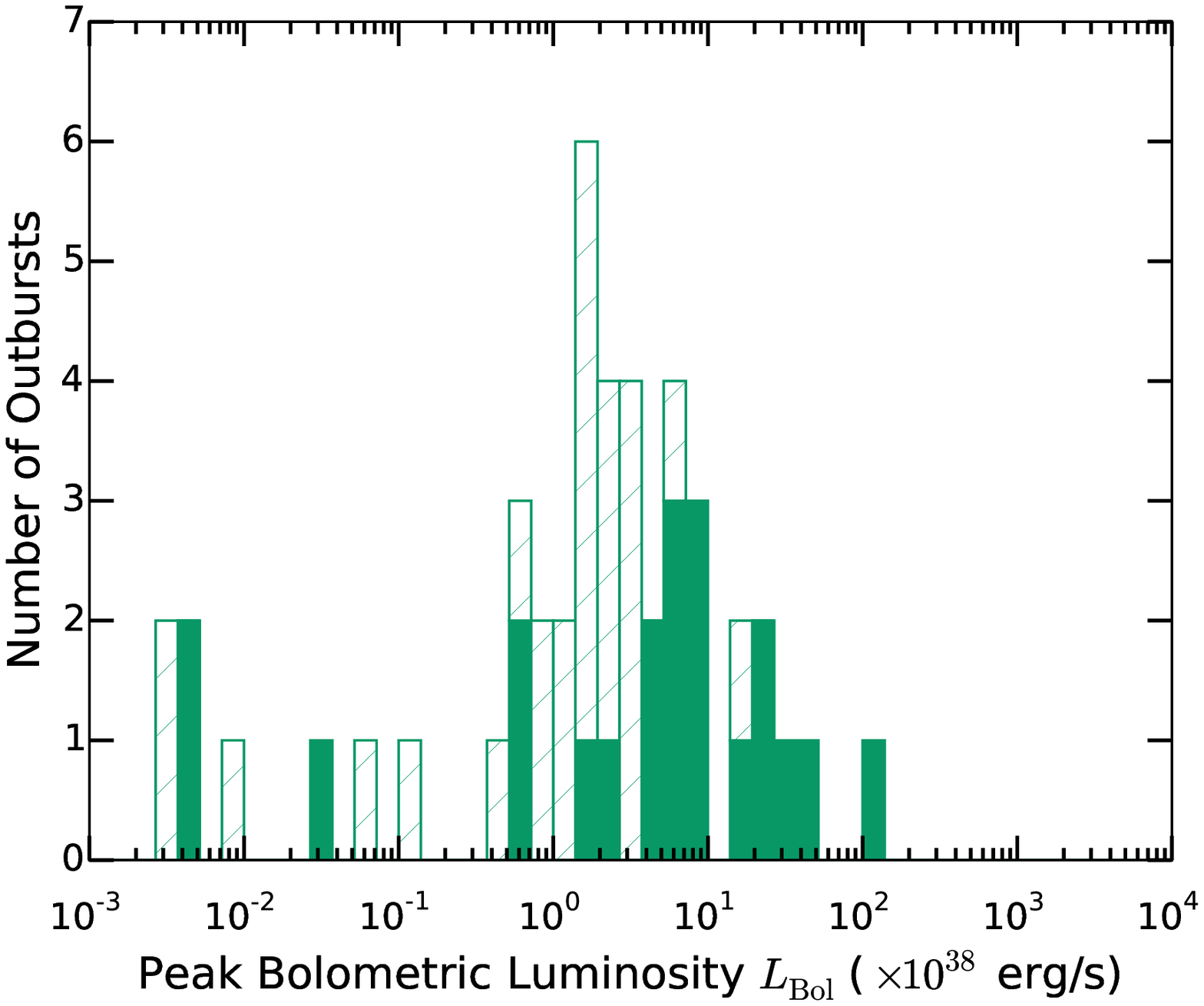}
    \caption{Distribution of the peak bolometric outburst luminosity for outbursts undergone by the transient (and long-term transient) Galactic BHXB population between 1996 January 6 -- 2015 May 14. The solid green and cross hatch regions represent outbursts from sources with and without known distance estimates, respectively. The histogram does not include outbursts that we have identified to contain days when our spectral fitting algorithm has failed. Taking into account only outbursts undergone by sources with an available distance estimate, we find a mean peak bolometric luminosity of $1.40 \times 10^{39}$ erg/s.}%
    \label{fig:outpeak1}%
\end{figure}

Third, we determine the fraction of time over the last 19 years that a source spends in each state, particularly the persistent sources.
Doing so, in combination with analysis of the state transitions occurring (see Tables \ref{table:statetransitions} and \ref{table:statetransitions2} and Section~\ref{subsec:transition}) has allowed us to separate the 10 persistent sources, for which data is available, into four separate classes based on long term behavioral characteristics as follows:
(i) mainly HCS behavior: the source spends $\gtrsim70$\% of the time it was in an X-ray bright state in the HCS and routinely undergoes attempted hard-soft transitions (i.e., ``incomplete'' state transitions), in which the source only reaches as soft as the IMS;
(ii) mainly SDS behavior: the source spends $\gtrsim70$\% of the time it was in an X-ray bright state in the SDS and routinely undergoes a combination of attempted hard-soft and soft-hard (`incomplete'') transitions only reaching as soft/hard as the IMS;
(iii) mainly IMS behavior: the source spends $\gtrsim70$\% of the time it was in an X-ray bright state in the IMS and routinely undergoes attempted intermediate-hard transitions; and 
(iv) anomalous/high $L_{X}$ behavior: while the source appears to spend the majority of its time repeatedly undergoing
the basic ``turtlehead'' pattern of behavior, it routinely shows peak luminosities at high fractions of Eddington, suggesting the presence of a high luminosity state such as an SPL state or perhaps a more complicated situation (e.g., ``heartbeat'' states of GRS 1915+105; \citealt{neil11}).
1E 1740.7$-$2942, SS 433, and 4U 1956+350 fall under case (i);
4U 1957+115 falls under case (ii);
GRS 1758$-$258 and Swift J1753.5$-$0127 fall under case (iii); and 
GRS 1915+105, LMC X-1, LMC X-3, and Cyg X-3 fall under case (iv).

Lastly, in Figure \ref{fig:outpeak1} we plot the distribution of peak outburst luminosities for the transient (and long-term transient) population, not including those outbursts that we have identified to contain days when our spectral fitting algorithm has failed (see Section~\ref{subsubsec:spectra} and outbursts marked with a ``*'' in Table \ref{table:lumdata}). 
Eliminating outbursts undergone by sources with no available distance estimates yields a range in peak outburst luminosity of $4.0 \times10^{35} - 1.0\times 10^{40}$ ${\rm erg s^{-1}}$ and a mean peak outburst luminosity of $1.4 \times 10^{39}$ ${\rm erg s^{-1}}$. Note that the one outburst peaking above $10^{40}$ ${\rm erg s^{-1}}$ is the 1997/1998 outburst of GS1354$-$64.
However, the distance to this source is poorly constrained (25--61 kpc).
If we were to place this system at our assumed standard Galactic value (i.e., a uniform distribution between 2 and 8 kpc) its peak luminosity would be on the order of $10^{38}$ ${\rm erg s^{-1}}$. 

\begin{figure}[]%
\plotone{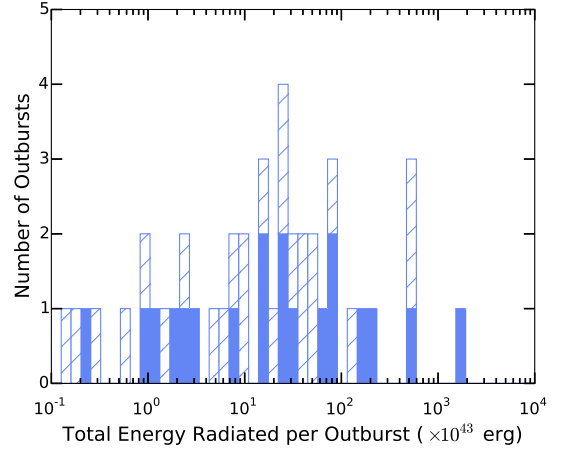}
    \caption{Distribution of the total X-ray energy released during outburst for outbursts undergone by the transient (and long-term transient) Galactic BHXB population between 1996 January 6 -- 2015 May 14. The light blue and cross hatch regions represent outbursts from sources with and without known distance estimates, respectively. The histogram only includes those outbursts that we have identified to not contain days when our spectral fitting algorithm has failed. Taking into account only outbursts undergone by sources with an available distance estimate, we find the mean energy released during outburst to be $2.0 \times 10^{46}$ erg. }%
    \label{fig:outflu1}%
\end{figure}

\subsection{Total Energy Radiated During Outburst}
\label{subsec:etotal}
We have estimated the total X-ray energy released during outburst for each of the 92 classified algorithm detected outbursts as well as the activity in the 2 long-term transient sources.
We define the total radiated energy as the $L_{\rm bol}$ (see Section \ref{subsubsec:spectra})  integrated over the duration of an outburst and calculate the quantity by first finding the weighted mean of $L_{\rm bol}$ during outburst then multiplying it by the total duration of the outburst.
As there are asymmetric errors on our $L_{\rm bol}$ estimates, we calculate this weighted mean iteratively.
Figure \ref{fig:outflu1} shows the distribution of total radiated energy.
Including only those sources that have a distance measurement available (31 sources; see Table \ref{table:binaryBH}) and only those outbursts that we have identified to not contain days when our spectral fitting algorithm has failed (see Section \ref{subsubsec:spectra} and outbursts marked with a ``*'' in Table \ref{table:lumdata}), we find a mean and range of total radiated energy during outburst for the transient population to be $2.0 \times 10^{46}$ erg and $8.8 \times10^{41} - 3.8\times 10^{47}$ erg, respectively.

\subsection{X-ray Luminosity Functions (XLFs)}
\label{subsec:xlfs}

Using the methods described in Section~\ref{subsubsec:lfmdot}, and given bolometric X-ray luminosity $L_{\rm bol}$ for a source on any given day, we have obtained the empirical XLFs for 46 transient sources and the 10 persistent sources for which sufficient data is available.

In Figures \ref{fig:tXLF1}--\ref{fig:tXLF82} (located in Appendix) luminosity data have been arranged into 31 bins between $0$ and $10^{41} \, \rm{erg}\, \rm{s}^{-1}$, where any values below $10^{34}\, \rm{erg}\, \rm{s}^{-1}$ are placed in the lowest bin.
The errors on each bin are quoted as 1$\sigma$ Gehrels errors. 
All luminosity data has been differentiated into accretion states where blue, red, yellow, and grey represent the HCS, SDS, IMS, and days in which we are unable to define state with available data, respectively.

As discussed in Section \ref{subsubsec:spectra}, there are days during an outburst of a given source when our spectral fitting algorithm fails, resulting in a large uncertainty in the derived luminosity for that day.
While we have identified 47 transient outbursts and multiple occurrences in the persistent sources in which this problem occurs (see Table \ref{table:lumdata}), the number of days when the algorithm fails during an outburst or long-term bright state is in many cases either only a small fraction of the outburst duration as a whole or a small fraction of the total 19 year outburst period of the source (i.e., in those sources with short recurrence times).
As this problem would be difficult to correct, we include all data in each source XLF and note that some source XLFs may not be representative of the true luminosity distribution for that individual source.

One may expect the exclusively ``hard-only'' outburst source XLFs to only exhibit one peak located at lower luminosities, associated with the HCS to which the source has been observed to remain in for the duration of outburst periods.
Of the 12 sources that have been observed to undergo exclusively ``hard-only'' outbursts, 5 exhibit this behavior (e.g., XTE J1118+480, SAX J1711.6$-$3808, XMMSL1 J171900.4$-$353217, GRS 1737$-$37, and IGRJ 17285$-$2922).
GS 1354$-$64 also displays a singular feature.
While this feature may peak at $\sim 10^{39} {\rm erg \, s^{-1}}$, we note that the upper bound on the poorly constrained distance estimate (61 kpc) is the most likely cause of these excessively high luminosity values.
Unfortunately, our ability to properly analyze the remaining 6 sources is hindered by either (i) uncertainties in derived bolometric luminosities caused by our spectral fitting algorithm failing, presumably leading to a scattered contribution at higher luminosities in addition to the expected low luminosity peak in the XLFs (e.g., Swift J1357.2$-$0933, and MAXI J1836$-$194), and (ii) not enough data available for a source to determine what state the source was in on a significant number of days during outburst (e.g., XTE J0421+560 and IGRJ17379$-$3747).

In addition, we also observe the appearance of significant hard state contributions in the higher luminosity regime ($>10^{38} {\rm erg s^{-1}}$) in a few sources (e.g., GX 339$-$4, GRS 1739$-$278, H1743$-$322).
We find that the appearance of this features correlates with sources that undergo outbursts that  peak in the HCS and display the ``dragon horn'' feature in their HIDs.
Given that these sources have been shown (through spectra/timing studies in the literature) to enter the SPL (or high luminosity state) state during outburst  (i.e., see \citealt{bo98,mo09,mc09}), this behavior is perhaps indicative of the SPL state extending into the hard state regime (see discussion in Section ~\ref{subsec:transition}).
A second alternative is that this contribution at higher luminosities could be caused (at least in part) by the days when our spectral fitting algorithm fails.

Lastly, in the case of the persistent XLFs, we find that the features exhibited correlate with each of the four classes we have defined above based on luminosity and temporal evolution observed.
Specifically,
(i) 1E 1740.7$-$2942, SS 433, and 4U1956+350, which spend most of their time in the HCS, exhibit a prominent HCS peak component at low luminosities ($\sim10^{37} \rm{erg} \rm{s}^{-1}$),
(ii) GRS 1758$-$258 and Swift J1753.5$-$0127, which show mainly IMS behavior exhibit a dominate IMS contribution at lower luminosities ($\sim10^{37} \rm{erg} \rm{s}^{-1}$),
(iii) 4U 1957+115, which undergoes mainly SDS behavior exhibits a single peak feature at $\sim10^{37} \rm{erg} \rm{s}^{-1}$ and
(iv) GRS 1915+105, LMC X-1, LMC X-3, and Cyg X-3, which exhibit anomalous/high $L_X$ behavior, display a single peak feature between $\sim10^{38} \rm{erg} \rm{s}^{-1}$ and $\sim10^{39} \rm{erg} \rm{s}^{-1}$.

We must also take into account our knowledge of where the spectral fitting algorithm fails in our analysis of the persistent source XLFs.
In both cases (i) and (ii), in addition to the observed peaks, we also observe the presence of a higher luminosity ``tail''.
This feature can be seen in 1E 1740.7$-$2942, Swift J1753.5$-$0127, SS 433, and GRS 1758$-$258.
Given that 4U 1956+350, a source in which our spectral fitting algorithm has never failed, is the only source in these two groups that does not exhibit this feature in the XLF, we conclude that it is probable that this higher luminosity ``tail'' may (at least in part) be an artificial feature occurring as a result of incorrect bolometric luminosity estimates.

\begin{figure*}%
\plottwo{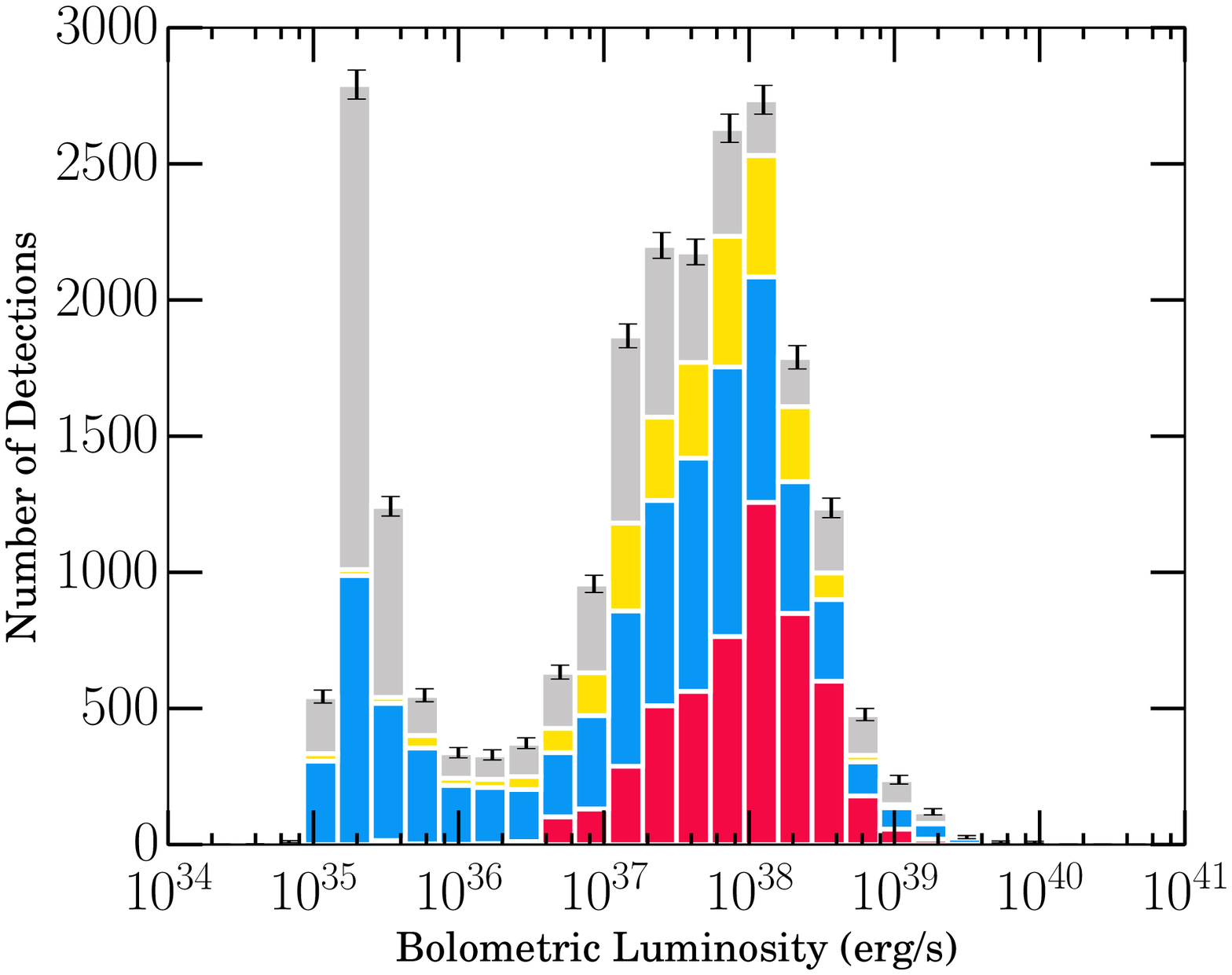}{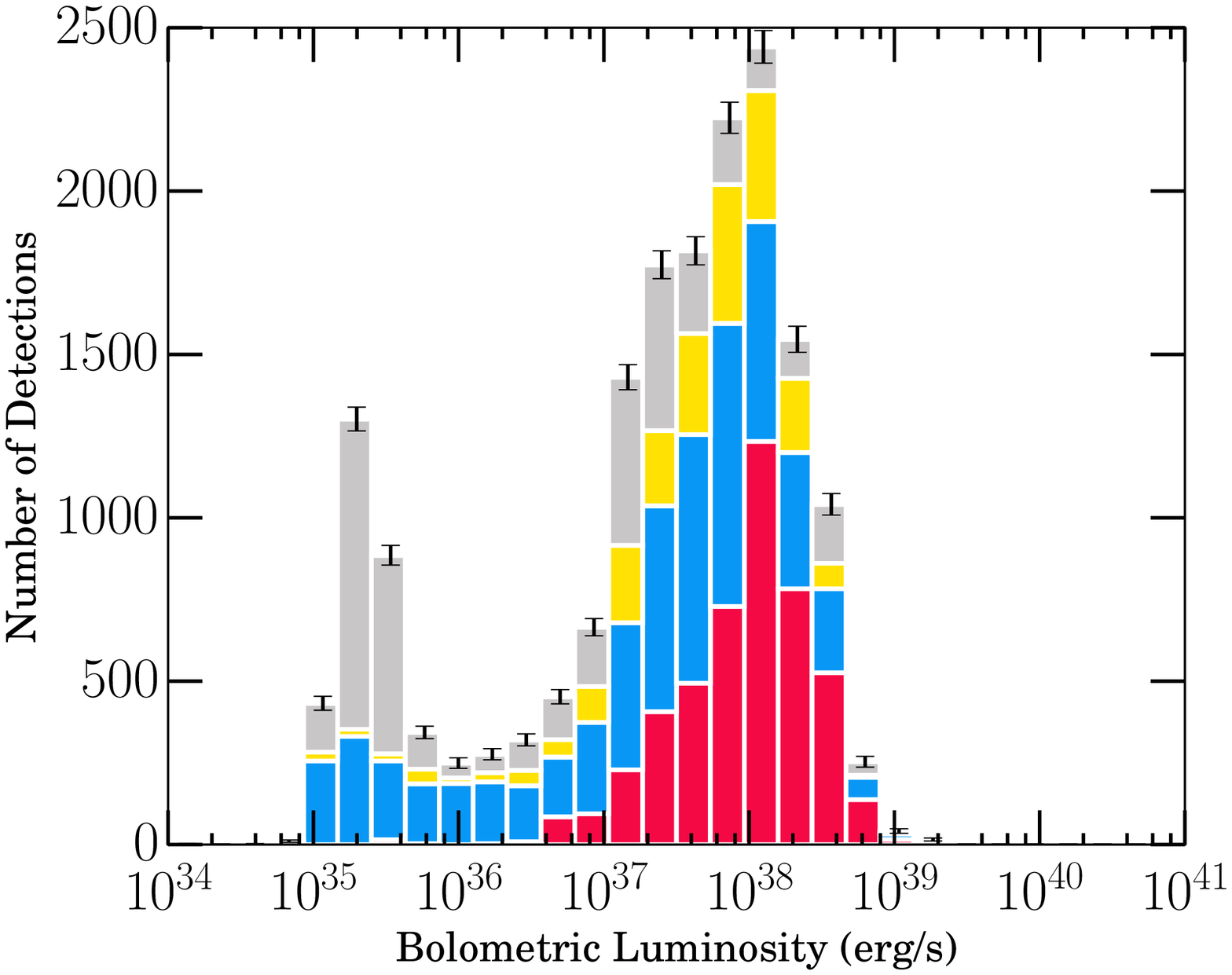}
\plottwo{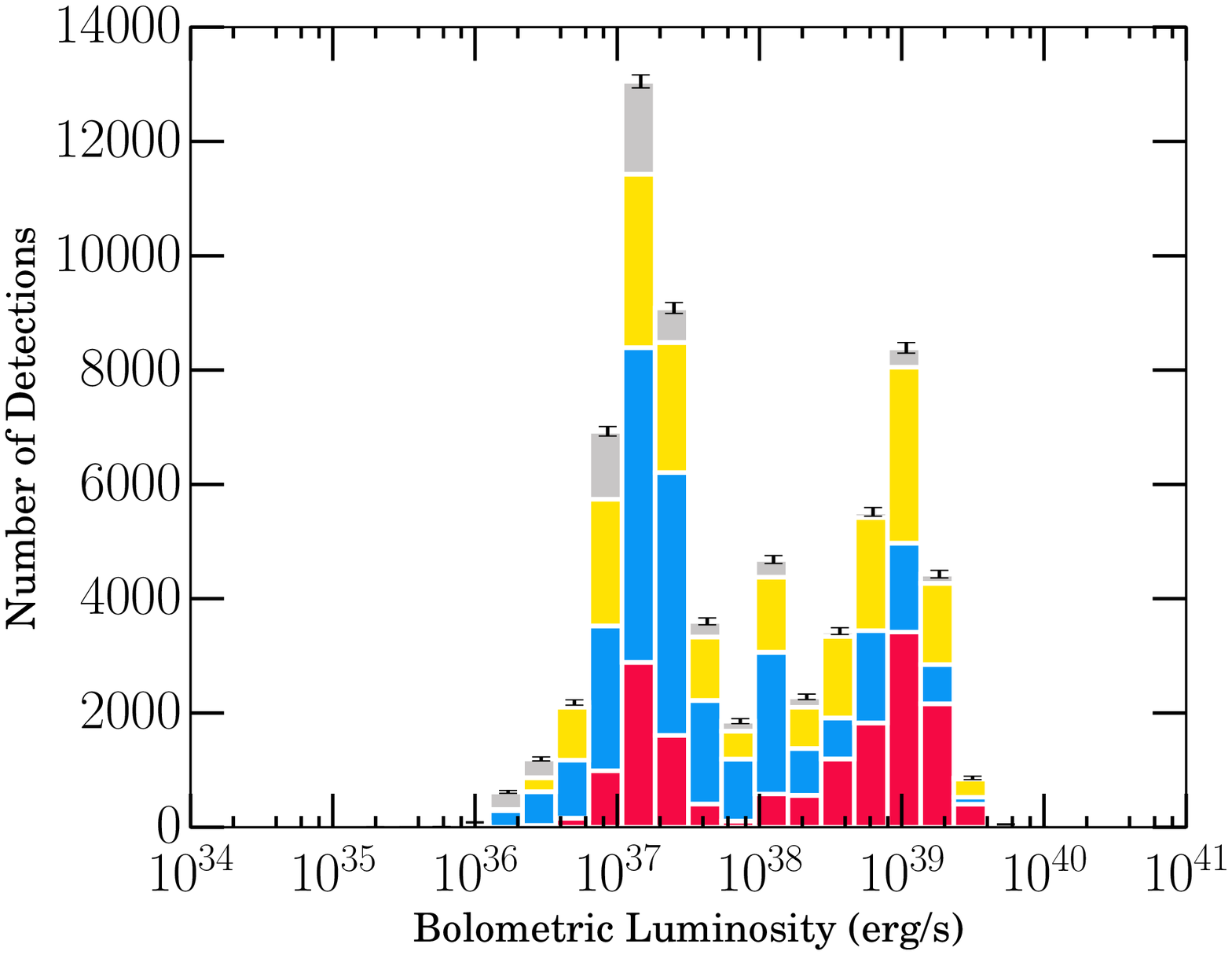}{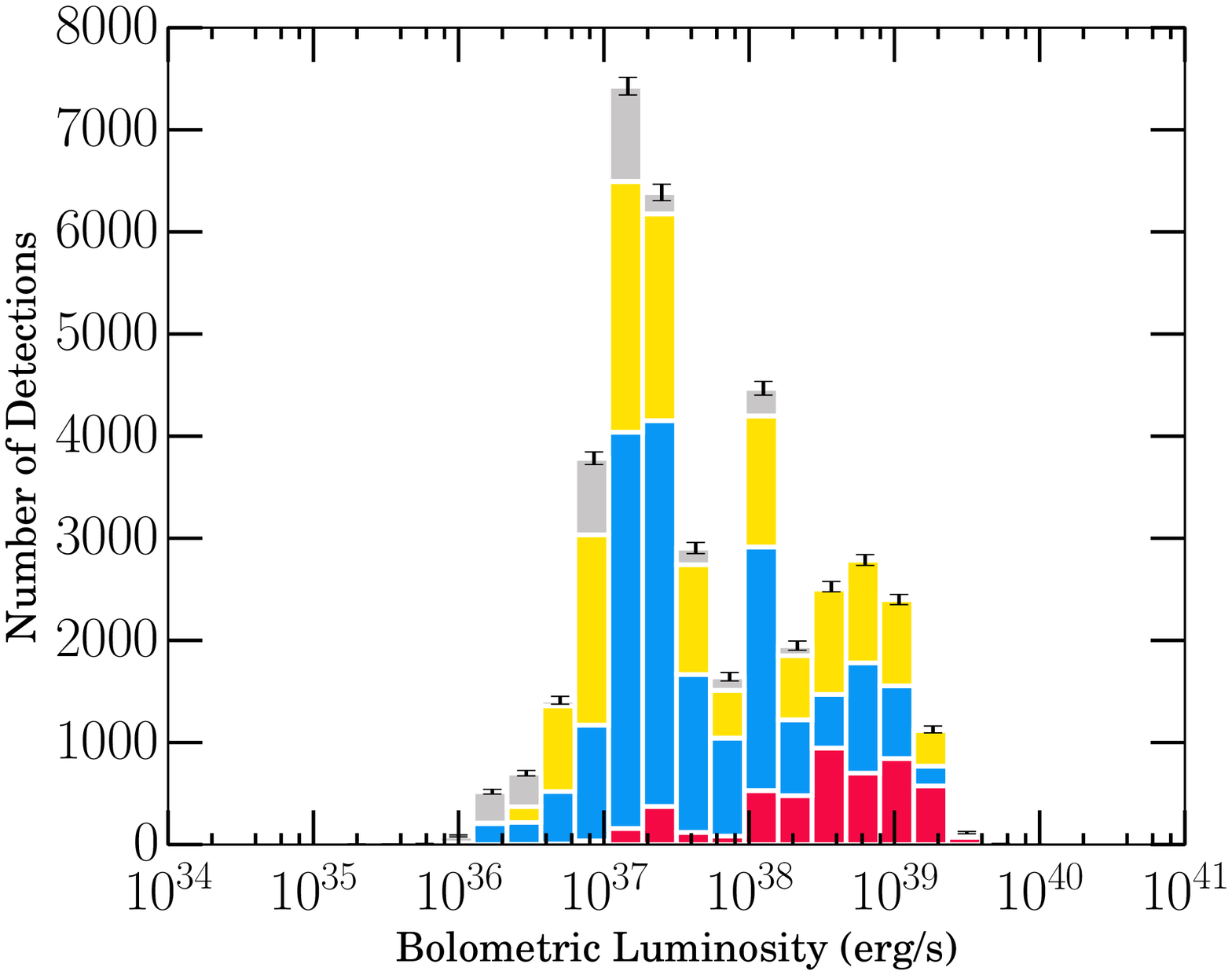}
\plottwo{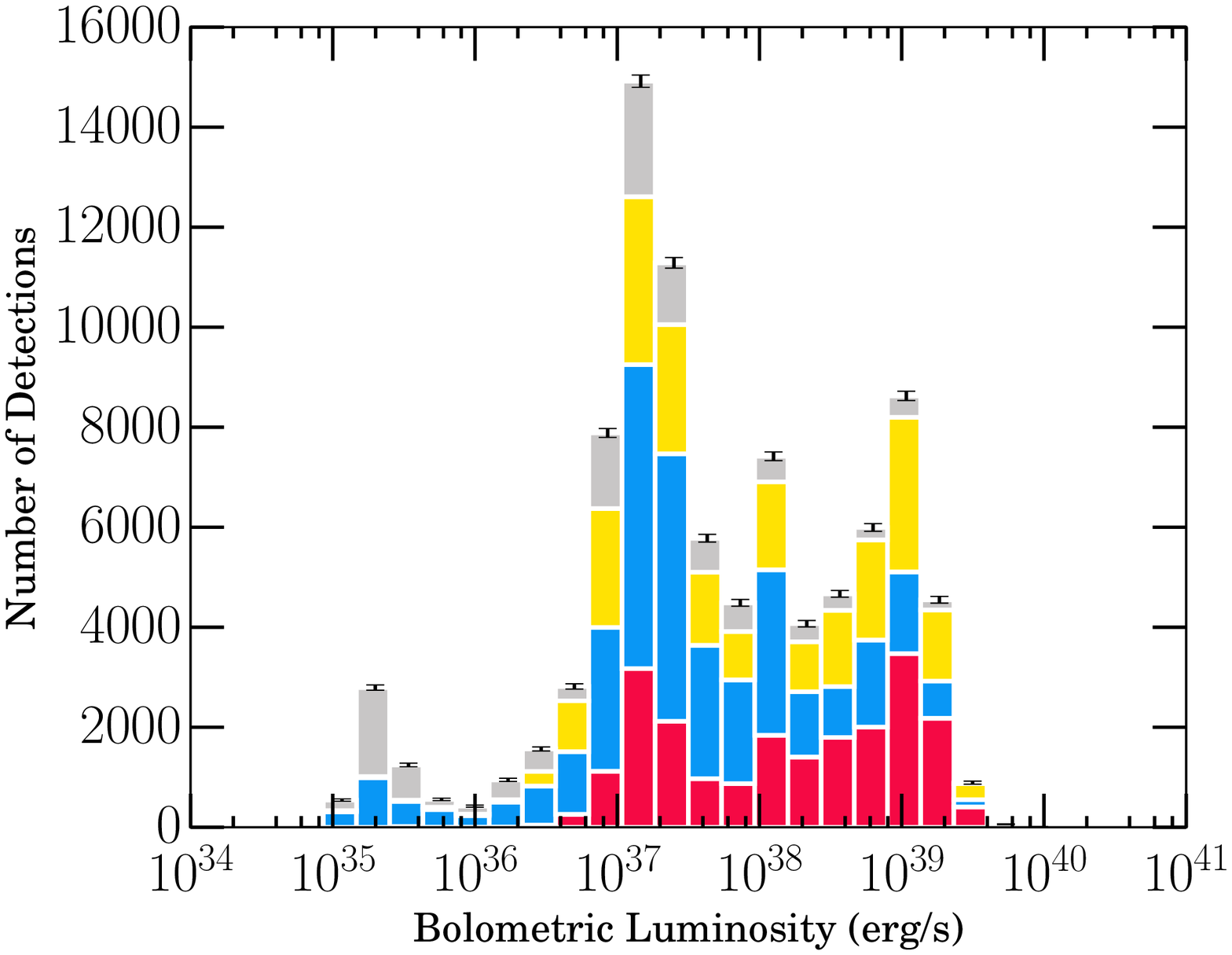}{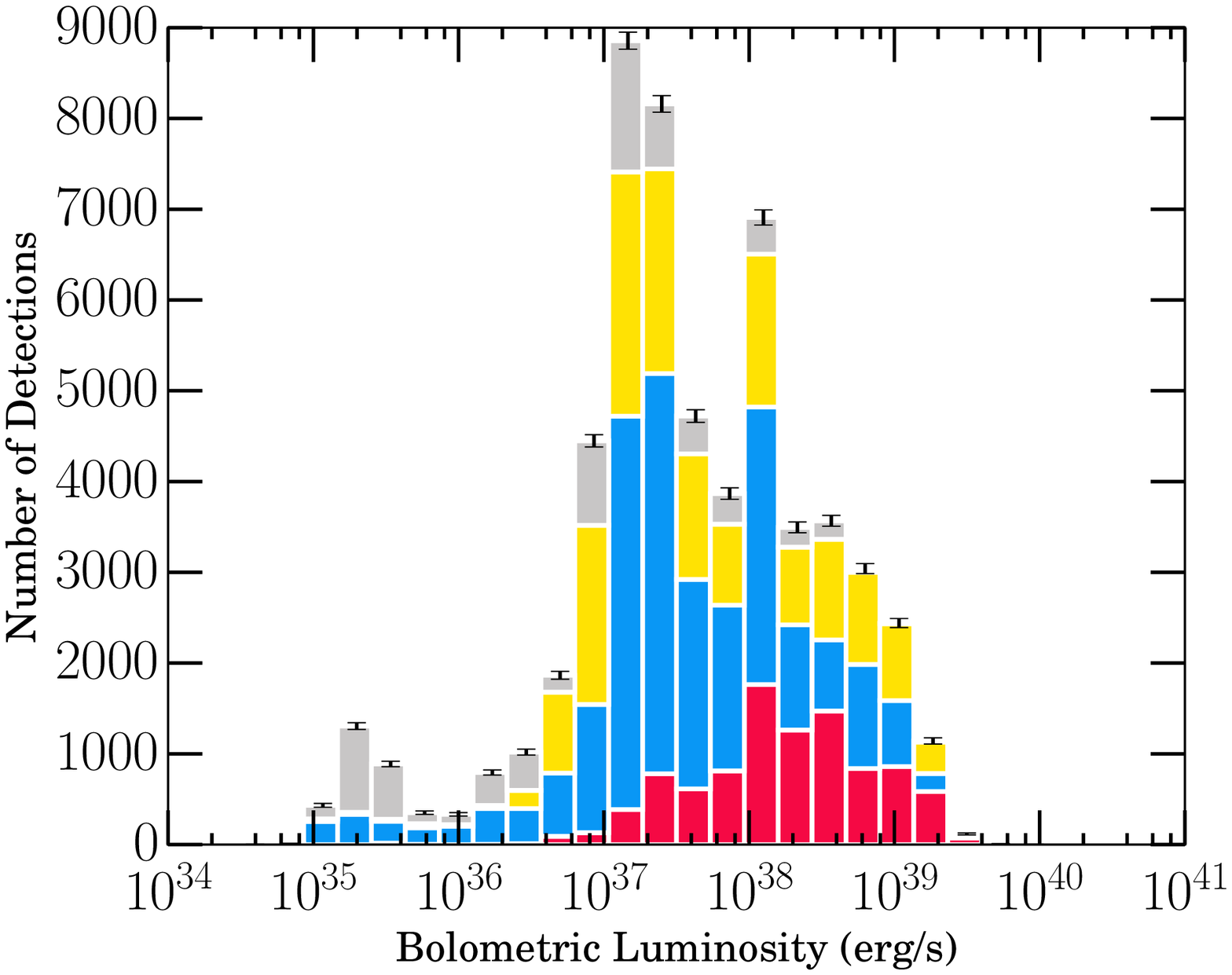}
    \caption{A combined XLF over time for the Galactic transient BHXB population (top), persistent BHXB population (middle), and total Galactic population (bottom). The XLFs in the \textit{left column}  were created from the individual source XLFs of all sources in our sample and the XLFs in the \textit{right column} were created from the individual source XLFs of only those sources that belong to BH classes A or B with distance estimates $<$ 10kpc. Luminosity data have been arranged into 31 bins between $0$ and $10^{41} \, \rm{erg}\, \rm{s}^{-1}$, where any values below $10^{34}\, \rm{erg}\, \rm{s}^{-1}$ are placed in the lowest bin. The 1$\sigma$ Gehrels errors are shown for each bin. Luminosity data have been split into accretion states where red represents the SDS, blue is the HCS, yellow is the IMS, and grey represents days when we were unable to define the state of the source.}%
    \label{fig:poplum1a}%
\end{figure*}

We have also created combined XLFs over time for the transient population, persistent population, and the entire Galactic BHXB population as a whole by combining each individual source XLF (See Figure \ref{fig:poplum1a}) . The population XLFs are rendered for two cases:  (i) including the entire source population in our sample, and (ii) including those sources that belong to BH classes A or B (i.e., securely classified BHs or BHCs) with distance estimates $<$10 kpc. 

In the case of the transient sources, we observe the appearance of a double-peaked profile, a complete deviation from the power-law type distribution found for the entire XRB population of the Galaxy in previous studies (e.g., \citealt{grimm2002}).
We note that even when we only use measurements in a single energy band (as to directly compare to the \citealt{grimm2002} result), this double-peaked profile is still present.
By splitting all the luminosity data into accretion states, we find one soft state contribution peaking at  $\sim10^{38}{\rm erg s^{-1}}$, and two significant hard state contributions, one peaking between $\sim 10^{35} {\rm erg s^{-1}}$ and $\sim 10^{36} {\rm erg s^{-1}}$ and, another that appears to peak only one bin below the maximum of the soft state contribution.
This implies that the bi-modal distribution exhibited by the transient population as a whole may likely be the result of the ``turtlehead'' behavior characteristic of BHXB systems.
We note that that while the small hard state contribution at $>10^{39} {\rm erg s^{-1}}$ may be indicative of a SPL state it may also (at least in part) be due to the failure of spectral fitting algorithm in some cases.

In addition, we note that, despite all that has been discussed, we may also be dealing with incompleteness at lower luminosities due to the limited sensitivity of the all-sky and scanning instruments.
In analyzing the XLFs for the individual transient sources it becomes clear that the limiting luminosity at which the number of detections for a given source begin to fall off is highly dependent on distance.
In the case of nearby sources (e.g., XTEJ1650$-$500 at $\sim$2.6 kpc, GRO J1655$-$40 at $\sim$ 3.2 kpc, and XTEJ1752$-$223 at $\sim3.5$ kpc) we have a significant amount of detections at luminosities as low as a few times $10^{35} \, {\rm erg \, s^{-1}}$. In comparison, detections in Galactic Center sources (e.g., 4U 1543$-$475 at $\sim$7.5 kpc, XTEJ1859+226 at $\sim$8 kpc, and H 1743$-$322 at $\sim$8--10 kpc) appear to fall off at luminosities between $\sim8\times 10^{35}-8 \times10^{36} \, {\rm erg \, s^{-1}}$.
Despite this possible bias, the existence of a bimodal profile in the XLF, as opposed to a single power-law distribution, is still robust.

Lastly, we note that this bi-modal distribution is still present even when we do not include those sources that have undergone outbursts that the spectral modelling algorithm has failed to accurately estimate bolometric luminosity on a select number of days.
This implies that the problem we have with our spectral modelling algorithm is not the cause of the peaked behavior we observe.

For the persistent sources, we also observe a bi-modal profile. 
However, as we are including the LMC sources in our persistent sample, we must take into account the effect distance will have on the total XLF before making any solid claims.
We are not sensitive to the lower luminosity XRBs in the Magellanic Clouds, resulting in our XLF of the LMC being artificially truncated at a few times $10^{38} {\rm erg s^{-1}}$.
As such, a strong argument can be made that the double peaked profile in the persistent sources is the result of the sample being divided into two segments at two distances, and thus two different luminosity cutoffs (at a few times $10^{36} {\rm erg s^{-1}}$ and a few times $10^{38} {\rm erg s^{-1}}$), rather than the result of BHXB hysteresis.
However, when we consider only class A and B systems at $<10$ kpc, the bi-modal profile is still present, suggesting that the distance effect may not in fact be causing the bimodal shape.
In addition, we note that this bimodal profile could also be caused naturally by random clumping of the very small number of persistent sources in our sample, which in general tend to stay roughly at the same X-ray luminosity over time. Lastly, we note that while an argument can be made for the exclusion of SS 433 in this type of population analysis, given its unique behaviour when compared to other persistent XRBs (see Section \ref{subsubsec:SS433s}), we find excluding this source has little impact on the observed bi-model profile, and therefore include the source in the presented population XLFs and note that the luminosity profile of SS 433 presented here may not be the true distribution.

\subsection{Mass-Transfer History}
\label{subsec:mdothist}
We have derived the time averaged bolometric luminosity and long-term mass transfer rates making use of the method described in Section~\ref{subsubsec:lfmdot} for 46 transient sources and 10 persistent sources for which sufficient data is available to us during the time period.
If a source has undergone repeated outbursts in the past 19 years, an $\Dot{M}$ estimate is given in Table \ref{table:mdotsum}. 
If a transient source has only undergone one outburst in the last 19 years, only an upper limit estimate on the $\Dot{M}$ for the source is presented in Table \ref{table:mdotsum} (indicated by a ``$<$'').

The resulting values, coupled with a summary of the prominent outburst behavior, are presented in Table \ref{table:mdotsum}.
In the case of the transient sources, the outburst behavior column indicates whether the source has undergone only ``successful'', only ``hard-only'', or a combination of ``successful'' and ``hard-only'' outbursts over the last 19 years.
In the case of persistent sources, it indicates one of the four long-term behavioral characteristics discussed above in Sections~\ref{subsec:peaklum} and \ref{subsec:xlfs}. 

As discussed in Section \ref{subsubsec:spectra}, there are days during outburst of a source when our spectral fitting algorithm fails, resulting in a large uncertainty in the derived luminosity for that day.
We have identified 47 outbursts in which this problem occurs (see outbursts marked with a ``*'' in Table \ref{table:lumdata}). Of these outbursts, 25 still display the expected trend in disk fraction (soft and hard states corresponding to high and low disk fraction, respectively), only with the addition of a few outliers often occurring when a source is very soft or very hard.
While the other 20 outbursts display a $>50$\% failure rate, the outbursts within this group either
(i) belong to a rapidly recurring transient source (i.e., $\gtrsim10$ outbursts in the last 19 years), or 
(ii) have total outburst periods amounting to $<5$\% of the past 19 years for which we calculate $\Dot{M}$ over.
As such, we believe this problem will not significantly affect our derived long term mass transfer rates, even though this may be a possible source of error that has not been accounted for in our calculations. See Section~\ref{subsec:mdot_porb} for further discussion on the long-term mass transfer rates derived for the Galactic population.

\subsection{Algorithm Data Products} 
\begin{figure*}%
\plotone{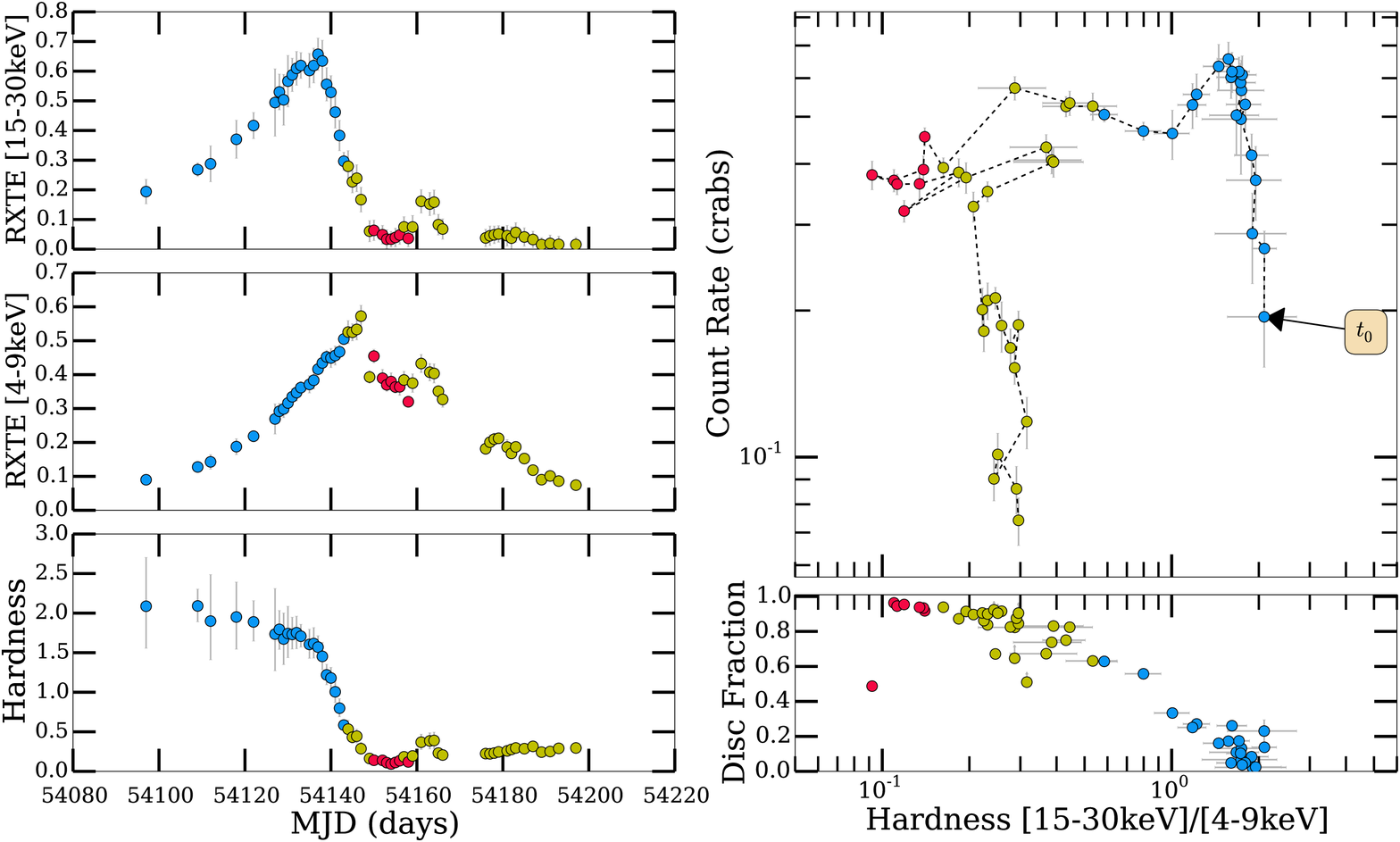}
\caption{Algorithm data product for the 2006/2007 ``successful'' outburst of GX 339$-$4 with RXTE/HEXTE and RXTE/PCA. Colours represent accretion states: HCS (blue), SDS (red) and IMS (yellow).}
    \label{fig:dataproducts}%
    \label{fig:dataproductsa}%
%\end{figure*}
%\begin{figure*}%
\ \\ \ 
\plotone{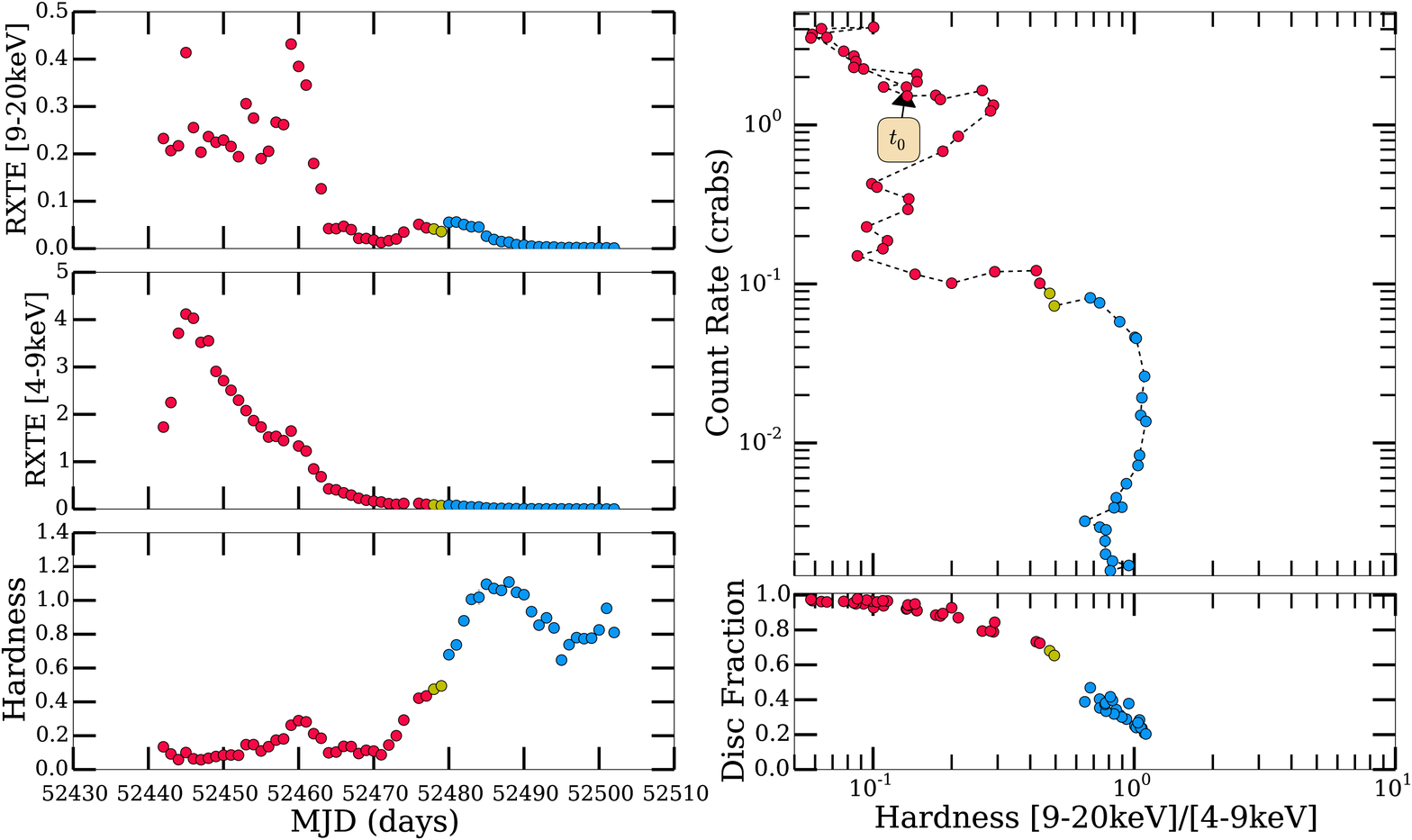}
 \caption{Algorithm data product for the 2002 ``successful'' outburst of 4U 1543$-$475 with RXTE/PCA (bottom). Colours represent accretion states: HCS (blue), SDS (red) and IMS (yellow).}%
    \label{fig:dataproductsb}%
\end{figure*}

\begin{figure*}%
\plotone{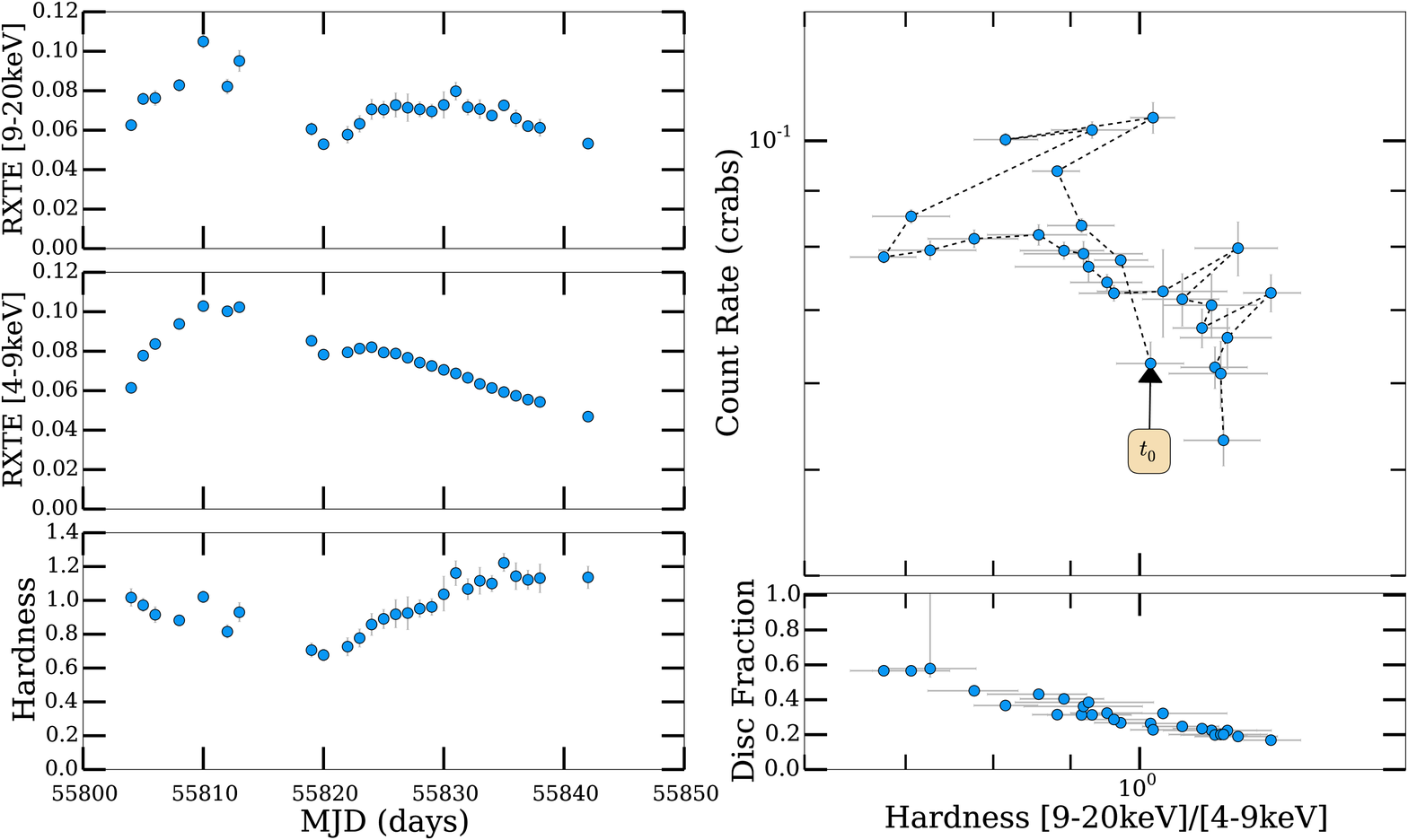}
\caption{Algorithm data product for the 2011/2012 ``hard-only'' outburst of MAXI J1836$-$194 with RXTE/PCA. Colours represent accretion states: HCS (blue), SDS (red) and IMS (yellow). The unusually large disk fraction in the hard state in this figure and Figure~\ref{fig:dataproductfb} indicates a situation where our spectral fitting algorithm has failed. See Section~\ref{subsubsec:spectra} for thorough discussion.}%
    \label{fig:dataproductf}%
    \label{fig:dataproductfa}%
%\end{figure*}
%\begin{figure*}%
\ \\ \ \\ \ \\
\plotone{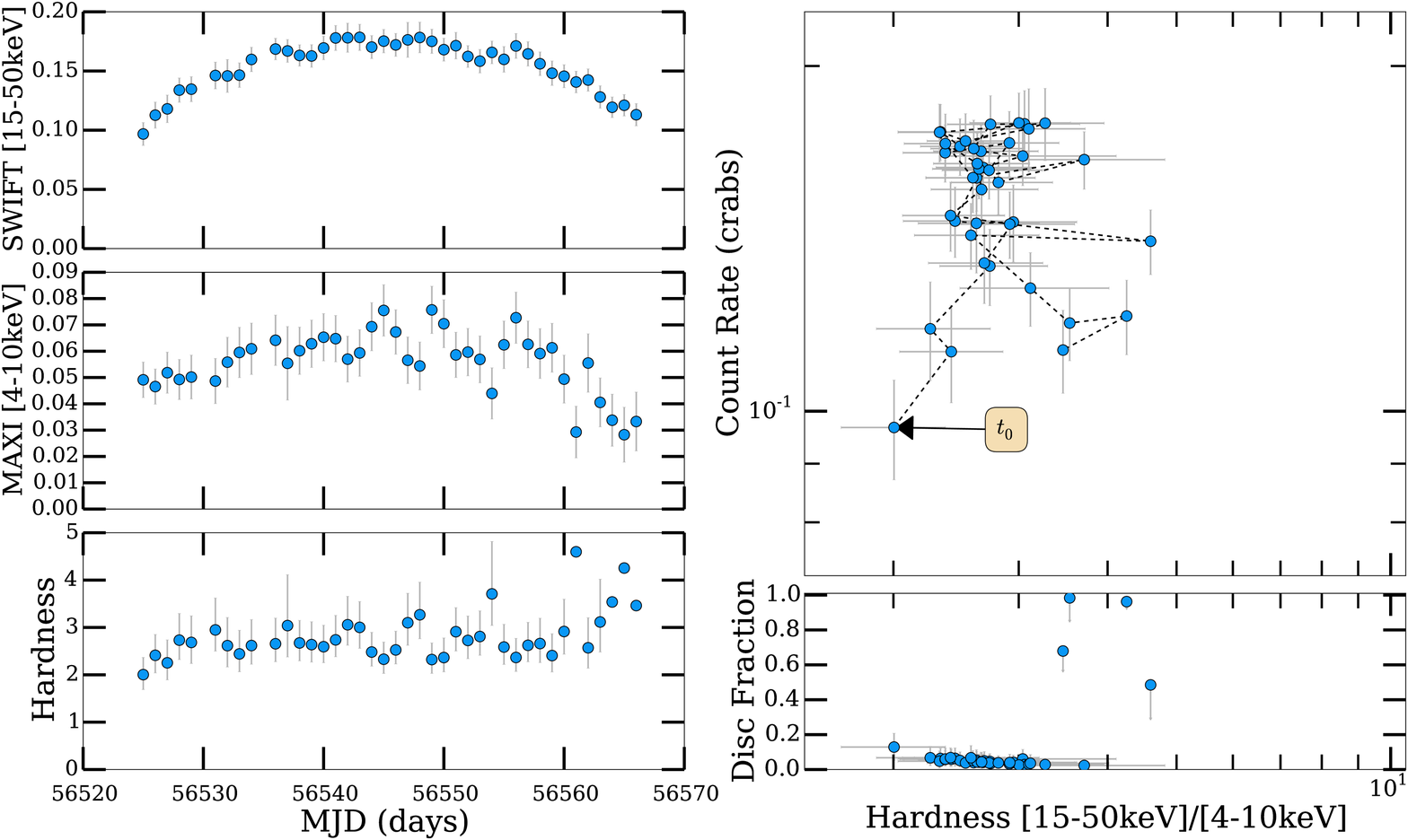}
 \caption{Algorithm data product for the 2013 ``hard-only'' outburst of GX 339$-$4 with Swift/BAT and MAXI/GSC. Colours represent accretion states: HCS (blue), SDS (red) and IMS (yellow).The unusually large disk fraction in the hard state in this figure and Figure~\ref{fig:dataproductfa} indicates a situation where our spectral fitting algorithm has failed. See Section~\ref{subsubsec:spectra} for thorough discussion.}%
     \label{fig:dataproductfb}%
\end{figure*}

\begin{figure*}%
\plotone{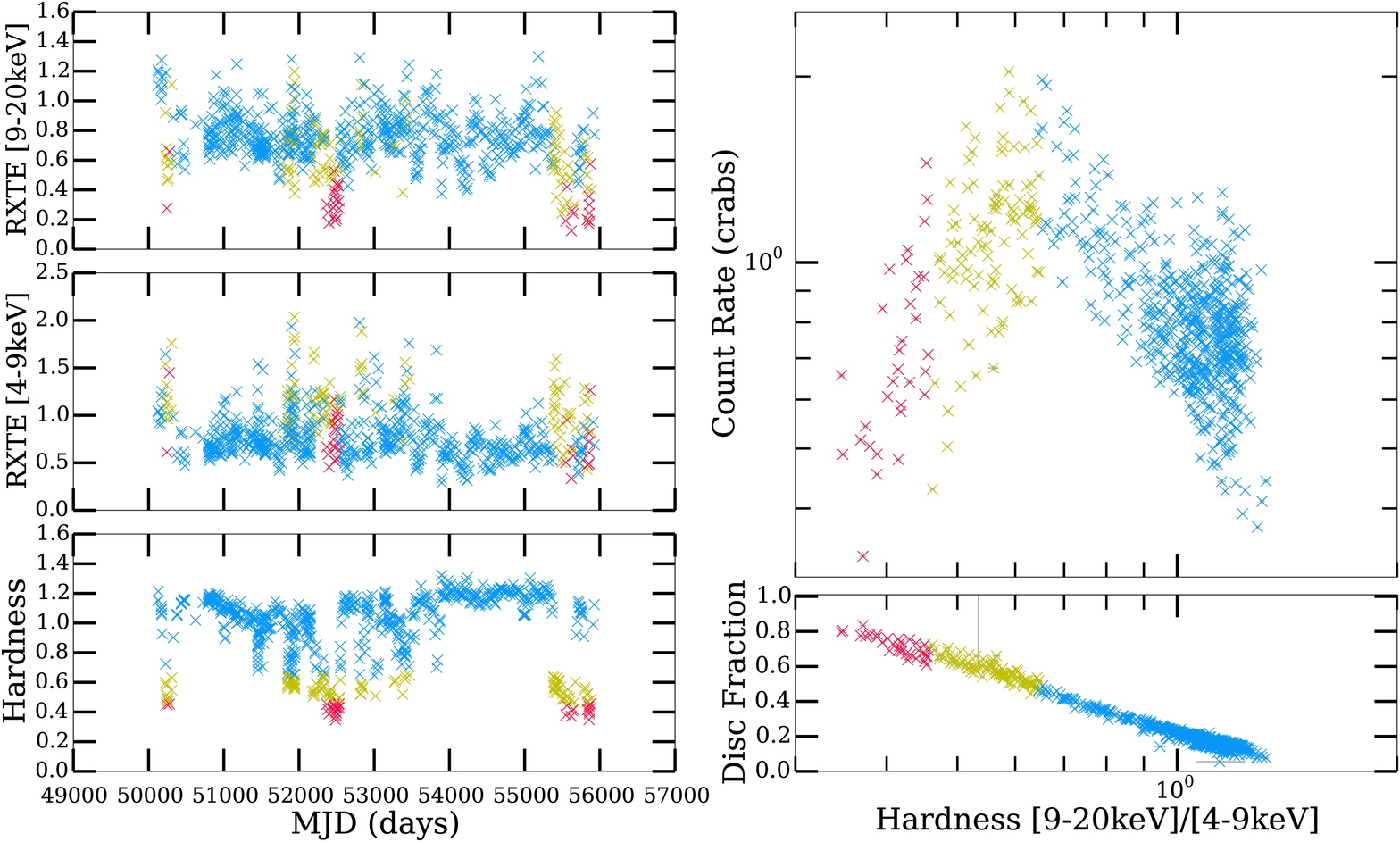}
  \caption{Algorithm data product for the long-term activity of 4U 1956+350 (Cyg X-1) with RXTE/PCA. Colours represent accretion states: HCS (blue), SDS (red) and IMS (yellow).}
    \label{fig:dataproductp}%
    \label{fig:dataproductpa}%
%\end{figure*}
%\begin{figure*}%
\ \\ \ \\ \ \\
\plotone{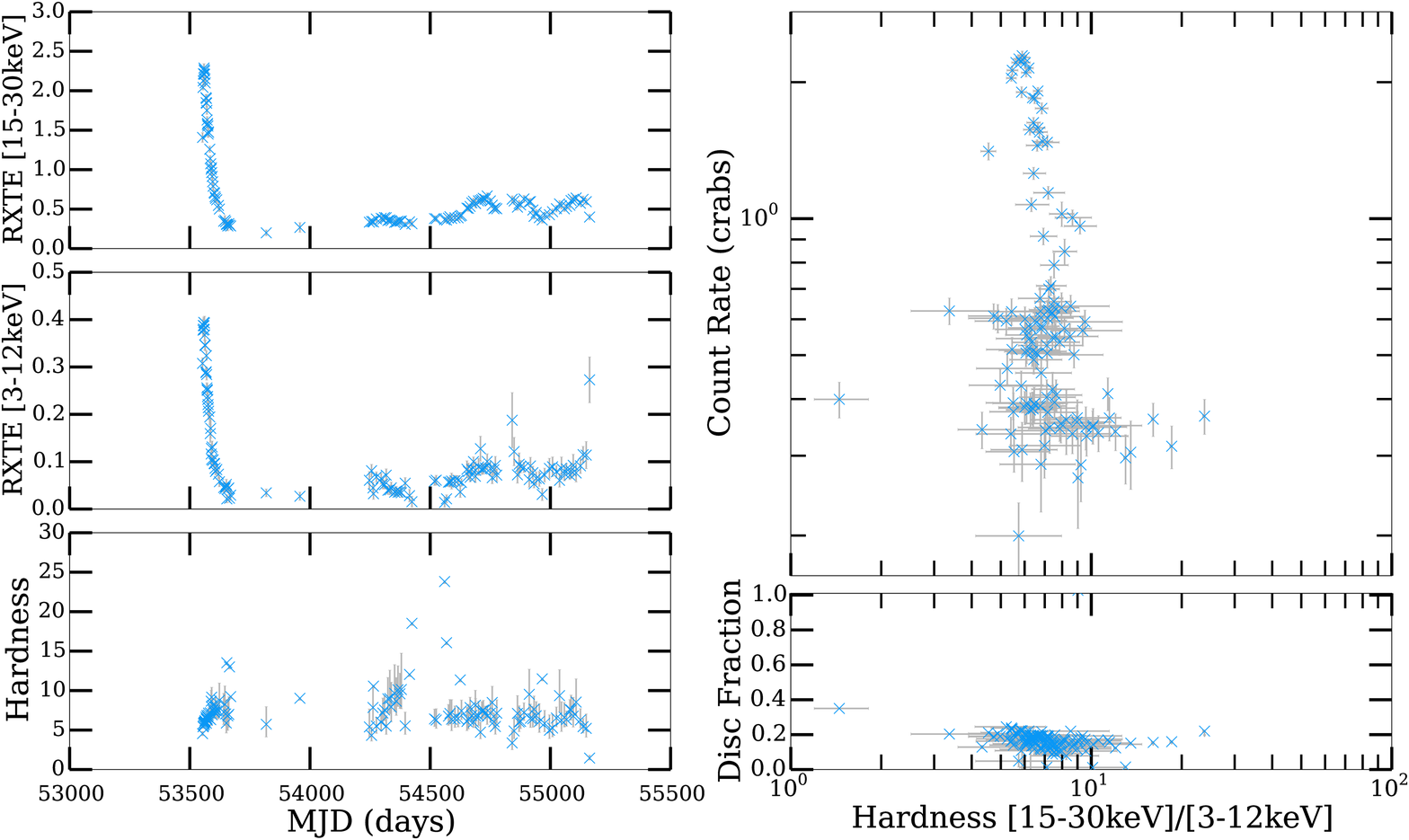}
 \caption{Algorithm data product for the long-term activity of Swift J1753.5$-$0127 with RXTE/HEXTE and RXTE/ASM  between 1996 and 2012. Colours represent accretion states: HCS (blue), SDS (red) and IMS (yellow).}%
    \label{fig:dataproductpb}%
\end{figure*}

\begin{turnpage}
\begin{figure*}%
\epsscale{0.87}
\plotone{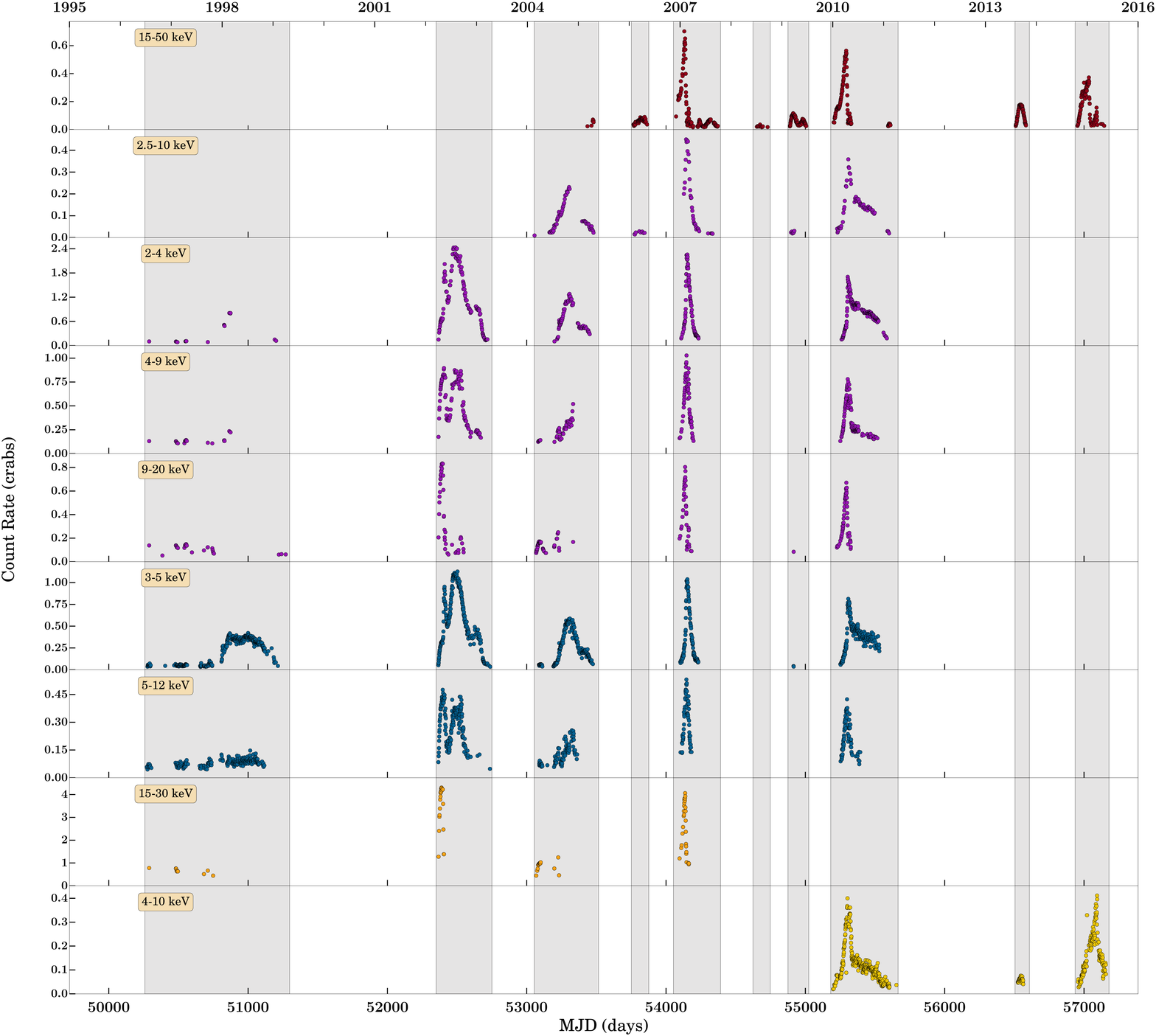}
  \caption{Long-Term Light Curve of the transient BHXB GX 339$-$4. Shaded grey regions span individual outbursts. Colours represent individual instruments: Swift/BAT (red), RXTE/PCA (purple), RXTE/ASM (blue), RXTE/HEXTE (orange), and MAXI/GSC (yellow) from top to bottom.}%
    \label{fig:GXlc}%
\end{figure*}
\end{turnpage}

\begin{turnpage}
\begin{figure}%
\epsscale{0.87}
\plotone{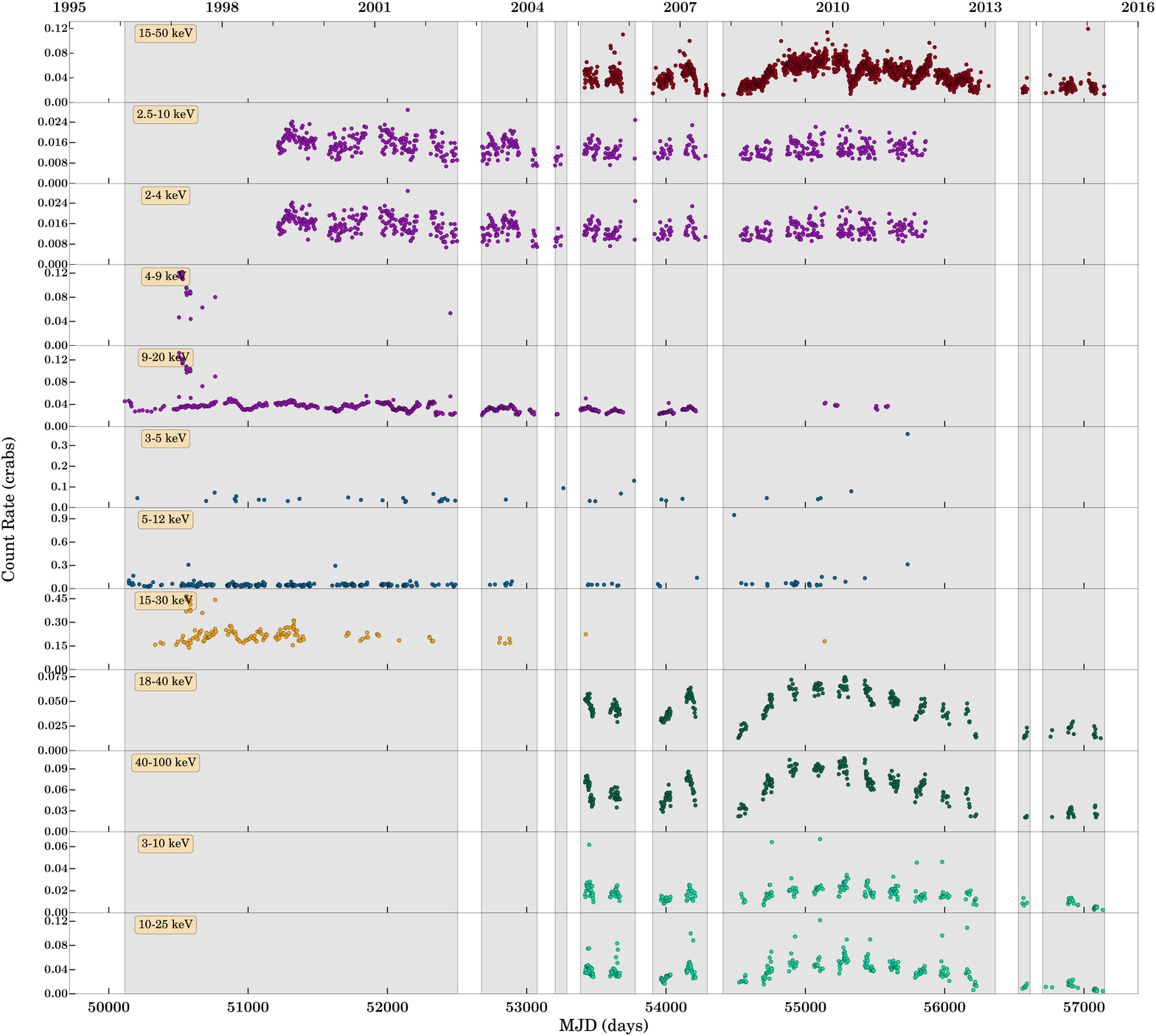}
  \caption{Long-Term Light Curve of the persistent BHXB 1E 1740.7$-$2942. Shaded grey regions span individual outbursts. Colours represent individual instruments: Swift/BAT (red), RXTE/PCA (purple), RXTE/ASM (blue), RXTE/HEXTE (orange), INTEGRAL/ISGRI (dark green), and INTEGRAL/\mbox{JEM-X} (light green) from top to bottom. }%
    \label{fig:1Elc}%
\end{figure}
\end{turnpage}

In Figures \ref{fig:dataproducts}--\ref{fig:1Elc} we present a sample of the detailed data products, created via the algorithm.
The full set of data products is available on the WATCHDOG website (http://astro.physics.ualberta.ca/WATCHDOG/).
The sample includes example data products detailing two ``successful'' and two ``hard-only'' transient outbursts, and the long-term activity of two persistent sources.
These data products come in two separate forms: long-term light curves depicting the structure of the outbursts undergone in each available energy band and an analysis package for each detected transient outburst or period of long-term persistent activity.
The long-term light curves have been color coded by instrument, with INTEGRAL/ISGRI in dark green, INTEGRAL/JEM-X in light green, MAXI/GSC in yellow, RXTE/ASM in blue, RXTE/HEXTE in orange, RXTE/PCA in purple and, Swift/BAT in red. The outburst analysis package presented 
includes the following:
(i) individual hard and soft band light curves for the outburst;
(ii) the evolution of $H_{X}$ over the duration of the outburst;
(iii) the complete HID of the outburst; and
(iv) the evolution of disk fraction $d_{f}$ as a function of $H_{X}$ throughout the outburst.
Note that in the HIDs the ``intensity'' on any particular day is the maximum of the hard and soft bands and that within the analysis package each data point is color coded with respect to the accretion state the system was in on that particular day (see Section~\ref{subsubsec:classify} for details), where blue represents the HCS, red represents the SDS, and yellow represents the IMS.

\section{Discussion}

\subsection{The relationship between average mass transfer rate and orbital period}
\label{subsec:mdot_porb}

\tabletypesize{\scriptsize}
\renewcommand{\thefootnote}{\alph{footnote}}
\renewcommand\tabcolsep{15pt}
\begin{longtable*}{lcccl}

\caption{\centerline{ Mass Transfer History of the Galactic BHXB Population from 1996--2015}}  \\

\hline \hline \\[-2ex]
   \multicolumn{1}{l}{Source Name} &
   \multicolumn{1}{c}{Source$^a$} &
     \multicolumn{1}{c}{${<L_{\rm Bol}>}_{ t}$$^b$} &
     \multicolumn{1}{c}{${<\dot{ M}>}_{ t}$$^c$} &
    \multicolumn{1}{l}{Outburst}  \\
      &\multicolumn{1}{c}{Type} & \multicolumn{1}{c}{$(\times 10^{36}$ ergs/s)} & \multicolumn{1}{c}{$(\times 10^{-9} \, M_{\odot}/\rm{yr}$)}&\multicolumn{1}{l}{Behaviour}\\[0.5ex] \hline
   \\[-4ex]
\label{table:mdotsum}
\endfirsthead
\multicolumn{5}{c}{{\tablename} \thetable{} -- Continued} \\[0.5ex]
\hline \hline \\[-2ex]
   \multicolumn{1}{l}{Source Name} &
   \multicolumn{1}{c}{Source$^a$} &
     \multicolumn{1}{c}{${<L_{\rm Bol}>}_{ t}$$^b$} &
     \multicolumn{1}{c}{${<\dot{ M}>}_{ t}$$^c$} &
    \multicolumn{1}{l}{Outburst}  \\
      &\multicolumn{1}{c}{Type} & \multicolumn{1}{c}{$(\times 10^{36}$ ergs/s)} & \multicolumn{1}{c}{$(\times 10^{-9} \, M_{\odot}/\rm{yr}$)}&\multicolumn{1}{l}{Behaviour}\\[0.5ex] \hline
   \\[-4ex]
\endhead
  \\ \hline \\[-1.8ex]  
             \multicolumn{5}{p{0.8\columnwidth}}{\hangindent=1ex NOTE 1.-- This table is also available in machine readable format online at the Astrophysical Journal and on the WATCHDOG website - http://astro.physics.ualberta.ca/WATCHDOG/.}\\
      \multicolumn{5}{p{0.8\columnwidth}}{\hangindent=1ex $^a$Indicates whether the source type is transient (T) or persistent (P). }\\
        \multicolumn{5}{p{0.8\columnwidth}}{\hangindent=1ex $^b$The time averaged bolometric luminosity calculated between 1996 January 6 -- 2015 May 14.} \\
          \multicolumn{5}{p{0.8\columnwidth}}{\hangindent=1ex $^c$The average mass-transfer rate calculated between 1996 January 6 -- 2015 May 14. If a source has only one detected outburst, the $\Dot{M}$ is considered an upper limit (indicated by $<$).} \\
\endlastfoot
XTEJ0421+560 & T & ${0.2578}^{+0.0035}_{-0.0033}$ & $<{0.04552}^{+0.00062}_{-0.00059}$ & hard-only\\[0.045cm]
4U0538$-$641 & P & ${1134}^{+13.97}_{-10}$ & ${200.2}^{+2.5}_{-1.8}$ & ``turtlehead''/anomalous high $L_X$\\[0.045cm]
4U0540$-$697 & P & ${1290}^{+172}_{-140}$ & ${229}^{+30.4}_{-25}$ & ``turtlehead''/anomalous high $L_X$\\[0.045cm]
XTEJ1118+480 & T & $0.0447 \pm 0.0013$ & ${0.00790}^{+0.00023}_{-0.00022}$ & hard-only\\[0.045cm]
MAXIJ1305$-$704 & T & ${3.71}^{+0.46}_{-0.32}$ & $<{0.654}^{+0.081}_{-0.057}$ & undefined\\[0.045cm]
SWIFTJ1357.2$-$0933 & T & ${0.202}^{+0.055}_{-0.042}$ & $<{0.0356}^{+0.0097}_{-0.0075}$ & hard-only\\[0.045cm]
GS1354$-$64 & T & ${37.6}^{+1.8}_{-2.0}$ & $<{6.64}^{+0.33}_{-0.35}$ & hard-only\\[0.045cm]
SWIFTJ1539.2$-$6227 & T & ${0.947}^{+0.080}_{-0.062}$ & $<{0.167}^{+0.014}_{-0.011}$ & successful only\\[0.045cm]
MAXIJ1543$-$564 & T & ${1.18}^{+0.16}_{-0.12}$ & $<{0.208}^{+0.028}_{-0.021}$ & successful only\\[0.045cm]
4U1543$-$475 & T & ${4.345}^{+0.030}_{-0.034}$ & $<{0.7669}^{+0.0053}_{-0.0060}$ & successful only\\[0.045cm]
XTEJ1550$-$564 & T & $12.00 \pm 0.48$ & ${2.119}^{+0.085}_{-0.084}$ & combined\\[0.045cm]
4U1630$-$472 & T & ${40.34}^{+0.76}_{-0.66}$ & ${7.12}^{+0.13}_{-0.12}$ & successful only\\[0.045cm]
XTEJ1637$-$498 & T & ${0.00085}^{+0.00030}_{-0.00027}$ & $<{0.000150}^{+0.000052}_{-0.000048}$ & undefined\\[0.045cm]
XTEJ1650$-$500 & T & ${0.3147}^{+0.0044}_{-0.0050}$ & $<{0.05556}^{+0.00078}_{-0.00089}$ & successful only\\[0.045cm]
XTEJ1652$-$453 & T & ${1.16}^{+0.18}_{-0.12}$ & $<{0.205}^{+0.031}_{-0.021}$ & successful only\\[0.045cm]
GROJ1655$-$40 & T & ${16.185}^{+0.072}_{-0.067}$ & ${2.857}^{+0.013}_{-0.012}$ & successful only\\[0.045cm]
MAXIJ1659$-$152 & T & ${2.00}^{+0.13}_{-0.12}$ & $<{0.354}^{+0.023}_{-0.021}$ & successful only\\[0.045cm]
GX339$-$4 & T & ${87.6}^{+1.8}_{-1.4}$ & ${15.46}^{+0.31}_{-0.24}$ & combined\\[0.045cm]
IGRJ17091$-$3624 & T & ${3.36}^{+0.30}_{-0.26}$ & ${0.594}^{+0.053}_{-0.046}$ & successful only\\[0.045cm]
SAXJ1711.6$-$3808 & T & ${0.444}^{+0.031}_{-0.029}$ & $<{0.0783}^{+0.0055}_{-0.0051}$ & hard-only\\[0.045cm]
SWIFTJ1713.4$-$4219 & T & ${0.00230}^{+0.00012}_{-0.00014}$ & $<{0.000406}^{+0.000022}_{-0.000025}$ & undefined\\[0.045cm]
XMMSL1J171900.4$-$353217 & T & ${0.0670}^{+0.0018}_{-0.0019}$ & ${0.01183}^{+0.00032}_{-0.00033}$ & hard-only\\[0.045cm]
XTEJ1720$-$318 & T & ${0.750}^{+0.053}_{-0.052}$ & $<{0.1324}^{+0.0093}_{-0.0092}$ & successful only\\[0.045cm]
XTEJ1727$-$476 & T & ${0.1170}^{+0.0069}_{-0.0071}$ & $<{0.0207}^{+0.0012}_{-0.0013}$ & successful only\\[0.045cm]
IGRJ17285$-$2922 & T & ${0.039}^{+0.037}_{-0.038}$ & ${0.0069}^{+0.0066}_{-0.0068}$ & hard-only\\[0.045cm]
IGRJ17379$-$3747 & T & ${0.037}^{+0.027}_{-0.021}$ & ${0.0065}^{+0.0048}_{-0.0036}$ & hard-only\\[0.045cm]
GRS1737$-$31 & T & ${0.358}^{+0.012}_{-0.013}$ & $<{0.0632}^{+0.0021}_{-0.0022}$ & hard-only\\[0.045cm]
GRS1739$-$278 & T & ${4.769}^{+0.029}_{-0.031}$ & ${0.8419}^{+0.0051}_{-0.0054}$ & successful only\\[0.045cm]
1E1740.7$-$2942 & P & ${55.6}^{+3.9}_{-3.1}$ & ${9.81}^{+0.70}_{-0.55}$ & mainly HCS \\[0.045cm]
SWIFTJ174510.8$-$262411 & T & ${1.7708}^{+0.0040}_{-0.0041}$ & $<{0.31259}^{+0.00070}_{-0.00072}$ & hard-only\\[0.045cm]
IGRJ17454$-$2919 & T & ${0.0119}^{+0.0014}_{-0.0012}$ & $<{0.00211}^{+0.00024}_{-0.00020}$ & undefined\\[0.045cm]
H1743$-$322 & T & ${85.1}^{+5.8}_{-4.3}$ & ${15.03}^{+1.03}_{-0.77}$ & combined\\[0.045cm]
XTEJ1748$-$288 & T & ${0.866}^{+0.022}_{-0.023}$ & $<{0.1528}^{+0.0039}_{-0.0040}$ & hard-only\\[0.045cm]
IGRJ17497$-$2821 & T & ${0.562}^{+0.113}_{-0.083}$ & $<{0.099}^{+0.020}_{-0.015}$ & combined\\[0.045cm]
SLX1746$-$331 & T & $1.783 \pm 0.046$ & $0.3147 \pm 0.0082$ & successful only\\[0.045cm]
XTEJ1752$-$223 & T & ${1.411}^{+0.088}_{-0.073}$ & $<{0.249}^{+0.015}_{-0.013}$ & successful only\\[0.045cm]
SWIFTJ1753.5$-$0127 & P & ${13.50}^{+0.49}_{-0.38}$ & $<{2.383}^{+0.086}_{-0.068}$ & combined\\[0.045cm]
XTEJ1755$-$324 & T & $0.398 \pm 0.019$ & $<{0.0703}^{+0.0033}_{-0.0034}$ & successful only\\[0.045cm]
GRS1758$-$258 & P & ${63.9}^{+4.9}_{-3.5}$ & ${11.28}^{+0.87}_{-0.61}$ & mainly IMS \\[0.045cm]
XTEJ1812$-$182 & T & ${0.384}^{+0.046}_{-0.038}$ & ${0.0678}^{+0.0081}_{-0.0066}$ & combined\\[0.045cm]
IGRJ18175$-$1530 & T & ${0.0073}^{+0.0032}_{-0.0028}$ & $<{0.00129}^{+0.00057}_{-0.00050}$ & undefined\\[0.045cm]
XTEJ1817$-$330 & T & ${2.02}^{+0.20}_{-0.16}$ & $<{0.356}^{+0.036}_{-0.027}$ & successful only\\[0.045cm]
XTEJ1818$-$245 & T & ${0.175}^{+0.095}_{-0.081}$ & $<{0.031}^{+0.017}_{-0.014}$ & successful only\\[0.045cm]
SAXJ1819.3$-$2525 & T & ${13.10}^{+0.42}_{-0.36}$ & ${2.312}^{+0.074}_{-0.064}$ & combined\\[0.045cm]
MAXIJ1836$-$194 & T & ${5.91}^{+0.62}_{-0.50}$ & $<{1.043}^{+0.110}_{-0.088}$ & hard-only\\[0.045cm]
SWIFTJ1842.5$-$1124 & T & ${2.06}^{+0.55}_{-0.48}$ & $<{0.363}^{+0.098}_{-0.084}$ & successful only\\[0.045cm]
XTEJ1856+053 & T & ${1.310}^{+0.088}_{-0.069}$ & ${0.231}^{+0.015}_{-0.012}$ & successful only\\[0.045cm]
XTEJ1859+226 & T & ${4.26}^{+0.15}_{-0.14}$ & $<{0.752}^{+0.026}_{-0.025}$ & successful only\\[0.045cm]
XTEJ1908+094 & T & ${7.09}^{+1.01}_{-0.96}$ & ${1.25}^{+0.18}_{-0.17}$ & successful only\\[0.045cm]
SWIFTJ1910.2$-$0546 & T & ${2.92}^{+0.33}_{-0.26}$ & $<{0.516}^{+0.058}_{-0.047}$ & successful only\\[0.045cm]
SS433 & P & ${26.6}^{+2.0}_{-1.6}$ & ${4.70}^{+0.35}_{-0.29}$ & mainly HCS \\[0.045cm]
GRS1915+105 & P & ${899.41}^{+0.47}_{-0.55}$ & $<{158.767}^{+0.083}_{-0.098}$ & ``turtlehead''/anomalous high $L_X$\\[0.045cm]
4U1956+350 & P & ${21.965}^{+0.014}_{-0.018}$ & ${3.8774}^{+0.0024}_{-0.0032}$ & mainly HCS\\[0.045cm]
4U1957+115 & P & ${16.06}^{+0.20}_{-0.14}$ & ${2.835}^{+0.035}_{-0.025}$ & mainly SDS\\[0.045cm]
XTEJ2012+381 & T & $0.506 \pm 0.014$ & $<0.0893 \pm 0.0025$ & successful only\\[0.045cm]
4U2030+40 & P & ${206.05}^{+1.00}_{-0.74}$ & ${36.37}^{+0.18}_{-0.13}$ & ``turtlehead''/anomalous high $L_X$\\[-1.84ex]
\end{longtable*}
\vspace*{1ex} \  \\
\vspace*{1ex}
\renewcommand\tabcolsep{5pt}
\renewcommand{\thefootnote}{\alph{footnote}}


\clearpage

\clearpage

\begin{figure*}[h]
\epsscale{0.87}
\plotone{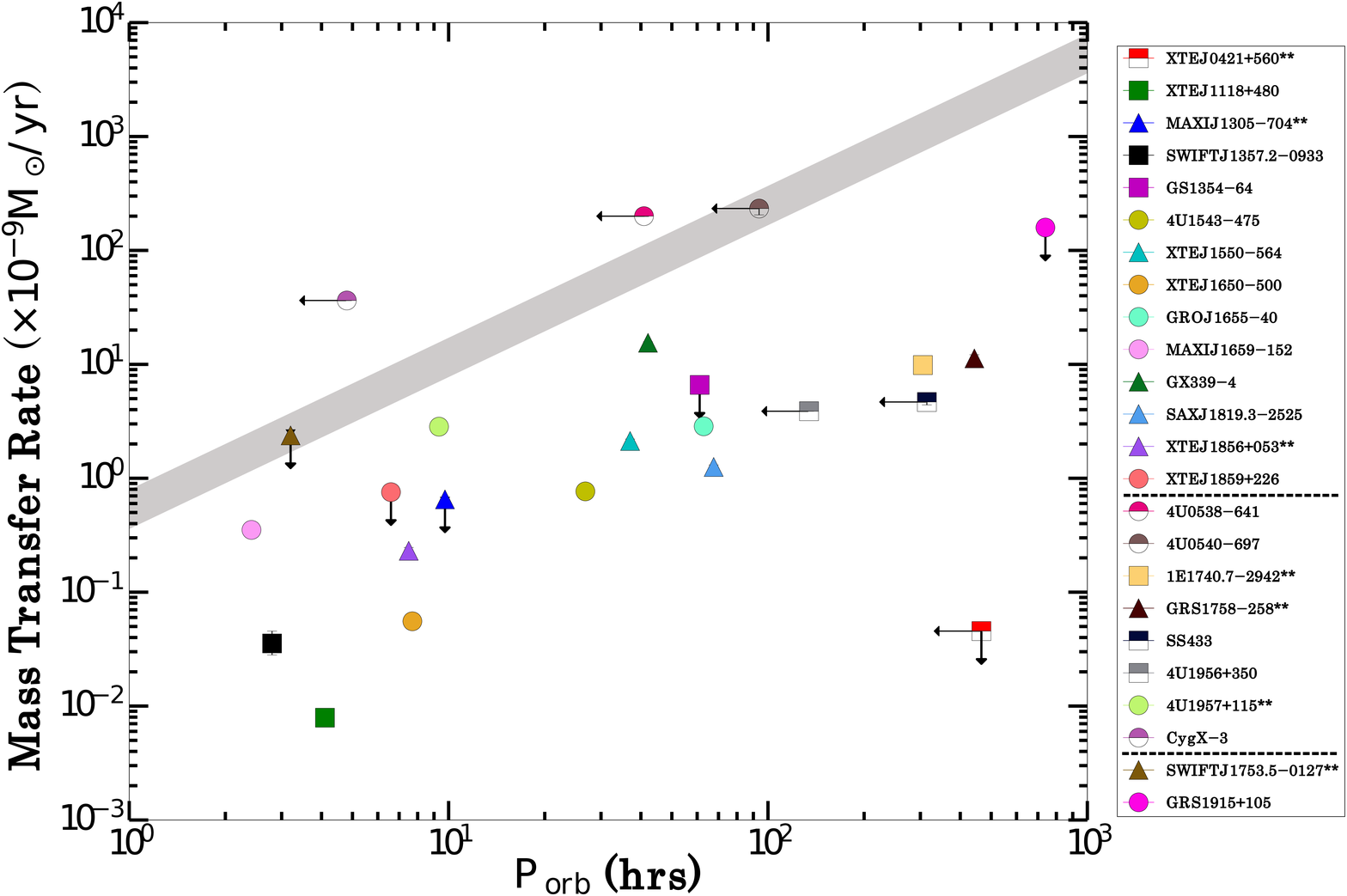}
\ \\ \ \\
\epsscale{0.87}
\plotone{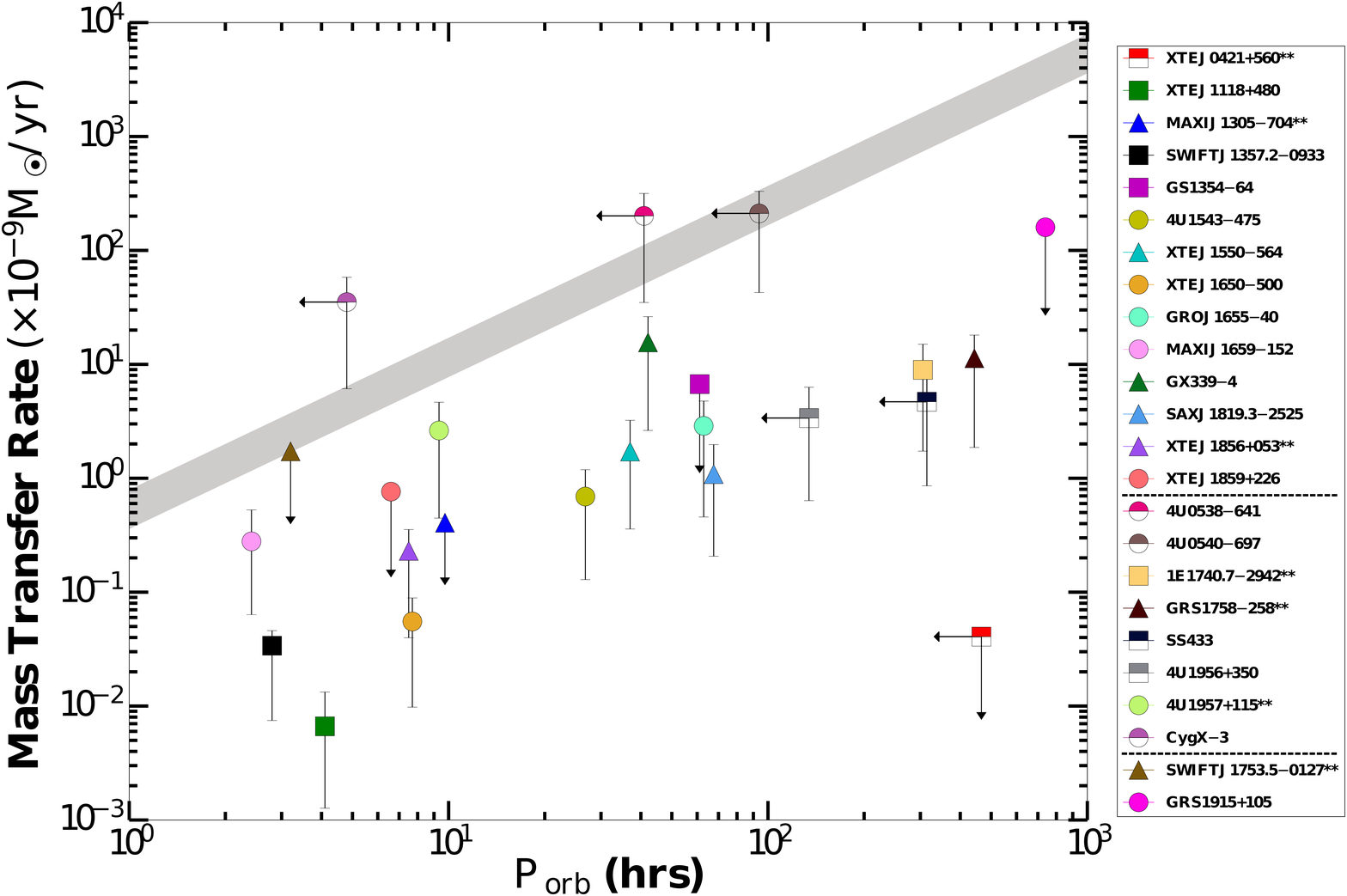}
\caption{{Long term mass-transfer rates vs.\ orbital period for Galactic BH/BHC systems. Colours represent individual sources (see legend). Sources without known distance estimates are indicated by a ``**'' symbol in the legend. The legend is split by system type: transient, persistent, and long-term transient. LMXBs are represented by filled shapes and HMXBs are represented by shapes with colored hats. The shaded grey region plots the critical accretion rate for an irradiated disk around a 5--$15 \, M_{\odot}$ BH according to the disk instability model with irradiation, providing a theoretical distinction between transient (below) and persistent (above) systems. The mass-transfer rate estimates of sources that have undergone one outburst are denoted as upper limits. Leftward facing arrows on the HMXBs are not upper limits on $P_{\rm orb}$. They indicate that these systems transfer mass via a stellar wind, resulting in the radius of their accretion disks likely being smaller than the one would expect at their orbital period if they were Roche lobe overflow sources. Shape denotes behavior: exclusively ``hard-only'' or mainly HCS (squares), exclusively ``successful'' or long-term ``turtlehead'' pattern (circles), combination ``successful'' and ``hard-only'' or exhibiting mainly IMS and/or incomplete state transition behavior  (triangles). (\textit{Top}) Mass-transfer rates calculated assuming a fixed accretion efficiency $\eta=0.1$. Error bars are too small to see. (\textit{Bottom}) Mass-transfer rates calculated assuming the full theoretical range of accretion efficiency, $\eta\sim0.06-0.40$. Errors are quoted to the 1$\sigma$ confidence level. }}

\label{figure:mdotporb}
\end{figure*}

We have analyzed the relationship between $\Dot{M}$ and $P_{\rm orb}$ for the Galactic population. Figure \ref{figure:mdotporb} presents the $(P_{\rm orb}, \Dot{M})$ diagram for the 14 transient and 10 persistently accreting sources (note that we classify the two long-term transient sources, GRS 1915+105 and Swift J1753.5$-$0127, as persistent) from our sample for which the orbital period is known (see Section 2 or Table \ref{table:binaryBH} for references). 
This plot distinguishes between (i) LMXBs (filled shapes) and HMXBs (shapes with colored hats), (ii) outburst behavior by shape: exclusively ``hard-only'' or mainly HCS (squares), exclusively ``successful'' or long-term ``turtlehead'' pattern (circles), combination ``successful'' and ``hard-only'' or exhibiting mainly IMS (incomplete state transition) behavior  (triangles), (iii) sources that have undergone multiple outbursts versus those with only lower limits on recurrence time (upper limit arrows) and (iv) sources with and without known distance estimates (the later indicated by ** in legend).

We note that all our $\Dot{M}$ estimates are in agreement with previous work down by \citet{coriat2012}, with the exception of XTE J1118+480 and GS 1354$-$64. In the case of XTE J1118+480, the discrepancy is most likely the result of \citet{coriat2012} using a different accretion efficiency prescription when $L_X<0.01 \, L_{\rm edd}$, a luminosity regime in which this source remains in throughout outburst (i.e., $L_{\rm peak}\sim0.004 \, L_{\rm edd}$). In the case of GS 1354$-$64, the discrepancy is due to the use of a different distance estimate. While we use the full 25--61 kpc range, \citet{coriat2012} use the lower limit of 25 kpc.

Generally, theory predicts a correlation, where a larger $P_{\rm orb}$ should correspond to a larger $\Dot{M}$ \citep{pod02}. This expectation is based on the predictions of angular momentum loss mechanisms, magnetic braking for relatively short-orbit systems \citep{verbunt81}, and nuclear evolution timescales for longer-orbit systems \citep{webbink83}. See \citet{king88} and \citet{king1996} for a review on the mechanisms driving mass transfer in these binary systems. 

However, we observe the appearance of numerous outliers and a great deal of scatter, in both transient and persistent sources, implying that the mass transfer rates presented here may in fact be systematically mis (under) estimating the true mass transfer from the companion.

First, we observe that the exclusively ``hard-only'' outburst transient sources appear to have significantly lower average 
$\Dot{M}_{\rm BH}$ then those sources that exhibit exclusively ``successful'' or a mix of ``successful'' and ``hard-only'' behavior. Performing a two sample KS-test between (i) the exclusively ``hard-only'' outburst transient sources and those transients that have never undergone any ``hard-only'' outbursts and (ii) the persistent sources that either remain in the HCS or IMS for the majority of the time or exhibit ``incomplete'' state transition behavior and those which regularly undergo the typical ``turtlehead'' pattern or spend the majority of time in the SDS yields p-values of 0.028 and 0.082, respectively. 
While this implies marginal evidence that we can reject the null hypothesis that either set of paired data arise from the same parent distribution, the combination of these independent tests provide clear statistical evidence of a difference between the long-term mass transfer rates exhibited by those sources which have been observed to undergo ``hard-only'' outburst behavior and ``incomplete'' transitions in persistent sources in comparison to those that undergo ``turtlehead'' patterned outbursts and persistent state transition behavior.

Second, we find many persistent sources appear to occupy the region of the $(P_{\rm orb}, \Dot{M})$ diagram reserved for transient sources (i.e., below the critical accretion rate for an irradiated disk around the BH; \citealt{kingrit8}). This group of systems includes (i) GRS 1915+105,  one of the long-term transient sources that we treat as persistent, and (ii) those sources that have been observed to exhibit mainly HCS, mainly IMS, and/or incomplete state transition behavior and have mass transfer rates that are apparently too low to sustain a persistent flux (i.e., 1E 1740.7$-$2942, GRS 1758$-$258, SS 433, Cyg X-1, and 4U 1957+115). 

All these sources must be considered carefully on a case-by-case basis to understand the situation. Cyg X-1 is an HMXB, meaning it transfers mass via a stellar wind, resulting in the radius of the accretion disk likely being smaller than one would expect at its orbital period if it was a Roche lobe overflow source. Moreover, it is also worth considering that the long period, apparently persistent sources 1E 1740.7$-$2942, GRS 1758$-$258, and SS 433 may actually be long-term transients like GRS 1915+105. As discussed by \citealt{coriat2012}, while these sources are usually classified as persistent within the literature (see relevant subsections in Section~\ref{sec:sample} for references), the validity of this classification is brought into question by both their large orbital periods, which in principle could allow for decade long outbursts, and their similar position to GRS 1915+105 in the $(P_{\rm orb}, \Dot{M})$ diagram.
Lastly, it is also possible that 1E 1740.7$-$2942, GRS 1758$-$258, SS 433, and 4U 1957+115 may be putting significant amounts of accretion energy into powering an outflow in the form of a relativistic jet, which has been observed in SS 433 \citep{hjon81,ver87,fej88,ver93}, 1E 1740.7$-$2942 \citep{mirbel2}, GRS 1758$-$258 \citep{rod92,pott6} and GRS 1915+105 \citep{mirbel4}, or an accretion disk wind, as observed in 4U 1957+115, GS 1758$-$258, and GRS 1915+105 \citep{ponti12}.

We consider, in more detail, two possibilities to explain (i) why some persistent sources lie well below the irradiated disk stability line and (ii) the scatter in the transient sources.
First, the scatter could imply a change in efficiency between the two regimes (e.g., more advection of energy during the hard state). However, given that (i) models that suggest this (e.g., ADAFs; \citealt{ny94}) show that the brightest hard states have only a minor reduction in luminosity due to advection ($\Dot{M}_{\rm BH} \propto \eta$), and (ii) there is an observed absence of a clear luminosity change during spectral transitions in these types of systems (e.g., \citealt{macar05}), the difference in accretion efficiencies between the hard and soft states (at the transition luminosity) is most likely minimal, ruling out the idea that radiative efficiency changes between the soft and hard states could effectively alter the observed mass transfer rates.

Second, the scatter may also be a result of significant mass (and energy) loss via outflows present in BHXB systems.
Specifically in the case of ``hard-only'' outbursts, the question becomes where all the accreted material (or more specifically its energy) going, if it is not contributing to the accretion luminosity. Thus, we consider how significant a role the the compact, steady, relativistic plasma jet (an outflow known to arise only in the hard state and not seen in the soft state; \citealt{fbg04}) could play. While some material will of course leave in the jet, the amount of material lost to the outflow is most  likely not as much as has been found for 

\afterpage{
\renewcommand{\thefootnote}{\alph{footnote}}
\renewcommand\tabcolsep{5pt}
\tabletypesize{\footnotesize}
\begin{longtable}{lcccr}

\caption{Linear Model fit results for $L_{\rm bol,peak}$ vs. $P_{\rm orb}$}  \\

\hline \hline \\[-2ex]
   \multicolumn{1}{c}{Fit ID} &
   \multicolumn{1}{c}{Slope ($m$)} &
     \multicolumn{1}{c}{Intercept ($b$)} &
          \multicolumn{1}{c}{$\chi^2/$dof} &
                    \multicolumn{1}{c}{$P_{\rm null}$$^a$} \\[0.5ex] \hline
   \\[-1.8ex]
\endfirsthead

  \\[-1.8ex] \hline \\[-1.0ex] 
      \multicolumn{5}{l}{{Note -- best fit to $\log(L_{\rm bol,peak}/L_{\rm edd})=m \log(P_{\rm orb})+b$}} \\   
      \multicolumn{5}{l}{{$^a$ null hypothesis probability}} \\   
      \multicolumn{5}{l}{$^b$ Subsample with known distance.} \\  
      \multicolumn{5}{l}{$^c$ Subsample including the brightest outburst of each source.} \\ 
      \multicolumn{5}{p{0.8\columnwidth}}{\hangindent=1ex$^d$Subsample including the brightest outburst of each source with a known distance.}\\ \ \\
\endlastfoot
all data &$0.21_{-0.066}^{+0.046}$&$-0.43_{-0.12}^ {+0.031}$&29.7/19&0.06 \\[0.1cm] 
$d$ known$^b$&$0.25_{-0.82}^{+0.17}$&$-0.47_{-0.49}^ {+0.060}$&8.0/12&0.92 \\[0.1cm] 
brightest all data$^c$ &$0.22_{-0.097}^{+0.051}$&$-0.44_{-0.15}^ {+0.039}$&24.9/12&0.02 \\[0.1cm] 
brightest $d$ known$^d$&$0.31_{-0.082}^{+0.12}$&$-0.50_{-0.30}^ {+0.074}$&4.8/9&0.86\\[-2ex]
\label{table:linearfit}
\end{longtable}
     \renewcommand{\thefootnote}{\arabic{footnote}}
     \renewcommand\tabcolsep{3pt}
}

\noindent accretion disk winds. However, what the jet may transport more effectively than mass is energy. It is thought that at lower Eddington luminosities (a regime associated with ``hard-only'' outbursts) a larger fraction of the energy released from the accreted material goes into the kinetic (and magnetic) energy of the jet rather than being radiated away (e.g., \citealt{fender2003}), effectively resulting in a smaller contribution to accretion luminosity (and hence a lower $\Dot{M}_{\rm BH}$) than would be the case if the jet was not present.

Since we also observe outliers that correspond to sources that routinely spend significant periods of time in the soft state, we also consider the opposite situation, namely significant outflows that exist in the soft state but are not observed in the hard state. Originally predicted by the early works on accretion disk theory \citep{ss73}, the presence of winds from the outer accretion disk have been observed (exclusively) in the soft accretion state  \citep{neil09,ponti12} of many Galactic BH systems \citep{lee2002,mil04,mi06,mi06c,miller2008,king2012,neil12,diaztrigo14}, indicating that these systems can drive outflows in forms other then jets \citep{diaz11}. For a recent review on accretion disk winds see \citet{neilsen13}.

\citet{ponti12} have estimated the wind outflow rate, $\Dot{M}_{\rm wind}$, in the majority of the sources in which a wind has been detected, to be at least twice the $\Dot{M}_{\rm BH}$. In addition, a few exceptional cases at high Eddington ratio (e.g., the ``heartbeat'' states of GRS 1915+105 and IGR J17091$-$3624; \citealt{neil09} and \citealt{king2012}), show mass loss rates in excess of 10--20 $\Dot{M}_{\rm BH}$. 
In particular, it seems quite plausible that 4U1957+115, which is a persistent BH with a short orbital period that spends substantial time in the SDS, may be losing a large fraction of the mass passing through its outer disk, and thus have an actual mass transfer rate consistent with the irradiated disk stability criterion.

Unfortunately, full calculations of kinetic energy, mass, and momentum flux for the jets and winds have yet to be done \citep{fendgal14}. However, recently  \citet{fender15} have been able to provide basic estimates for both the mass flow and radiative and kinetic feedback over the course of an individual outburst of GX339$-$4.  While we recognize that these estimates are based on a number of assumptions, they are consistent with the ideas postulated above. Namely that, (i) the jet kinetic power exceeds the radiative luminosity during the early and late stages of an outburst (when the source is in the jet-dominated hard state), and (ii) the disk winds (present when the source is in the soft state) are the mechanism responsible for the majority of the mass lost (to the environment) throughout outburst.

\afterpage
{  
\renewcommand{\thefootnote}{\alph{footnote}}
\renewcommand\tabcolsep{2.0pt}
\tabletypesize{\footnotesize}
\begin{longtable}{lcccr}

\caption{Linear Saturation Model fit results for $L_{\rm bol,peak}$ vs. $P_{\rm orb}$}  \\

\hline \hline \\[-2ex]
   \multicolumn{1}{c}{Fit ID} &
   \multicolumn{1}{c}{Slope ($m$)} &
     \multicolumn{1}{c}{Intercept ($b$)} &
          \multicolumn{1}{c}{$\chi^2/$dof} &
                    \multicolumn{1}{c}{$P_{\rm null}$$^a$} \\[0.5ex] \hline
   \\[-1.8ex]
\endfirsthead

  \\[-1.8ex] \hline \\[-1.0ex] 
        \multicolumn{5}{l}{{Note -- best fit found when simultaneously fitting a pure linear }} \\   
                \multicolumn{5}{l}{{model for $P_{\rm orb}>10$ hrs and a constant for $P_{\rm orb}<10$ hrs.  }} \\   
      \multicolumn{5}{l}{{$^a$null hypothesis probability}} \\   
                  \multicolumn{5}{l}{$^b$Subsample with known distance.} \\  
                  \multicolumn{5}{l}{$^c$Subsample including the brightest outburst of each source.} \\  
                        \multicolumn{5}{p{0.8\columnwidth}}{\hangindent=1ex$^d$Subsample including the brightest outburst of each source with a known distance.}    
\endlastfoot

all data &$0.47_{-0.077}^{+0.079}$&$-1.26_{-0.072}^ {+0.071}$&60.6/19&$3.0\times10^{-12}$\\[0.1cm] 
d known$^b$&$0.25_{-1.00}^{+0.198}$&$-0.57_{-0.45}^ {+0.16}$&7.0/15&0.96\\[0.1cm]
brightest all data$^c$ &$0.21_{-0.10}^{+0.046}$&$-0.43_{-0.0837}^ {+0.048}$&25.3/12&0.01 \\[0.1cm] 
brightest d known$^d$&$0.29_{-0.063}^{+0.15}$&$-0.48_{-0.34}^ {+0.047}$&4.8/9&0.85\\[-2ex]
\label{table:linearfit2}
\end{longtable}
\renewcommand\tabcolsep{5.0pt}
     \renewcommand{\thefootnote}{\arabic{footnote}}
}

\begin{figure*}
\epsscale{0.80}
\plotone{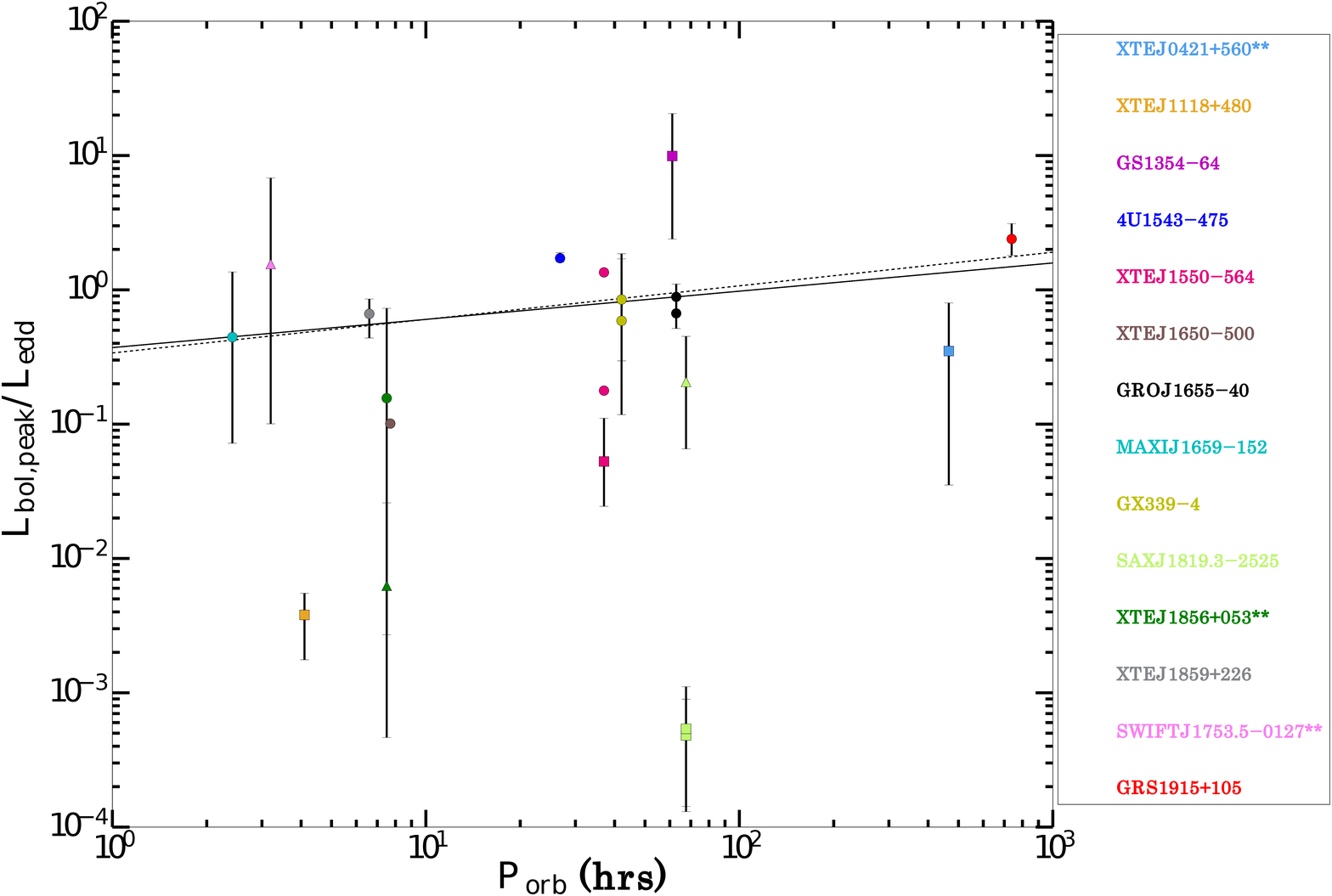}
\ \\ \ \\
\epsscale{0.80}
\plotone{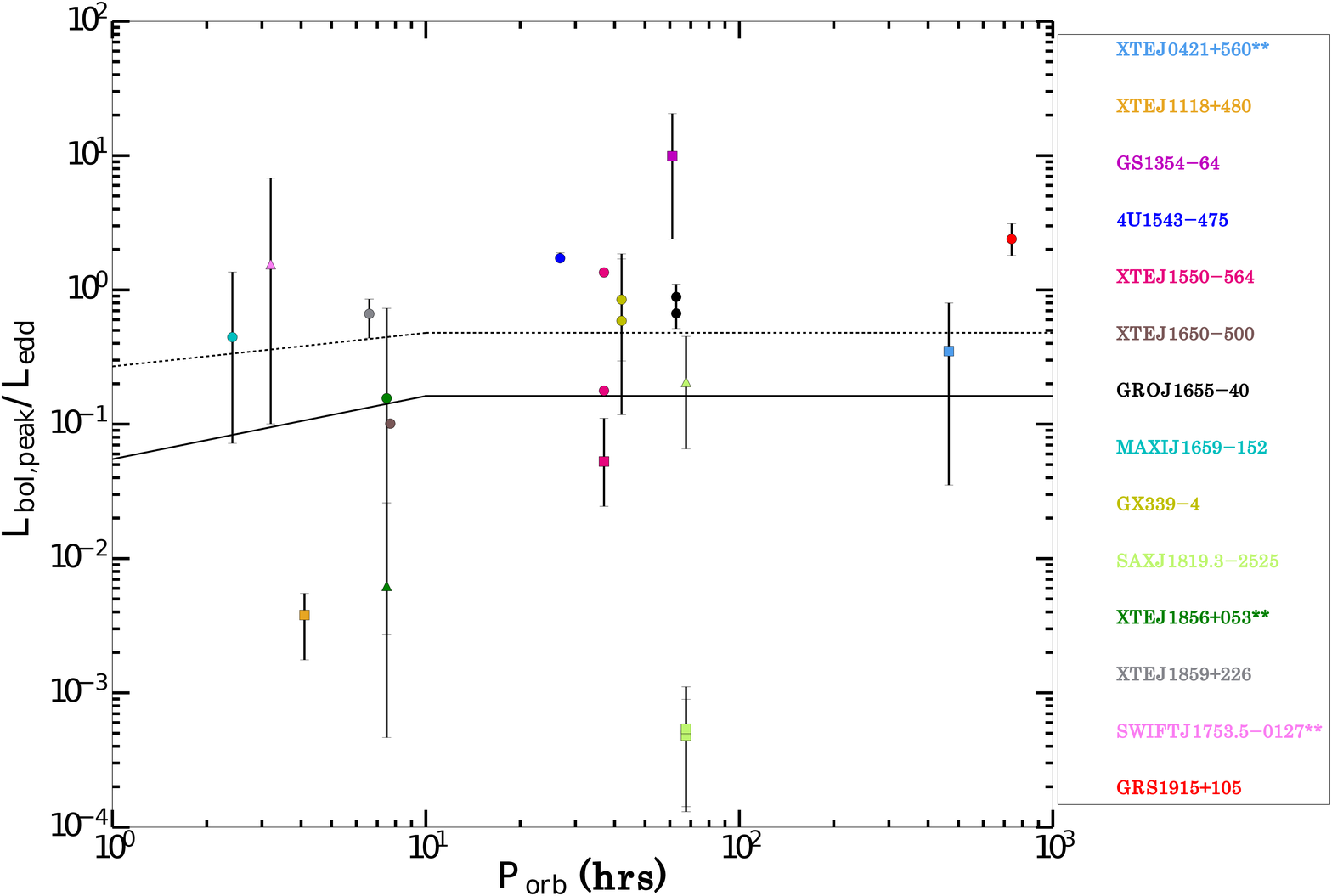}
\caption{{Peak bolometric outburst luminosity in Eddington units as a function of orbital period for transient and long-term transient Galactic BH/BHC systems. Colours represent individual sources (see legend). Sources without known distance estimates are indicated by a ``**'' symbol in the legend. Errors are quoted to the 1$\sigma$ confidence level. Shapes denote outburst behavior: exclusively ``hard-only'' (squares), exclusively ``successful'' or long-term ``turtlehead'' pattern (circles), ``intermediate'' or exhibiting ``incomplete"" state transition behavior  (triangles). (\textit{Top}) Displays the best fit linear model including all outbursts (solid black line) and only those outbursts from sources with distance estimates (dashed black line). (\textit{Bottom}) Displays the best fit linear saturation model  including all outbursts (solid black line) and only those outbursts from sources with distance estimates (dashed black line). See Tables \ref{table:linearfit} and \ref{table:linearfit2} for best fit parameters.}}
\label{figure:lpeakporb}
\end{figure*}

\begin{figure*}
\epsscale{0.80}
\plotone{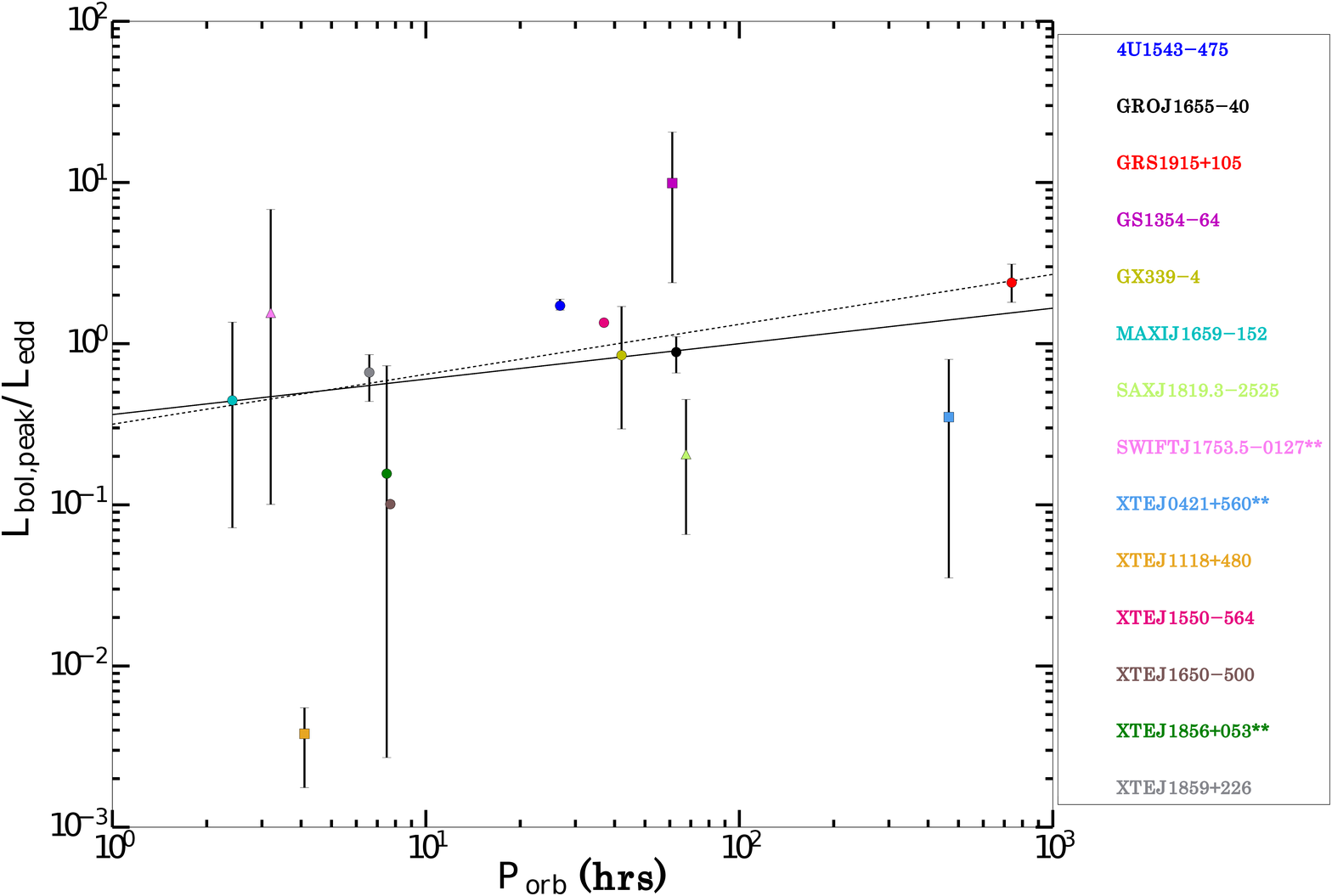}
\ \\ \ \\
\epsscale{0.80}
\plotone{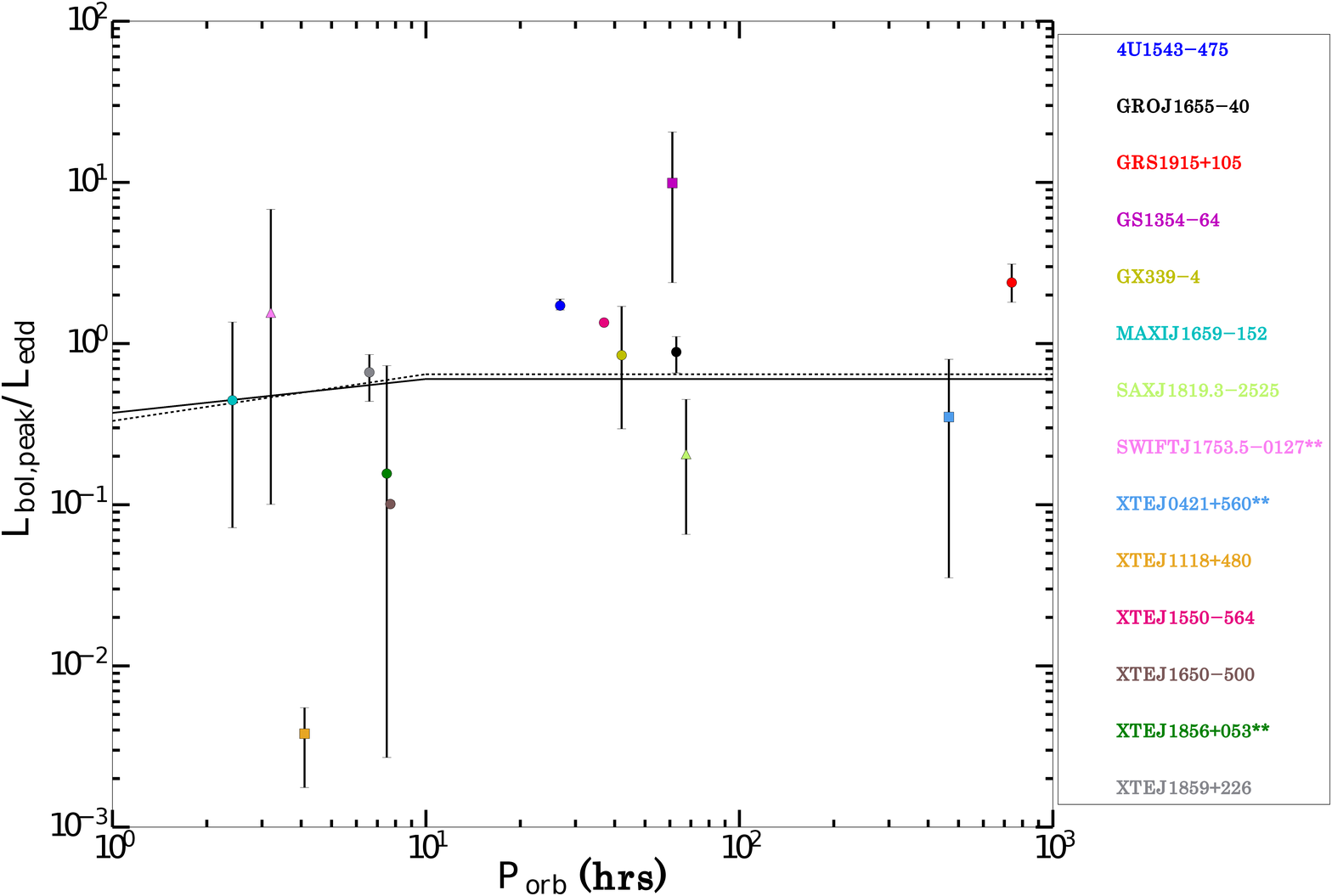}
\caption{{Peak bolometric outburst luminosity in Eddington units as a function of orbital period including only the brightest outburst from each transient or long-term transient Galactic BH/BHC system. Colours represent individual sources (see legend). Sources without known distance estimates are indicated by a ``**'' symbol in the legend. Errors are quoted to the 1$\sigma$ confidence level. Shapes denote outburst behavior: exclusively ``hard-only'' (squares), exclusively ``successful'' or long-term ``turtlehead'' pattern (circles), ``intermediate'' or exhibiting ``incomplete" state transition behavior  (triangles). (\textit{Top}) Displays the best fit linear model including all outbursts (solid black line) and only those outbursts from sources with distance estimates (dashed black line). (\textit{Bottom}) Displays the best fit linear saturation model including all outbursts (solid black line) and only those outbursts from sources with distance estimates (dashed black line). See Tables \ref{table:linearfit} and \ref{table:linearfit2} for best fit parameters.}}
\label{figure:lpeakporbbright}
\end{figure*}

\subsection{The relationship between peak outburst luminosity and orbital period}
\label{subsec:peaklum_porb}
A positive correlation between $P_{\rm orb}$ and peak outburst luminosity for LMXBs has been established in previous studies.
\citep[e.g., ][]{shab98,port2004,wu2010}.
In Figure \ref{figure:lpeakporb} we plot peak outburst luminosity versus $P_{ \rm orb}$ for all the outbursts detected in the 12 recurrent transient and 2 long-term transient (GRS1915+105 and Swift J1753.5$-$0127) sources from our sample for which the orbital period is known (see Section~\ref{sec:sample} or Table \ref{table:binaryBH} for references). In addition, given that it is the brightest outburst of each source that will tell us the most about the intrinsic properties of the binary, we also plot peak outburst luminosity vs. $P_{ \rm orb}$, for the brightest outburst detected in each of the 12 recurrent transient and 2 long-term transient (GRS1915+105 and Swift J1753.5$-$0127) sources from our sample for which the orbital period is known, in Figure \ref{figure:lpeakporbbright}. Note that we do not include those outbursts that we have identified to 
contain days when our spectral fitting algorithm has failed, due to the resulting uncertainty in the bolometric luminosities derived from this process (see Sections \ref{subsubsec:spectra} and \ref{subsec:peaklum} for discussion and outbursts marked with a ``*'' in Table \ref{table:lumdata}), in any analysis presented in this Section.

Before comparing our results to past results, we performed non-parametric (Kendall-Tau) tests to determine whether there is evidence for a positive correlation between $P_{\rm orb}$ and peak outburst luminosity. As our data in Figures \ref{figure:lpeakporb} \& \ref{figure:lpeakporbbright} include several subsamples, we performed analysis on both the full sample, as well as the individual subsamples of the ``successful'', and ``hard-only'' outbursts. To account for the errors in the peak luminosity, we performed Monte Carlo simulations in logarithmic space assuming a normal distribution and the errors we measured for each outburst. We performed this analysis for all outbursts (Figure \ref{figure:lpeakporb}) and the brightest outbursts (Figure \ref{figure:lpeakporbbright}). Contrary to past claims, we do not find definitive evidence of a positive correlation. In particular, the ``hard-only'' outbursts do not show any correlation, which affects both this subsample and the complete sample. However, we find potential agreement with past results for the ``successful outburst'' sample. In the case where we consider all outbursts, this subsample had a median Kendall-Tau correlation of $\tau = 0.26$ with only a 23\% chance that $\tau$ is consistent with 0.  When we only consider the brightest outburst in each source, this sample had a median $\tau = 0.39$ with only a 14\% chance that $\tau$ is consistent with 0. We note that in both cases, the median results from the Monte Carlo simulations are more conservative than the results one gets if the error in luminosity is ignored. If we ignore luminosity errors, we measure $\tau = 0.39$ with only a 7.9\% chance that $\tau$ is consistent with 0 and $\tau = 0.44$ with only a 9.5\% chance that $\tau$ is consistent with 0 for the entire  ``successful outburst'' sample and the brightest ``successful outburst'' sample, respectively. Given that past results did not consider as broad a suite of errors as we considered and did not use non-parametric tests, this might explain why our result only marginally suggests a correlation. Despite this, we consider the potential correlations that other studies have found in more detail below.

First, following the procedures of \citet{port2004} and \citet{wu2010}, we attempt to fit the data on a logarithmic scale. Using the MCMC algorithm described in Sections 3.7.5 and 3.7.6 we attempt to fit two different models to the data. We first fit all the data as well as only the data from sources with available distance estimates with a linear relation (resulting in a single power-law model of luminosity versus orbital period). The fit results are presented in Table \ref{table:linearfit}. We note that our best fit slopes in all four cases are smaller then than \citet{wu2010} result. 

Next, following \citet{port2004}, we fit all the data as well as only the data from sources with available distance estimates with a linear model  that saturates at a constant for long $P_{\rm orb}$ (resulting in a single power-law model of luminosity versus orbital period at low orbital period and a flat model at high orbital period). This model is motivated by theoretical studies that have suggested that the amount of mass accumulated in the disk during an outburst is related to the orbital period, with a cutoff at longer periods (see \citealt{meyerH2000} and \citealt{meyer4}). As discussed in \citet{wu2010}, if the $P_{\rm orb}$ at which the break occurs is not constrained, the best fit for this parameter will be exceedingly large, essentially making the model identical to the basic linear relation discussed above. For this reason, we do not attempt to fit for the break $P_{\rm orb}$. Instead, following \citet{port2004} we choose to fix the break $P_{\rm orb}$ at a value of 10 hours, as suggested by the theory.
 
The best fit parameters found by fitting the linear saturation model are presented in Table \ref{table:linearfit2}. Comparing the resulting $\chi^2$ of the linear saturation model to those obtained for the pure linear relation, we find that in the case involving all outbursts from only the sources with known distance estimates, the goodness of fit improves when adding a break and saturation luminosity at long orbital periods, on par with the result found in \citet{port2004}. While the $\chi^2$ decreases for the same numbers of degrees of freedom compared to the linear model, Monte Carlo simulations indicate that this difference is not statistically significant enough to favor the saturation model over the linear model. Similarly, Monte Carlo simulations indicate that the improved $\chi^2$ when fitting all outbursts and only the brightest outbursts of the entire sample of sources is not statistically significant enough to favor the linear model over the saturation model in either case. 
We note that in the two cases involving only sources with known distance estimates, we find a best fit saturation luminosity (constant for the $P_{\rm orb}>10$ hrs) consistent with the results found by \citet{port2004}. However, in all four cases, our best fit slope is significantly smaller then the results found by \citet{port2004}. 
 
We chose to consider all outbursts to be consistent with past studies, but have additionally fit the relations with respect to the brightest outburst of each source. From a theoretical perspective, we expect any relation between orbital period and outburst luminosity to break down for low luminosity sources, as in these outbursts it is more likely that much a smaller fraction of the accretion disk has been accreted. However, this choice does not provide sufficient statistical leverage to disentangle which model is preferred in our current sample.

Second, given that outbursts from transient systems are far from phenomenologically identical (see \citealt{chen97}, \citealt{wu2010}, and this study) and the fact that we have numerous outliers in our dataset, the question becomes whether or not ``hard-only'' outbursts have statistically different peak outburst luminosities when compared to ``successful'' outbursts. Performing a two sample KS-test comparing these two sample groups yields p-values of 0.002 and 0.069 for all data and for only outbursts from sources with known distance estimates, respectively. 

The former test provides clear statistical evidence that the peak outburst luminosities of  ``hard-only''  and ``successful'' outbursts do not arise from the same parent distribution, while the latter subsample only provides suggestive statistical evidence of the same.

Third, we find that all ``hard-only'' outbursts (with the exception of GS 1354$-$64) have significantly lower (sub-Eddington) peak luminosities when compared to ``successful'' outbursts 
from systems at the same orbital period. This observation is expected as it is the inevitable result of these systems never reaching the high luminosity soft state associated with radiatively efficient accretion. In the case of GS 1354$-$64, the higher then expected peak luminosity maybe due to uncertainty in the poorly constrained distance estimate ($25-61$ kpc). If we were to place GS 1354$-$64 at our assumed standard Galactic value (i.e., a uniform distribution between 2 and 8 kpc), $L_{\rm bol,peak}<0.1\, L_{\rm edd}$, on par with the behavior we see from the other ``hard-only'' outbursts in our sample.
 
Lastly, it has been postulated that the low peak outburst luminosities associated with short period BH LMXBs could potentially cause them to remain in the low luminosity hard state (as peak outburst luminosity drops near the limit for radiatively inefficient accretion), rather than entering the high luminosity soft state expected for radiatively efficient accretion. This implies that short period systems may be more prone to ``hard-only'' outbursts then the longer period systems \citep{meyer4,maccarone2013,knev14}. 

Currently there are few known BH systems that belong to the short $P_{ \rm orb}$ ($<5$ hours) regime (See \citealt{port2004} and this work). There are only two possible explanations for why these types of systems are missing from the observed sample: (i) they are too faint to detect with current instrumentation, or (ii) they do not exist. However, given that binary evolution indicates that these systems should not only exist but also dominate the total BHXB number counts (e.g., see \citealt{knev14}), we favor case (i) and reason that we need a more-sensitive all-sky monitor to detect their faint (and likely ``hard-only'') outbursts \citep{maccarone2015}.

\section{Summary}

Taking advantage of the current suite of more sensitive all-sky, narrow field and scanning X-ray instruments on board INTEGRAL, MAXI, RXTE, and Swift, which have made an in-depth exploration of the transient X-ray Universe possible, we have established a comprehensive database of BH (and BHC) XRB activity over the last 19 years. This database collates observable properties from the literature with the X-ray light curves to enable quantitative classification of these accreting stellar mass BHs.
We have assembled our database by running data from seven separate instruments through a custom pipeline composed of a comprehensive algorithm built to discover, track, and quantitatively classify outburst behavior.
Until the 2015 May 14 cutoff for our database, we have detected 132 transient outbursts, tracked and classified behavior occurring in 47 transient and 10 persistently accreting black holes, and performed a statistical study on a number of outburst properties across the Galactic population, including outburst detection rates, duration, recurrence rates, duty cycles, total energy radiated, peak luminosity and state transitions.

We have found that not only are our current suite of more sensitive X-ray instruments in space detecting a greater number of sources, an estimated $\sim$4--12 transient outbursts per year, more than a factor of three larger than in the pre-RXTE era \citep{chen97} alone, but also that $38^{+6.0}_{-5.6}$ \% of the outbursts detected in the last 19 years do not complete the typical ``turtlehead'' pattern, never transitioning from the HCS to the SDS. This ``hard-only'' behavior is not just limited to recurrent transients but can also be observed in a number of long-term transient and persistently accreting BH systems as well. In their case this takes the form of long continuous periods spent in the HCS or periodic ``incomplete'' state transitions (i.e., attempted hard-soft state transitions in which the source only reached as far as the IMS before transitioning back to the hard state).

This ``hard-only'' behavior is neither a rare nor recent phenomena. Through an extensive literature search, we find a near constant appearance of these ``hard-only'' outbursts over the last $\sim50$ years. 
This finding, paired with the fact that ``hard-only'' outbursts tend to have peak Eddington scaled luminosities that are $\lesssim0.11 \, L_{\rm edd}$ (a regime at or below that where we expect the transition from the hard to the soft state to happen) may indicate that the ``hard-only'' behavior simply involves the mass-transfer rate onto the BH remaining at a low level throughout the outburst, below the (yet unknown) instability that triggers the change to the soft state.

Given that we find a substantial fraction of the total transient BH outburst sample over the past two decades is represented by ``hard-only'' outbursts, we have considered the ramifications the larger number of these so-called ``hard-only'' outbursts (and the entire population of observed BHXB outbursts) have on the luminosity function and mass-transfer history of the Galactic BHXB population.

First, we observe the appearance of a bi-modal distribution present in the luminosity function for the entire transient population, presumably indicative of the cyclic ``turtlehead'' patterns of temporal evolution in BHXBs.  This is a complete deviation from the power-law type distribution found for the entire XRB population of the Galaxy in previous studies (e.g., \citealt{grimm2002}). In addition, we find that the features present in the individual source XLFs correlate with outburst classifications and temporal evolution observed in the sources. In particular we find the appearance of a single peak contribution at low luminosities ($\sim10^{35}-10^{37} {\rm erg s^{-1}}$) in the exclusively ``hard-only'' outburst and persistently HCS sources, a prominent peak at lower luminosity ($\sim10^{37} {\rm erg s^{-1}}$) that is dominated by intermediate state detections in those source that exhibit mainly ``incomplete'' state transitions, and a dominant contribution at high luminosities ($>10^{38} {\rm erg s^{-1}}$), that is found in those sources that exhibit the typical ``turtlehead'' pattern (i.e., a significant soft state).

Second, we observe numerous outliers (including both transient and persistent sources) from the theoretically expected correlation in the $(P_{\rm orb}, \Dot{M})$ diagram \citep{pod02}. There exist a number of explanations for these outliers, including but not limited to, different binary evolution paths (i.e., not all objects at one $P_{\rm orb}$ have the same history), differing local conditions at the first Lagrangian point (e.g., star spots), irradiation-induced mass transfer cycles (e.g., see \citealt{podsiadlowski91,harpaz91,buningritter2004}), uncertainties in distance and inclination (and thus beaming),  a significant change in accretion efficiency existing between accretion states (e.g., more advection of energy during the hard state; see \citealt{knev14}), the possibility that some long period apparently persistent sources are actually long-term transients, or the possibility that a significant amount of accreted matter (or energy from this matter) may be removed from a system via a substantial outflow arising either exclusively in the hard state or exclusively in the soft state (i.e., relativistic jet vs. accretion disk wind) before it has a chance to fall through the disk and contribute to the accretion luminosity, simultaneously implying that the observational difference in $\Dot{M}$ estimates is in fact physical and providing further indirect observational evidence signifying the importance of outflows to accretion theory.

In this study we have considered the last three options. Given that previous studies (e.g., \citealt{macar05}) have ruled out the possibility that the bright hard states of BHXBs have a significantly different radiative efficiency than the soft states at similar luminosities, we favor a combination of the last two scenarios to explain the scatter in the observed mass transfer rates from the expected correlation. We conclude that our inferred $\Dot{M}_{\rm BH}$ estimates may in fact only be lower limits on the true mass transfer rates from the companions in these systems.

Third, while we find that fitting a linear saturation model (see \citealt{meyerH2000} and \citealt{meyer4}) between $P_{ \rm orb}$ and peak outburst luminosity does improve the goodness of fit when the distance to a source is known, when compared to fitting a pure linear model \citep{shab98}, the improvement is only marginal. As such, we can not favor either model over the other. For the pure linear model, we find best fit slopes, taking into account all outbursts and only outbursts from sources with distance estimates, that are significantly smaller then previous studies (e.g., \citealt{wu2010}). For the linear saturation model, we find best fit saturation luminosities for the $P_{\rm orb}>10$ hrs regime consistent with the results found by \citet{port2004} and best fit slopes for the $P_{\rm orb}\leq10$ hrs regime  that are significantly smaller then the results found by \citet{port2004}.

Overall, this all-sky study has allowed us to probe the wide and varying array of outburst behavior exhibited by Galactic BHXBs, its impact on the physical observables of individual systems alone, and thus the universal properties of the Galactic population as a whole. With our results we have demonstrated that enumerating the frequency at which outbursts occur, tracking outburst properties across the population and quantitatively classifying the wide range of behavior exhibited during outburst will be critical to furthering our understanding of the physical mechanisms driving mass-transfer in binary BH systems and a key step toward filling in the many gaps in our knowledge of how BHXBs form, accrete and evolve.

\section{Acknowledgements}
BET would like to thank Hans Krimm for sharing his extensive knowledge of the Galactic BH population and his enthusiastic interest in the project, Erik Rosolowsky for helping develop the MCMC algorithms, Serena Repetto for catching multiple mistakes in the long list of references and her useful comments on the manuscript, Ann Hornschemeier for her interest in this project and positive encouragement during the writing of the manuscript, and Alex Tetarenko for her help creating the name and logo of the Database.  BET, GRS, and COH acknowledge support by NSERC Discovery Grants. COH additionally acknowledges support by an Alexander von Humboldt fellowship, and is grateful to the Max Planck Institute for Radioastronomy in Bonn for their hospitality. The authors would also like to thank the anonymous scientific and statistical referees for there detailed comments that have significantly improved the manuscript. This research has made use of data, software, and/or web tools obtained from the High Energy Astrophysics Science Archive Research Center (HEASARC), a service of the Astrophysics Science Division at NASA/GSFC and of the Smithsonian Astrophysical Observatory's High Energy Astrophysics Division, and the MAXI data provided by RIKEN, JAXA and the MAXI team. This work has also made extensive use of NASA's Astrophysics Data System (ADS).

\bibliographystyle{apj.bst}

\bibliography{BHXB_DB.bib}

\clearpage
\LongTables
\renewcommand{\thefootnote}{\alph{footnote}}
\renewcommand\tabcolsep{3pt}

\begin{landscape}

\tabletypesize{\scriptsize}
\begin{longtable}{lccccccccr}
\caption[Galactic BH/BHC Primary Source Information]{Galactic BH/BHC Primary Source Information}  \\
\hline \hline \\[-2ex]
   \multicolumn{1}{c}{Source Name$^a$} &
   \multicolumn{1}{c}{Optical Counterpart/$^a$} &
      \multicolumn{1}{c}{Class$^b$} &  
     \multicolumn{1}{c}{Type$^c$} &
     \multicolumn{1}{c}{RA (J2000)$^d$} &
    \multicolumn{1}{c}{DEC (J2000)$^d$} &
    \multicolumn{2}{c}{Discovery$^e$} &        
        \multicolumn{1}{c}{Outbursts$^f$} &   
   \multicolumn{1}{c}{References$^g$}\\
      &\multicolumn{1}{c}{Alternate Name(s)} && & & & \multicolumn{1}{c}{Year }&\multicolumn{1}{c}{Telescope} &  \multicolumn{1}{c}{Undergone} &\\[0.5ex] \hline
   \\[-4ex]
      \label{table:primaryBH}
\endfirsthead
\multicolumn{9}{c}{{\tablename} \thetable{} -- Continued} \\[0.5ex]
\hline \hline \\[-2ex]
   \multicolumn{1}{c}{Source Name$^a$} &
   \multicolumn{1}{c}{Optical Counterpart/$^a$} &
      \multicolumn{1}{c}{Class$^b$} &  
     \multicolumn{1}{c}{Type$^c$} &
     \multicolumn{1}{c}{RA (J2000)$^d$} &
    \multicolumn{1}{c}{DEC (J2000)$^d$} &
    \multicolumn{2}{c}{Discovery$^e$} &        
        \multicolumn{1}{c}{Outbursts$^f$} &   
   \multicolumn{1}{c}{References$^g$}\\
      &\multicolumn{1}{c}{Alternate Name(s)} && & & & \multicolumn{1}{c}{Year }&\multicolumn{1}{c}{Telescope} &  \multicolumn{1}{c}{Undergone} &\\[0.5ex] \hline
   \\[-2ex]
\endhead
  \\[-1ex] \hline  \\[-2ex]
  \multicolumn{10}{l}{{Continued on Next Page\ldots}} \\
\endfoot
  \\[-1ex] \hline \\[-1ex]                
\endlastfoot
XTE J0421+560&CI Cam&C&BHC,T,R&04 19 42.141 (0.0566\arcsec) &+55 59 57.70 (0.0678\arcsec)&1998&Rx&1&5,185--189,276--278,367\\[1.00ex]

GRO J0422+32&V518 Per&A&BH,T,R&04 21 42.79 (0.2\arcsec) &+32 54 27.1 (0.2\arcsec)&1992&CG&3&1,36--38,279\\
&Nova Per 1992 &&&&&&&\\[1.00ex]

4U 0538$-$641&LMC X-3&A&BH,P&05 38 56.299 (0.15\arcsec) &$-$64 05 03.00 (0.15\arcsec)&1971&U&-&1,11,12\\ [0.50ex]

4U 0540$-$697&LMC X-1&A&BH,P&05 39 38.839 (0.0015\arcsec) &$-$69 44 35.66 (0.0015\arcsec)&1971&U&-&1,12,13\\[1.00ex]

1A 0620$-$00&V616 Mon&A&BH,T,R,J&06 22 44.503 (0.18\arcsec) &$-$00 20 44.72 (0.10\arcsec)&1975&Ar&2&1,5,6,280--284,365\\
&Nova Mon 1975&&&&&&&\\
&Mon X$-$1&&&&&&&\\
\end{longtable}
\clearpage
\addtocounter{table}{-1}
\begin{longtable}{lccccccccr}
\multicolumn{10}{c}{{\tablename} \thetable{} -- Continued} \\[0.5ex]
\hline \hline \\[-2ex]
   \multicolumn{1}{c}{Source Name$^a$} &
   \multicolumn{1}{c}{Optical Counterpart/$^a$} &
      \multicolumn{1}{c}{Class$^b$} &  
     \multicolumn{1}{c}{Type$^c$} &
     \multicolumn{1}{c}{RA (J2000)$^d$} &
    \multicolumn{1}{c}{DEC (J2000)$^d$} &
    \multicolumn{2}{c}{Discovery$^e$} &        
        \multicolumn{1}{c}{Outbursts$^f$} &   
   \multicolumn{1}{c}{References$^g$}\\
      &\multicolumn{1}{c}{Alternate Name(s)} && & & & \multicolumn{1}{c}{Year }&\multicolumn{1}{c}{Telescope} &  \multicolumn{1}{c}{Undergone} &\\[0.5ex] \hline
   \\[-1ex]
\endhead
\multicolumn{10}{c}{{\tablename} \thetable{} -- Continued} \\[0.5ex]
\hline \hline \\[-2ex]
   \multicolumn{1}{c}{Source Name$^a$} &
   \multicolumn{1}{c}{Optical Counterpart/$^a$} &
      \multicolumn{1}{c}{Class$^b$} &  
     \multicolumn{1}{c}{Type$^c$} &
     \multicolumn{1}{c}{RA (J2000)$^d$} &
    \multicolumn{1}{c}{DEC (J2000)$^d$} &
    \multicolumn{2}{c}{Discovery$^e$} &        
        \multicolumn{1}{c}{Outbursts$^f$} &   
   \multicolumn{1}{c}{References$^g$}\\
      &\multicolumn{1}{c}{Alternate Name(s)} && & & & \multicolumn{1}{c}{Year }&\multicolumn{1}{c}{Telescope} &  \multicolumn{1}{c}{Undergone} &\\[0.5ex] \hline
   \\[-1ex]
\endfirsthead
  \\[-0.5ex] \hline \\[-1ex]                
\endlastfoot
\multicolumn{10}{p{0.95\columnwidth}}{\hangindent=1ex$^g${\bf References.} (cont.) --- [251]~\citet{cas4}, [252]~\citet{sri13b}, [253]~\citet{steph5}, [254]~\citet{r02a}, [255]~\citet{kar7}, [256]~\citet{kar8}, [257]~\citet{krimm10}, [258]~\citet{woods2}, [259]~\citet{int02b}, [260]~\citet{gog4}, [261]~\citet{jon4}, [262]~\citet{kr13a}, [263]~\citet{hy99}, [264]~\citet{rem98}, [265]~\citet{cam2}, [266]~\citet{v00}, [267]~\citet{rm06}, [268]~\citet{kit84}, [269]~\citet{harm92}, [270]~\citet{park04}, [271]~\citet{kale05}, [272]~\citet{krimm2014}, [273]~\citet{cui2000}, [274]~\citet{lep93}, [275]~\citet{barr92}, [276]~\citet{hj98a}, [277]~\citet{clark20}, [278]\citet{mi02004}, [279]~\citet{shrad4}, [280]~\citet{dav75}, [281]~\citet{dul79}, [282]~\citet{geld83}, [283]~\citet{gallo6}, [284]~\citet{owen76}, [285]~\citet{uem20}, [286]~\citet{fend01}, [287]~\citet{ball95}, [288]~\citet{br01}, [289]~\citet{mj11},[290]~\citet{hun02}, [291]~\citet{hann01}, [292]~\citet{cor02}, [293]~\citet{tomsick3}, [294]~\citet{karr3}, [295]~\citet{hje99}, [296]~\citet{calv9}, [297]~\citet{tav96}, [298]~\citet{harm95}, [299]~\citet{hjr95}, [300]~\citet{ting95}, [301]~\citet{vanh10}, [302]~\citet{corb0}, [303]~\citet{fend97}, [304]~\citet{fend99a}, [305]~\citet{hann98}, [306]~\citet{rup3a}, [307]~\citet{pand5}, [308]~\citet{rup5}, [309]~\citet{rup3c}, [310]~\citet{del94}, [311]~\citet{hj96}, [312]~\citet{dav76}, [313]~\citet{hj96b}, [314]~\citet{ana3}, [315]~\citet{reich92}, [316]~\citet{hei94}, [317]~\citet{mirbel2}, [318]~\citet{mjs12}, [319]~\citet{corb12}, [320]~\citet{corr13a}, [321]~\citet{rup3d}, [322]~\citet{brock13}, [323]~\citet{diaz13},[324]~\citet{fend5}, [325]~\citet{marti02}, [326]~\citet{rod92}, [327]~\citet{rup6b}, [328]~\citet{rup5b}, [329]~\citet{hj00b}, [330]~\citet{r12}, [331]~\citet{r14b}, [332]~\citet{brock2}, [333]~\citet{rup2}, [334]\citet{king12},[335]~\citet{hj88}, [336]~\citet{hj98b}, [337]~\citet{hanhj92}, [338]~\citet{ab79}, [339]~\citet{blun10}, [340]~\citet{fend00}, [341]~\citet{chak03}, [342]~\citet{chak05}, [343]~\citet{gies02}, [344]~\citet{eik8}, [345]~\citet{harm97}, [346]~\citet{aah97}, [347]~\citet{fend02}, [348]~\citet{fend97b}, [349]~\citet{fend98}, [350]~\citet{pool97}, [351]~\citet{rod97}, [352]~\citet{rod98}, [353]~\citet{rod95}, [354]~\citet{mirbel4}, [355]~\citet{millerj5}, [356]~\citet{fost96}, [357]~\citet{chat96}, [358]~\citet{braes71}, [359]~\citet{braes76}, [360]~\citet{hj73}, [361]~\citet{hj71}, [362]~\citet{tig4}, [363]~\citet{stir1}, [364]~\citet{fend00b}, [365]~\citet{kul99}, [366]~\citet{mj12}, [367]\citet{tychos2}, [368]~\citet{jain99}, [369]~\citet{bai95}, [370]~\citet{mait5}, [371]~\citet{vog99}, [372]~\citet{ueda7}, [373]~\citet{bm83}, [374]~\citet{mot98}, [375]~\citet{sam99}, [376]~\citet{garn99}, [377]~\citet{ken12}, [378]~\citet{dhaw00}, [379]~\citet{oka88}, [380]~\citet{mj9}, [381]~\citet{kap80}, [382]~\citet{bradt79}, [383]~\citet{marg78}, [384]~\citet{kit90}, [385]~\citet{delv91}, [386]~\citet{baiorz95}, [387]~\citet{krimm2009b}, [388]~\citet{ponti12}, [389]~\citet{mi06c}, [390]~\citet{miller2008}, [391]~\citet{mil04}, [392]~\citet{prat2009},[393]~\citet{zhang2015b}, [394]~\citet{millerjm2015}, [395]~\citet{negoroh15}, [396]~\citet{kuulkerse15}, [397]~\citet{yoshit2015}, [398]~\citet{tachibanay2014}, [399]~\citet{muno2005}, [400]~\citet{muno2005b}, [401]~\citet{belanger2005}, [402]~\citet{bower2005},  [403] \citet{giacconi66}, [404] \citet{hjalm2008}, [405] \citet{szostek2008}, [406] \citet{szostek2008b}, [407] \citet{koljonen2010}, [408] \citet{watanabe1994}, [409] \citet{marti2001}, [410] \citet{miod2001}, [411] \citet{zdz2013}, [412] \citet{hanson2000}, [413] \citet{vilhu2009}, [414] \citet{oos97}, [415] \citet{lucarelli10}, [416] \citet{williams10}, [417] \cite{casar14}, [418] \citet{casares2012}, [418] \citet{munar14}, [419] \citet{dzib2015}}\\
\end{longtable}
\clearpage
\end{landscape}

\renewcommand{\thefootnote}{\alph{footnote}}

\clearpage
\LongTables
\renewcommand{\thefootnote}{\alph{footnote}}
\renewcommand\tabcolsep{3pt}

\begin{landscape}
\tiny
\begin{longtable}{lcccccccr}
\caption[Galactic BH/BHC Binary Source Information]{Galactic BH/BHC Binary Source Information}  \\
\hline \hline \\[-2ex]
   \multicolumn{1}{c}{Source Name} &
   \multicolumn{1}{c}{Distance$^b$} &
     \multicolumn{1}{c}{$M_{\rm BH}$$^b$} &
     \multicolumn{1}{c}{$f(M)$} &
    \multicolumn{1}{c}{$i$} &
    \multicolumn{1}{c}{$q$} & 
        \multicolumn{1}{c}{Spectral} &          
        \multicolumn{1}{c}{$P_{\rm orb}$} &   
   \multicolumn{1}{c}{References$^c$}\\
      &\multicolumn{1}{c}{(kpc)} & \multicolumn{1}{c}{($M_{\odot}$) } &  \multicolumn{1}{c}{($M_{\odot}$) }&  \multicolumn{1}{c}{(deg) }& \multicolumn{1}{c}{($M_{2}/M_{\rm BH}$) }&\multicolumn{1}{c}{Type} &  \multicolumn{1}{c}{(hr)} &\\[0.5ex] \hline
   \\[-4ex]
         \label{table:binaryBH}
\endfirsthead
\multicolumn{9}{c}{{\tablename} \thetable{} -- Continued} \\[0.5ex]
\hline \hline \\[-2ex]
   \multicolumn{1}{c}{Source Name} &
   \multicolumn{1}{c}{Distance$^b$} &
     \multicolumn{1}{c}{$M_{\rm BH}$$^b$} &
     \multicolumn{1}{c}{$f(M)$} &
    \multicolumn{1}{c}{$i$} &
    \multicolumn{1}{c}{$q$} &   
      \multicolumn{1}{c}{Spectral} &        
        \multicolumn{1}{c}{$P_{\rm orb}$} &   
   \multicolumn{1}{c}{References$^c$}\\
      &\multicolumn{1}{c}{(kpc)} & \multicolumn{1}{c}{($M_{\odot}$) } &  \multicolumn{1}{c}{($M_{\odot}$) }&  \multicolumn{1}{c}{(deg) }& \multicolumn{1}{c}{($M_{2}/M_{\rm BH}$) }&\multicolumn{1}{c}{Type} &  \multicolumn{1}{c}{(hr)} &\\[0.5ex] \hline
   \\[-2ex]
\endhead
  \\[-2ex] \hline  \\[-1ex]
  \multicolumn{9}{l}{{Continued on Next Page\ldots}} \\
\endfoot
  \\[-2ex] \hline \\[-1.5ex]  
               \multicolumn{9}{p{0.7\columnwidth}}{\hangindent=1ex NOTE 1.-- This table is also available in machine readable format online at the Astrophysical Journal and on the WATCHDOG website - http://astro.physics.ualberta.ca/WATCHDOG/.}\\
    \multicolumn{9}{p{0.7\columnwidth}}{\hangindent=1ex NOTE 2. -- All error quantities are quoted to the 1$\sigma$ confidence level.} \\
    \multicolumn{9}{p{0.7\columnwidth}}{\hangindent=1ex$^a$ $M_{\rm BH}$ has been calculated based on improved measurements of the orbital parameters of the system. For details per source see the individual source sections.}\\
    \multicolumn{9}{p{0.7\columnwidth}}{\hangindent=1ex$^b$ When no acceptable estimates for distance or BH mass are available, $5\pm3$ kpc is assumed and the BH mass is sampled from the Ozel mass distribution \citep{oz10}.}\\
        \multicolumn{9}{p{0.7\columnwidth}}{\hangindent=1ex$^c${\bf References.} --- [1] \citet{kreid12}, [2] \citet{oz10}, [3] \citet{can10}, [4] \citet{lmxb07}, [5] \citet{van84}, [6] \citet{mart10}, [7] \citet{mr06}, [8] \citet{or14}, [9] \citet{cow92}, [10] \citet{or09}, [11] \citet{barr96b}, [12] \citet{or03}, [13] \citet{or98}, [14] \citet{or11}, [15] \citet{re11}, [16] \citet{cab9}, [17] \citet{hi05}, [18] \citet{fi95}, [19] \citet{har99}, [20] \citet{webb2}, [21] \citet{beer2}, [22] \citet{hjr95}, [23] \citet{shab99}, [24] \citet{or02}, [25] \citet{fi99}, [26] \citet{shab96}, [27] \citet{del94}, [28] \citet{mas96}, [29] \citet{mart97}, [30] \citet{mart98}, [31] \citet{eik01}, [32] \citet{roth2}, [33] \citet{gr01}, [34] \citet{harg4}, [35] \citet{or96}, [36] \citet{cas97}, [37] \citet{cas9}, [38] \citet{ca04}, [39] \citet{barr96b}, [40] \citet{har96}, [41] \citet{cas95}, [42] \citet{mj9}, [43] \citet{casc94}, [44] \citet{chc6}, [45] \citet{hi03}, [46] \citet{shap9}, [47] \citet{martin5}, [48] \citet{fi97}, [49] \citet{har97},[50] \citet{neil7}, [51] \citet{or01}, [52] \citet{gel6}, [53] \citet{or11b}, [54] \citet{hom6}, [55] \citet{rm06}, [56] \citet{gies82}, [57] \citet{cow83}, [58] \citet{gh2008}, [59] \citet{orosz4}, [60] \citet{corr11}, [61] \citet{jon4}, [62] \citet{mae6}, [63] \citet{mason85}, [64] \citet{thor87}, [65] \citet{cram87}, [66] \citet{bb2004}, [67] \citet{mar89}, [68] \citet{fab90}, [69] \citet{gies02}, [70] \citet{sm2002}, [71] \citet{shap10}, [72] \citet{shap9}, [73] \citet{ku13}, [74] \citet{corr13}, [75] \citet{cb09}, [76] \citet{sg07}, [77] \citet{shid13}, [78] \citet{st12}, [79] \citet{r14b}, [80] \citet{kr13}, [81] \citet{zur8}, [82] \citet{md13}, [83] \citet{mot98}, [84]\citet{pai07}, [85] \citet{si11}, [86] \citet{chenevez2014b}, [87] \citet{st05}, [88] \citet{hmxb06}, [89] \citet{tendulkar2014}, [90] \citet{chenevez14}, [91] \citet{gomez15}, [92] \citet{bars5}, [93] \citet{rob02}, [94] \citet{bart13}, [95] \citet{hy99}, [96] \citet{kar7}, [97] \citet{chat6}, [98] \citet{br5}, [99] \citet{mait5}, [100] \citet{star8}, [101] \citet{arm11}, [102] \citet{grein8}, [103] \citet{cac6}, [104] \citet{sala8}, [105] \citet{torr6}, [106] \citet{shab13}, [107] \citet{gh03}, [108] \citet{or02b}, [109] \citet{mj11a}, [110] \citet{kong12}, [111] \citet{hi04}, [112] \citet{z04}, [113] \citet{mac11}, [114] \citet{krimm2009b}, [115] \citet{co05}, [116] \citet{reid2014}, [117] \citet{white84}, [118] \citet{bm83}, [119] \citet{curran11},[120] \cite{macdonald2014},[121] \citet{muno2005}, [122] \citet{muno2005b}, [123] \citet{bower2005}, [124] \citet{vank96}, [125] \citet{predehl2000}, [126] \citet{ling2009}, [127] \citet{zdz2013},[128] \citet{williams10}, [129] \cite{casar14}, [130] \citet{casares2012}}\\                
\endlastfoot

XTE J0421+560&$5\pm3$&-&-&-&-&B0--2&465.84&92--94\\[0.500ex]

GRO J0422+32&2.49$\pm$0.3&3.69$\pm$0.41$^a$&1.19$\pm$0.02&63.7$\pm$5.2&0.116$\pm$0.08&M2(+2/-1)V&5.1&1,4,17--20,107\\[0.500ex]

4U 0538$-$641&48.1$\pm$2.2&6.95$\pm$0.33&2.3$\pm$0.3&69.65$\pm$0.56&0.532$\pm$0.096&B3V&40.9&8--10,57\\[0.500ex]

4U 0540$-$697&50$\pm$2.3&10.91$\pm$1.41&0.866$\pm$0.037&36.38$\pm$1.92&0.343$\pm$0.220&O7III&93.8&2,7,9--11,55\\[0.500ex]

1A 0620$-$00&1.06$\pm$0.12&6.60$\pm$0.25&3.1$\pm$0.04&51$\pm$0.9&0.060$\pm$0.004&K5V&7.8&1,3,4\\[0.500ex]

GRS 1009$-$45&3.82$\pm$0.27&5.95$\pm$0.89$^a$&3.17$\pm$0.12&62.0$\pm$5.1&0.137$\pm$0.015&G5--K7V&6.8&1,4,17,25,26\\
\end{longtable}
\clearpage
\end{landscape}

\renewcommand{\thefootnote}{\alph{footnote}}

\setcounter{LTchunksize}{50}

\clearpage
\LongTables
\renewcommand{\thefootnote}{\alph{footnote}}
\renewcommand\tabcolsep{3pt}

\begin{landscape}
\tabletypesize{\footnotesize}
\begin{longtable}{lccccccr}
\caption[Galactic BHXB Outburst History]{Galactic BHXB Outburst History}  \\
\hline \hline \\[-2ex]
   \multicolumn{1}{c}{Source Name} &
   \multicolumn{1}{c}{$t_{\rm beg}$$^a$} &
     \multicolumn{1}{c}{$t_{\rm end}$$^a$} &
     \multicolumn{1}{c}{Outburst} &
    \multicolumn{1}{c}{Literature$^b$} &
    \multicolumn{1}{c}{Algorithm $^b$} &        
        \multicolumn{1}{c}{Instrument $^y$} &   
   \multicolumn{1}{c}{References$^z$}\\
      && & \multicolumn{1}{c}{ID}&\multicolumn{1}{c}{Class} & \multicolumn{1}{c}{Class}&\multicolumn{1}{c}{Detection} & \\[0.5ex] \hline
   \\[-4ex]
      \label{table:outhistBH}
\endfirsthead
\multicolumn{8}{c}{{\tablename} \thetable{} -- Continued} \\[0.5ex]
\hline \hline \\[-2ex]
   \multicolumn{1}{c}{Source Name} &
   \multicolumn{1}{c}{$t_{\rm beg}$$^a$} &
     \multicolumn{1}{c}{$t_{\rm end}$$^a$} &
     \multicolumn{1}{c}{Outburst} &
    \multicolumn{1}{c}{Literature$^b$} &
    \multicolumn{1}{c}{Algorithm $^b$} &        
        \multicolumn{1}{c}{Instrument $^y$} &   
   \multicolumn{1}{c}{References$^z$}\\
      && & \multicolumn{1}{c}{ID}&\multicolumn{1}{c}{Class} & \multicolumn{1}{c}{Class}&\multicolumn{1}{c}{Detection} & \\[0.5ex] \hline
   \\[-1.8ex]
\endhead

  \\[-1.8ex] \hline  \\[-1.5ex]   
  \multicolumn{8}{l}{{Continued on Next Page\ldots}} \\
\endfoot

  \\[-1.8ex] \hline \\[-1.5ex]               
  \multicolumn{8}{l}{{Continued on Next Page\ldots}} \\
\endlastfoot

XTEJ0421+560 & 50885.0 & 50934.0 & 1998 & - &H&$\textbf{ASM}$,BATSE,EPIC,HEXTE,NFI,\textbf{PCA}&140--143\\[0.500ex]

  GROJ0422+32& - & -&1992  & H&-&BATSE&242,271--274\\
& - & - & 1993a & - &-&GIS,PSPC,SIS&270,272\\
& - & - & 1993b & - &-&GIS,PSPC,SIS&270,272\\[0.500ex]

4U0538$-$641&-&-& 1971--2015 & - &-&A-2,\textbf{ASM,BAT,GSC,HEXTE},LAC,ME,NFI&296--303,344,345\\
&  &  &  &&&SSS,UHURU,\textbf{PCA}&\\[0.500ex]

4U0540$-$697&-&-& 1971--2015 & - &-&A-2,\textbf{ASM,BAT,GSC,HEXTE},LAC,ME,NFI&296--303, 344,345\\
&  &  &  &&&SSS,UHURU,\textbf{PCA}&\\
\end{longtable}
\clearpage
\addtocounter{table}{-1}
\begin{longtable}{lccccccr}
\multicolumn{8}{c}{{\tablename} \thetable{} -- Continued} \\[0.5ex]
\hline \hline \\[-2ex]
   \multicolumn{1}{c}{Source Name} &
   \multicolumn{1}{c}{$t_{\rm beg}$$^a$} &
     \multicolumn{1}{c}{$t_{\rm end}$$^a$} &
     \multicolumn{1}{c}{Outburst} &
    \multicolumn{1}{c}{Literature$^b$} &
    \multicolumn{1}{c}{Algorithm $^b$} &        
        \multicolumn{1}{c}{Instrument $^y$} &   
   \multicolumn{1}{c}{References$^z$}\\
      && & \multicolumn{1}{c}{ID}&\multicolumn{1}{c}{Class} & \multicolumn{1}{c}{Class}&\multicolumn{1}{c}{Detection} & \\[0.5ex] \hline
   \\[-1.8ex]
\endfirsthead

\multicolumn{8}{c}{{\tablename} \thetable{} -- Continued} \\[0.5ex]
\hline \hline \\[-2ex]
   \multicolumn{1}{c}{Source Name} &
   \multicolumn{1}{c}{$t_{\rm beg}$$^a$} &
     \multicolumn{1}{c}{$t_{\rm end}$$^a$} &
     \multicolumn{1}{c}{Outburst} &
    \multicolumn{1}{c}{Literature$^b$} &
    \multicolumn{1}{c}{Algorithm $^b$} &        
        \multicolumn{1}{c}{Instrument $^y$} &   
   \multicolumn{1}{c}{References$^z$}\\
      && & \multicolumn{1}{c}{ID}&\multicolumn{1}{c}{Class} & \multicolumn{1}{c}{Class}&\multicolumn{1}{c}{Detection} & \\[0.5ex] \hline
   \\[-1.8ex]
\endhead

  \\[-1.8ex] \hline  \\[-1.5ex]   
  \multicolumn{8}{l}{{Continued on Next Page\ldots}} \\
\endfoot

  \\[-1.8ex] \hline \\[-1.5ex]               
  \multicolumn{8}{l}{{Continued on Next Page\ldots}} \\
\endlastfoot
4U1630$-$472 &-&-& 1969 & - &-& XC&5\\
&-&-& 1970/1971 & - &-& SSI,UHURU,XC&5,6\\
&-&-& 1972 & - &-& SSI,UHURU,XC&5,6\\
&-&-& 1974 & - &-& SSI,UHURU,XC&5,6\\
&-&-& 1976 & - &-& SSI,UHURU,XC&5,6\\
&-&-& 1977 & - &-& A-2,LASS,SSI&7\\
&-&-& 1978 & - &-& LASS&7\\
&-&-& 1979 & - &-& GIS,HRI,LAC,PSPC,SSS&8,10\\
\end{longtable}
\clearpage
\addtocounter{table}{-1}
\renewcommand\tabcolsep{3pt}
\begin{longtable}{lccccccr}
\multicolumn{8}{c}{{\tablename} \thetable{} -- Continued} \\[0.5ex]
\hline \hline \\[-2ex]
   \multicolumn{1}{c}{Source Name} &
   \multicolumn{1}{c}{$t_{\rm beg}$$^a$} &
     \multicolumn{1}{c}{$t_{\rm end}$$^a$} &
     \multicolumn{1}{c}{Outburst} &
    \multicolumn{1}{c}{Literature$^b$} &
    \multicolumn{1}{c}{Algorithm $^b$} &        
        \multicolumn{1}{c}{Instrument $^y$} &   
   \multicolumn{1}{c}{References$^z$}\\
      && & \multicolumn{1}{c}{ID}&\multicolumn{1}{c}{Class} & \multicolumn{1}{c}{Class}&\multicolumn{1}{c}{Detection} & \\[0.5ex] \hline
   \\[-1.8ex]
\endfirsthead

\multicolumn{8}{c}{{\tablename} \thetable{} -- Continued} \\[0.5ex]
\hline \hline \\[-2ex]
   \multicolumn{1}{c}{Source Name} &
   \multicolumn{1}{c}{$t_{\rm beg}$$^a$} &
     \multicolumn{1}{c}{$t_{\rm end}$$^a$} &
     \multicolumn{1}{c}{Outburst} &
    \multicolumn{1}{c}{Literature$^b$} &
    \multicolumn{1}{c}{Algorithm $^b$} &        
        \multicolumn{1}{c}{Instrument $^y$} &   
   \multicolumn{1}{c}{References$^z$}\\
      && & \multicolumn{1}{c}{ID}&\multicolumn{1}{c}{Class} & \multicolumn{1}{c}{Class}&\multicolumn{1}{c}{Detection} & \\[0.5ex] \hline
   \\[-1.8ex]
\endhead

  \\[-1.8ex] \hline  \\[-1.5ex]   
  \multicolumn{8}{l}{{Continued on Next Page\ldots}} \\
\endfoot

  \\[-1.8ex] \hline \\[-1.5ex]               
  \multicolumn{8}{l}{{Continued on Next Page\ldots}} \\
\endlastfoot
H 1705$-$250& - & - &1977  & - &-&A-3, Ariel V/ASM,SSI&233--236\\[0.500ex]

IGRJ17091$-$3624$^h$ & - & -& 1994& - & -&TTM&65\\
& - & -& 1996& - &- &ISGRI,WFC&66\\
& - & -& 2001& - &- &ISGRI,WFC&66\\
& - & -& 2003&S  &- &HEXTE,ISGRI,JEM-X,PCA&67,68\\
& 54267.0     & 54395.0 & 2007& S &-&\textbf{BAT},ISGRI,JEM-X,PCA,XRT&73\\
& 55590.0 & 56412.0 & 2011--2013 & S &S&\textbf{ASM,BAT,ISGRI}, JEM-X,\textbf{PCA},XRT&69,70\\[0.500ex]

IGRJ17098$-$3628$^i$& - & - & 2005--2007 & S&- &\textbf{BAT},EPIC,\textbf{ISGRI,PCA},RGS,XRT&226,227\\
\end{longtable}
\clearpage
\addtocounter{table}{-1}
\begin{longtable}{lccccccr}
\multicolumn{8}{c}{{\tablename} \thetable{} -- Continued} \\[0.5ex]
\hline \hline \\[-2ex]
   \multicolumn{1}{c}{Source Name} &
   \multicolumn{1}{c}{$t_{\rm beg}$$^a$} &
     \multicolumn{1}{c}{$t_{\rm end}$$^a$} &
     \multicolumn{1}{c}{Outburst} &
    \multicolumn{1}{c}{Literature$^b$} &
    \multicolumn{1}{c}{Algorithm $^b$} &        
        \multicolumn{1}{c}{Instrument $^y$} &   
   \multicolumn{1}{c}{References$^z$}\\
      && & \multicolumn{1}{c}{ID}&\multicolumn{1}{c}{Class} & \multicolumn{1}{c}{Class}&\multicolumn{1}{c}{Detection} & \\[0.5ex] \hline
   \\[-1.8ex]
\endfirsthead

\multicolumn{8}{c}{{\tablename} \thetable{} -- Continued} \\[0.5ex]
\hline \hline \\[-2ex]
   \multicolumn{1}{c}{Source Name} &
   \multicolumn{1}{c}{$t_{\rm beg}$$^a$} &
     \multicolumn{1}{c}{$t_{\rm end}$$^a$} &
     \multicolumn{1}{c}{Outburst} &
    \multicolumn{1}{c}{Literature$^b$} &
    \multicolumn{1}{c}{Algorithm $^b$} &        
        \multicolumn{1}{c}{Instrument $^y$} &   
   \multicolumn{1}{c}{References$^z$}\\
      && & \multicolumn{1}{c}{ID}&\multicolumn{1}{c}{Class} & \multicolumn{1}{c}{Class}&\multicolumn{1}{c}{Detection} & \\[0.5ex] \hline
   \\[-1.8ex]
\endhead

  \\[-1.8ex] \hline  \\[-1.5ex]   
  \multicolumn{8}{l}{{Continued on Next Page\ldots}} \\
\endfoot

  \\[-1.8ex] \hline \\[-1.5ex]               
  \multicolumn{8}{l}{{Continued on Next Page\ldots}} \\
\endlastfoot
CXOGC J174540.0$-$290031& - & - & 2004 &-  & -&EPIC,Chandra&497,498\\[0.500ex]

H1743$-$322 & - & - & 1977 &-  & -&A-4&58,59\\
& - & - & 1984 &-  & -&ME&44\\
 & - & - & 1996 &-  & -&TTM&45\\
& 52713.0 & 52945.0 & 2003 & S &S&\textbf{ASM},Chandra,\textbf{HEXTE},ISGRI,JEM-X,\textbf{PCA}&46--50\\
& 53182.0 & 53307.0 & 2004 & S &S &\textbf{ASM}.Chandra,HEXTE,ISGRI,JEM-X,\textbf{PCA}&51\\
& 53575.0 & 53654.0 & 2005 & S &S&\textbf{ASM,BAT},Chandra,HEXTE,\textbf{ISGRI,JEM-X,PCA}&51\\
& 54446.0 & 54513.0 & 2007/2008 & S &S&\textbf{ASM,BAT},Chandra,HEXTE,\textbf{ISGRI,JEM-X,PCA},XRT&55,56,351\\
\end{longtable}
\clearpage
\addtocounter{table}{-1}
\begin{longtable}{lccccccr}
\multicolumn{8}{c}{{\tablename} \thetable{} -- Continued} \\[0.5ex]
\hline \hline \\[-2ex]
   \multicolumn{1}{c}{Source Name} &
   \multicolumn{1}{c}{$t_{\rm beg}$$^a$} &
     \multicolumn{1}{c}{$t_{\rm end}$$^a$} &
     \multicolumn{1}{c}{Outburst} &
    \multicolumn{1}{c}{Literature$^b$} &
    \multicolumn{1}{c}{Algorithm $^b$} &        
        \multicolumn{1}{c}{Instrument $^y$} &   
   \multicolumn{1}{c}{References$^z$}\\
      && & \multicolumn{1}{c}{ID}&\multicolumn{1}{c}{Class} & \multicolumn{1}{c}{Class}&\multicolumn{1}{c}{Detection} & \\[0.5ex] \hline
   \\[-1.8ex]
\endfirsthead

\multicolumn{8}{c}{{\tablename} \thetable{} -- Continued} \\[0.5ex]
\hline \hline \\[-2ex]
   \multicolumn{1}{c}{Source Name} &
   \multicolumn{1}{c}{$t_{\rm beg}$$^a$} &
     \multicolumn{1}{c}{$t_{\rm end}$$^a$} &
     \multicolumn{1}{c}{Outburst} &
    \multicolumn{1}{c}{Literature$^b$} &
    \multicolumn{1}{c}{Algorithm $^b$} &        
        \multicolumn{1}{c}{Instrument $^y$} &   
   \multicolumn{1}{c}{References$^z$}\\
      && & \multicolumn{1}{c}{ID}&\multicolumn{1}{c}{Class} & \multicolumn{1}{c}{Class}&\multicolumn{1}{c}{Detection} & \\[0.5ex] \hline
   \\[-1.8ex]
\endhead

  \\[-1.8ex] \hline  \\[-1.5ex]   
  \multicolumn{8}{l}{{Continued on Next Page\ldots}} \\
\endfoot

  \\[-1.8ex] \hline \\[-1.5ex]               
  \multicolumn{8}{l}{{Continued on Next Page\ldots}} \\
\endlastfoot
SAXJ1819.3$-$2525$^r$ & -&- &1978  & - &- &SAI&408\\ 
& 51214.0 &    51605.0 &1999/2000  & - &I &\textbf{ASM,HEXTE},NFI,\textbf{PCA},WFC&94--97,104,105\\                   
&51615.0 &    51636.0& 2000a &-  &-&ASM,\textbf{PCA}&-\\                   
&51645.0 &    51697.0&2000b  &-  & -&ASM,\textbf{PCA}&-\\     
&51743.0 &    51949.0& 2000/2001 & - &-&ASM,\textbf{PCA}&98,99\\  
&51960.0 &    52560.0& 2001/2002 & - &-&\textbf{ASM,HEXTE,PCA}&98,99\\            
&52680.0 &    52761.0&2003  & - &- &ASM,\textbf{PCA}&-\\                    
&52776.0&     53147.0&2003/2004  &H &-&\textbf{ASM,HEXTE,PCA}&100--102\\
\end{longtable}
\clearpage
\addtocounter{table}{-1}
\begin{longtable}{lccccccr}
\multicolumn{8}{c}{{\tablename} \thetable{} -- Continued} \\[0.5ex]
\hline \hline \\[-2ex]
   \multicolumn{1}{c}{Source Name} &
   \multicolumn{1}{c}{$t_{\rm beg}$$^a$} &
     \multicolumn{1}{c}{$t_{\rm end}$$^a$} &
     \multicolumn{1}{c}{Outburst} &
    \multicolumn{1}{c}{Literature$^b$} &
    \multicolumn{1}{c}{Algorithm $^b$} &        
        \multicolumn{1}{c}{Instrument $^y$} &   
   \multicolumn{1}{c}{References$^z$}\\
      && & \multicolumn{1}{c}{ID}&\multicolumn{1}{c}{Class} & \multicolumn{1}{c}{Class}&\multicolumn{1}{c}{Detection} & \\[0.5ex] \hline
   \\[-1.8ex]
\endfirsthead

\multicolumn{8}{c}{{\tablename} \thetable{} -- Continued} \\[0.5ex]
\hline \hline \\[-2ex]
   \multicolumn{1}{c}{Source Name} &
   \multicolumn{1}{c}{$t_{\rm beg}$$^a$} &
     \multicolumn{1}{c}{$t_{\rm end}$$^a$} &
     \multicolumn{1}{c}{Outburst} &
    \multicolumn{1}{c}{Literature$^b$} &
    \multicolumn{1}{c}{Algorithm $^b$} &        
        \multicolumn{1}{c}{Instrument $^y$} &   
   \multicolumn{1}{c}{References$^z$}\\
      && & \multicolumn{1}{c}{ID}&\multicolumn{1}{c}{Class} & \multicolumn{1}{c}{Class}&\multicolumn{1}{c}{Detection} & \\[0.5ex] \hline
   \\[-1.8ex]
\endhead

  \\[-1.8ex] \hline  \\[-1.5ex]
  \multicolumn{8}{l}{{Continued on Next Page\ldots}} \\
\endfoot

  \\[-1.8ex] \hline \\[-1.5ex]          
  \multicolumn{8}{l}{{Continued on Next Page\ldots}} \\
\endlastfoot
GRS 1915+105& -&-&1992--2015  & - &-&\textbf{ASM,BAT},BATSE,Chandra,\textbf{GSC},IXAE&245--255\\
&  &  & & &  &PCA,PSPC,SIGMA,WATCH&365,366\\[0.500ex]

4U1956+350&-&-& 1971--2015 & - &-&A-2,A-4,Ariel V/ASM,\textbf{ASM,BAT},BATSE&310--326\\
&  &  &  &  &&BBXRT,CMA,GBM,GCXSE,GIS,\textbf{GSC},GPSC,HEXE&369--379\\
&  &  &  &  &&HEXTE,HRGRS,HSEM,LAC,LE&\\
&  &  &  &  &&ME,OSSE,PCA,SIS,Tenma,UHURU,VELA5B&\\[0.500ex]

4U1957+115&-&-& 1973--2015 & - &-&A-2,\textbf{ASM,BAT},Chandra,CMA,EPIC,GIS&304--309\\
&  &  &  &  &&\textbf{GSC},GSPC,HEXTE,LAC,LE,ME,PCA,PSPC&367,368\\
\end{longtable}
\clearpage
\addtocounter{table}{-1}
\begin{longtable}{lccccccr}
\multicolumn{8}{c}{{\tablename} \thetable{} -- Continued} \\[0.5ex]
\hline \hline \\[-2ex]
   \multicolumn{1}{c}{Source Name} &
   \multicolumn{1}{c}{$t_{\rm beg}$$^a$} &
     \multicolumn{1}{c}{$t_{\rm end}$$^a$} &
     \multicolumn{1}{c}{Outburst} &
    \multicolumn{1}{c}{Literature$^b$} &
    \multicolumn{1}{c}{Algorithm $^b$} &        
        \multicolumn{1}{c}{Instrument $^y$} &   
   \multicolumn{1}{c}{References$^z$}\\
      && & \multicolumn{1}{c}{ID}&\multicolumn{1}{c}{Class} & \multicolumn{1}{c}{Class}&\multicolumn{1}{c}{Detection} & \\[0.5ex] \hline
   \\[-1.8ex]
\endfirsthead

\multicolumn{8}{c}{{\tablename} \thetable{} -- Continued} \\[0.5ex]
\hline \hline \\[-2ex]
   \multicolumn{1}{c}{Source Name} &
   \multicolumn{1}{c}{$t_{\rm beg}$$^a$} &
     \multicolumn{1}{c}{$t_{\rm end}$$^a$} &
     \multicolumn{1}{c}{Outburst} &
    \multicolumn{1}{c}{Literature$^b$} &
    \multicolumn{1}{c}{Algorithm $^b$} &        
        \multicolumn{1}{c}{Instrument $^y$} &   
   \multicolumn{1}{c}{References$^z$}\\
      && & \multicolumn{1}{c}{ID}&\multicolumn{1}{c}{Class} & \multicolumn{1}{c}{Class}&\multicolumn{1}{c}{Detection} & \\[0.5ex] \hline
   \\[-1.8ex]
\endhead

  \\[-1.8ex] \hline  \\[-1.5ex]
  \multicolumn{8}{l}{{Continued on Next Page\ldots}} \\
\endfoot

  \\[-1.8ex] \hline \\[-1.5ex]            
  \multicolumn{8}{l}{{Continued on Next Page\ldots}} \\
\endlastfoot
\multicolumn{8}{p{0.95\columnwidth}}{\hangindent=1ex$^w$These outbursts were in progress as of May 14, 2015 (the cut-off date for our analysis).}\\
\multicolumn{8}{p{0.95\columnwidth}}{\hangindent=1ex$^x$These outbursts began after May 14, 2015, and therefore are not included in our analysis.}\\
\end{longtable}
\clearpage
\end{landscape}

%GRSMARK

\clearpage
\LongTables
\renewcommand{\thefootnote}{\alph{footnote}}
\renewcommand\tabcolsep{3pt}

\begin{landscape}
\tiny
\begin{longtable}{lccccccccccr}

\caption[Luminosity data for Galactic BHs \& BHCs Between 1996-2014]{Luminosity data for Galactic BHs \& BHCs Between 1996-2014}  \\

\hline \hline \\[-2ex]
   \multicolumn{1}{l}{Source Name} &
   \multicolumn{1}{c}{Outburst} &
     \multicolumn{1}{c}{$t_{\rm dur}$$^a$} &
     \multicolumn{1}{c}{$\int L dt$$^b$} &
    \multicolumn{1}{c}{$L_{\rm peak,tot}$$^c$} &
    \multicolumn{1}{c}{$L_{\rm tot,edd}$$^c$} &        
        \multicolumn{1}{c}{$L_{\rm peak,HCS}$$^d$} &   
            \multicolumn{1}{c}{$L_{\rm peak,SDS}$$^e$}&   
 \multicolumn{1}{c}{$t_{\rm HCS}$$^f$}&  
 \multicolumn{1}{c}{$t_{\rm SDS}$$^g$}& 
  \multicolumn{1}{c}{$t_{\rm IMS}$$^h$}&  
    \multicolumn{1}{r}{$t_{\rm und}$$^i$}\\                          
      &\multicolumn{1}{c}{ID} & \multicolumn{1}{c}{(days)} & \multicolumn{1}{c}{($\times 10^{43}$ erg)}&\multicolumn{1}{c}{($\times 10^{38}$ erg/s)} & \multicolumn{1}{c}{($L_{\rm peak,tot}/L_{\rm edd}$)}&\multicolumn{1}{c}{($\times 10^{38}$ erg/s)}&  \multicolumn{1}{c}{($\times 10^{38}$ erg/s)} & \multicolumn{1}{c}{(days)}& \multicolumn{1}{c}{(days)}& \multicolumn{1}{c}{(days)}& \multicolumn{1}{r}{(days)}\\[0.5ex] \hline
   \\[-4ex]
      \label{table:lumdata}
\endfirsthead

\multicolumn{12}{c}{{\tablename} \thetable{} -- Continued} \\[0.5ex]
\hline \hline \\[-2ex]
   \multicolumn{1}{l}{Source Name} &
   \multicolumn{1}{c}{Outburst} &
     \multicolumn{1}{c}{$t_{\rm dur}$$^a$} &
     \multicolumn{1}{c}{$\int L dt$$^b$} &
    \multicolumn{1}{c}{$L_{\rm peak,tot}$$^c$} &
    \multicolumn{1}{c}{$L_{\rm tot,edd}$$^c$} &        
        \multicolumn{1}{c}{$L_{\rm peak,HCS}$$^d$} &   
            \multicolumn{1}{c}{$L_{\rm peak,SDS}$$^e$}&   
 \multicolumn{1}{c}{$t_{\rm HCS}$$^f$}&  
 \multicolumn{1}{c}{$t_{\rm SDS}$$^g$}& 
  \multicolumn{1}{c}{$t_{\rm IMS}$$^h$}&  
    \multicolumn{1}{r}{$t_{\rm und}$$^i$}\\                          
      &\multicolumn{1}{c}{ID} & \multicolumn{1}{c}{(days)} & \multicolumn{1}{c}{($\times 10^{43}$ erg)}&\multicolumn{1}{c}{($\times 10^{38}$ erg/s)} & \multicolumn{1}{c}{($L_{\rm peak,tot}/L_{\rm edd}$)}&\multicolumn{1}{c}{($\times 10^{38}$ erg/s)}&  \multicolumn{1}{c}{($\times 10^{38}$ erg/s)} & \multicolumn{1}{c}{(days)}& \multicolumn{1}{c}{(days)}& \multicolumn{1}{c}{(days)}& \multicolumn{1}{r}{(days)}\\[0.5ex] \hline
   \\[-2ex]
\endhead

  \\[-1.8ex] \hline  \\[-1.5ex]
  \multicolumn{12}{l}{{Continued on Next Page\ldots}} \\
\endfoot

  \\[-1.8ex] \hline \\[-1.5ex]  
               \multicolumn{12}{l}{NOTE 1.-- This table is also available in machine readable format online at the Astrophysical Journal and on the WATCHDOG website -  }\\
                              \multicolumn{12}{l}{http://astro.physics.ualberta.ca/WATCHDOG/. }\\
    \multicolumn{12}{l}{NOTE 2. -- $L_{\rm peak,HCS}=$``-'' and $t>0$ or $L_{\rm peak,SDS}=$``-'' and $t>0$, indicate the minimum requirement of two different bands (one hard and one soft) for the spectral} \\
            \multicolumn{12}{l}{modelling process has not been met. A ``*'' beside the outburst year marks those outbursts which contain day in which our spectral modelling algorithm has} \\
                    \multicolumn{12}{l}{{\scriptsizeold{failed. As such the peak luminosities quoted in these cases may not be accurate. }}} \\
        \multicolumn{12}{l}{$^a$Indicates the duration of the outburst in days.} \\
         \multicolumn{12}{l}{$^b$Indicates the total amount of energy released during outburst.}\\
            \multicolumn{12}{l}{$^c$Indicates the peak (bolometric) luminosity reached during outburst.} \\
                \multicolumn{12}{l}{$^d$Indicates the peak (bolometric) luminosity reached when the source was in the (HCS).} \\
                    \multicolumn{12}{l}{$^e$Indicates the peak (bolometric) luminosity reached when the source was in the (SDS).} \\
                        \multicolumn{12}{l}{$^f$Indicates the total time the source spends in the HCS measured in days.} \\
                        \multicolumn{12}{l}{$^g$Indicates the total time the source spends in the SDS measured in days.} \\
                         \multicolumn{12}{l}{$^h$Indicates the total time the source spends in transition between the HCS and the SDS measured in days.} \\
                        \multicolumn{12}{l}{$^i$Indicates the total amount of time we do not have sufficient data to determine state of the source.}

\endlastfoot
XTEJ0421+560 & 1998 & 50 & 5.2 $\pm$ 1.2 & ${3.2}^{+5.2}_{-2.6}$ & ${0.32}^{+0.53}_{-0.29}$ & ${3.2}^{+5.2}_{-2.6}$ & - & 4 & 0 & 4 & 42\\[0.125ex]
 &  &  &  &  &  &  &  &  &  &  & \\[0.500ex]
 4U0538$-$641 & 1996-1999* & 1131 & 9830 $\pm$ 130 & ${41}^{+20}_{-18}$ & ${4.5}^{+2.0}_{-1.8}$ & ${16.1}^{+5.5}_{-4.9}$ & ${41}^{+20}_{-18}$ & 127 & 867 & 123 & 14\\[0.125ex]
 & 1999/2000* & 326 & 2904 $\pm$ 64 & ${37}^{+16}_{-13}$ & ${4.1}^{+1.4}_{-1.3}$ & ${19}^{+11}_{-10}$ & ${37}^{+16}_{-13}$ & 123 & 117 & 56 & 30\\[0.125ex]
 & 2000/2001* & 390 & 3658 $\pm$ 86 & ${73}^{+43}_{-44}$ & ${7.8}^{+4.4}_{-4.2}$ & ${73}^{+43}_{-44}$ & ${14.3}^{+8.1}_{-6.7}$ & 281 & 41 & 44 & 24\\[0.125ex]
 & 2001-2003* & 585 & 4850 $\pm$ 98 & ${23.4}^{+8.8}_{-8.0}$ & ${2.55}^{+0.85}_{-0.77}$ & ${18.0}^{+5.8}_{-5.3}$ & ${22.5}^{+8.1}_{-6.7}$ & 263 & 92 & 224 & 6\\[0.125ex]
 & 2003* & 255 & 2097 $\pm$ 73 & ${18.3}^{+7.2}_{-6.5}$ & ${2.05}^{+0.65}_{-0.73}$ & ${18.0}^{+8.5}_{-6.7}$ & - & 170 & 0 & 69 & 16\\[0.125ex]
 & 2004-2011* & 2794 & 25110 $\pm$ 240 & $39 \pm 21$ & ${4.5}^{+2.0}_{-2.2}$ & $39 \pm 21$ & ${38}^{+27}_{-23}$ & 56 & 2269 & 336 & 133\\
\end{longtable}
\clearpage
\end{landscape}

\clearpage

\renewcommand\tabcolsep{3pt}
\begin{longtable}{lcccccr}

\caption{State Transitions occurring in the Transient Galactic BHXB Population Between 1996--2015}  \\

\hline \hline \\[-2ex]
   \multicolumn{1}{l}{Source Name} &
   \multicolumn{1}{c}{Outburst} &
     \multicolumn{1}{c}{Stage} &
     \multicolumn{1}{c}{State(s)} &
    \multicolumn{1}{c}{$t_{\rm beg}$$^a$} &
    \multicolumn{1}{c}{$t_{\rm end}$$^a$} &        
        \multicolumn{1}{r}{$t_{\rm dur}$$^b$} \\   
      &\multicolumn{1}{c}{ID} &&&\multicolumn{1}{c}{(MJD)} & \multicolumn{1}{c}{(MJD)}&\multicolumn{1}{r}{(days)}\\[0.5ex] \hline
   \\[-1.8ex]
\endfirsthead

\multicolumn{7}{c}{{\tablename} \thetable{} -- Continued} \\[2.25ex]
\hline \hline \\[-2ex]
   \multicolumn{1}{l}{Source Name} &
   \multicolumn{1}{c}{Outburst} &
     \multicolumn{1}{c}{Stage} &
     \multicolumn{1}{c}{State(s)} &
    \multicolumn{1}{c}{$t_{\rm beg}$$^a$} &
    \multicolumn{1}{c}{$t_{\rm end}$$^a$} &        
        \multicolumn{1}{r}{$t_{\rm dur}$$^b$} \\   
      &\multicolumn{1}{c}{ID} &&&\multicolumn{1}{c}{(MJD)} & \multicolumn{1}{c}{(MJD)}&\multicolumn{1}{r}{(days)}\\[0.5ex] \hline
   \\[-1.8ex]
\endhead

  \\[-1.8ex] \hline  \\[0.12183380128ex]
  \multicolumn{7}{l}{{Continued on Next Page\ldots}} \\
\endfoot

  \\[-1.8ex] \hline \\   
      \multicolumn{7}{l}{{\scriptsizeold{NOTE. -- This is an excerpt of the full table. A portion is shown here for guidance}}} \\
                  \multicolumn{7}{l}{{\scriptsizeold{regarding its form and content. Table \ref{table:statetransitions} is published in its entirety (in machine }}} \\
            \multicolumn{7}{l}{{\scriptsizeold{readable format) online at the Astrophysical Journal and the WATCHDOG website -}}} \\
                        \multicolumn{7}{l}{{\scriptsizeold{http://astro.physics.ualberta.ca/WATCHDOG/.    }}} \\
      \multicolumn{7}{l}{{\scriptsizeold{$^a$The start and end times of a particular stage in an outburst, given in MJD.}}} \\
        \multicolumn{7}{l}{{\scriptsizeold{$^b$Duration of the particular outburst stage in days.}}} \\

\endlastfoot
GX339$-$4 & 1996-1999 & rise & - & 50259 & 50272 & 13\\
 &  & rise & HCS & 50272 & 50612 & 340\\
 &  & transition & HCS $\rightarrow$ IMS $\rightarrow$ HCS & 50612 & 50692 & 80\\
 &  & rise & HCS & 50692 & 50711 & 19\\
 &  & decline & HCS & 50711 & 50815 & 104\\
 &  & transition & HCS $\rightarrow$  SDS & 50815 & 50830 & 15\\
 &  & decline & SDS & 50830 & 50921 & 91\\
 &  & transition & SDS $\rightarrow$ IMS $\rightarrow$ SDS & 50921 & - & 1\\
\end{longtable}

 \renewcommand{\thefootnote}{\arabic{footnote}}

\clearpage

\renewcommand\tabcolsep{3pt}
\begin{longtable}{lcccccr}

\caption{State Transitions occurring in the Persistent Galactic BHXB Population Between 1996--2015}  \\

\hline \hline \\[-2ex]
   \multicolumn{1}{l}{Source Name} &
   \multicolumn{1}{c}{Outburst} &
     \multicolumn{1}{c}{Stage} &
     \multicolumn{1}{c}{State(s)} &
    \multicolumn{1}{c}{$t_{\rm beg}$$^a$} &
    \multicolumn{1}{c}{$t_{\rm end}$$^a$} &        
        \multicolumn{1}{r}{$t_{\rm dur}$$^b$} \\   
      &\multicolumn{1}{c}{ID} &&&\multicolumn{1}{c}{(MJD)} & \multicolumn{1}{c}{(MJD)}&\multicolumn{1}{r}{(days)}\\[0.5ex] \hline
   \\[-1.8ex]
\endfirsthead

\multicolumn{7}{c}{{\tablename} \thetable{} -- Continued} \\[2.25ex]
\hline \hline \\[-2ex]
   \multicolumn{1}{l}{Source Name} &
   \multicolumn{1}{c}{Outburst} &
     \multicolumn{1}{c}{Stage} &
     \multicolumn{1}{c}{State(s)} &
    \multicolumn{1}{c}{$t_{\rm beg}$$^a$} &
    \multicolumn{1}{c}{$t_{\rm end}$$^a$} &        
        \multicolumn{1}{r}{$t_{\rm dur}$$^b$} \\   
      &\multicolumn{1}{c}{ID} &&&\multicolumn{1}{c}{(MJD)} & \multicolumn{1}{c}{(MJD)}&\multicolumn{1}{r}{(days)}\\[0.5ex] \hline
   \\[-1.8ex]
\endhead

  \\[-1.8ex] \hline  \\[0.12183380128ex]
  \multicolumn{7}{l}{{Continued on Next Page\ldots}} \\
\endfoot

  \\[-1.8ex] \hline \\   
      \multicolumn{7}{l}{{\scriptsizeold{NOTE. -- This is an excerpt of the full table. A portion is shown here for guidance}}} \\
                  \multicolumn{7}{l}{{\scriptsizeold{regarding its form and content. Table \ref{table:statetransitions2} is published in its entirety (in machine }}} \\
            \multicolumn{7}{l}{{\scriptsizeold{readable format) online at the Astrophysical Journal and the WATCHDOG website - }}} \\
                        \multicolumn{7}{l}{{\scriptsizeold{http://astro.physics.ualberta.ca/WATCHDOG/.    }}} \\
      \multicolumn{7}{l}{{\scriptsizeold{$^a$The start and end times of a particular stage in an outburst, given in MJD.}}} \\
        \multicolumn{7}{l}{{\scriptsizeold{$^b$Duration of the particular outburst stage in days.}}} \\
 
\endlastfoot
4U1956+350 & 1996-2015 & rise & HCS & 50087 & 50206 & 119\\
 &  & transition & HCS $\rightarrow$ IMS $\rightarrow$ HCS & 50206 & 50226 & 20\\
 &  & rise & HCS & 50226 & 50229 & 3\\
 &  & transition & HCS $\rightarrow$  SDS & 50229 & 50240 & 11\\
 &  & rise & SDS & 50240 & 50249 & 9\\
 &  & transition & SDS $\rightarrow$ IMS $\rightarrow$ SDS & 50249 & 50269 & 20\\
 &  & rise & SDS & 50269 & 50303 & 34\\
 &  & transition & SDS $\rightarrow$  HCS & 50303 & 50322 & 19\\
\end{longtable}

 \renewcommand{\thefootnote}{\arabic{footnote}}
\tabletypesize{\scriptsize}
\clearpage

\clearpage

\begin{appendix}

The luminosity functions of individual sources are presented in this appendix.

\clearpage

\begin{figure*}%
\epsscale{0.85}
\plottwo{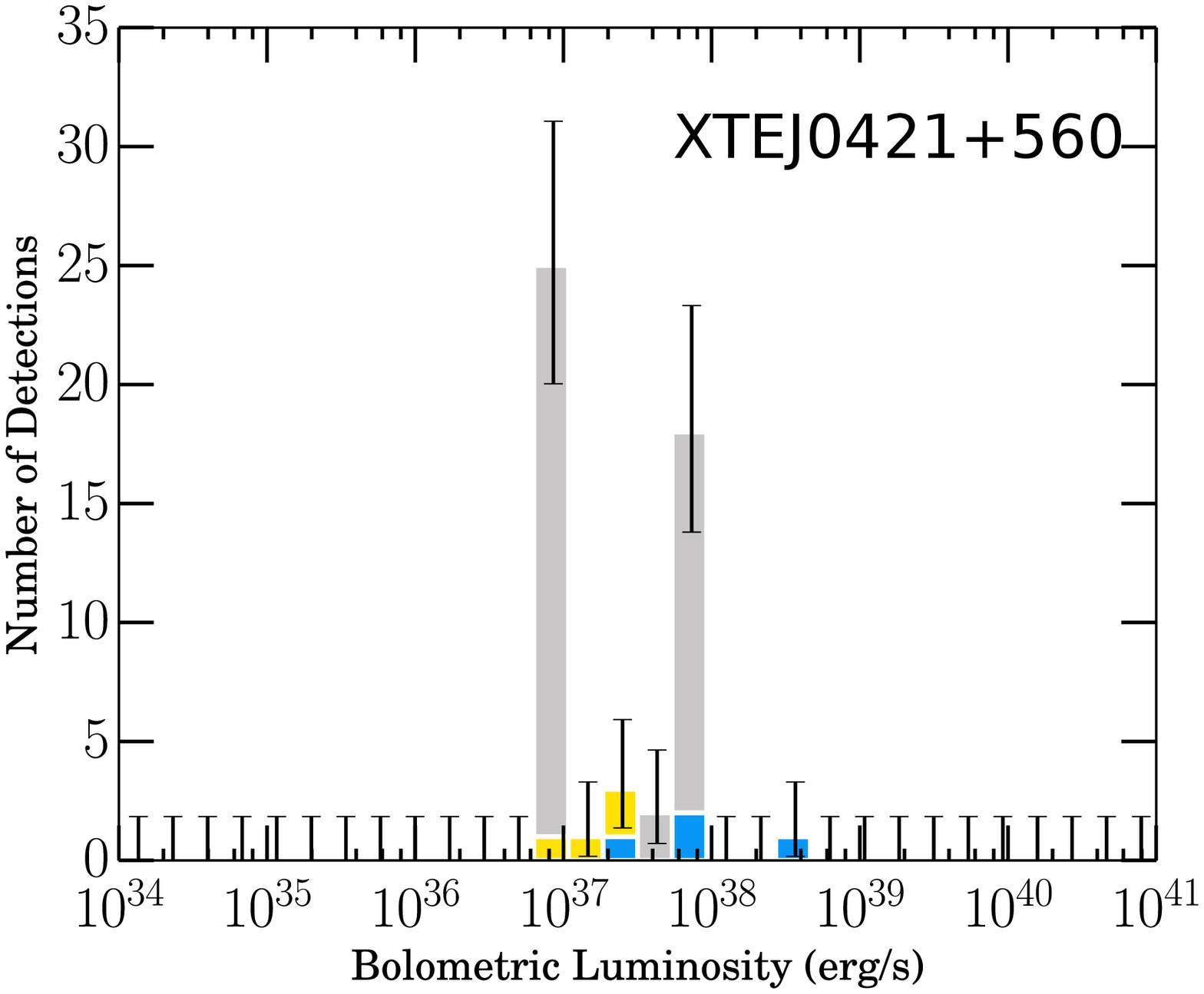}{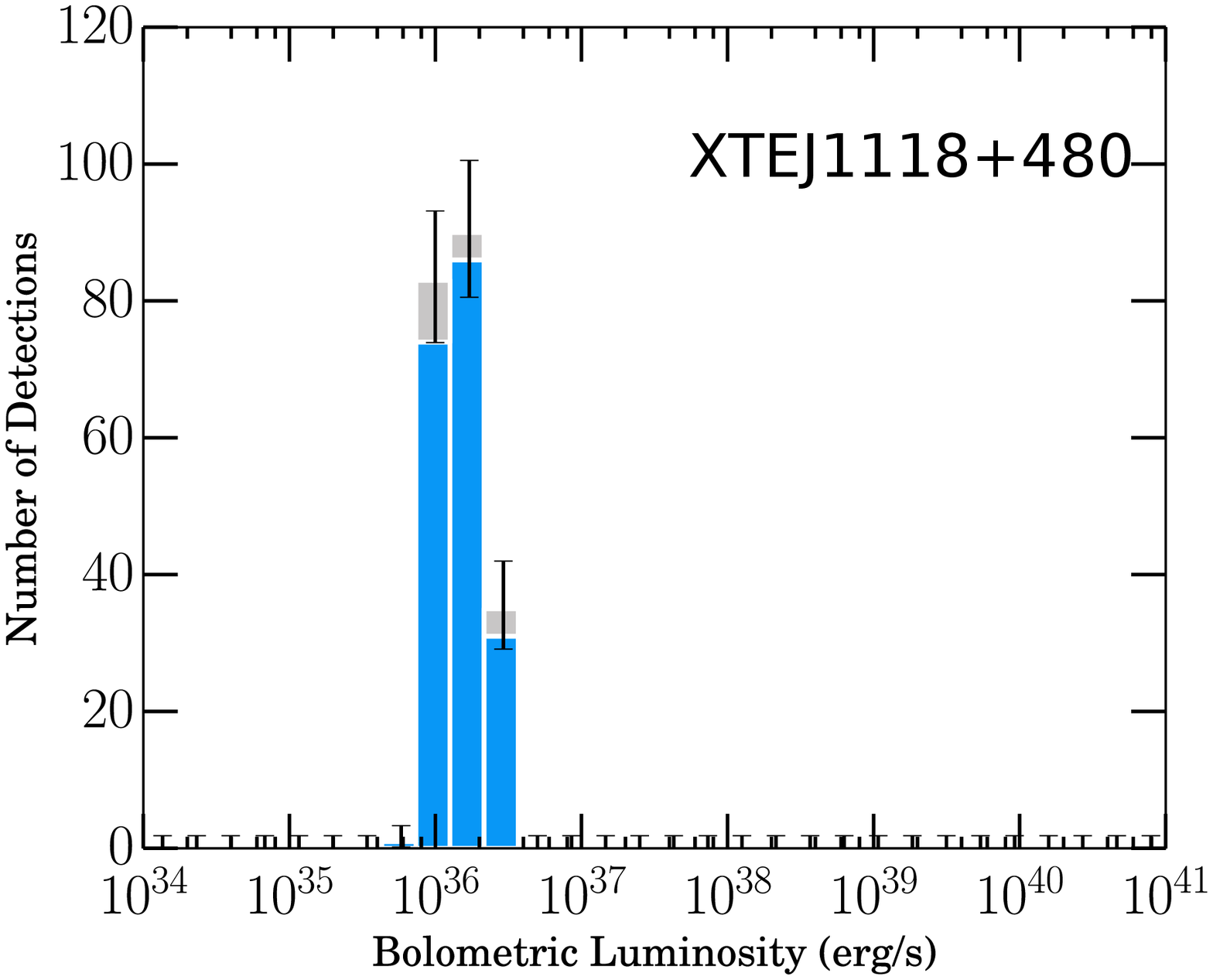}
\plottwo{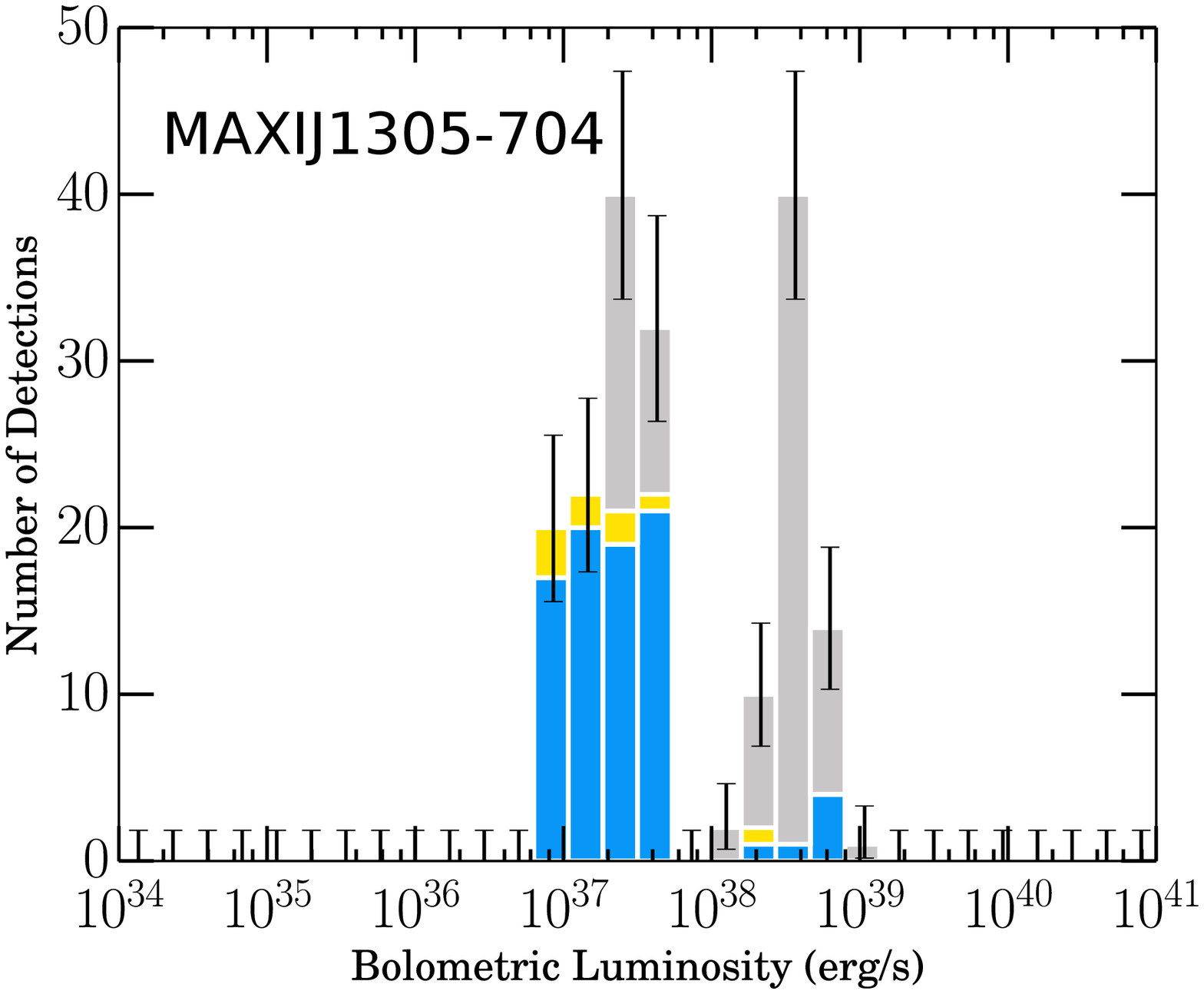}{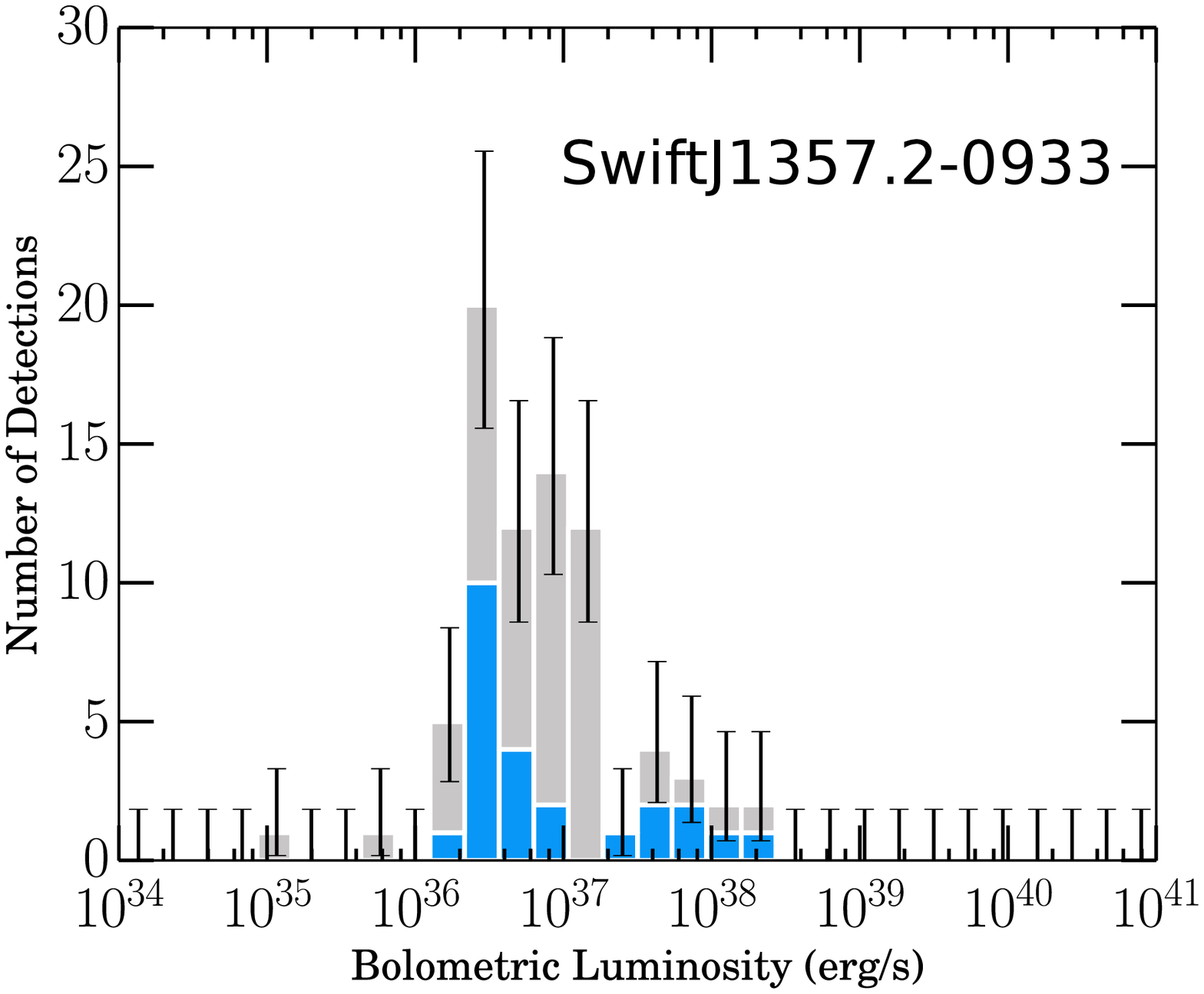}
\plottwo{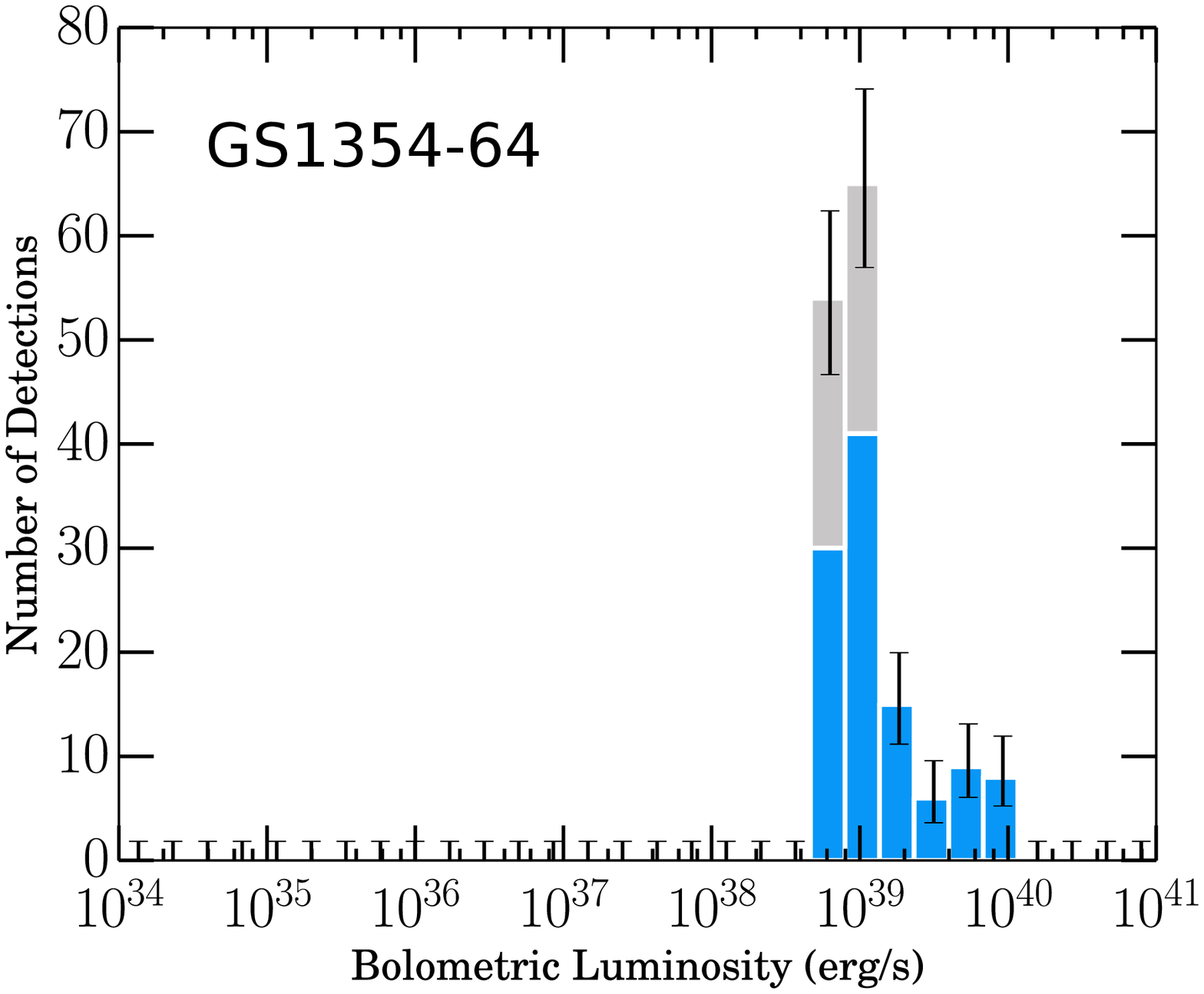}{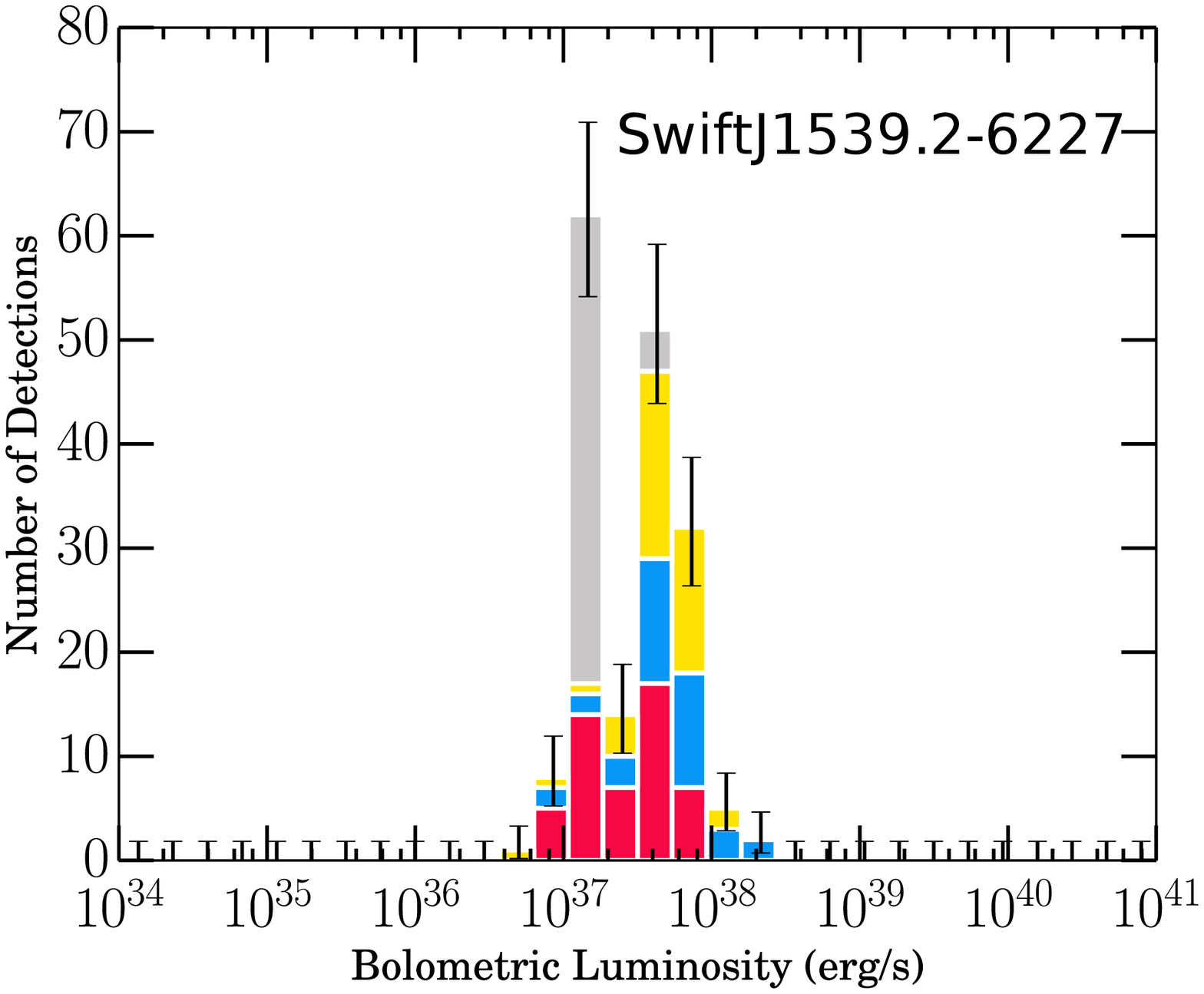}
\plottwo{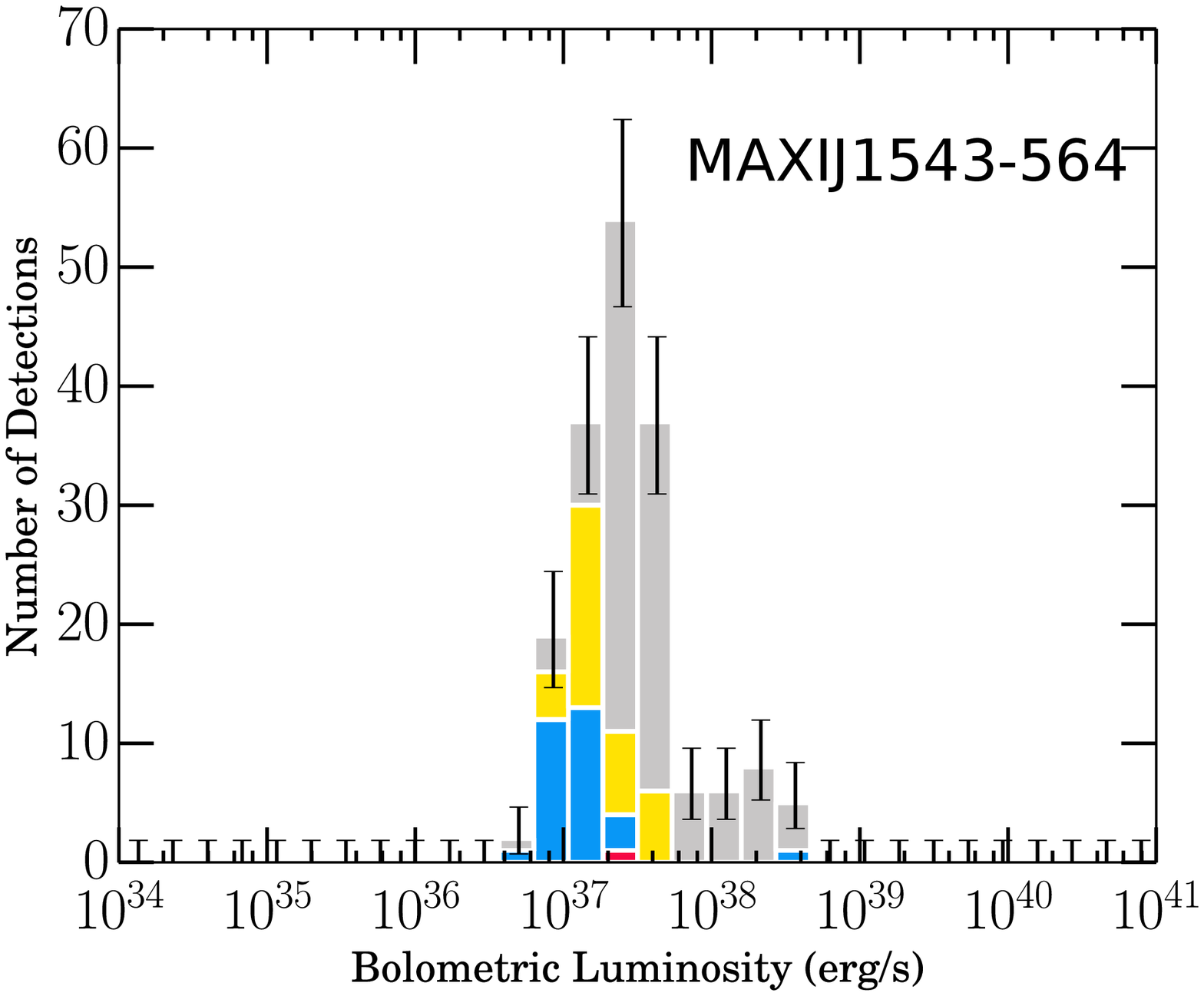}{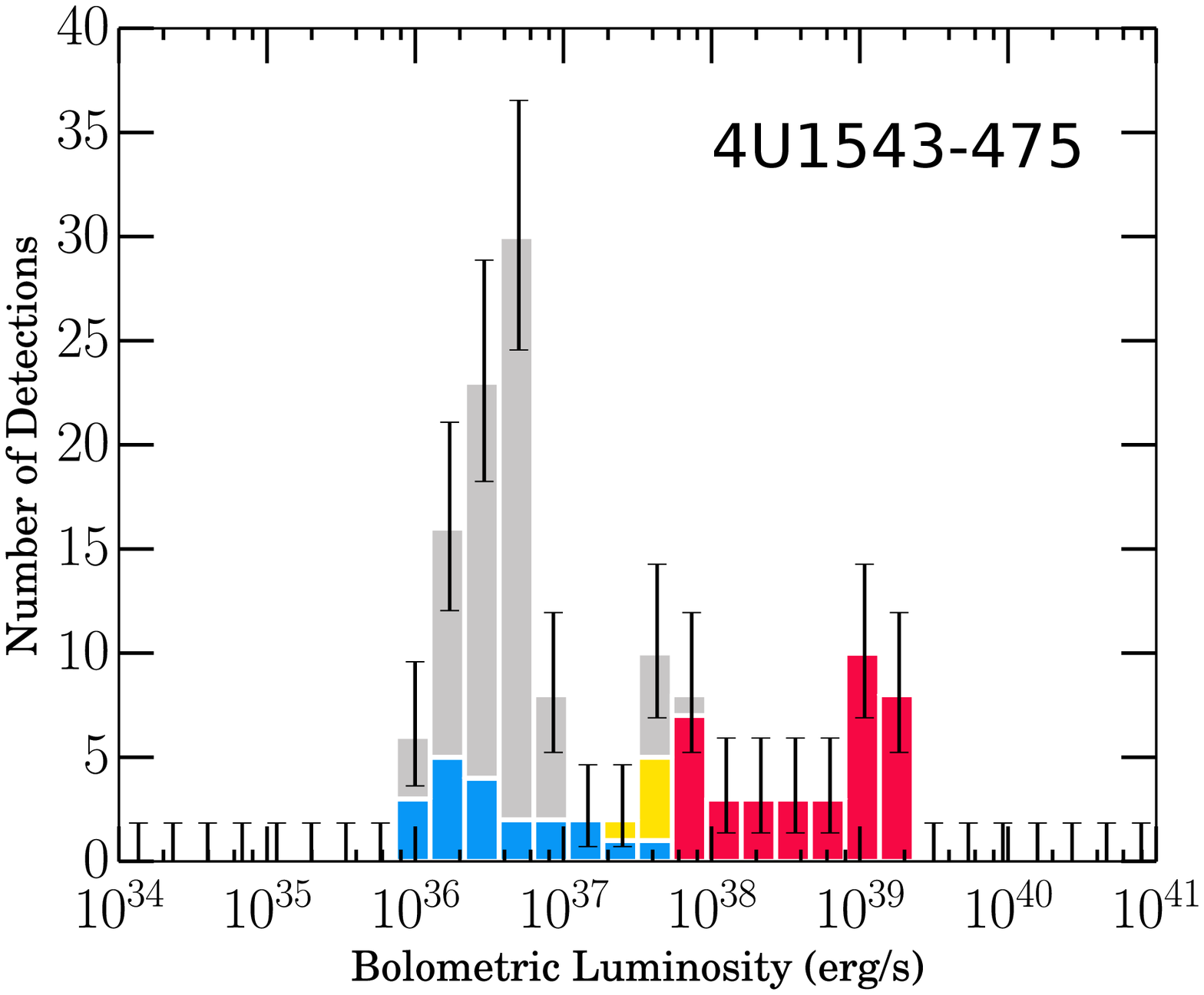}
    \caption{Transient XLFs color coded by state. HCS (blue), SDS (red), IMS (yellow), and unable to determine state with data available (grey).}%
    \label{fig:tXLF1}%
\end{figure*}
\afterpage{\clearpage}

\addtocounter{figure}{-1}
\begin{figure*}%
\epsscale{0.85}
\plottwo{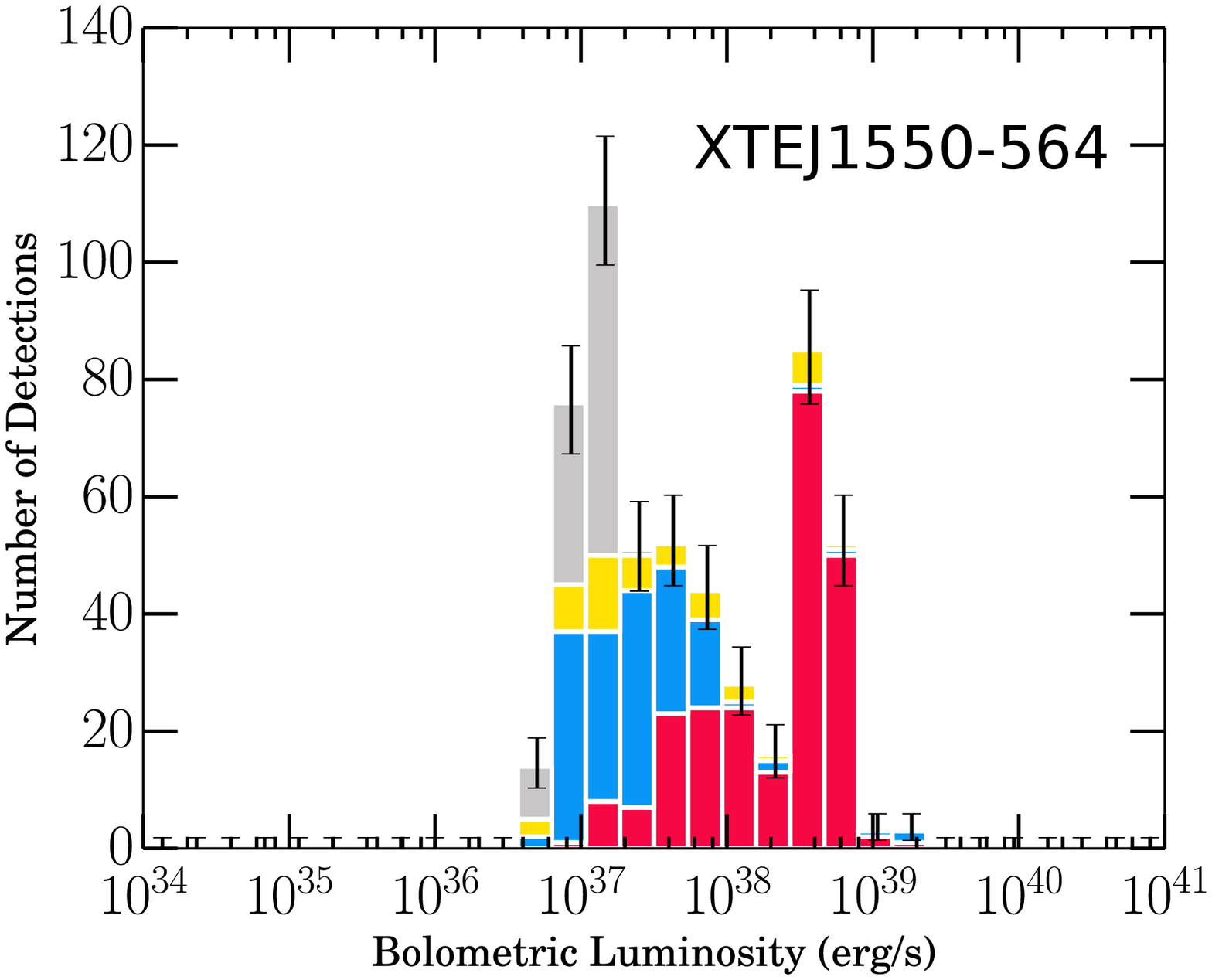}{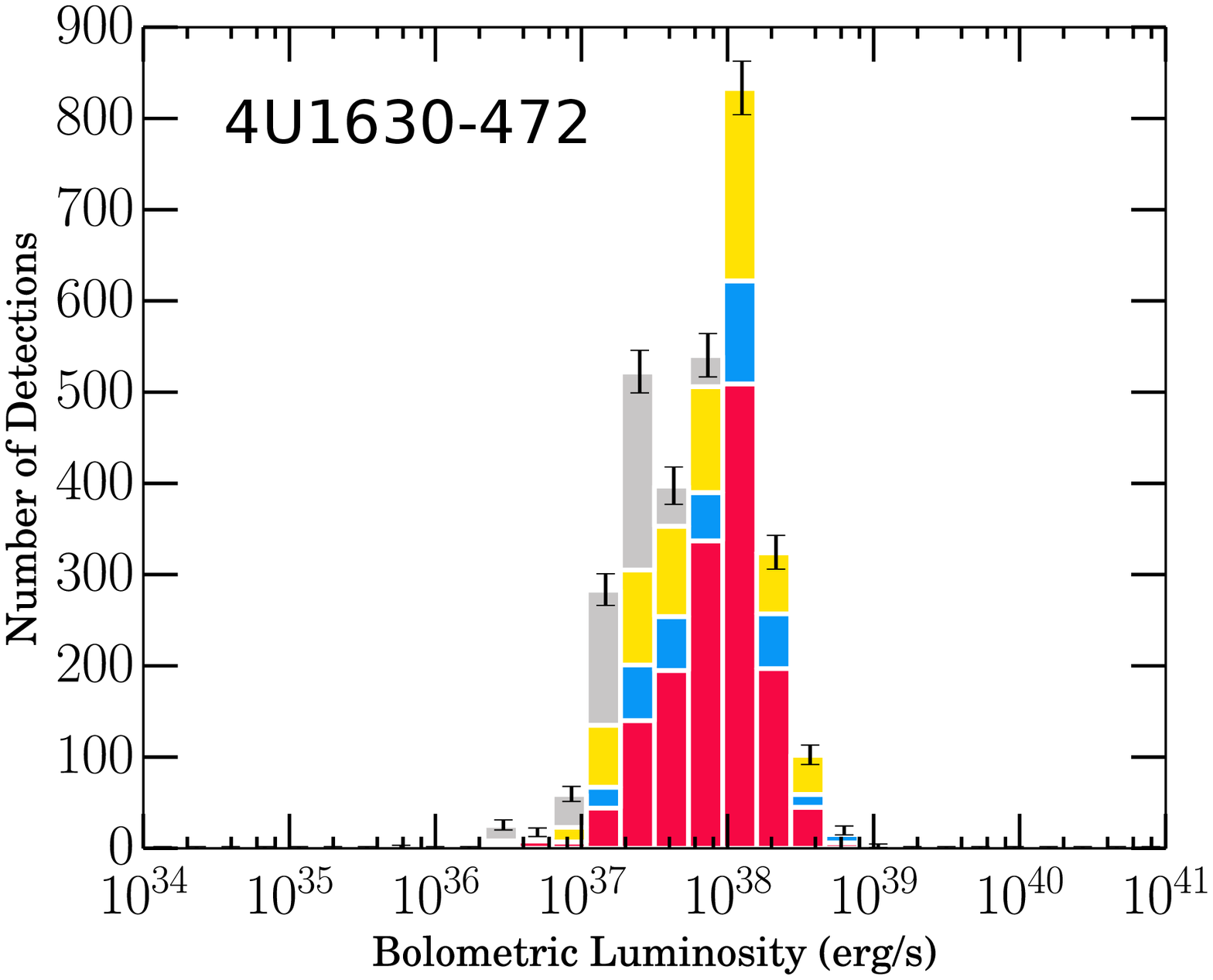}
\plottwo{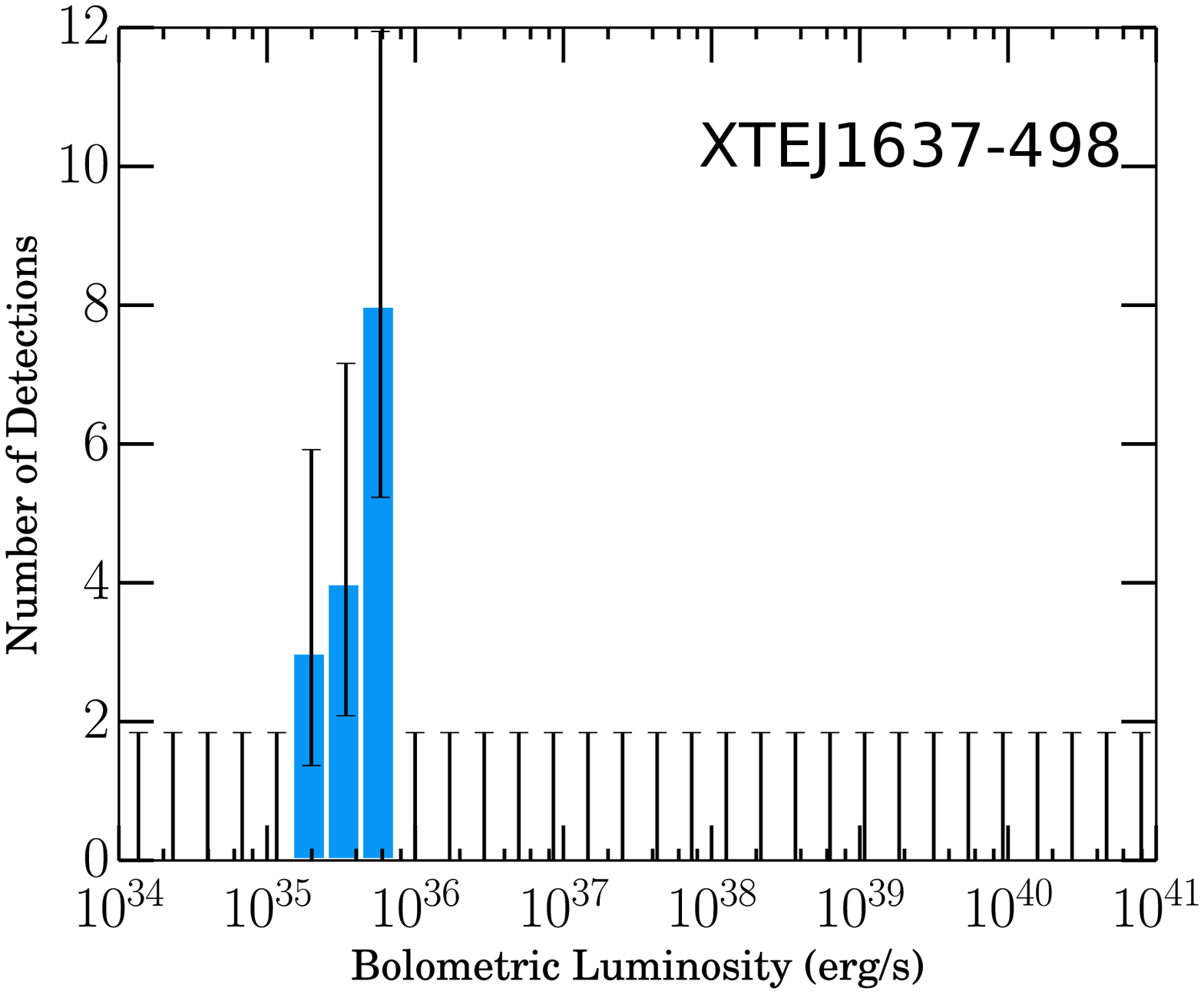}{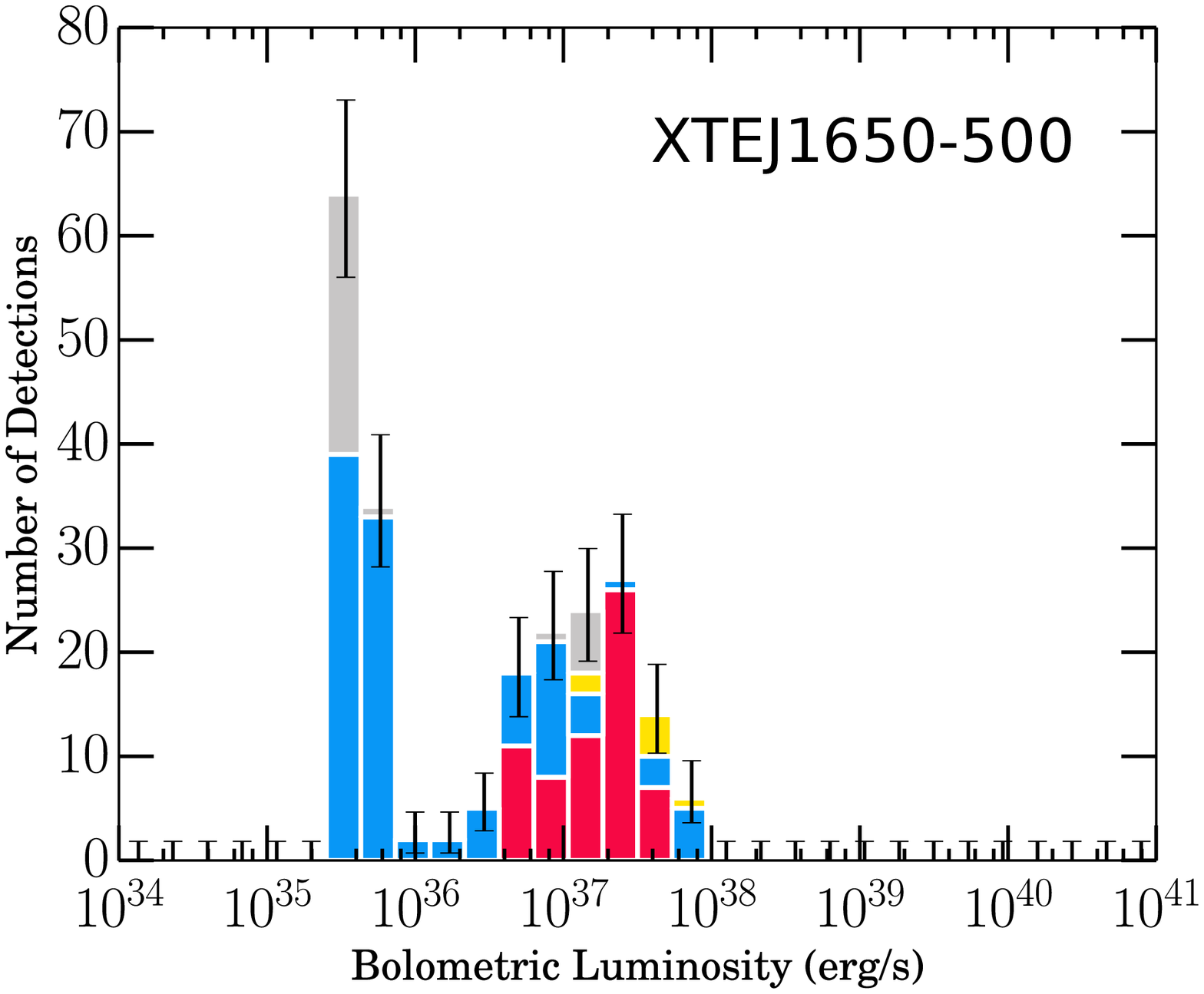}
\plottwo{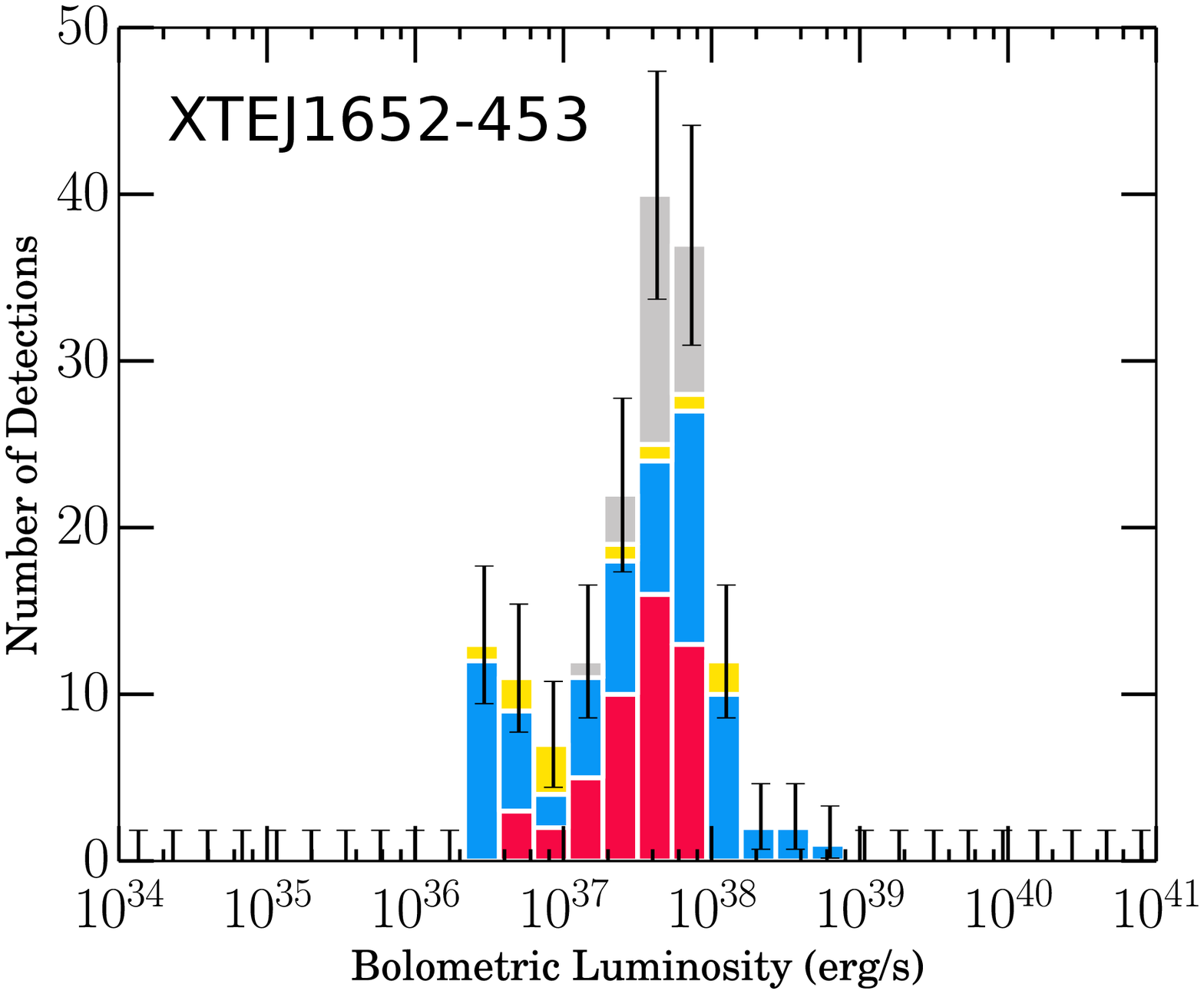}{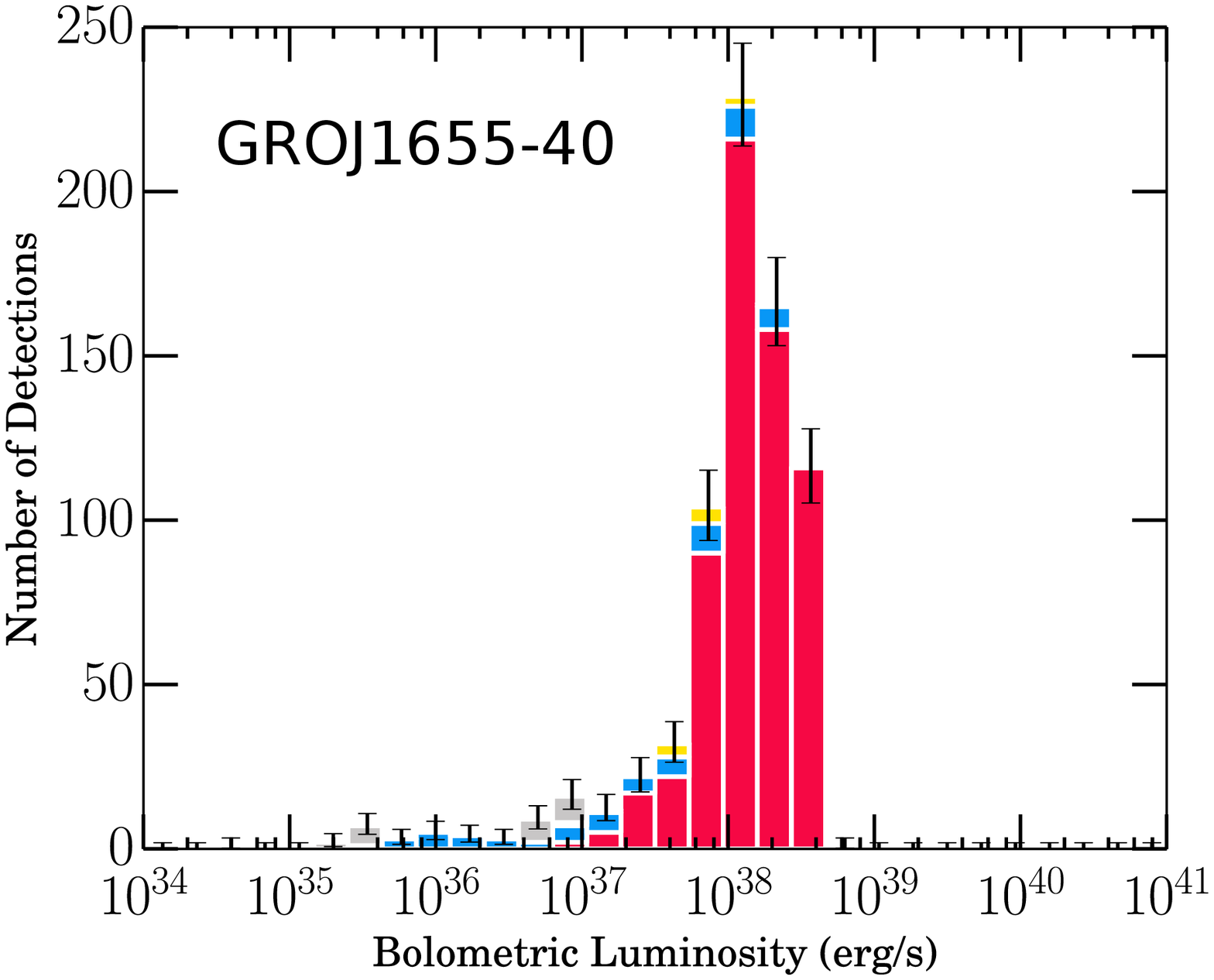}
\plottwo{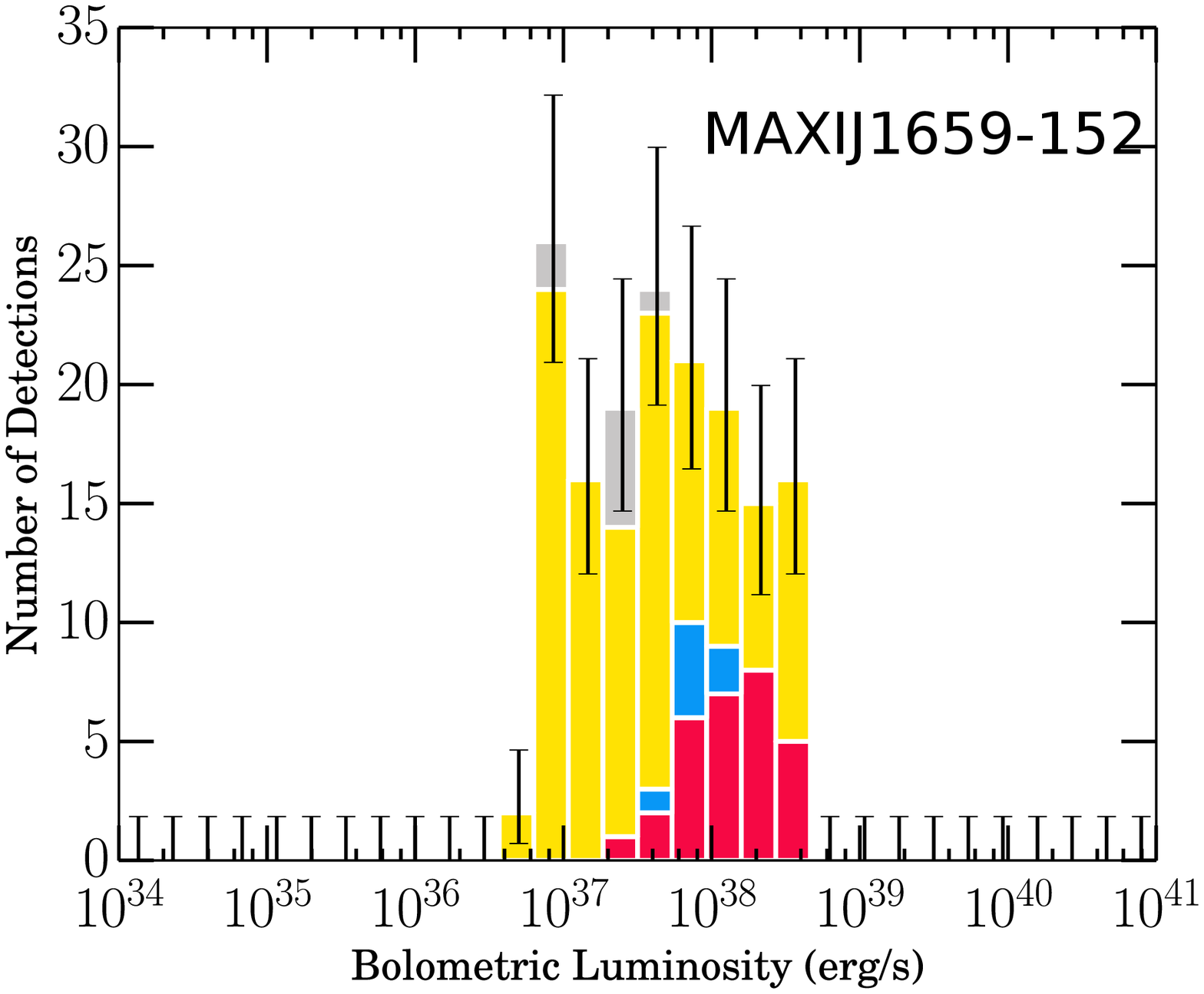}{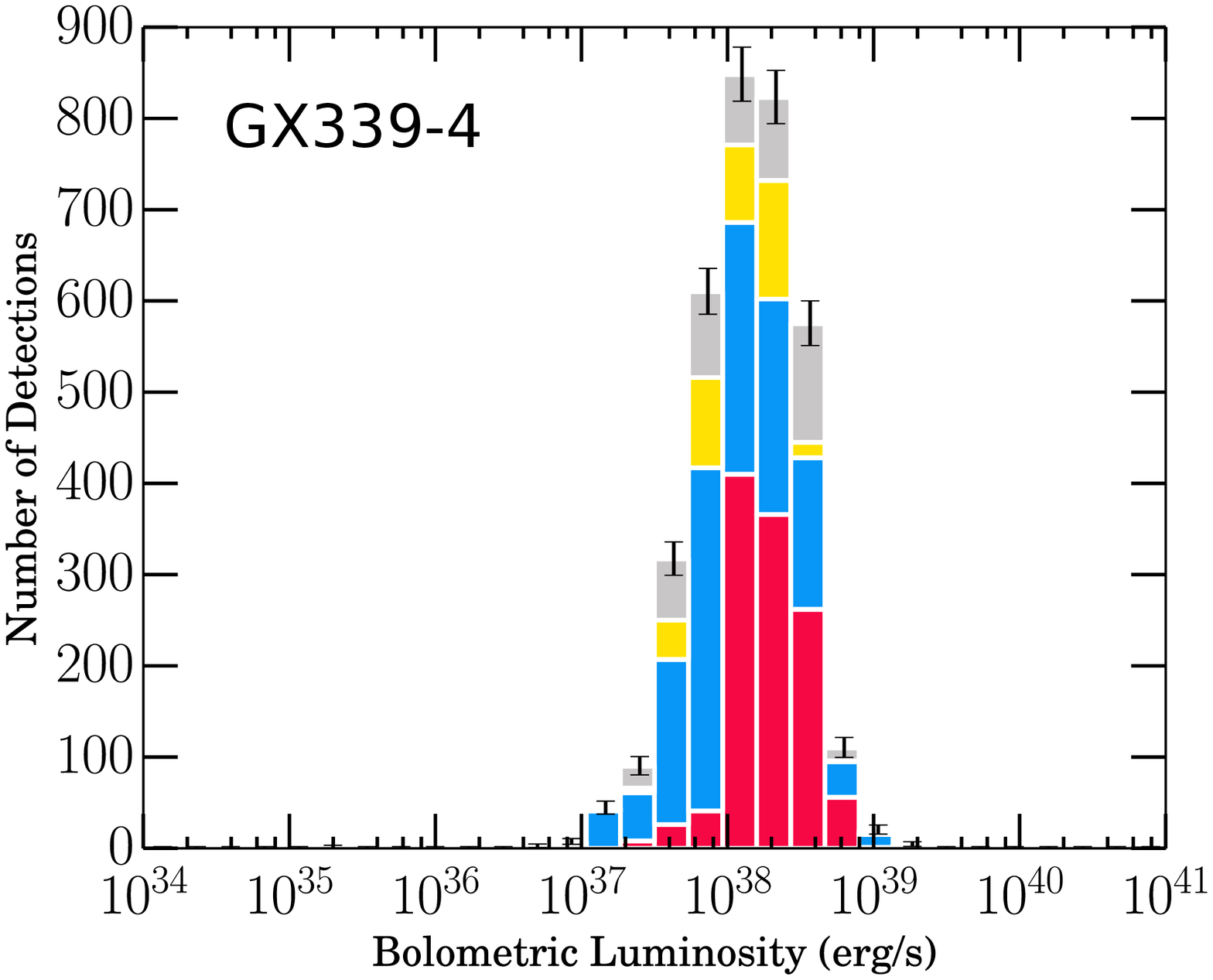}
    \caption{{\bf (cont.)} Transient XLFs color coded by state. HCS (blue), SDS (red), IMS (yellow), and unable to determine state with data available (grey). }%
    \label{fig:tXLF22}%
\end{figure*}
\afterpage{\clearpage}

\addtocounter{figure}{-1}
\begin{figure*}%
\epsscale{0.85}
\plottwo{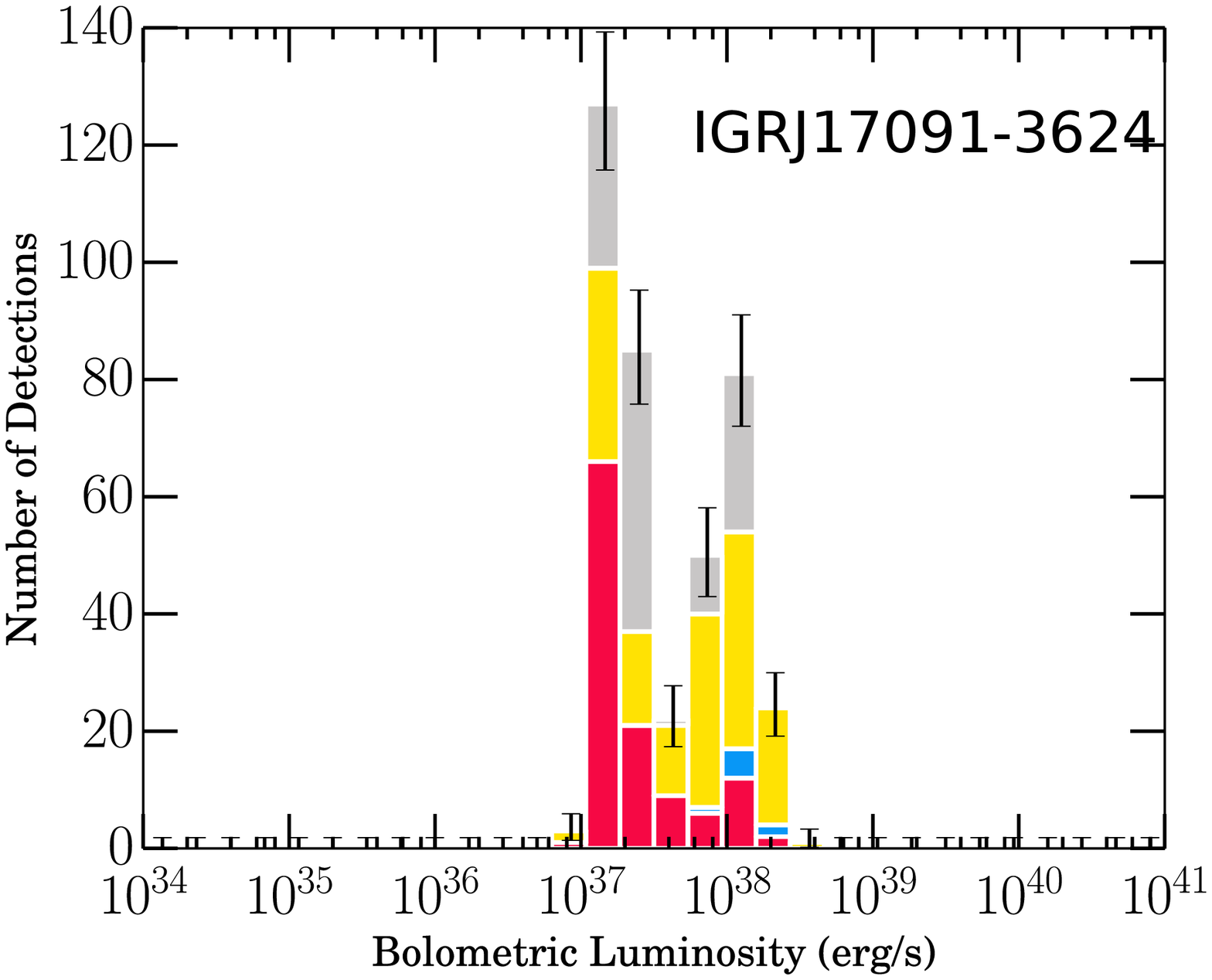}{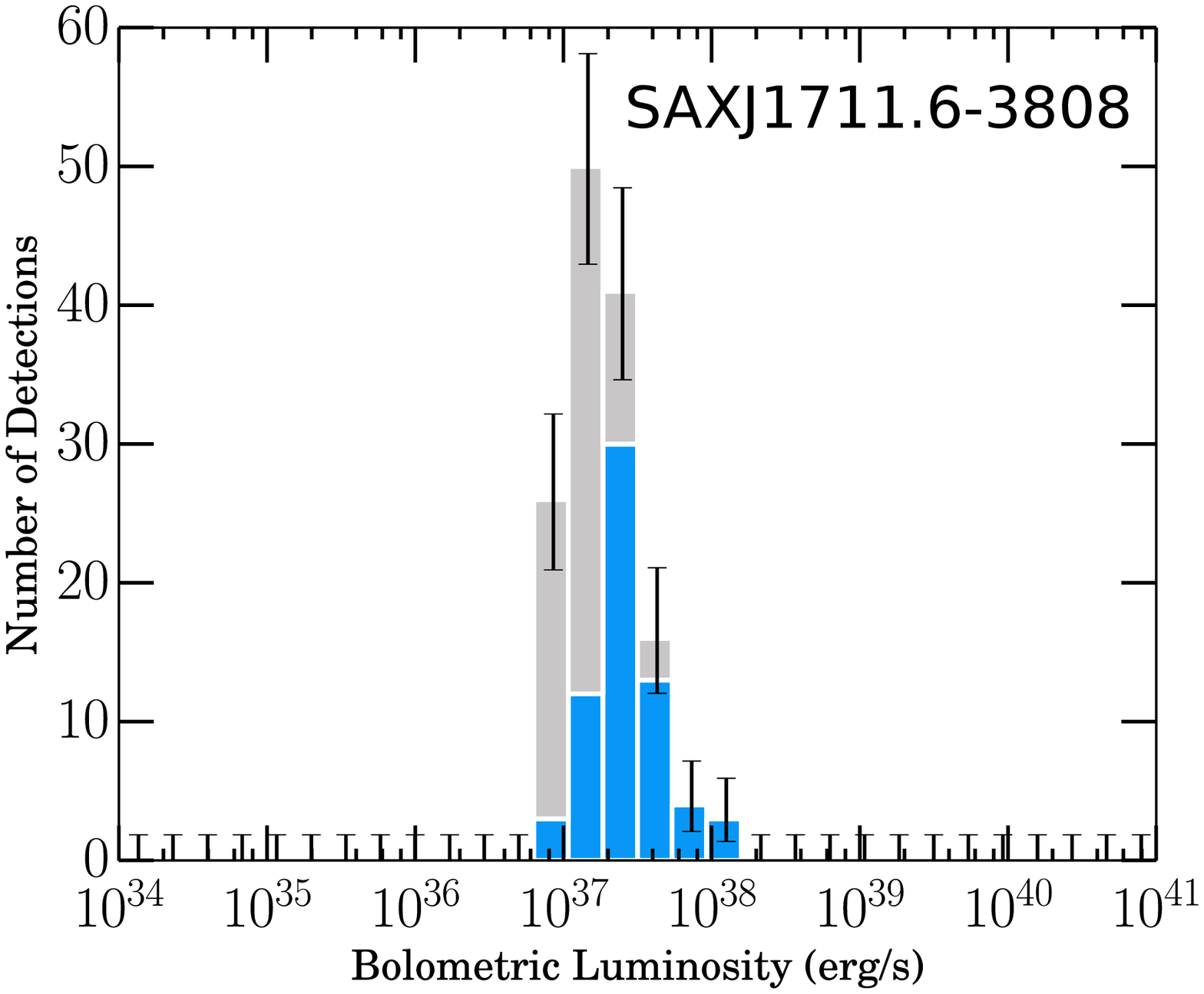}
\plottwo{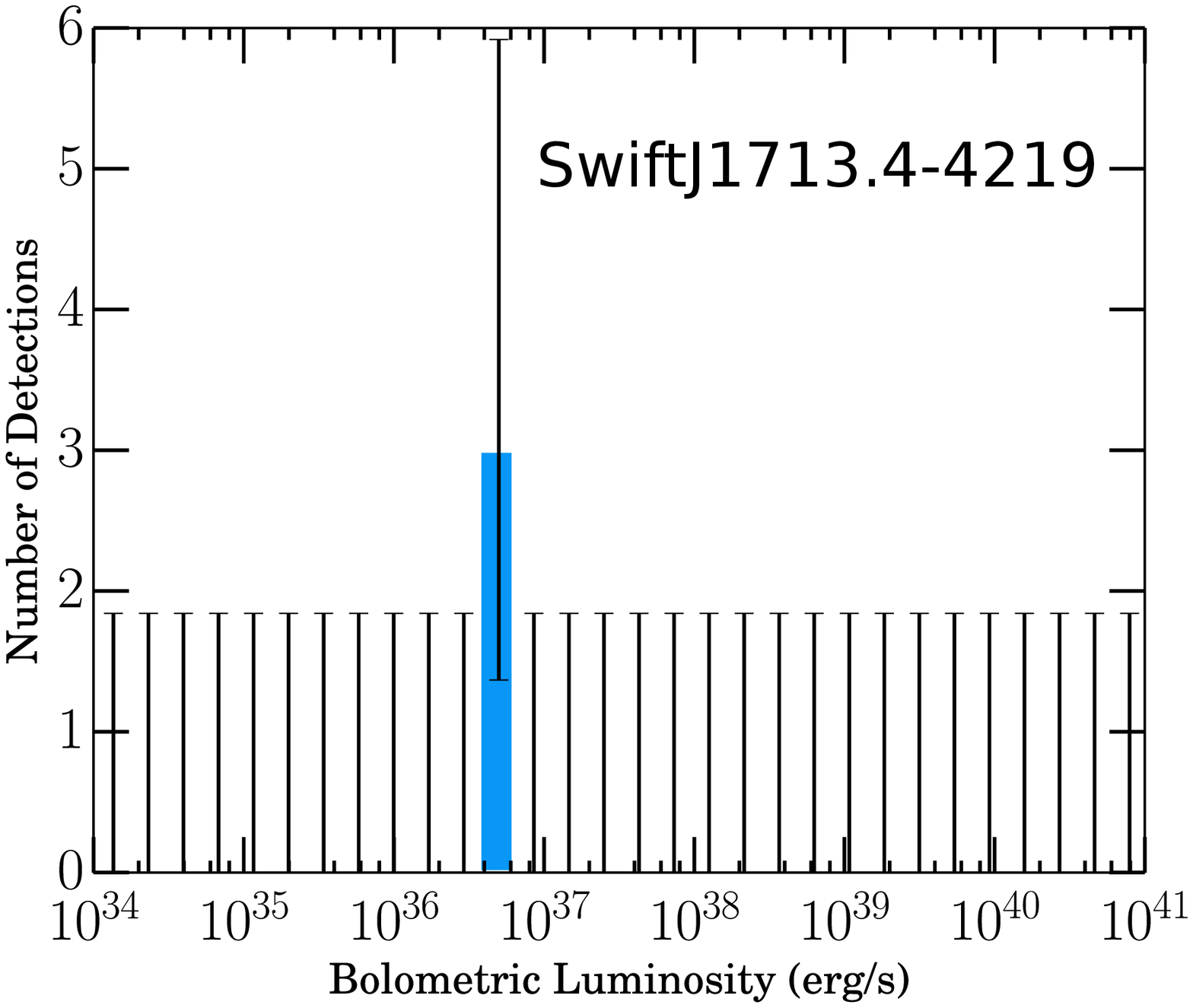}{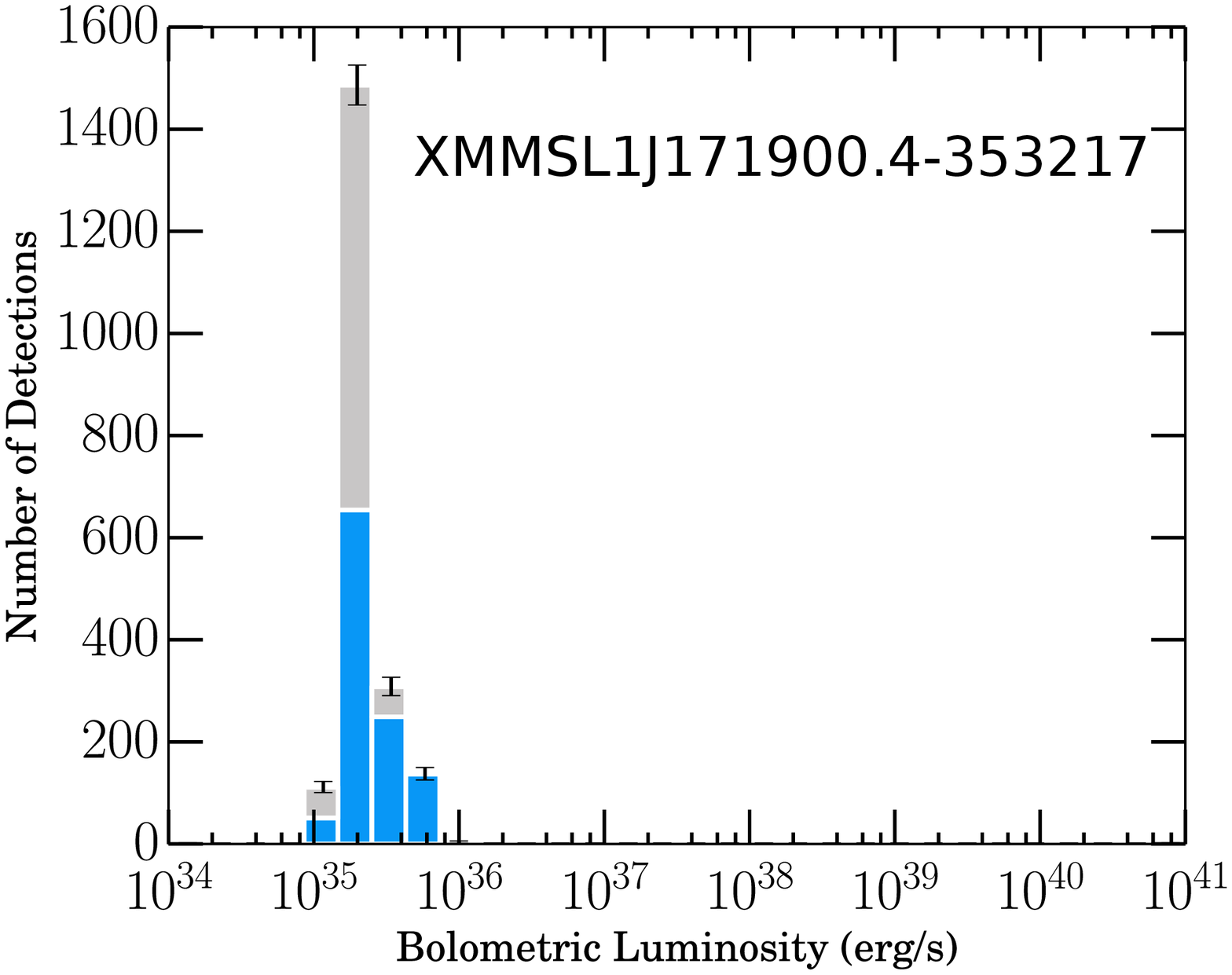}
\plottwo{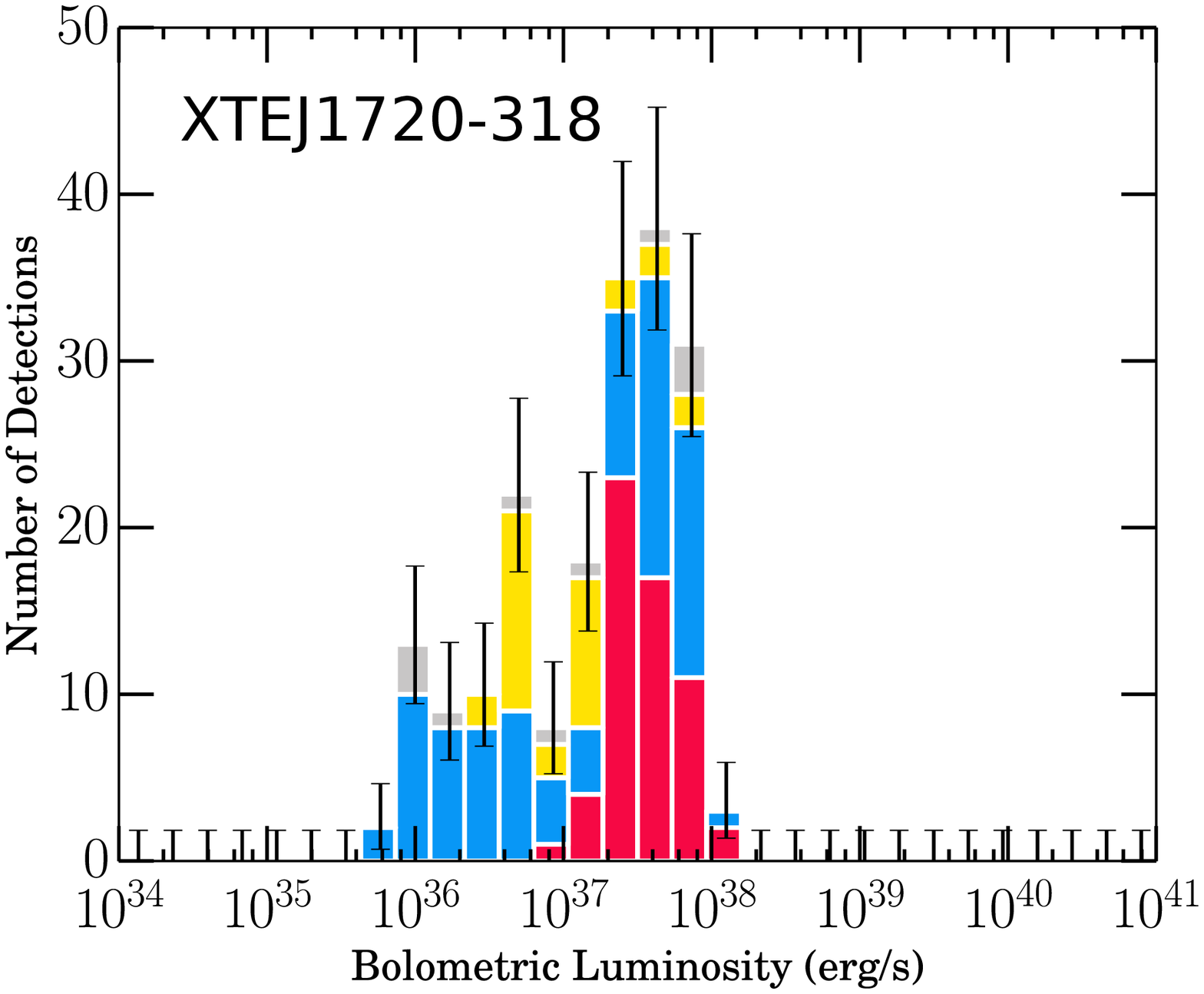}{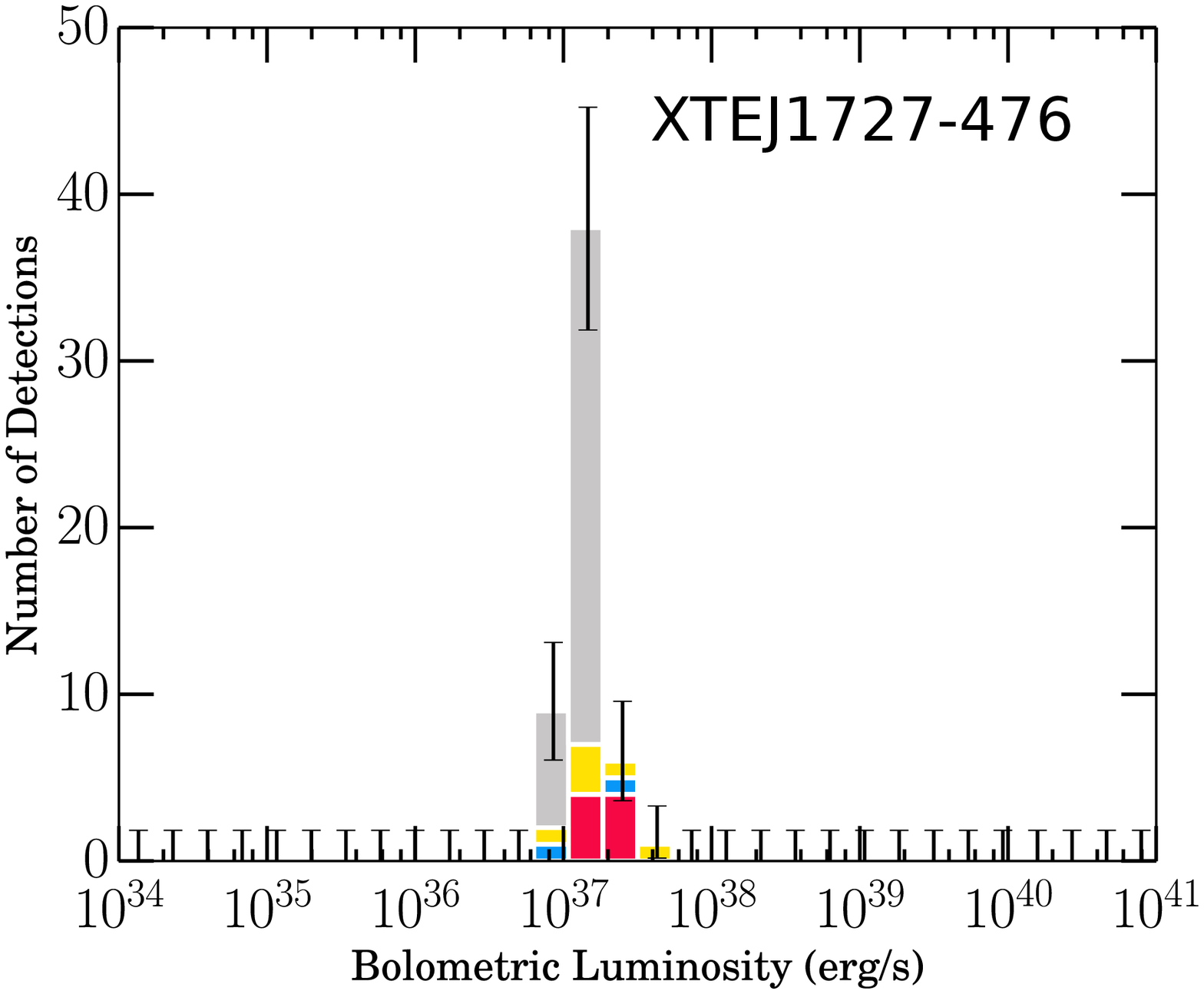}
\plottwo{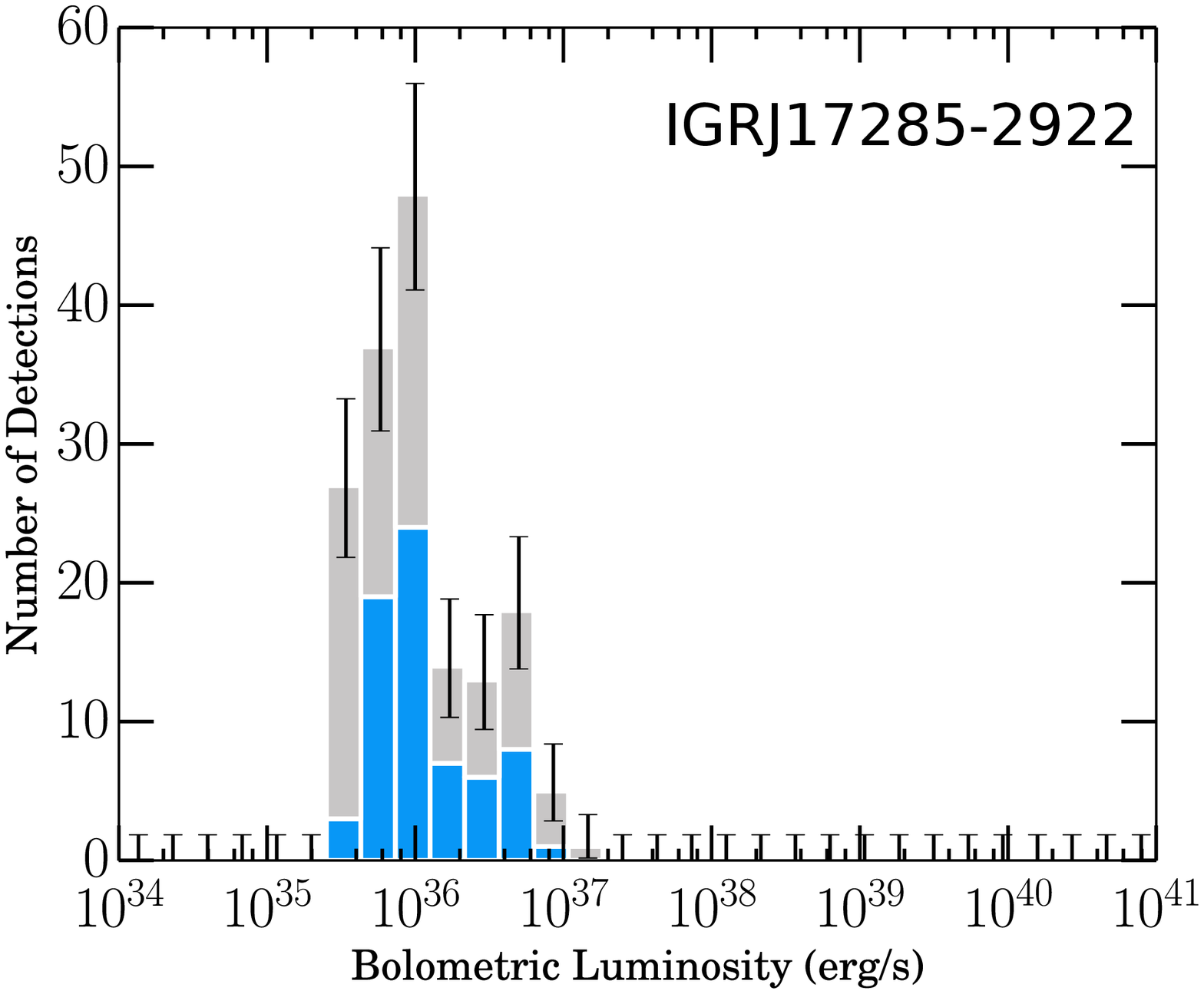}{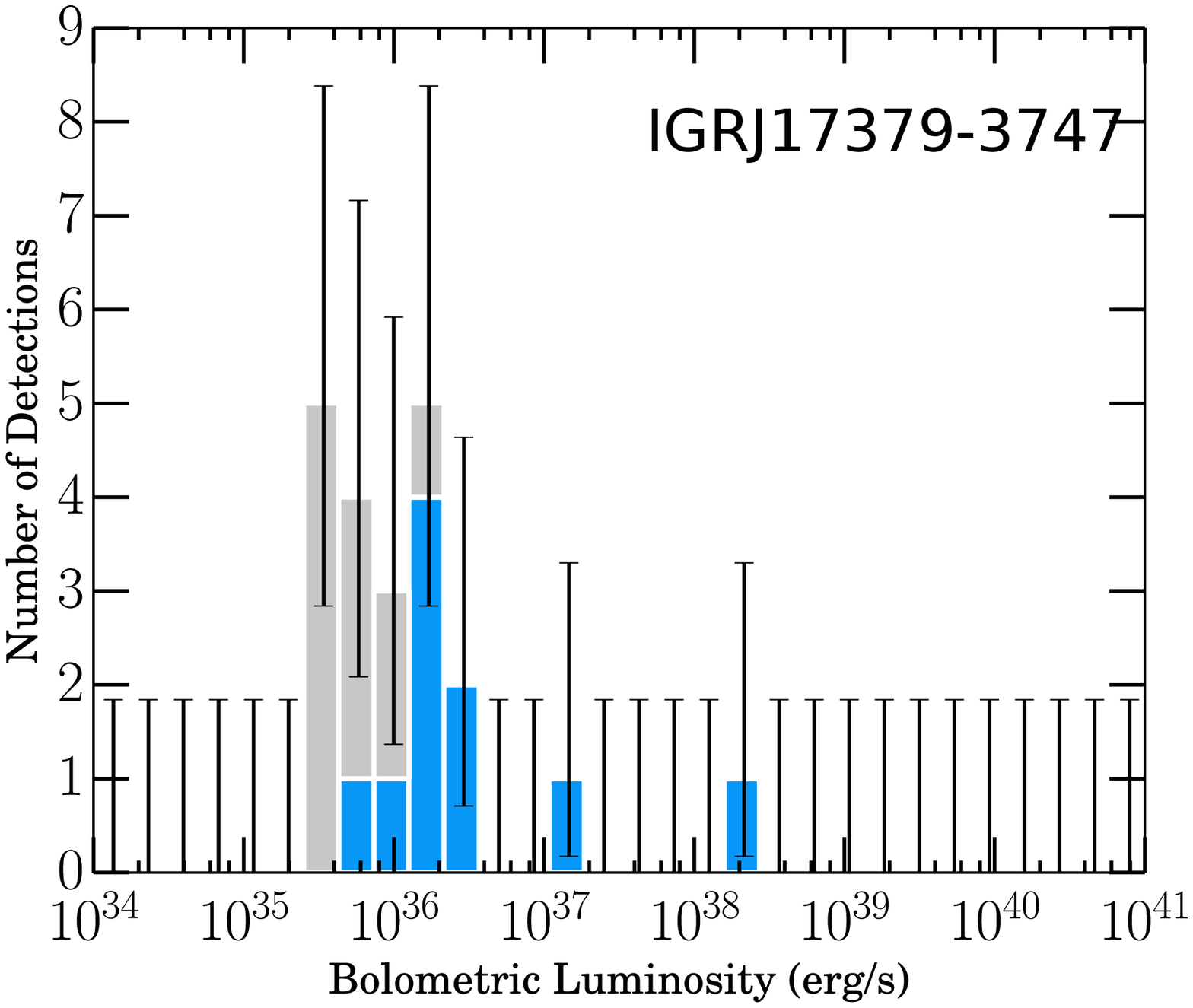}
    \caption{{\bf (cont.)} Transient XLFs color coded by state. HCS (blue), SDS (red), IMS (yellow), and unable to determine state with data available (grey).  }%
    \label{fig:tXLF32}%
\end{figure*}
\afterpage{\clearpage}

\addtocounter{figure}{-1}
\begin{figure*}%
\epsscale{0.85}
\plottwo{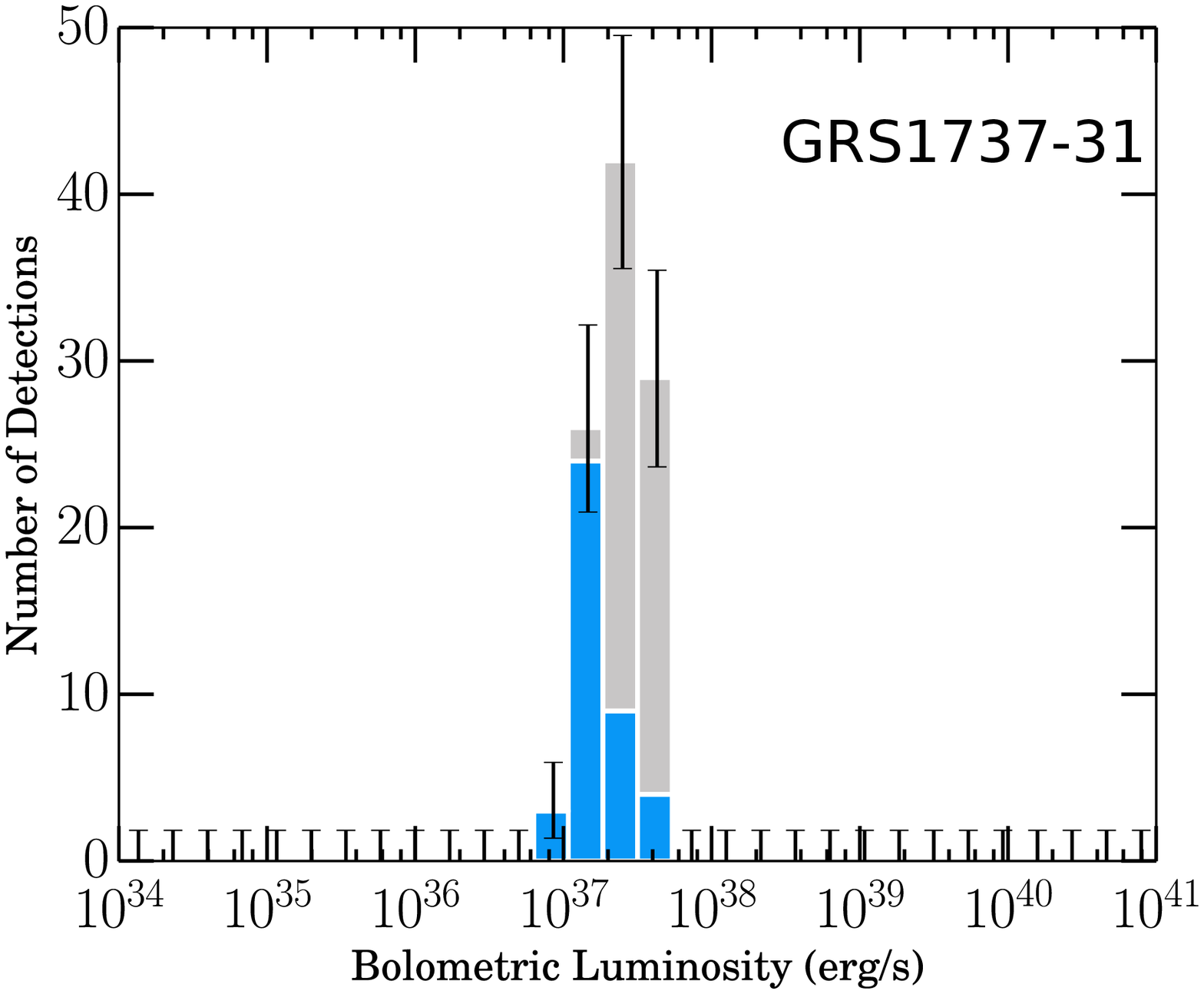}{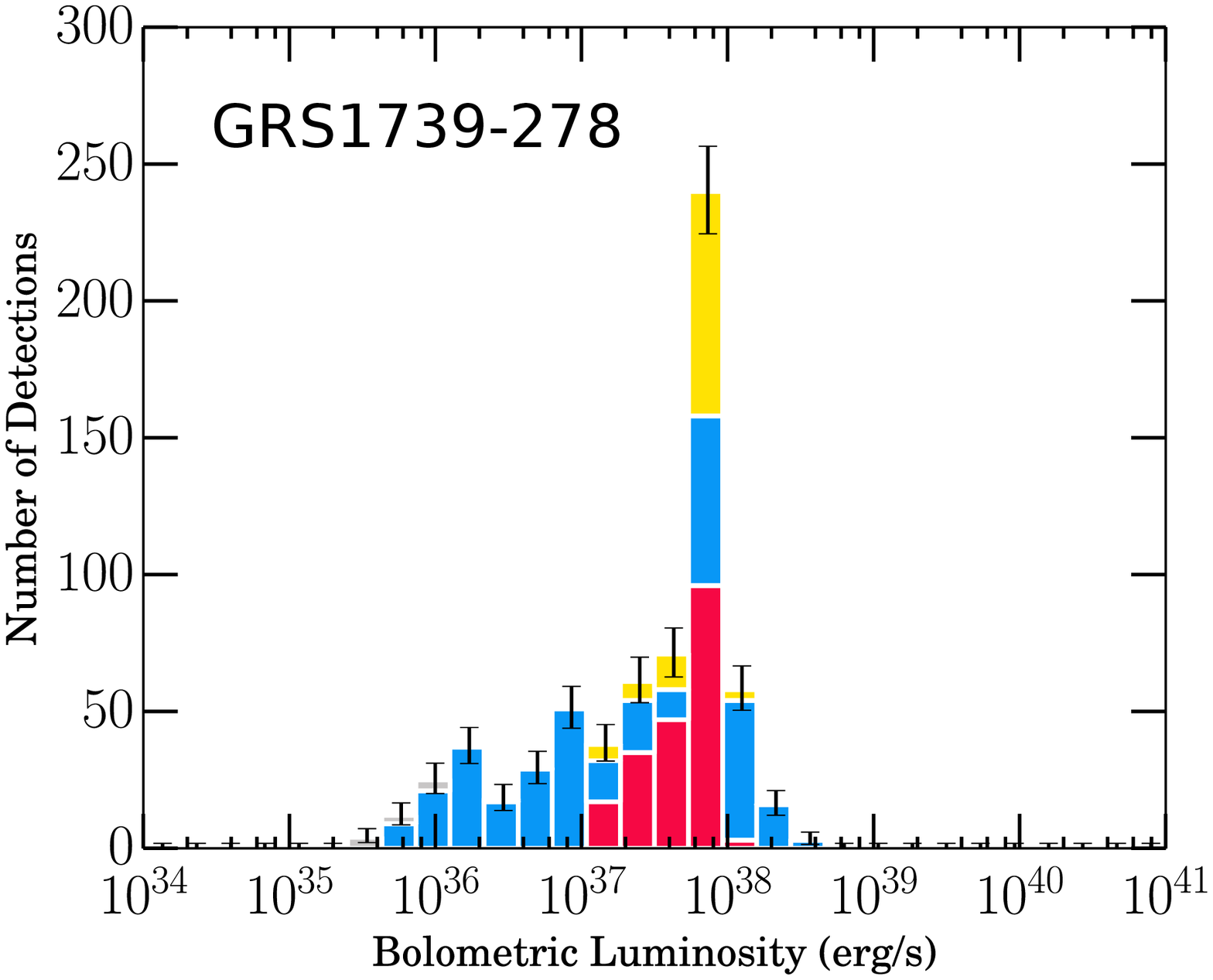}
\plottwo{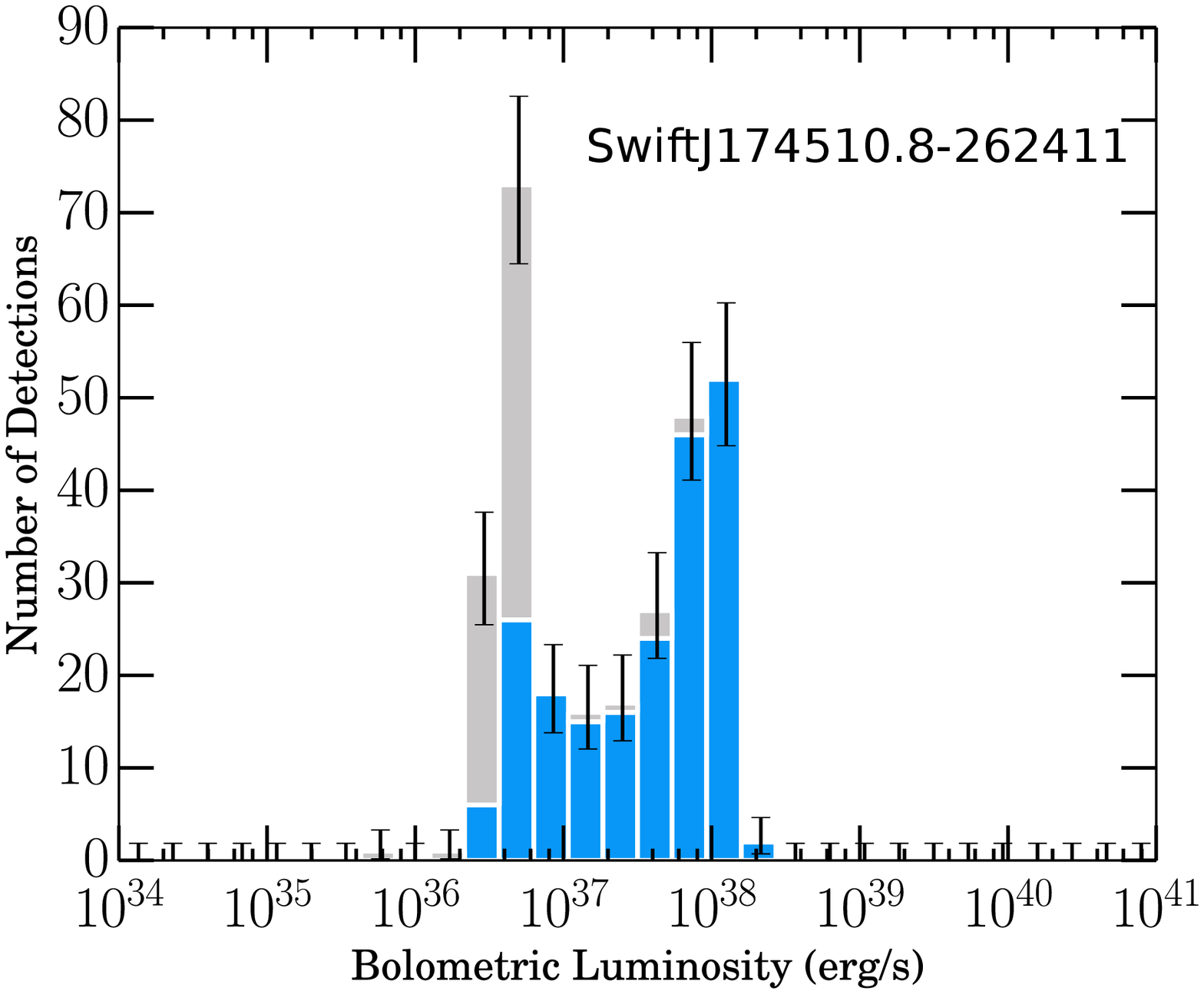}{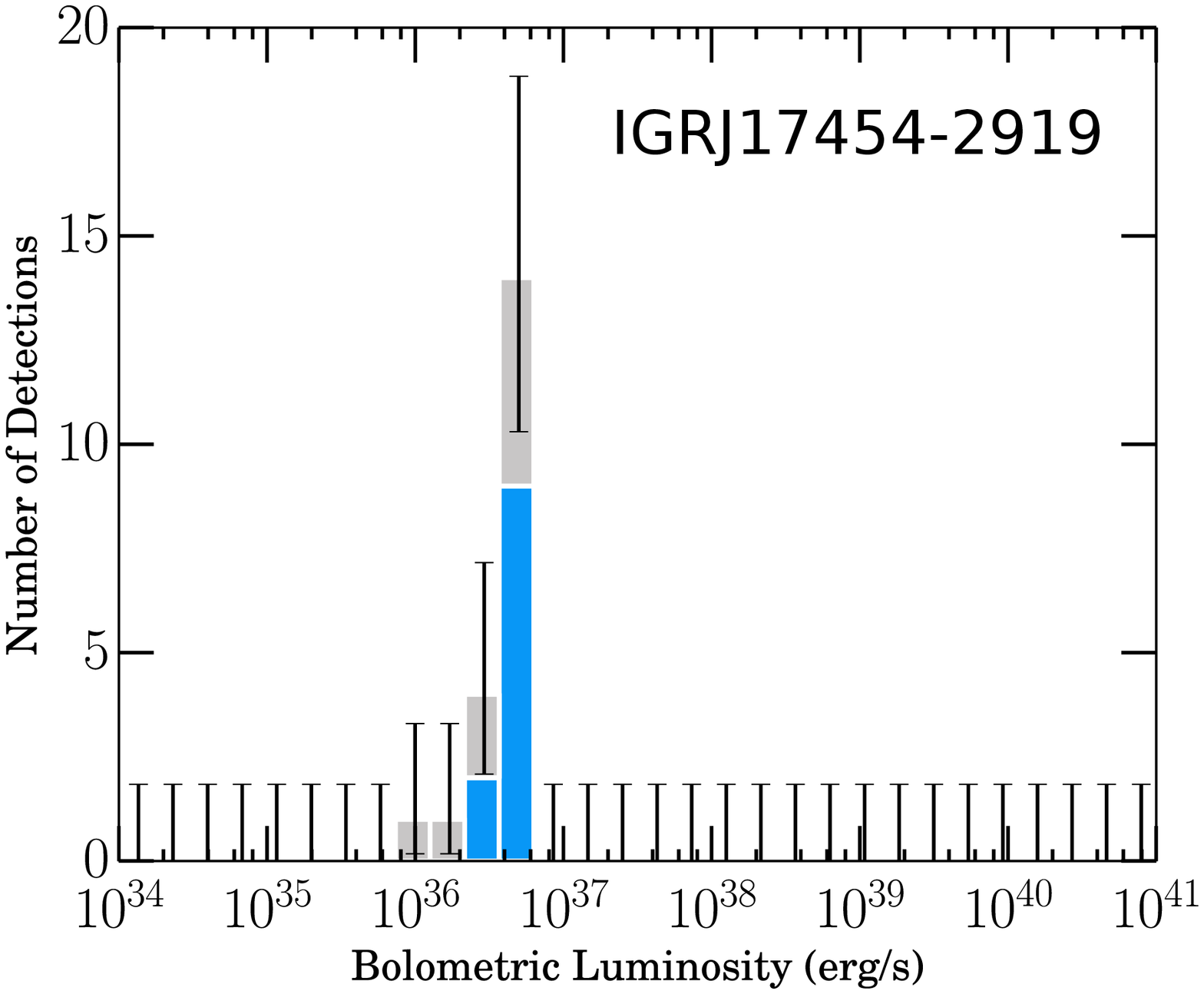}
\plottwo{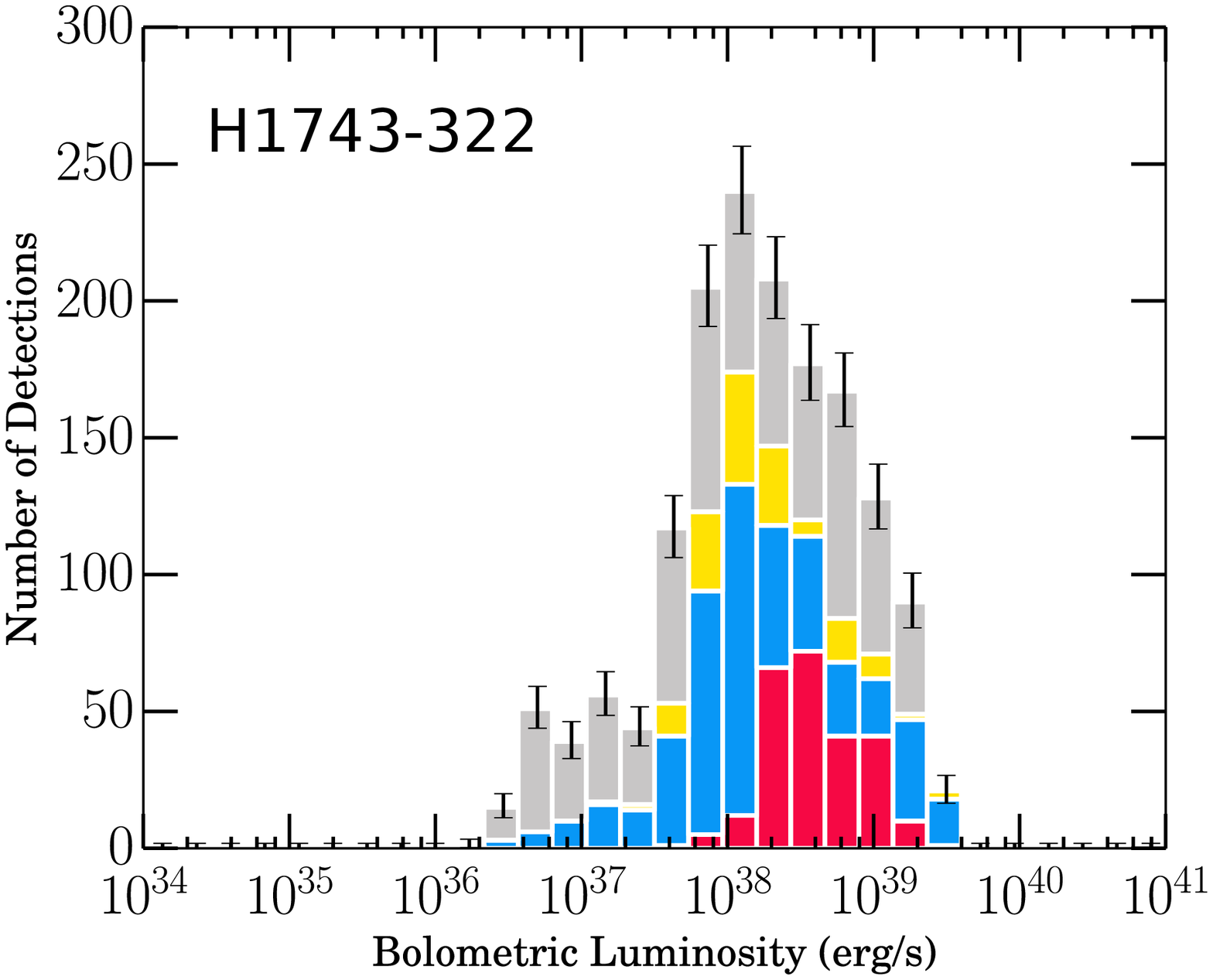}{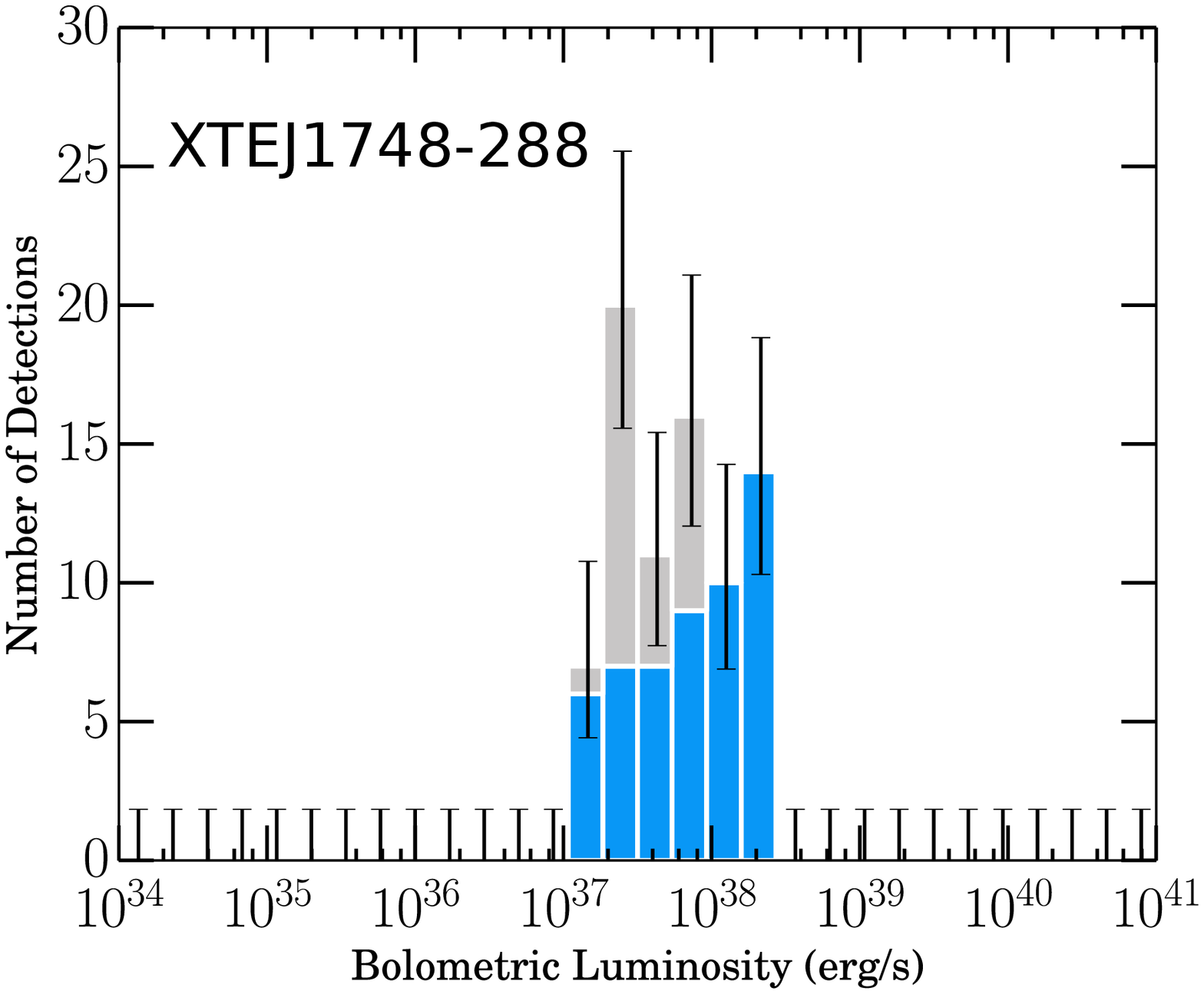}
\plottwo{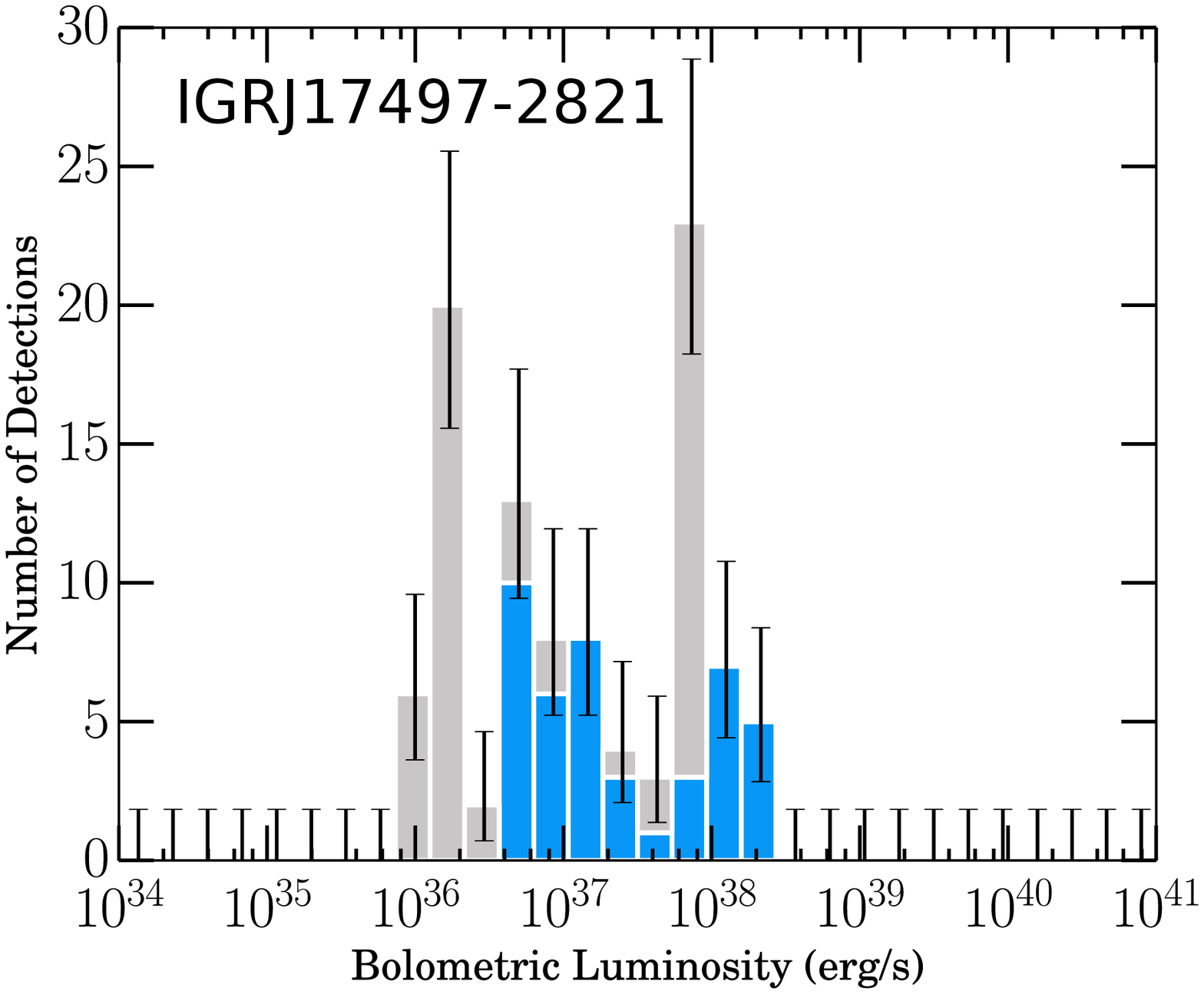}{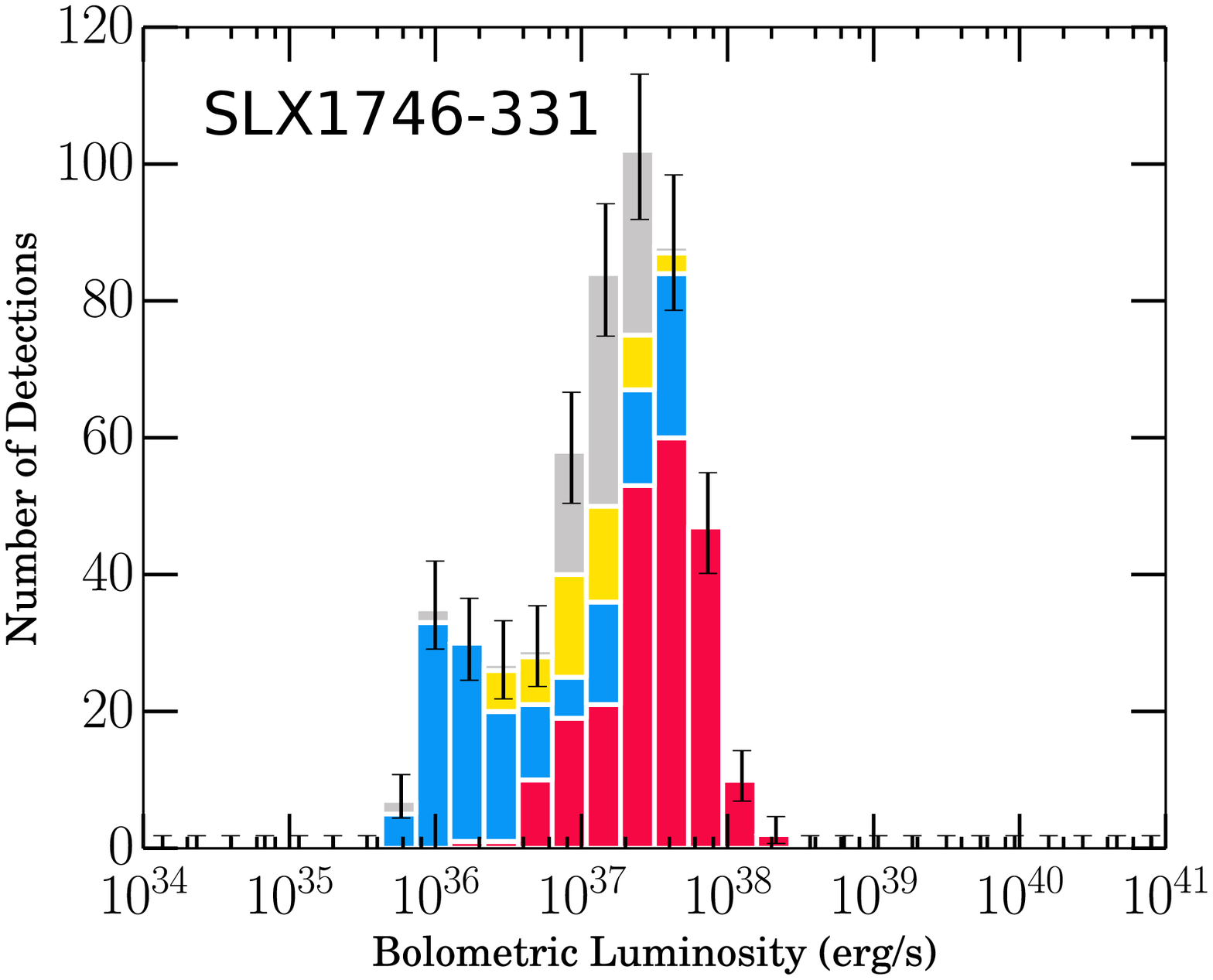}
    \caption{{\bf (cont.)} Transient XLFs color coded by state. HCS (blue), SDS (red), IMS (yellow), and unable to determine state with data available (grey).}%
    \label{fig:tXLF42}%
\end{figure*}
\afterpage{\clearpage}

\addtocounter{figure}{-1}
\begin{figure*}%
\epsscale{0.85}
\plottwo{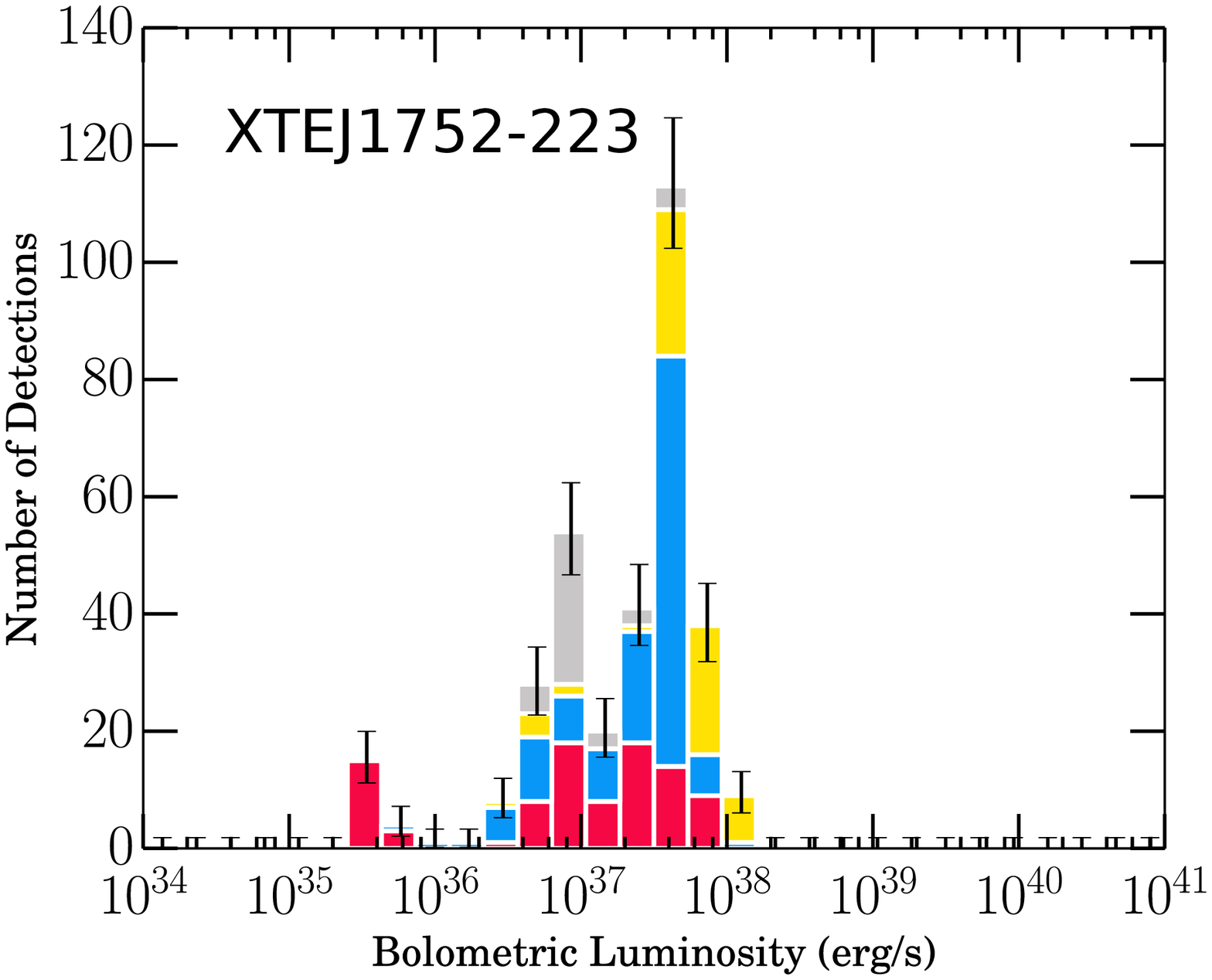}{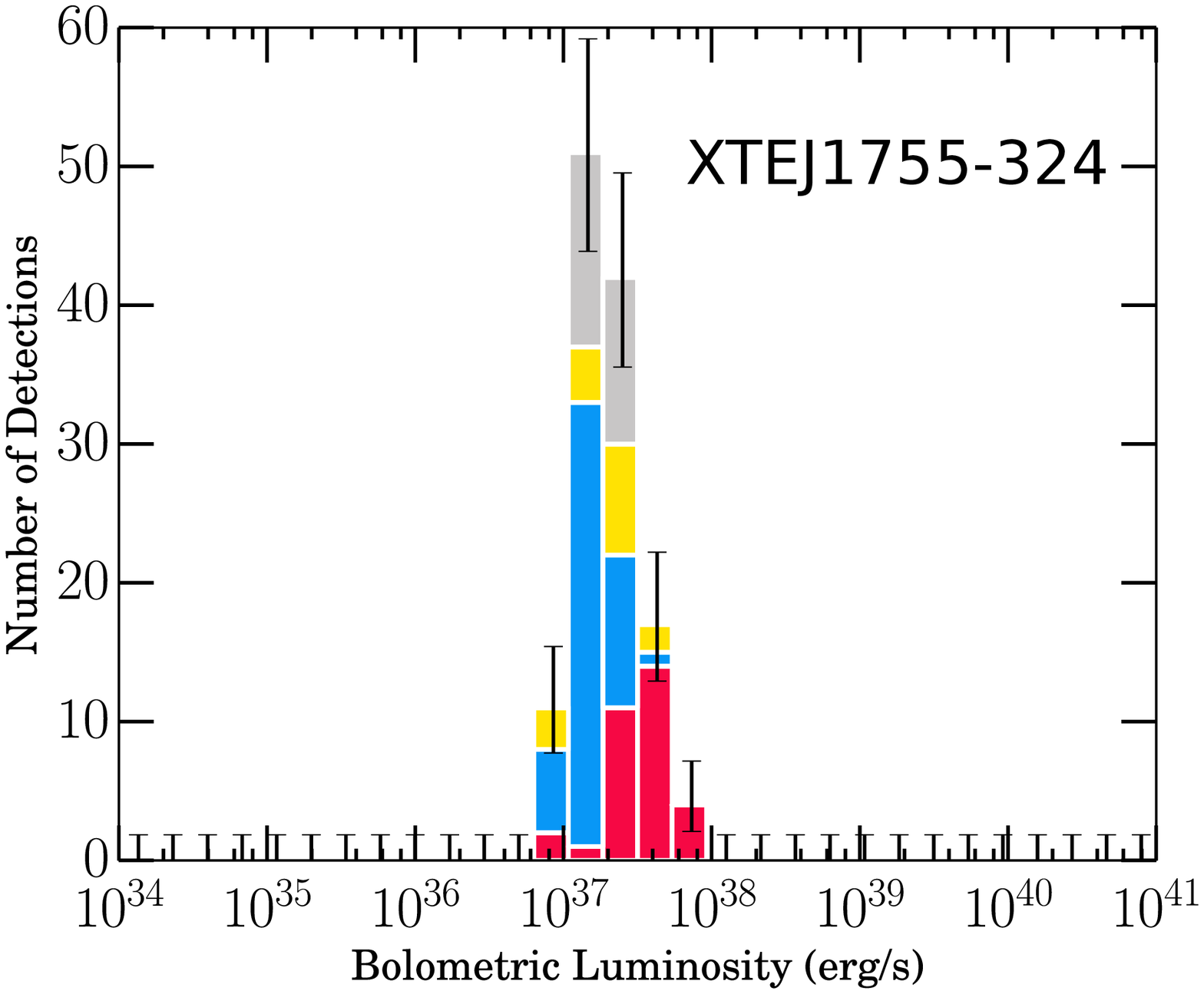}
\plottwo{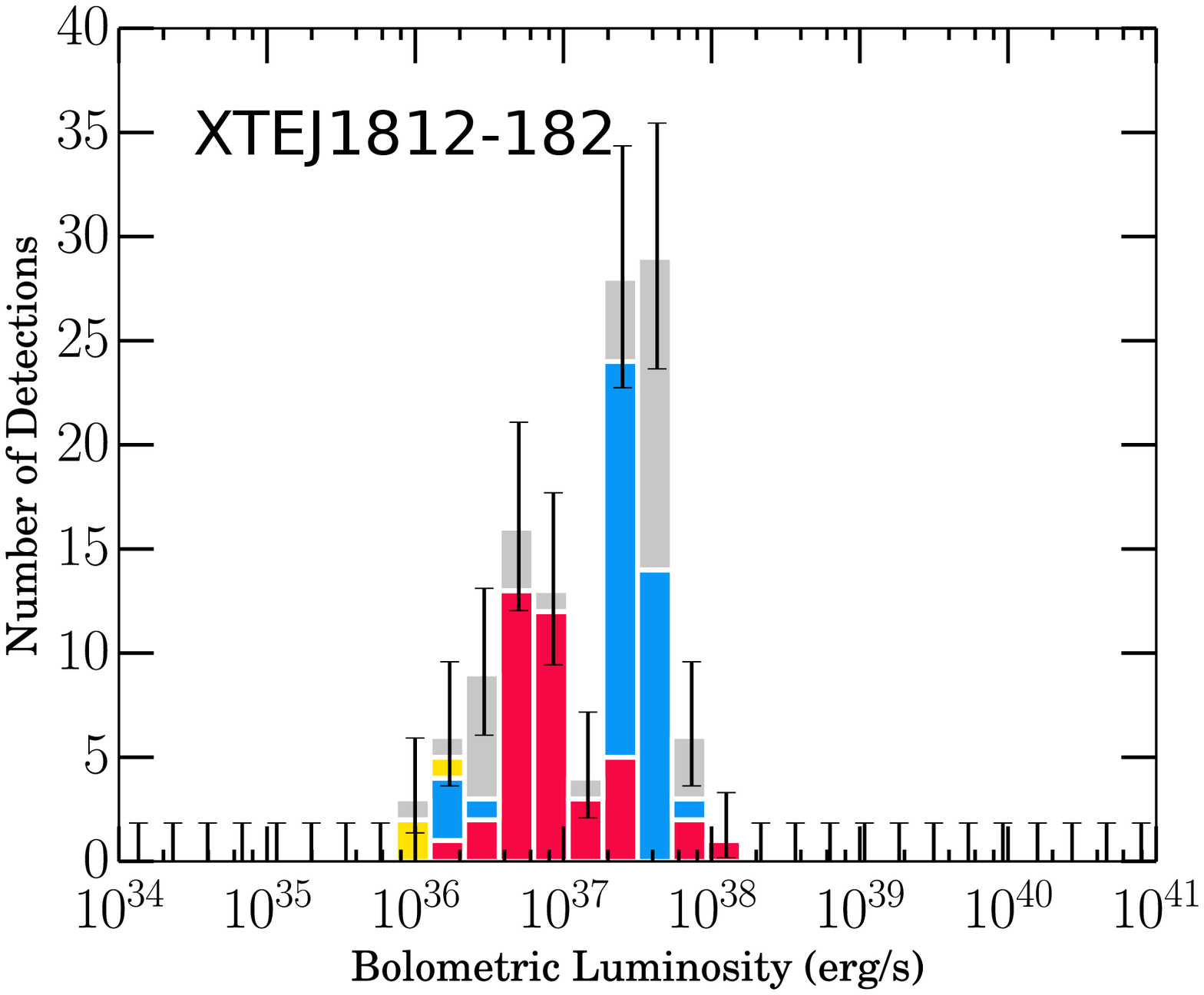}{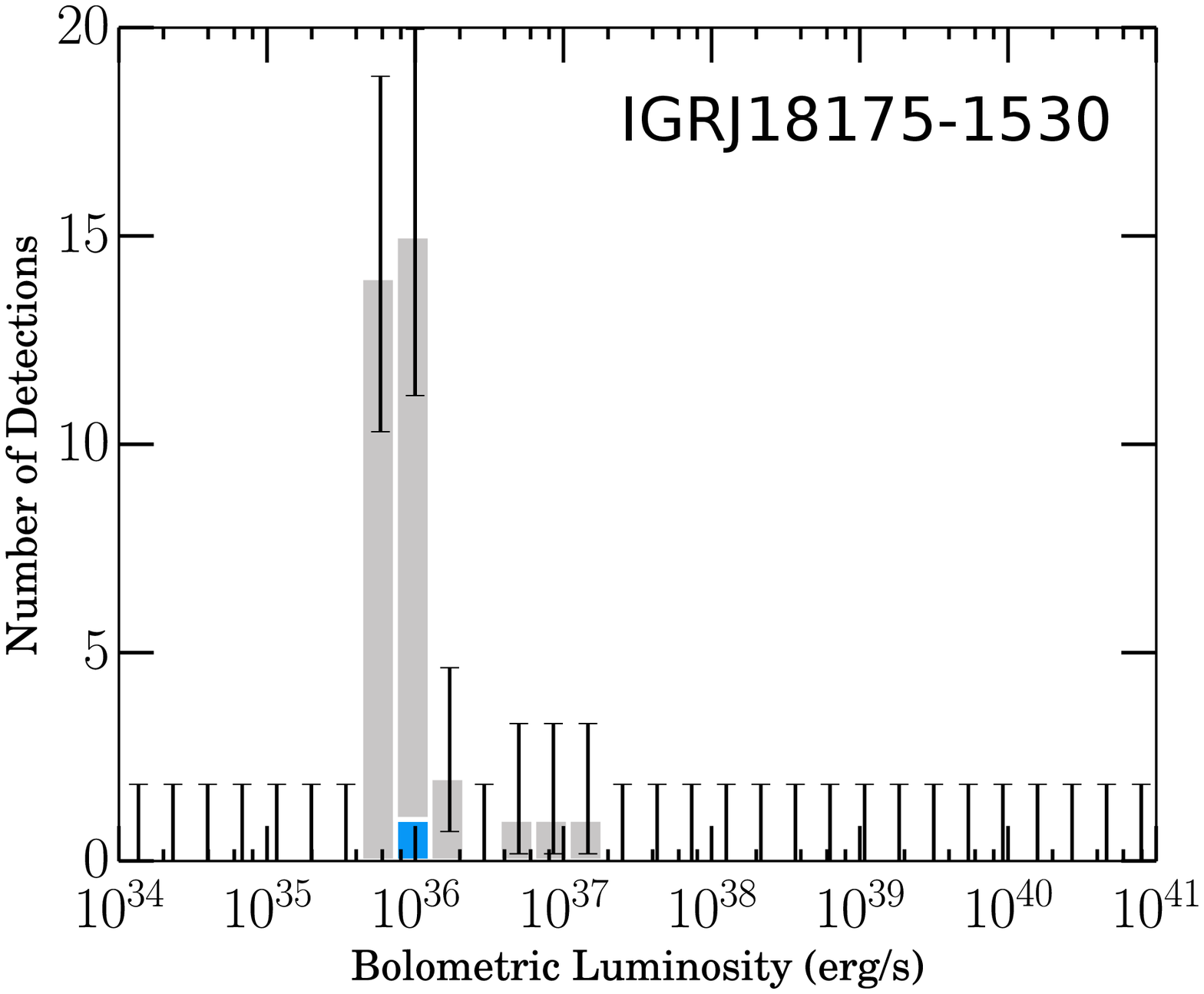}
\plottwo{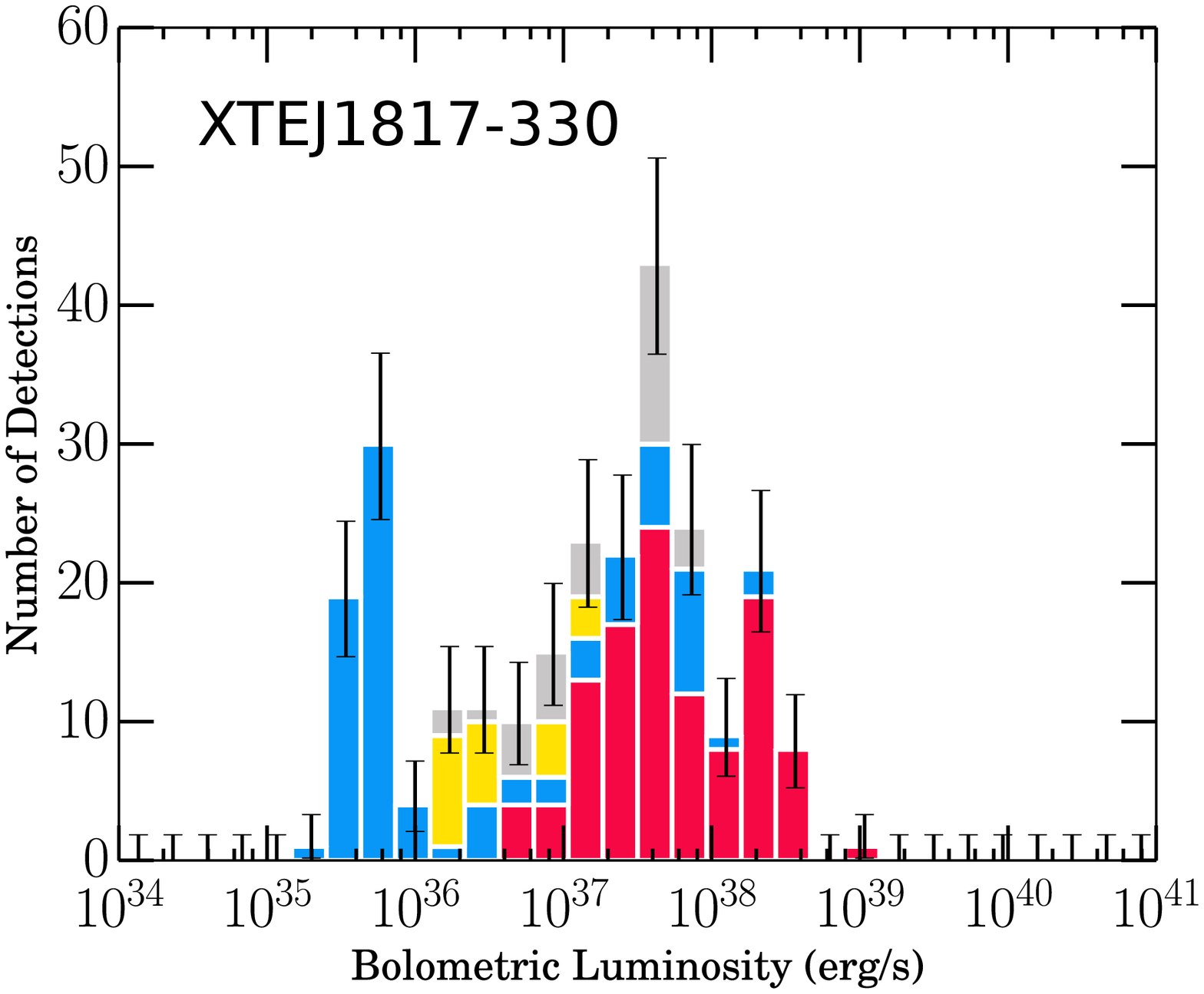}{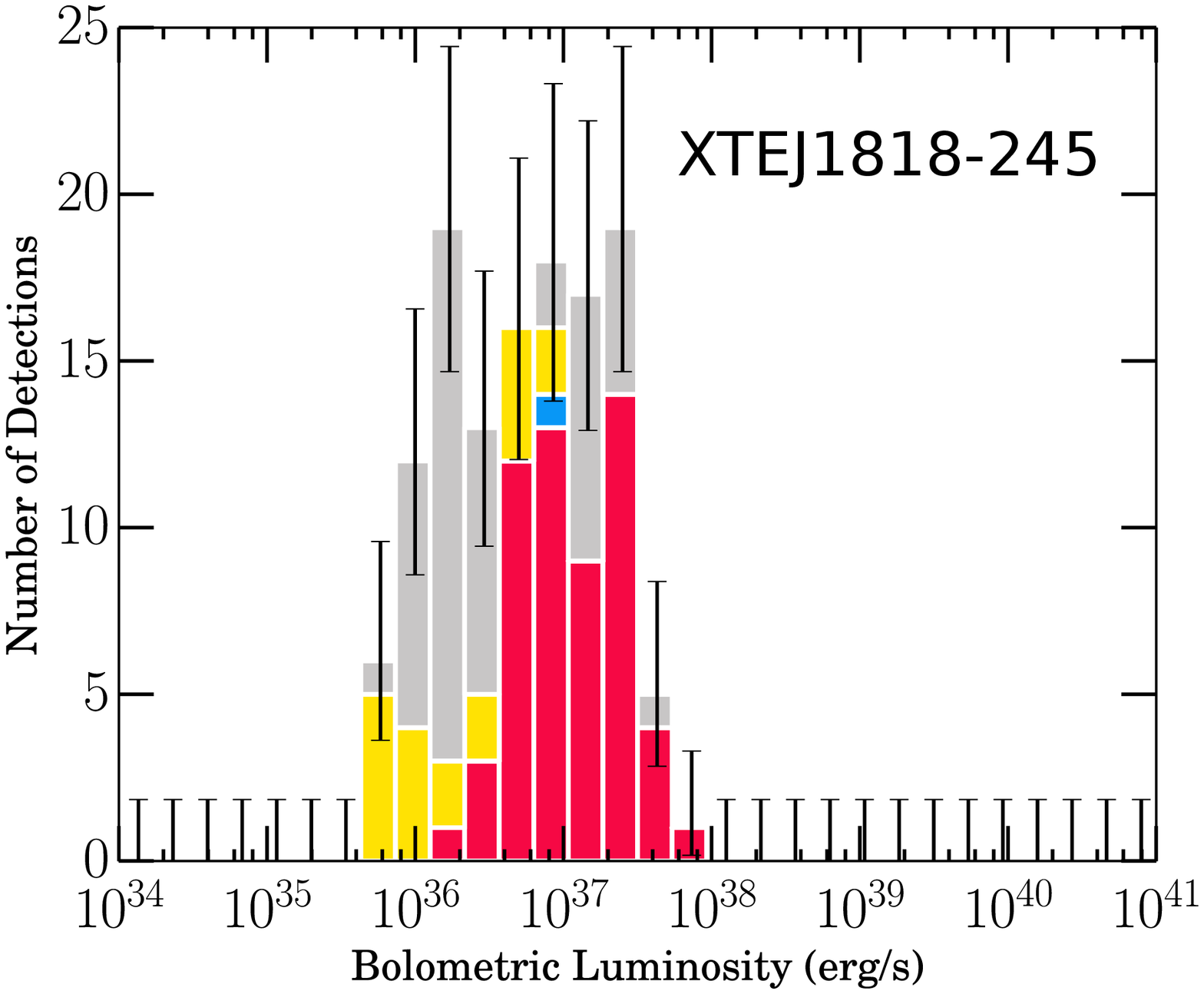}
\plottwo{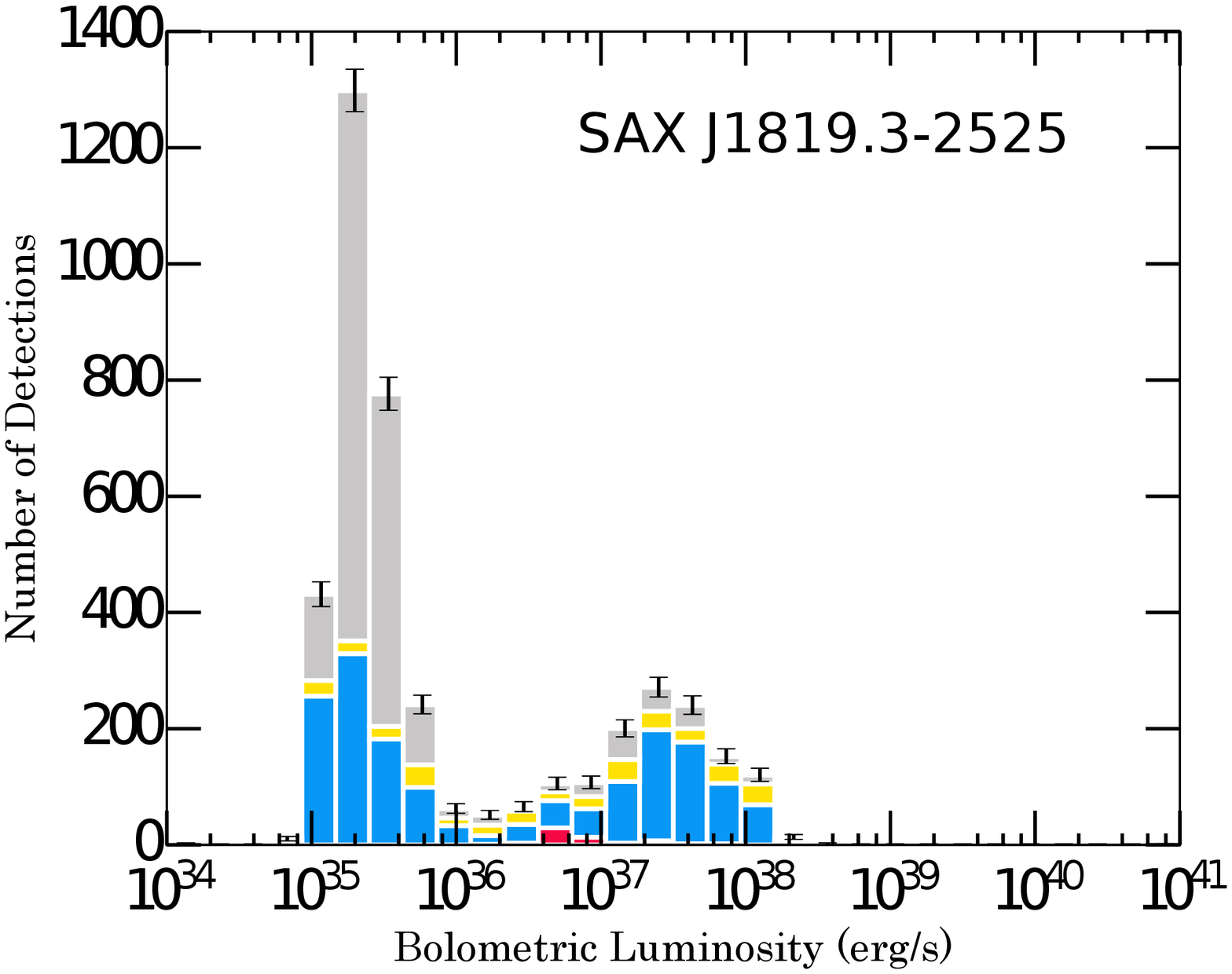}{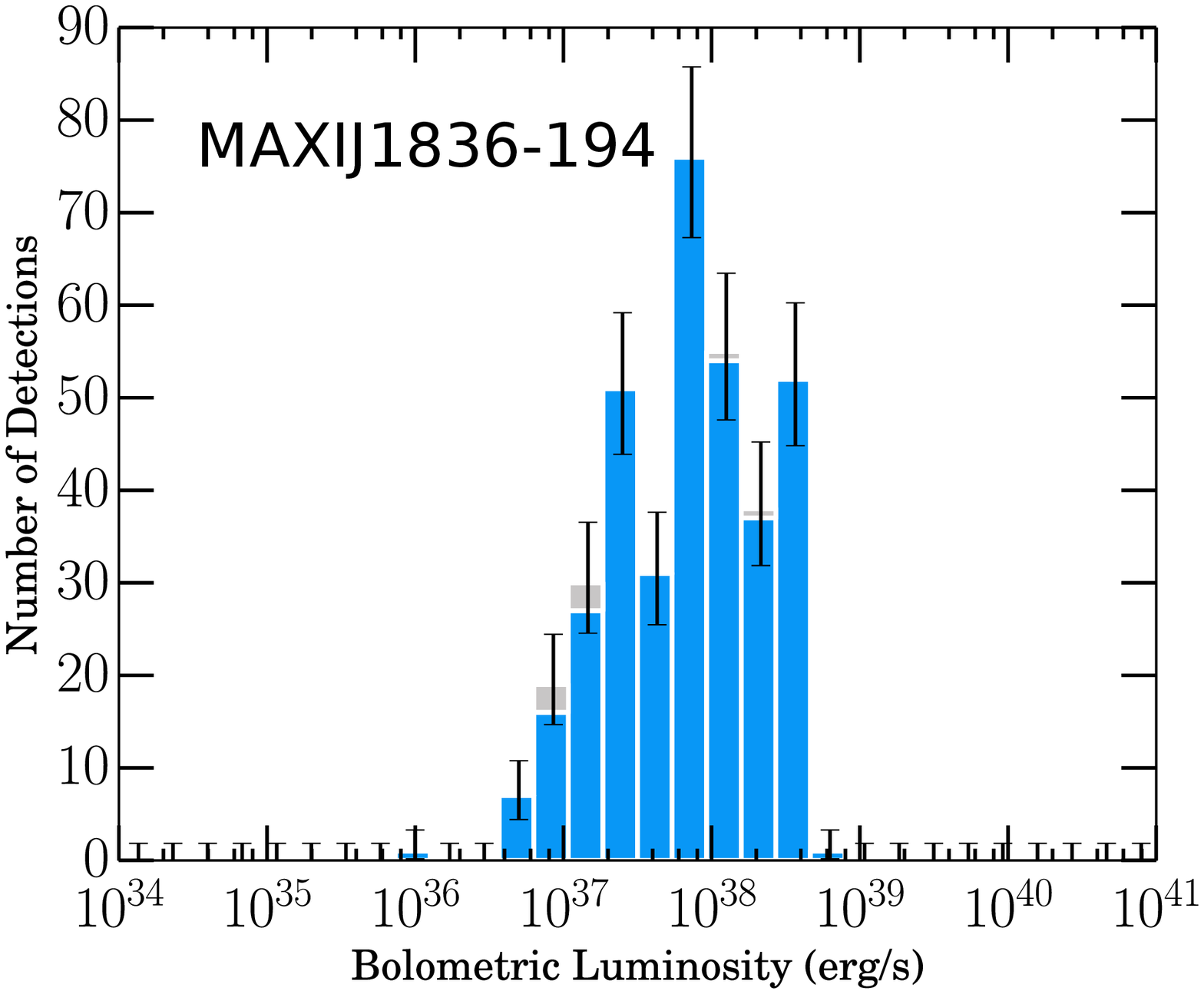}
    \caption{{\bf (cont.)} Transient XLFs color coded by state. HCS (blue), SDS (red), IMS (yellow), and unable to determine state with data available (grey).}%
    \label{fig:tXLF52}%
\end{figure*}
\afterpage{\clearpage}

\addtocounter{figure}{-1}
\begin{figure*}%
\epsscale{0.85}
\plottwo{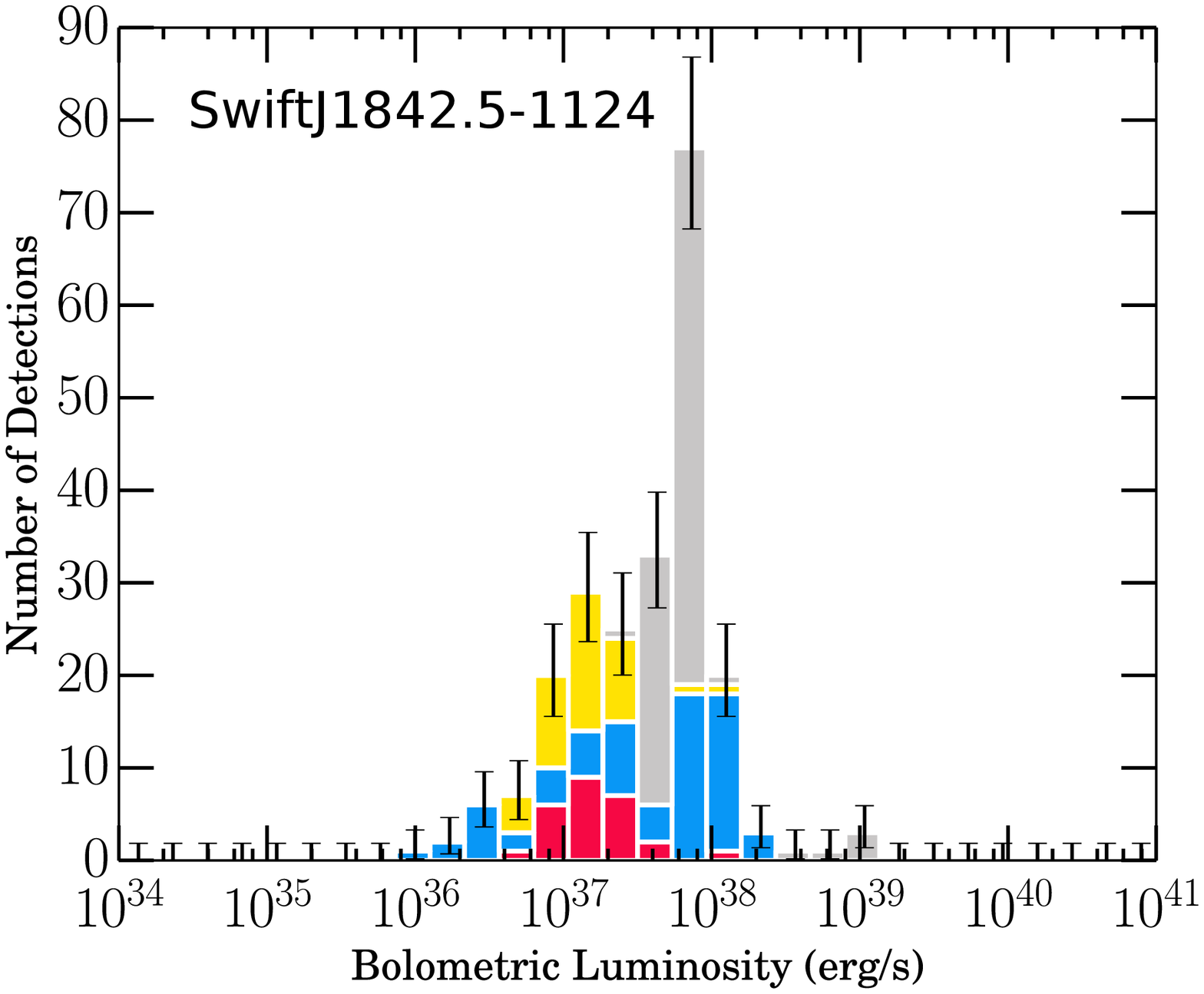}{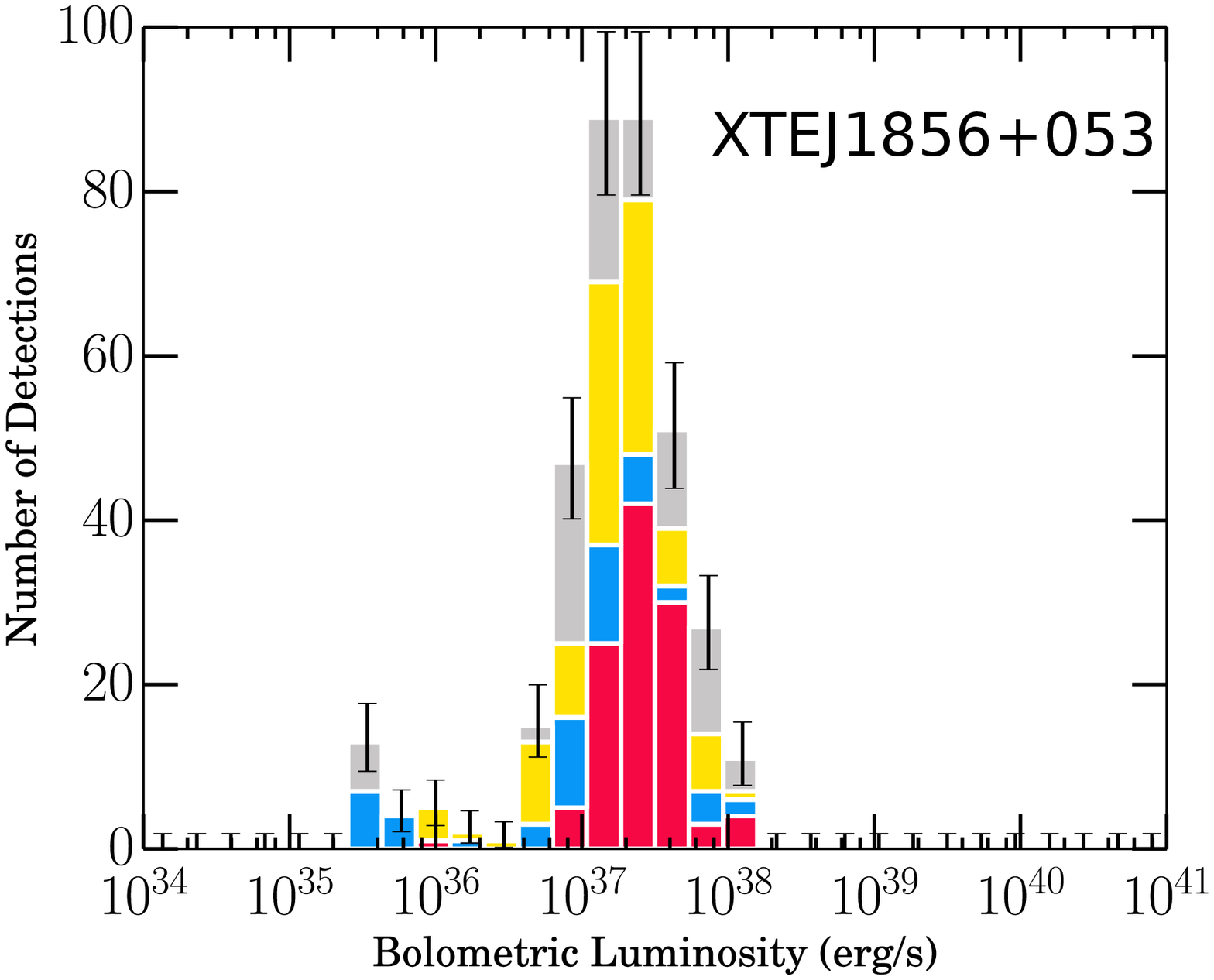}
\plottwo{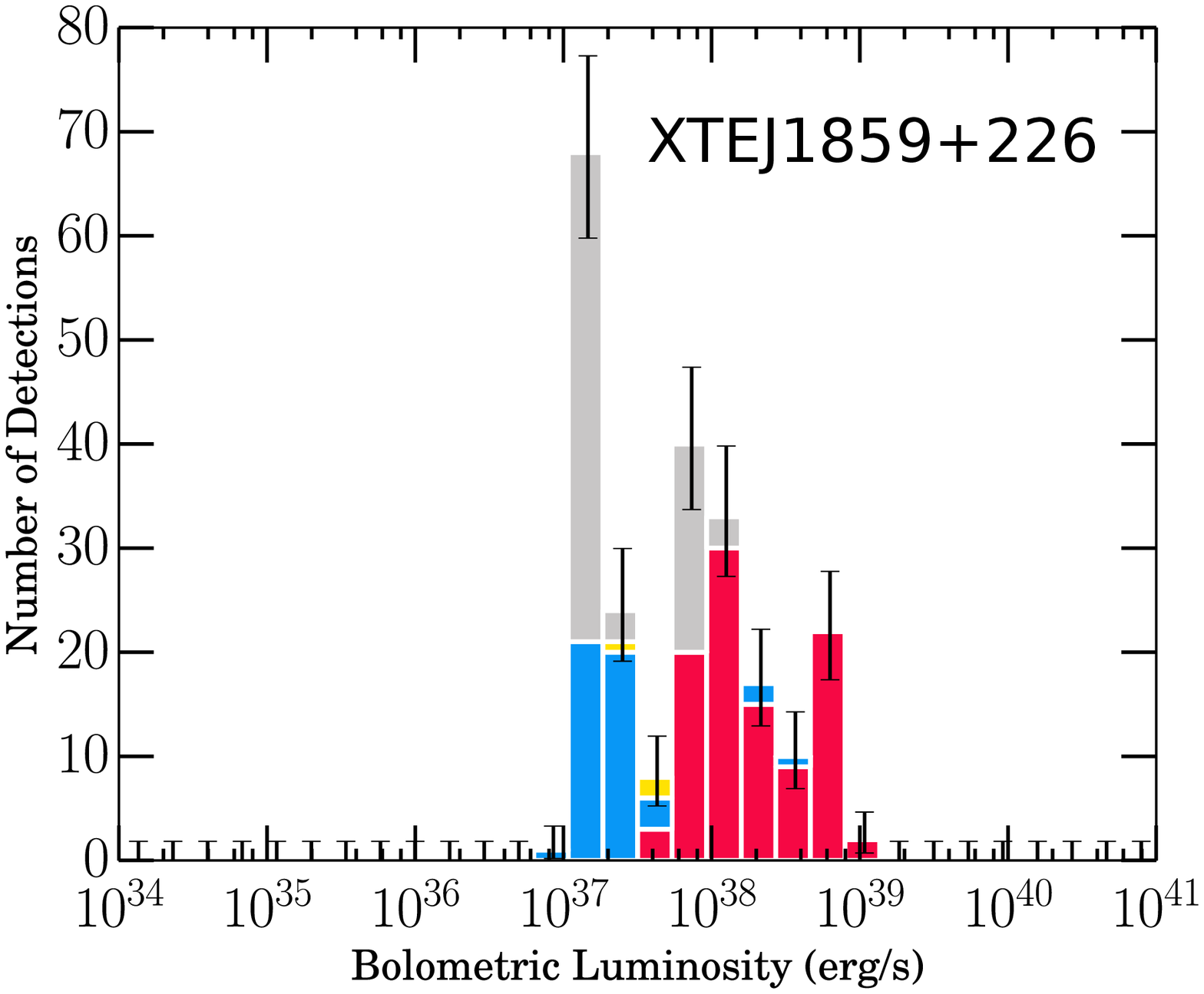}{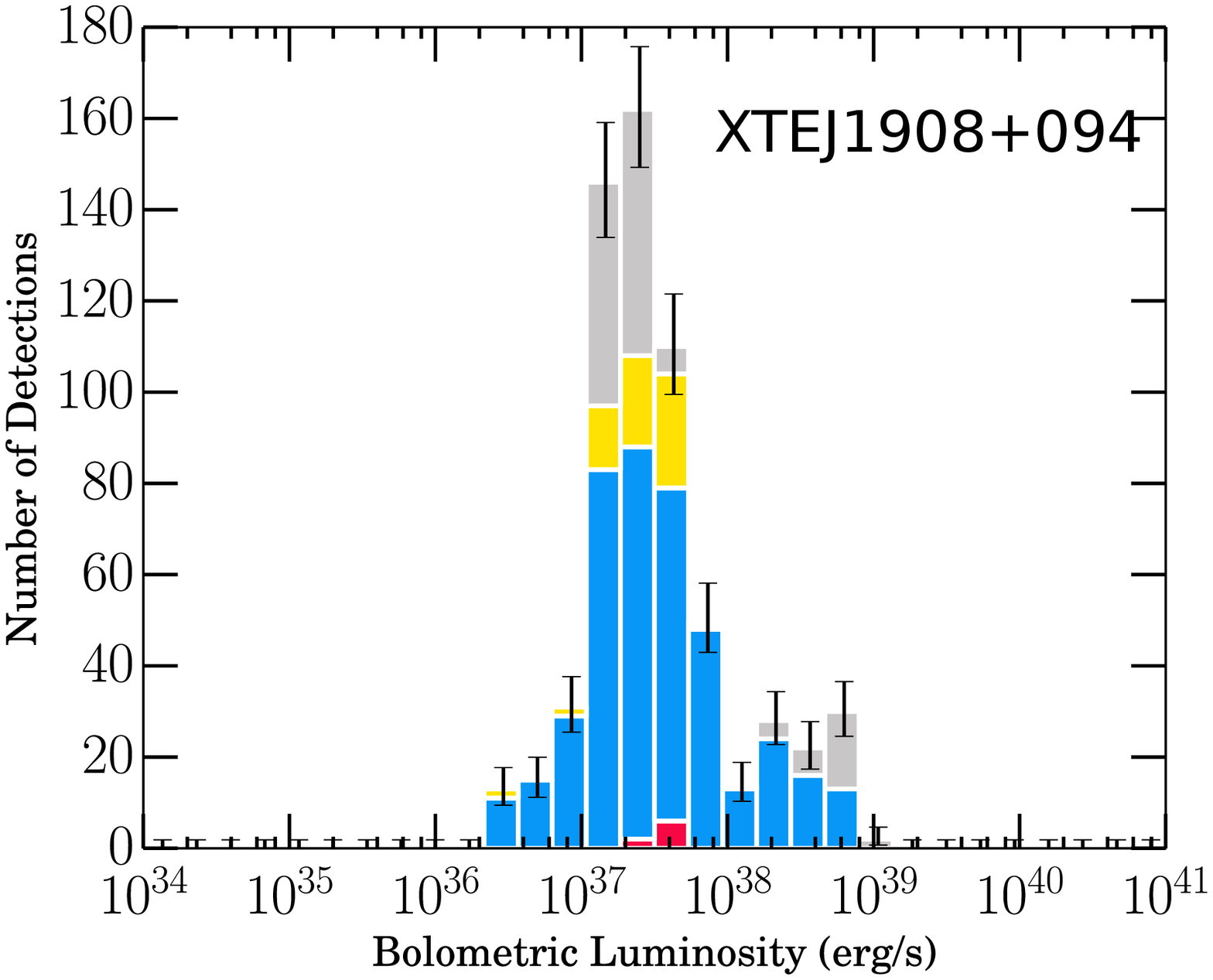}
\plottwo{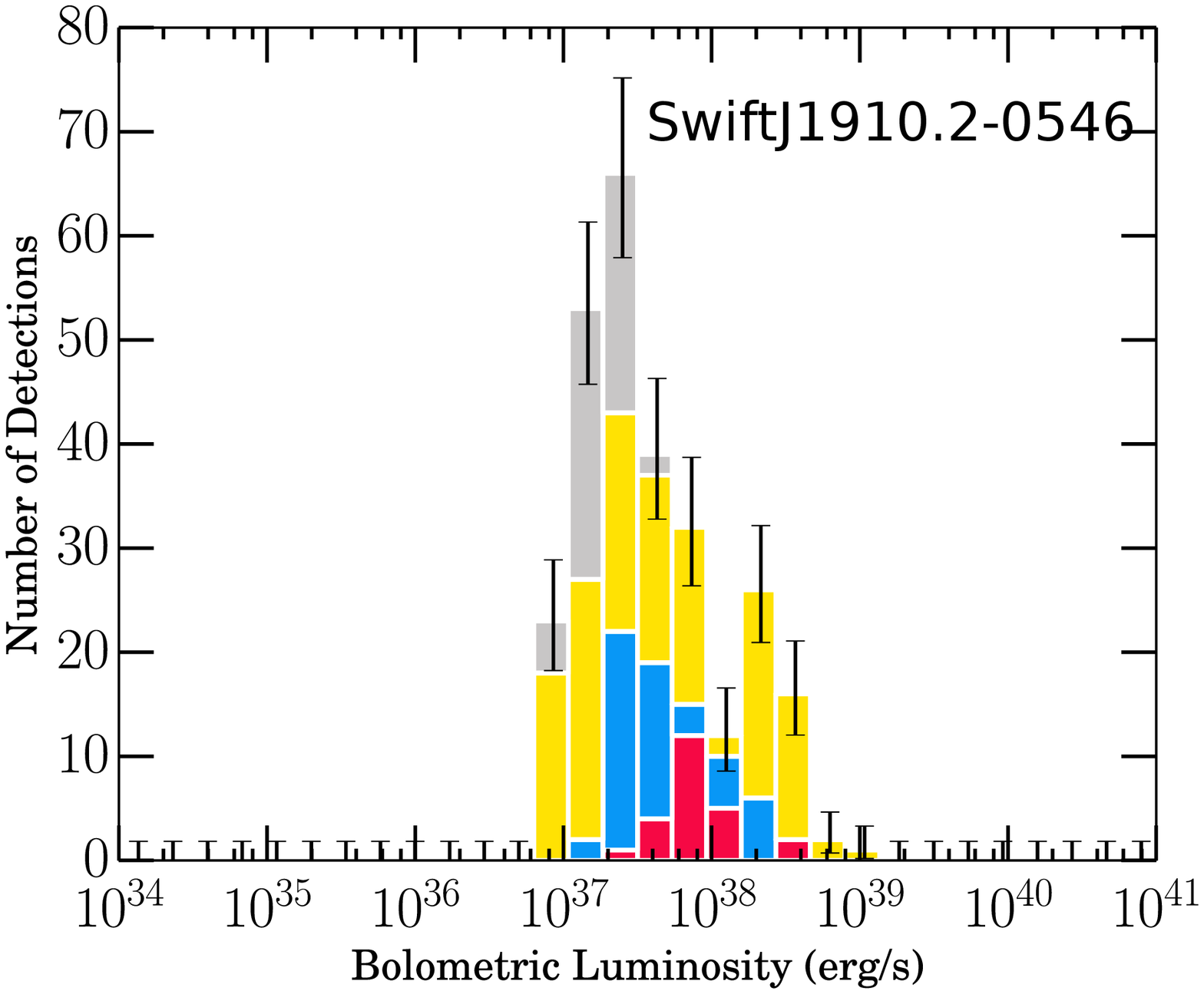}{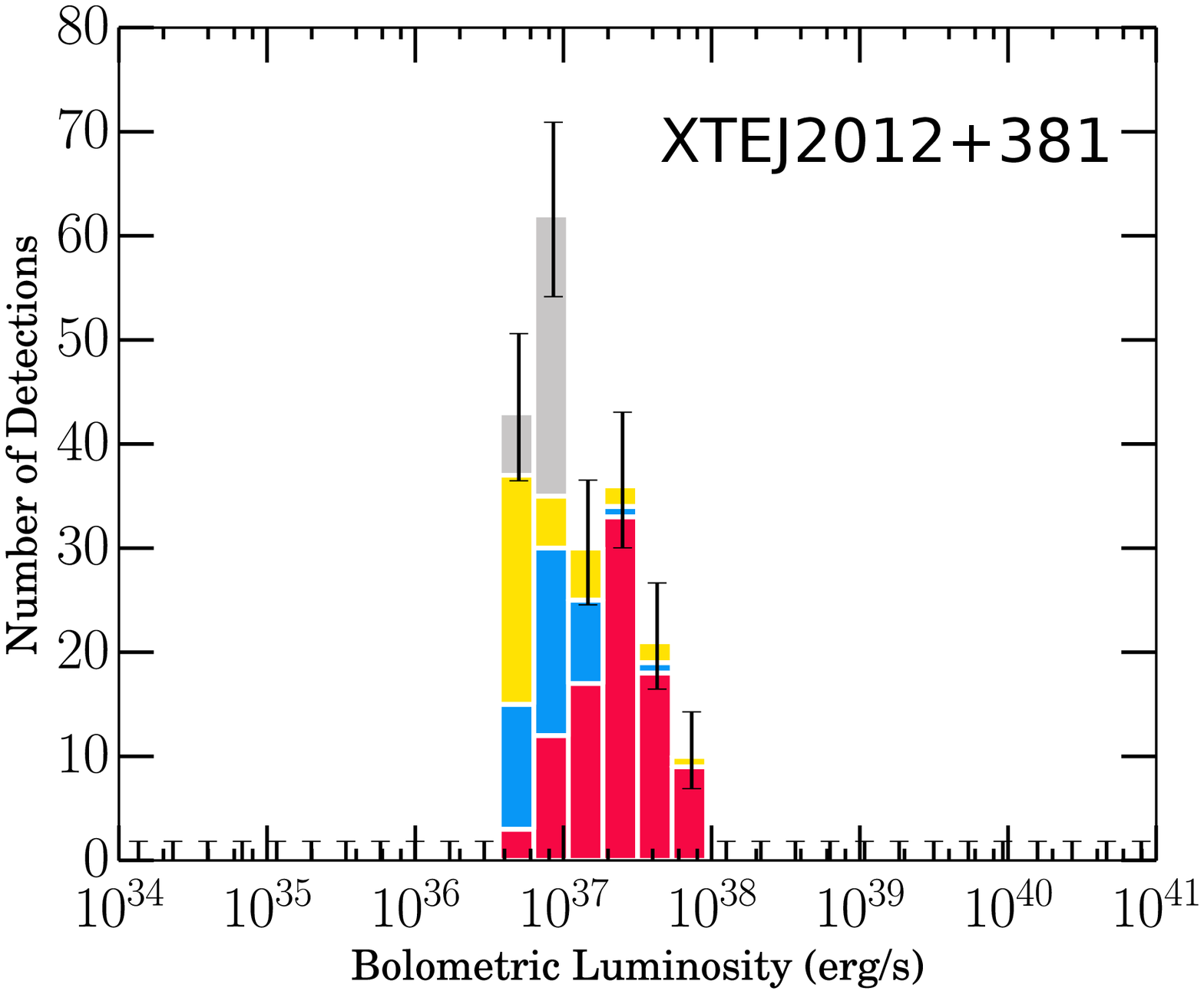}
  \caption{{\bf (cont.)} Transient XLFs color coded by state. HCS (blue), SDS (red), IMS (yellow), and unable to determine state with data available (grey).}%
    \label{fig:tXLF62}%
\end{figure*}
\afterpage{\clearpage}

\begin{figure*}%
\epsscale{0.85}
\plottwo{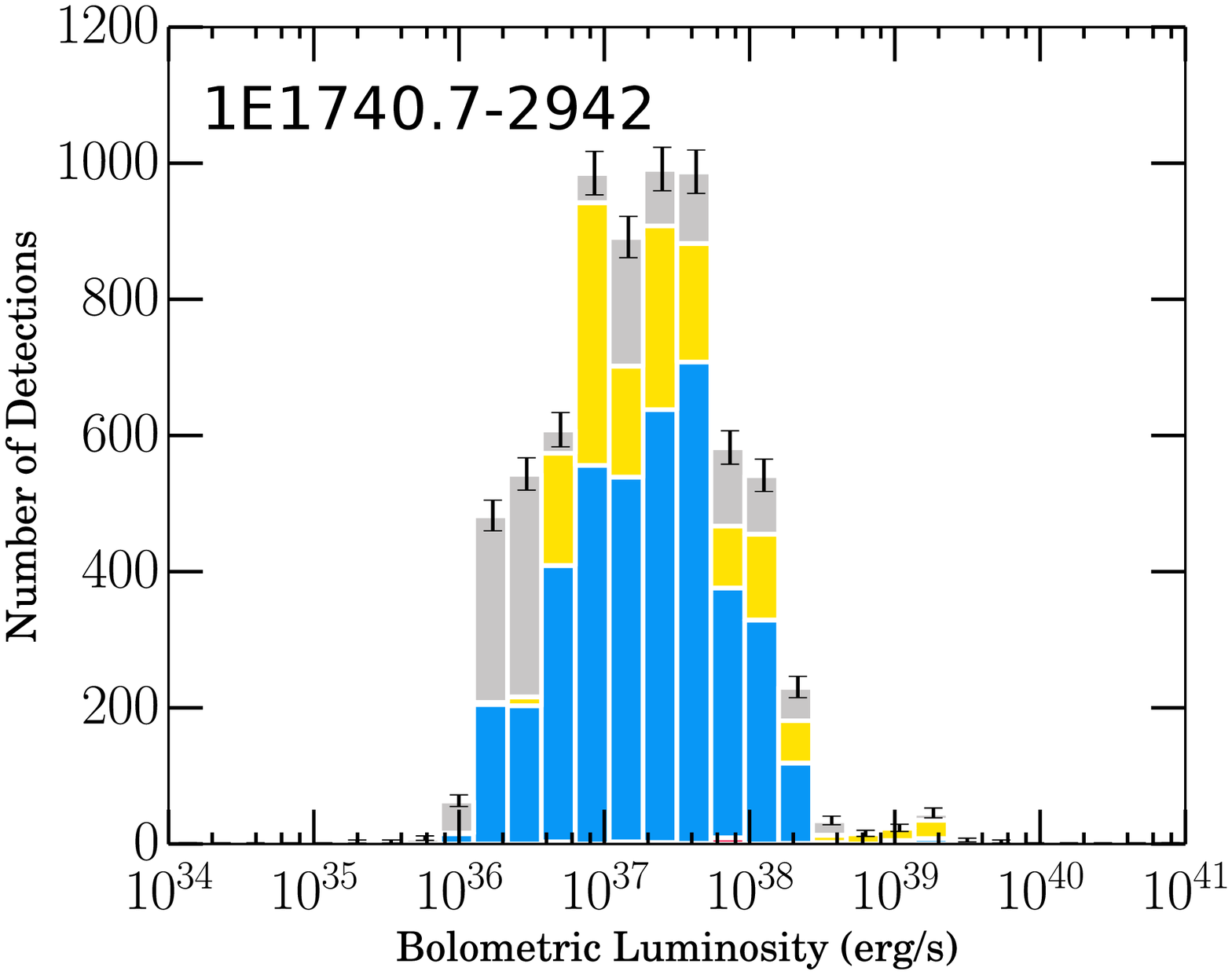}{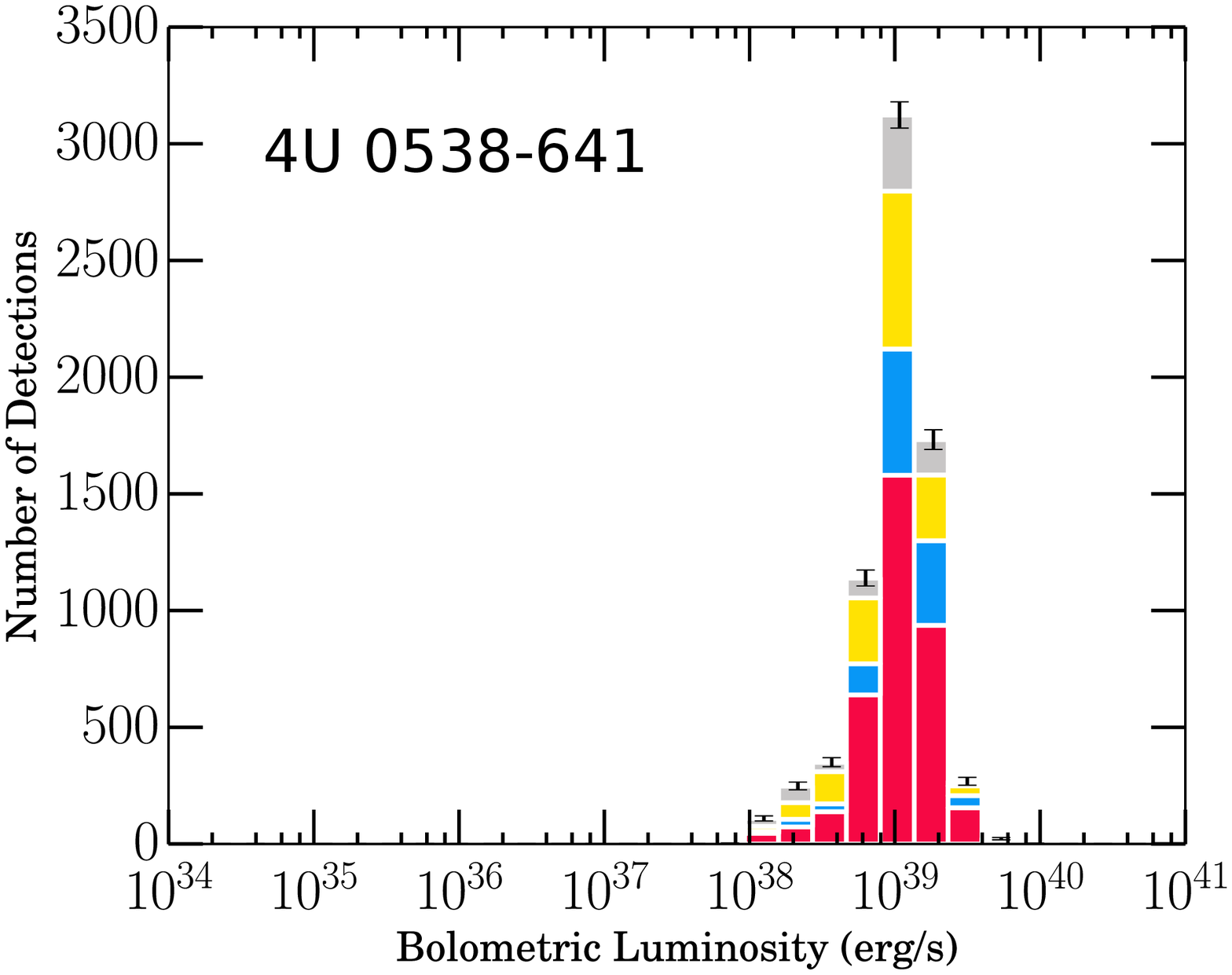}
\plottwo{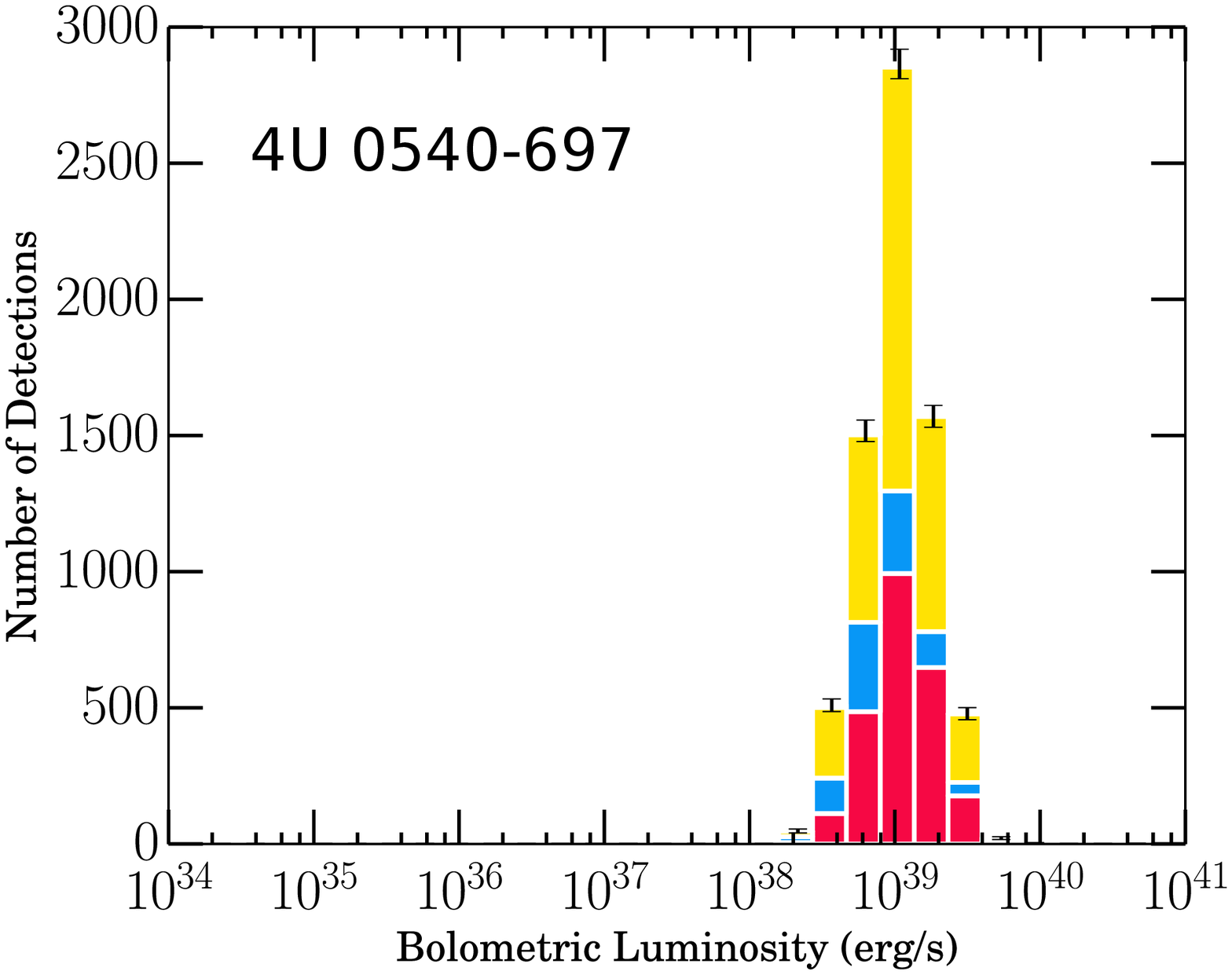}{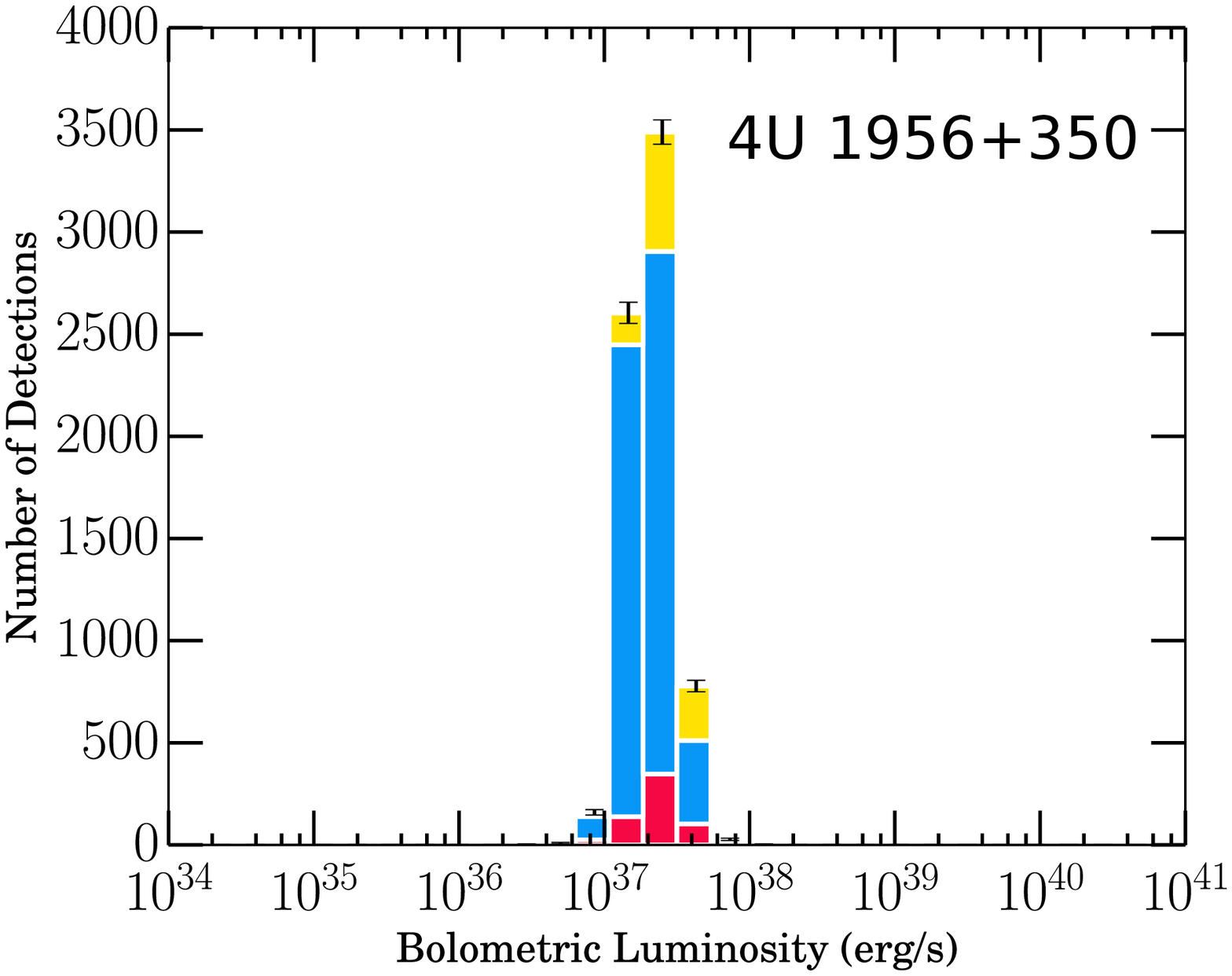}
\plottwo{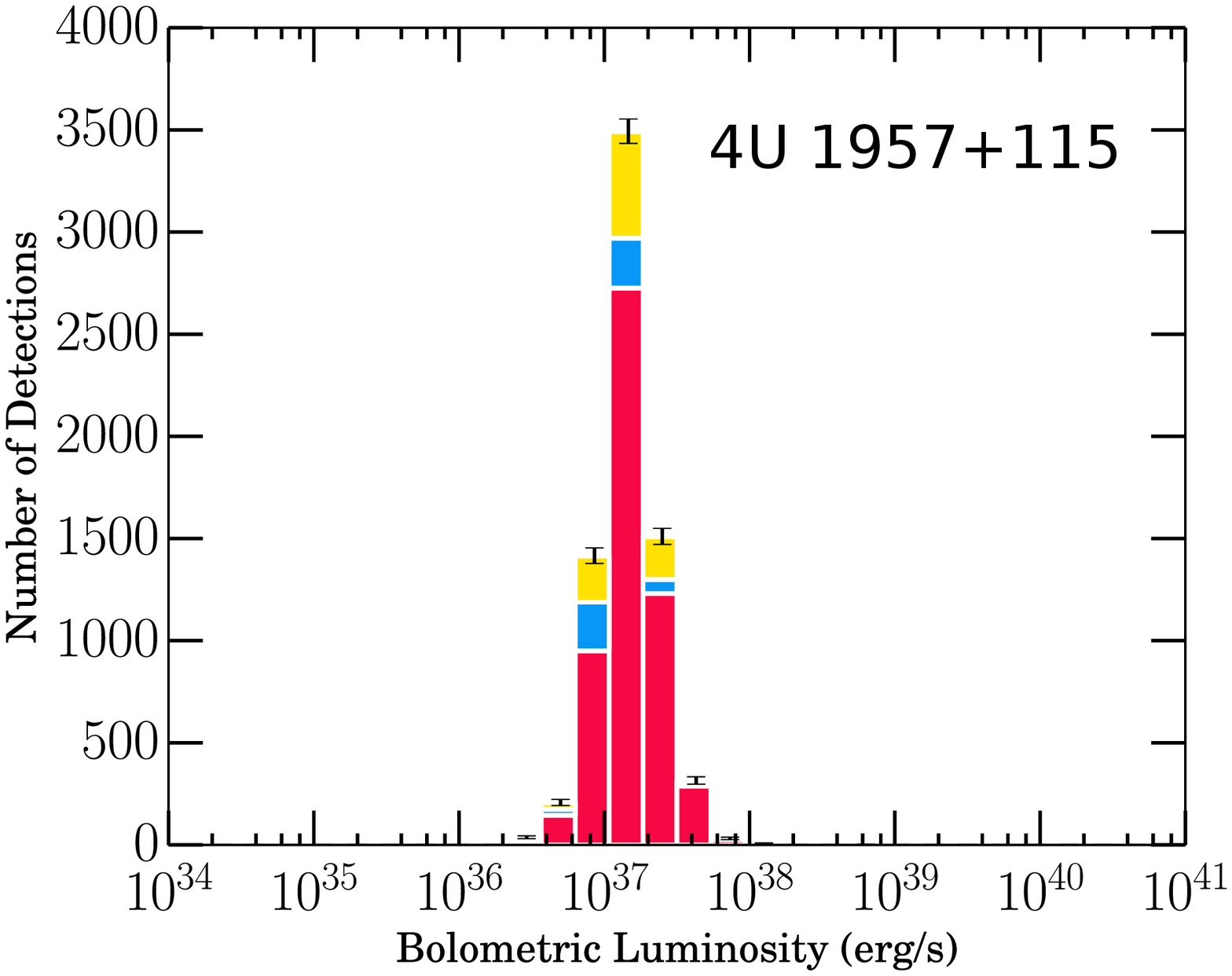}{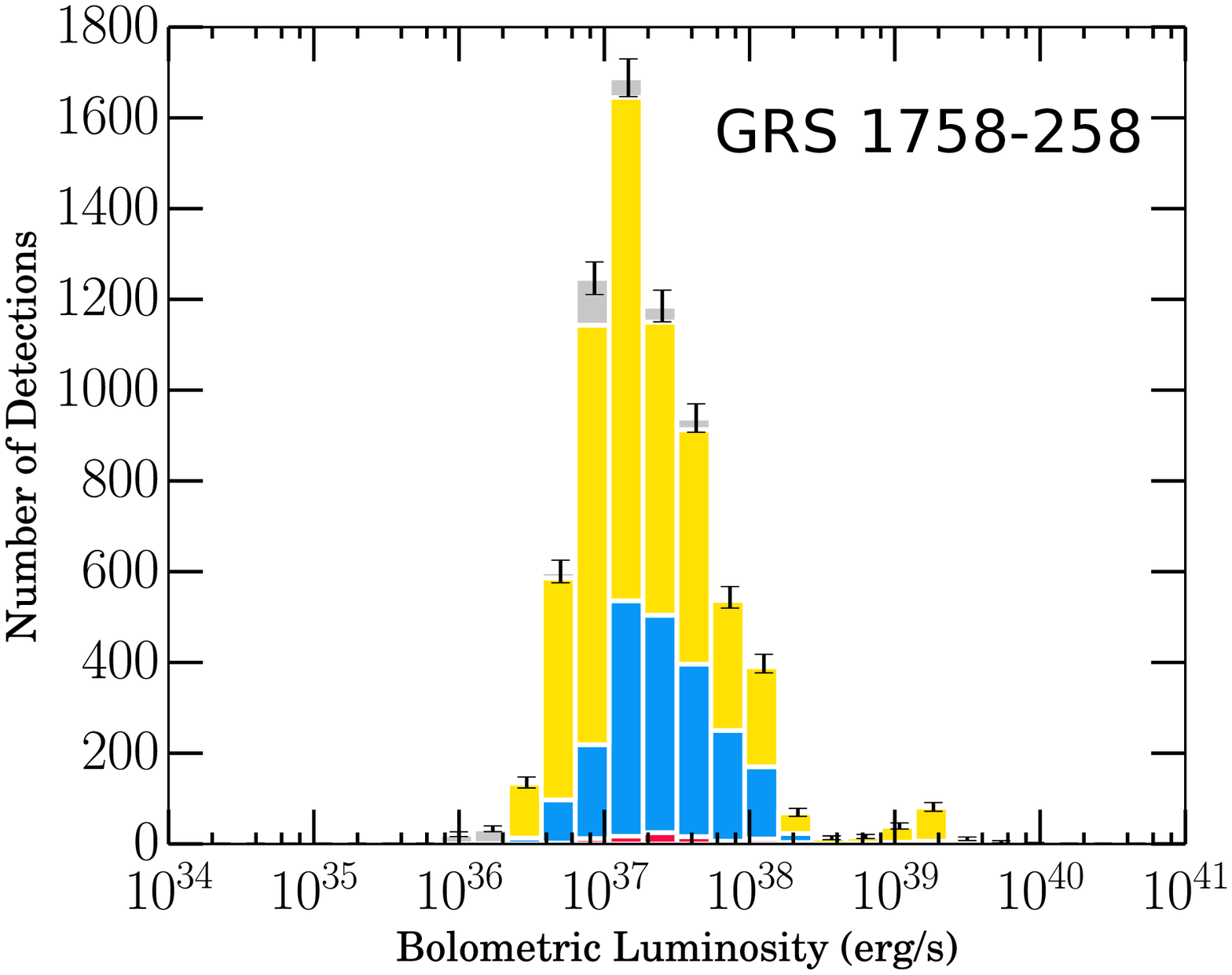}
\plottwo{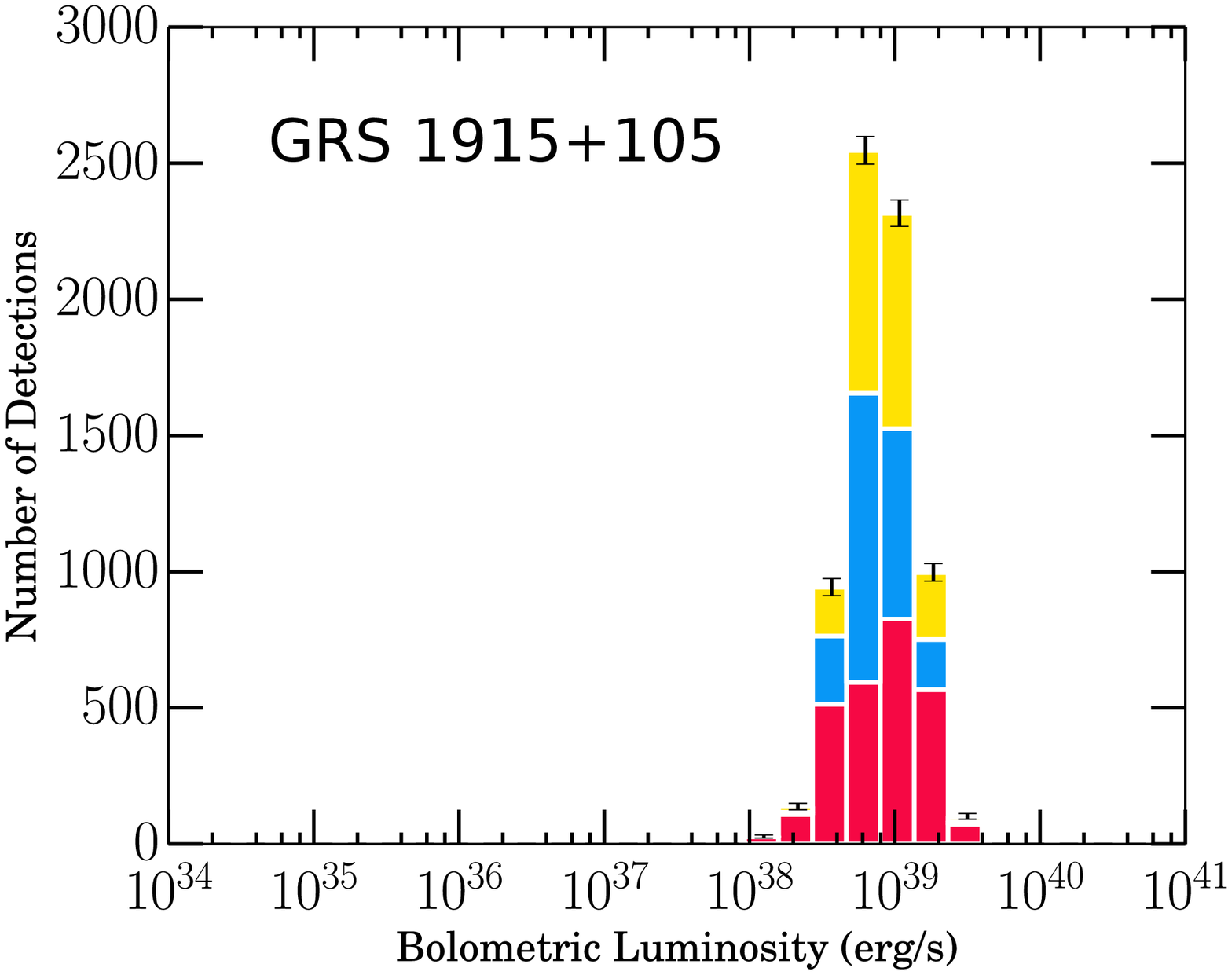}{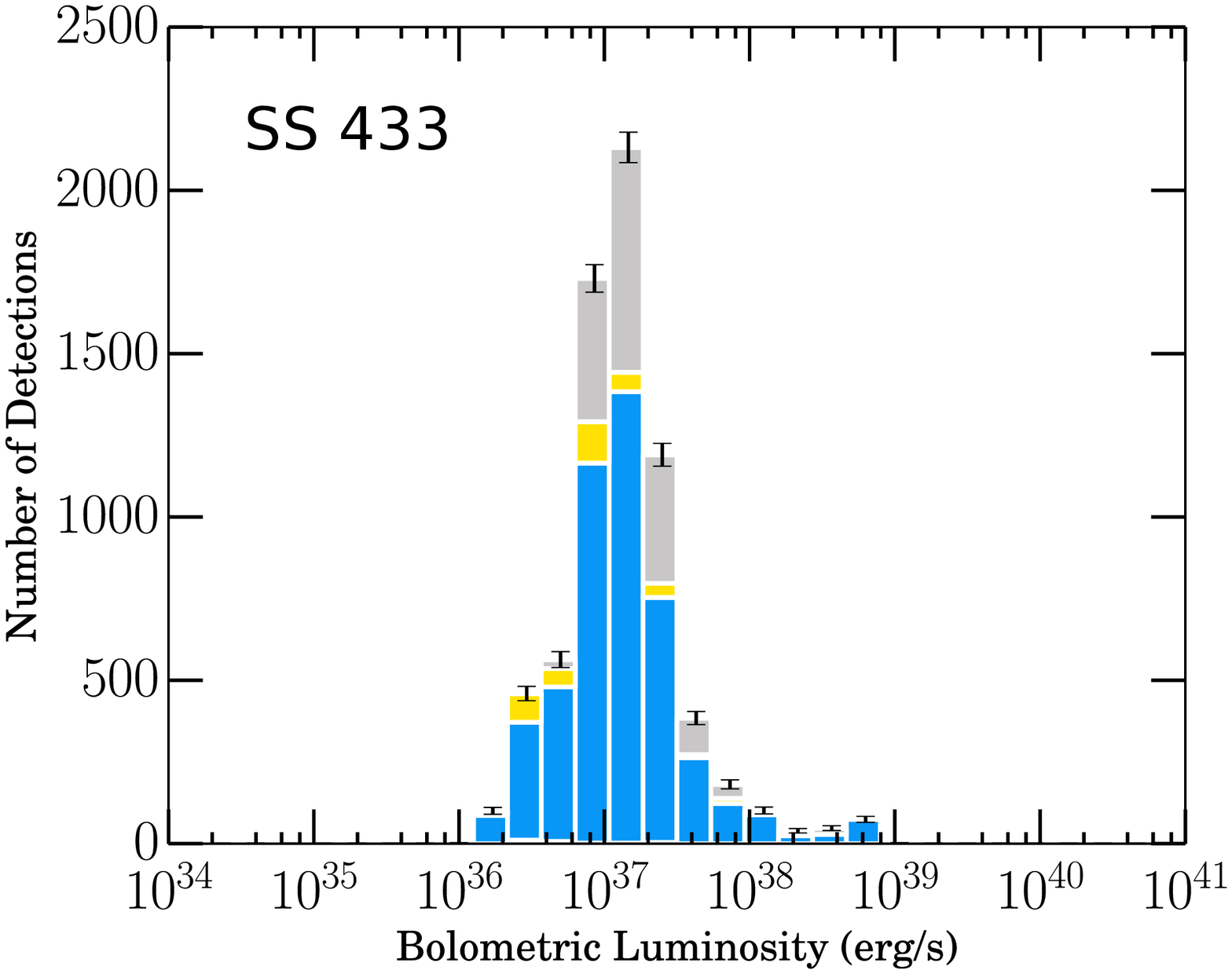}
  \caption{Persistent XLFs color coded by state. HCS (blue), SDS (red), IMS (yellow), and unable to determine state with data available (grey).}%
    \label{fig:tXLF72}%
\end{figure*}
\afterpage{\clearpage}

\addtocounter{figure}{-1}
\begin{center}
\begin{figure*}%
\epsscale{0.85}
\plottwo{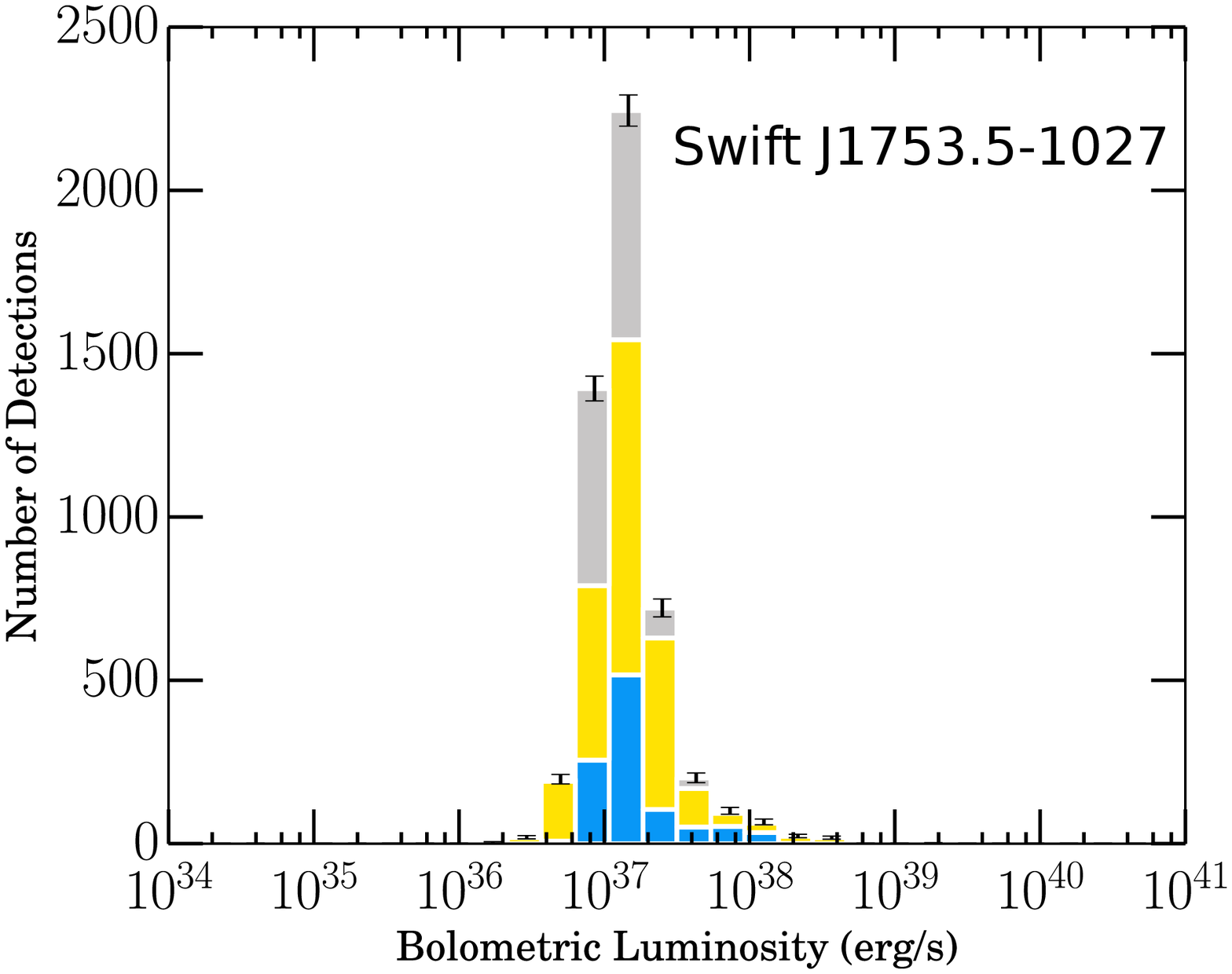}{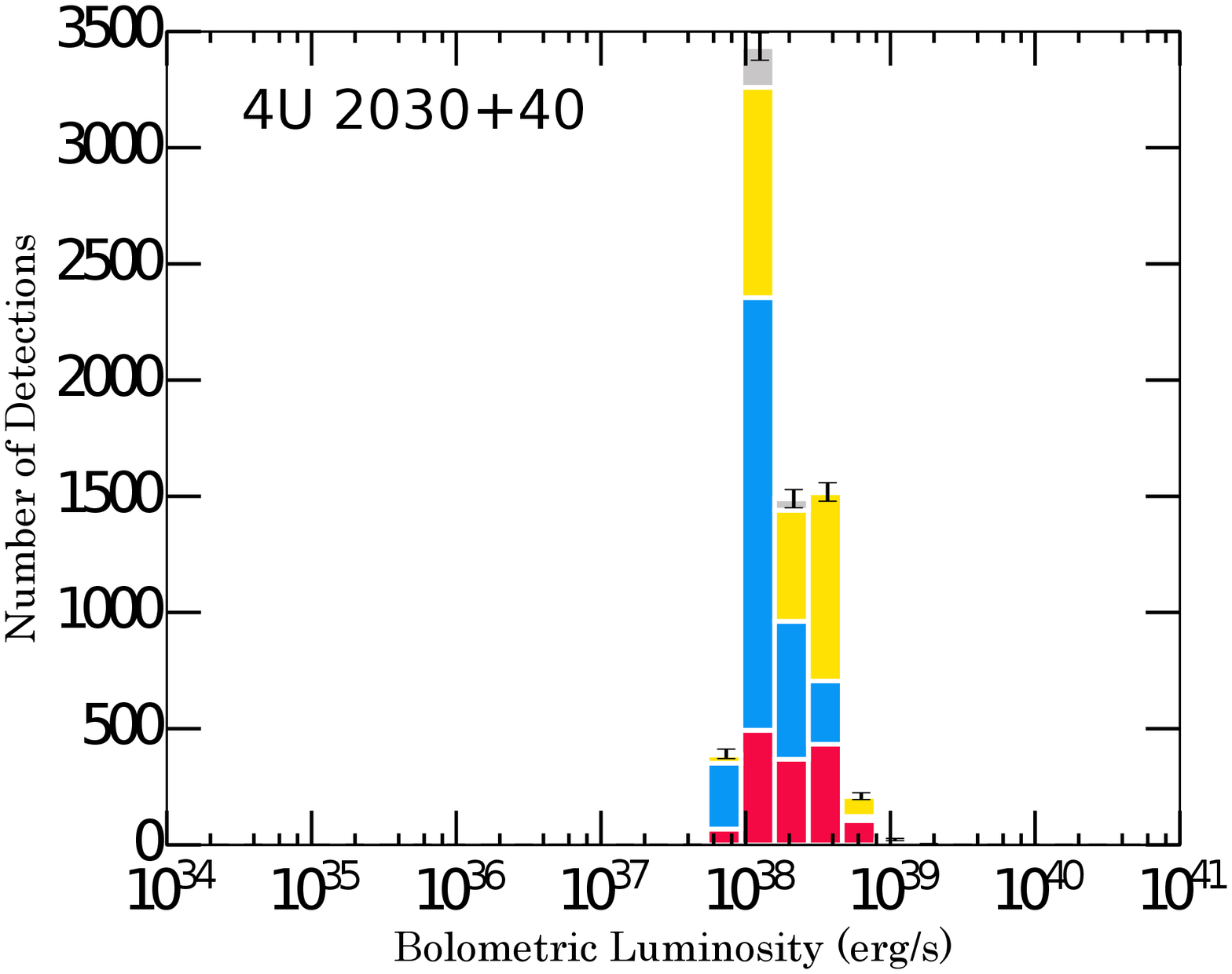}
  \caption{{\bf (cont.)} Persistent XLFs color coded by state. HCS (blue), SDS (red), IMS (yellow), and unable to determine state with data available (grey). }%
    \label{fig:tXLF82}%
\end{figure*}
\end{center}

\end{appendix}
\end{document}